\documentstyle[12pt,epsf,cite,amsfonts,amssymb,amsbsy,eufrak]{article}
\newcommand{\Rsub}{\rm\scriptscriptstyle}
\def\slashchar#1{\setbox0=\hbox{$#1$}           
   \dimen0=\wd0                                 
   \setbox1=\hbox{/} \dimen1=\wd1               
   \ifdim\dimen0>\dimen1                        
      \rlap{\hbox to \dimen0{\hfil/\hfil}}      
      #1                                        
   \else                                        
      \rlap{\hbox to \dimen1{\hfil$#1$\hfil}}   
      /                                         
   \fi}                                         %

\begin{document}
\thispagestyle{empty}
\begin{center}
{\Large\sc
Baryons with two heavy quarks}\\
\normalsize
\rm
\vspace*{1cm}
{\sf V.V.Kiselev, A.K.Likhoded,}\\[2mm]
{\sl Russian State Research Center "Institute for High Energy Physics"}\\
{\sl Protvino, Moscow Region 142284 Russia}\\[3mm]
\vspace*{8mm}
\begin{abstract}
We consider general physical characteristics of doubly heavy baryons:
the spectroscopy in the framework of potential approach and sum rules of QCD,
mechanisms of production in various interactions on the basis of fragmentation
model with account of preasymptotic corrections caused by higher twists over
the transverse momentum of baryon, inclusive decays and lifetimes in the
operator product expansion over the inverse powers of heavy quark masses as
well as the exclusive decays in the sum rules of QCD. We generalize the methods
developed in the effective theory of heavy quarks towards the description of
systems with two heavy quarks and a single light quark. The calculations are
presented for the masses, decay widths and yields of baryons with two heavy
quarks in the running and planned experimental facilities. We discuss the
prospects of search for the baryons and possibilities of experimental
observation. The most bright physical effects concerning these baryons as well
as their position in the system for the theoretical description of heavy quark
dynamics are considered.
\end{abstract}

\end{center}
\setcounter{page}{0}
\newpage
\makeatletter
\def\thesection {Introduction.}
\def\thsection {\arabic{section}.}
\def\thesubsection {\thsection\arabic{subsection}.}
\def\thesubsubsection {\thesubsection\arabic{subsubsection}.}
 \@addtoreset{equation}{section}
  \@addtoreset{table}{section}
 \@addtoreset{figure}{section}
  \def\thetable{\arabic{table}}
  \def\thefigure{\arabic{figure}}

\def\theequation{\arabic{equation}}

\def\section{\@startsection {section}{1}{\z@}{-3.5ex plus -1ex minus
 -.2ex}{2.3ex plus .2ex}{\large\bf}}
\def\subsection{\@startsection{subsection}{2}{\z@}{-3.25ex plus -1ex minus
 -.2ex}{1.5ex plus .2ex}{\large\bf}}
\def\subsubsection{\@startsection{subsubsection}{3}{\z@}{-3.25ex plus
 -1ex minus -.2ex}{1.5ex plus .2ex}{\sl\bf}}
\long\def\@makecaption#1#2{
 \vskip 10pt
 \setbox\@tempboxa\hbox{\underline{#1.} \small \sf #2}
 \ifdim \wd\@tempboxa >\hsize \underline{#1.} \small \sf #2\par \else \hbox
to\hsize{\hfil\box\@tempboxa\hfil}
 \fi}
\def\l@section#1#2{\addpenalty{\@secpenalty} \addvspace{1.0em plus 1pt}
\@tempdima 
1in \begingroup
 \parindent \z@ \rightskip \@pnumwidth
 \parfillskip -\@pnumwidth
 \bf \leavevmode \advance\leftskip\@tempdima \hskip -\leftskip 
#1\nobreak\hfil
\nobreak\hbox to\@pnumwidth{\hss #2}\par
 \endgroup}

\makeatother

\tableofcontents
\newpage
\section{}
After the precision investigations of neutral intermediate $Z$ boson at LEP
(CERN) and observation of $t$ quark at FNAL were done, the study of electroweak
interactions in the sector of heavy quarks is a task among the most actual
problems in the physics of fundamental particles, since in the framework of
standard model \cite{sm} such the searches can result in a complete picture of
effects responsible for the unreversability of time at energies below the scale
of electroweak symmetry breaking. A comparative analysis of processes with
violation of combined CP parity inverting both the charges (C) and space
orientation (P) in decays of hadrons containing the heavy quarks becomes
accessible for quite precise measurements due to putting into action
specialized facilities Belle (KEK) and BaBar (SLAC) along with the upgraded
detectors of CDF and D0 at FNAL. The results of such experiments probably allow
us to add an essential missing chain-cell in the standard model of
interactions, i.e. a complete description of charged currents for three quark
generations \cite{stone}. This study is the most important problem along with
both the observation of scalar Higgs particles providing the mechanism of
electroweak symmetry breaking, and the investigation of neutrino currents.

The pricision study of electroweak properties of heavy quarks make rise a deep
theoretical problem, that is the description of dynamics for the strong
interactions of quarks, since the strong forces cause the formation of bound
states, hadrons: mesons and baryons, and the observable quantities like
characteristics of rare decays with the CP violation are assigned to the bound
states, so that we need definite and reliable representations about the
relation of measured properties with the parameters of heavy quark interactions
themselves. In this way, fine effects of electroweak physics can be extracted
under a high precision description for the dominant contributions of quantum
chromodynamics (QCD). In this respect, we deal with a general problem on
quantitative understanding the quark confinement in QCD, that can be
efficiently investigated not only in the spectroscopy and processes with the
production and decays of exotic hybrid and glueball states \cite{glueball}, but
also in the study of hadrons with heavy quarks. In practice, the measured
quantities of asymmetries in the decays of heavy hadrons, for instance, are
expressed in termes of functions parametrically depending on both the primary
characteristics assigned to charged and flavor changing neutral currents of
quarks, and the hadronic matrix elements of quark operators, which usually
cannot be straightforwardly determined from a wide set of various experimental
data. So, we need a sound theoretical anlaysis of such the matrix elements in
QCD.

A more complete list of heavy quark bound states under study provides a more
wide region for the variation of conditions, wherein the forces of QCD act on
the heavy quarks, so that the theoretical methods for the description of
hadrons containing the heavy quarks should be more accomplished in order to get
a consistent understanding of various quark systems. In this way, we see a new
field of activity is baryons, containing two heavy quarks. So, the theoretical
predictions for the properties of such the baryons are of interest and actual.
The doubly heavy baryons naturally continue the list of long-lived heavy
hadrons with both a single heavy quark ($D$, $B$ mesons and и $\Lambda_c$,
$\Sigma_c$, $\Xi_c$, $\Omega_c$, $\Lambda_b$ baryons) and two heavy quarks (the
$B_c$ meson). With respect to the character of strong interactions, these
baryons could have common features with the heavy quarkonia $\bar c c$ and
$\bar b b$. In practice, we could expect experimental observation of doubly
heavy baryons in the searches at the modern hadron colliders with high
lumonosities (Tevatron, LHC), since the yield of such baryons should be of the
same order of magnitude as for the doubly heavy meson containing the quarks of
different flavors, the $B_c$ meson\footnote{See the review on the physivs of 
$B_c$ in \cite{bc-rev}.}, while the methods for the registration of rare decays
with heavy particles recently get a high efficiency due to a development in the
technique of vertex detectors as was successfully exhibited in the first
experimental observation of $B_c$ by the CDF collaboration \cite{cdf}.

Constructing the theoretical methods for the description of QCD dynamics with
the heavy quarks is based on a clear physical definition: the quark $Q$ is
ascribed heavy, if its mass $m_Q$ is much greater than the scale of heavy quark
confinement in the bound state, so $m_Q\gg \Lambda_{QCD}$. Thus, considering
the problem of strong interactions with the heavy quarks, i.e. calculating
hadronic matrix elements of quark operators, we could involve a small parameter
$\Lambda_{QCD}/m_Q \ll 1$, which could serve for the development of formal
approximate methods.

So, first of all, in hard processes with virtualities of the order of heavy
quark masses, say, in the production of heavy quarks, the QCD coupling constant
is small, $\alpha_s \sim \frac{1}{\ln m_Q/\Lambda_{QCD}}$, and the usual
technique of {\sf perturbative theory} in powers of $\alpha_s$ is justified.
Another productive method is the {\sf operator product expansion} (OPE) in
inverse powers of heavy quark mass. In such the approach a calculation of
hadronic matrix element for a quark operator leads to summing up the matrix
elements of operators, whose properties assume a hierarchy in terms of small
parameter $\Lambda_{QCD}/m_Q \ll 1$, i.e. a suppression of some contributions
by powers of $\Lambda_{QCD}/m_Q$, since the interaction in the bound state
containing the heavy quark is characterized by the energies close to
$\Lambda_{QCD}$ determining the inverse size of the hadron. This expansion
gives a complete description of hadronic systems containing a single heavy
quark. If there are two heavy quarks in the hadron, then along with the scale
of nonperturbative interactions another energetic scale is a momentum transfer
in the coulomb interaction with a virtuality of $\mu \sim \alpha_s m_Q$, so
that a relative velocity for two heavy quarks moving inside the hadron, $v$, is
determined by a comparatively small value of coupling constant in QCD, and
$v\sim \alpha_s$, where the coupling constant is taken at the scale of
virtualities prescribed to the coulomb interactions. Thus, in the heavy
quarkonium $\bar Q Q'$, say, there is an additional small parameter, that is
the relative velocity of nonrelativistic quarks $v$, which can be used in the
OPE in order to calculate the hadronic matrix elements. Following such the
approach, we underline three methods for the evaluation of quantities
characteristic for the bound states with heavy quarks:

\begin{description}
\item[-]
the {\it operator product expansion} in the inverse powers of heavy quark mass
in QCD for calculations of inclusive width and lifetimes of heavy hadrons,
where terms correcting the leading approximation are given by some external
parameters \cite{vs};
\item[-]
{\it sum rules} of QCD and nonrelativistic QCD (NRQCD) for both two-point
correlators of quark currents in spectroscopic calculations and three-point
correlators in estimations of exclusive decay modes \cite{SVZ};
\item[-]
{\it potential models} used for the evaluation of exclusive characteristics of
hadrons containing the heavy quarks \cite{RQ}. 
\end{description}
Let us stress, first, that the sum rules are based on the operator product
expansion, too. However, the only external parameters of sum rules are
fundamental quantities such as the masses of heavy quarks, the normalization of
QCD coupling constant and the quark-gluon condensates in contrast to inclusive
estimates in the general OPE, wherein we should use, for instance, the value of
heavy quark binding energy in the hadron, the average square of heavy quark
momentum etc. Second, in the expansion over the inverse powers of heavy quark
mass as well as over the relative velocity of heavy quarks inside the hadron an
important role is played by both the perturbation theory and the
renormalization group relations, which are necessary for the calculation of
Wilson coefficients in the operator product expansion, since these coefficients
enter the factors in front of operators or matrix elements under
consideration.

It is worth to emphasize that in the method of OPE for the heavy quarks we
could consider an actual operator, say, a current of weak decay or a product of
currents as it does in sum rules of QCD, with the consequent expansion, while
the approach of effective field theory could be useful, too, wherein the
starting point of construction is a formal expansion of QCD lagrangian itself
for the heavy quarks. Then we can isolate a leading term in the effective
lagrangian and treat the rest of terms as perturbations in the expansion. In
this way the definition of leading term is determined by the character of the
problem, i.e. by the actual convergency in the estimation of physical
quantities calculated in the effective lagrangian. So, for the hadrons with a
single heavy quark the effective theory of heavy quarks (HQET) \cite{HQET} was
developed. In this theory we can neglect the binding energy of heavy quark
inside the hadron to the leading order, therefore, the kinetic energy of heavy
quark is the perturbation. It is important to note that, first, in the leading
order of HQET the effective lagragian possesses the following symmetries: {\it
a}) the heavy quarks with identical velocities equal to the velocity of hadron,
where the quarks are bound, can be permuted, that implies the heavy flavor
symmetry, b) the heavy quark spin is decoupled from the interaction with
low-virtual gluons, since the current is given by the quark velocity $v_\nu$,
that implies the spin symmetry. Second, the leading term of effective
lagrangian provides the renormalization group behaviour different from the full
QCD. Particularly, the currents being conserved in full QCD and, hence, having
zero anomalous dimensions become divergent after the transition to the fields
of effective theory. The same note concerns to the correction terms of
effective lagrangian: the corresponding Wilson coefficients in the effective
theory have nonzero anomalous dimensions, too. Thus, we deal with the
situation, wherein we need an infinite number of normalization conditions for
the anomalous Wilson coefficients in the renormalization group in order to
correctly define the theory. This problem has a clear physical reason, since in
the effective theory constructed for the fields with low virtualities one
should introduce a cut off in the ultraviolet region at the scale of the order
of heavy quark mass because at high virtualities the assumptions made in the
derivation of the theory are not correct. Since the effective theory was
derived from the full QCD, we have to constructively put the effective
lagrangian equal to the lagrangian of full QCD at a scale $\mu_{hard}$ about
the heavy quark mass $m_Q$. This fact implies that in the given order over the
coupling constant of QCD we have to calculate the effective action of QCD with
account of corresponding loop corrections and to transform it by the expansion
in inverse powers of heavy quark mass, so that the expanded lagrangian should
be equalized to the effective lagrangian calculated in the effective theory in
the same order of coupling constant with the anomalous Wilson coefficients at
the scale $\mu_{hard}$, that results in the matching conditions for the unknown
constans of integration in the renormalization group equations for the Wilson
coefficients in the effective theory. After the matching is done, we remove an
ambiguity in the choice of finite contraterms in the renormalization of
effective lagrangian, and it becomes definite at $\mu$ below the scale of
matching with full QCD $\mu_{hard}$. Thus, for the hadrons containing  single
heavy quark along with a light one the scheme of effective theory HQET is
cosistently constructed.

For the heavy quarkonia composed of heavy quark and heavy anti-quark another
physical situation takes place. Indeed, the coulomb interaction of
nonrelativistic heavy quarks leads to that the kimetic energy is of the same
order of magnitude as the potential energy, while one could naively expect that
the kinetic term ${\bf p}^2/2 m_Q$ should be suppressed by the heavy quark
mass. However, since in the coulomb exchange we have $p\sim \alpha_s m_Q$, such
the suppression of kinetic energy does not take place, and the wave functions
of heavy quarkonia depend on the quark masses, i.e. they are flavor-dependent.
In the formal approach of effective theory for the nonrelativistic quarks in
the heavy quarkonium the leading term of lagrangian is defined with account of
the kinetic energy, so that we deal with the nonrelativistic QCD (NRQCD)
\cite{NRQCD}. In NRQCD compared with HQET, the spin symmetry of leading order
effective lagragian survives, while there is no heavy flavor symmetry because
the contribution of kinetic energy explicitely depends on the quark masses. In
the same manner as in HQET, the Wilson coefficients in the effective lagrangian
of NQRCD have to be matched with the full QCD at a scale about the heavy quark
mass, so that, in general, these coefficients have anomalous dimensions
different from those of HQET, since the kinetic energy results in the
ultraviolet behaviour of quark propagators, which is different from the
behaviour in HQET.

The potential approach is also based on the operator product expansion. So, the
static potential means the expansion of effective action of QCD for two
infinitely heavy sources $j$ posed at a fixed distance $r$, so that the
effective action has the form of $\Gamma(j) = - V(r) \cdot T$, where $T\to
\infty$ is a time of the sources are switched on. For actual problems the long
time interval means that virtualities of external gluon fields interacting with
the heavy quarks are much less than the inverse distance, i.e. $\mu \sim
\frac{1}{T} \ll \frac{1}{r} \sim m_Q v$. This constraint for the consistency of
potential approach can be expressed in terms of effective theory, which is
called the potential nonrelativistic QCD (pNRQCD) \cite{pNRQCD}. The
construction of pNRQCD follows the matching of NQRCD action with the effective
action of supersoft fields treated in the multipole expansion of QCD with
nonrelativistic heavy quarks at $\mu \sim v\cdot m_Q$. The leading order
includes the kinetic energy as well as the Wilson coefficient meaning the
static potential depending on the distance between the quarks, $r$. Thus, the
potential approach also has the status of OPE, and the static approximation for
the potential is determined by the covergency of this expansion in pNRQCD.

\begin{figure}[t]
\hspace*{3.5cm}
\epsfxsize=10cm \epsfbox{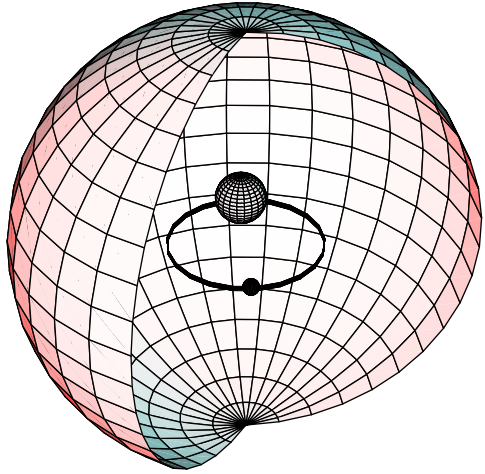}
\caption{The character of strong interactions in the doubly heavy baryon
$\Xi_{bc}$: the compton lengths of quarks $\lambda_Q=1/m_Q$, the size of heavy
diquark $r_{bc}\sim 1/(v\cdot m_Q)$ and the scale of nonperturbative
confinement of light quark $r_{QCD}=1/\Lambda_{QCD}$ are arranged by
$\lambda_b\approx \frac{1}{3}\, \lambda_c\approx \frac{1}{9}\, r_{bc}\approx
\frac{1}{27}\, r_{QCD}$.}
\label{dynia}
\end{figure}

In this respect, the baryons with two heavy quarks especially are of interest
for the theoretical consideration, since for their description one should
develop and use a combined approach involving the features of HQET, NRQCD and
pNRQCD because in this systems the interaction of light quark with the heavy
quarks is essential as well as the interaction between the heavy quarks (see
Fig. \ref{dynia}).

In this review we solve the problem on the description of bound baryonic states
$QQ'q=\Xi_{QQ'}$ with two heavy quarks $Q$, $Q'$ and a light quark $q$ on the
basis of factorizing the interactions with various virtualities determined by
the following: 
\begin{description}
\item[-]
the confinement scale $\Lambda_{QCD}$ for the nonperturbative interactions of
both the heavy quarks with the light one and the quarks with the quark-lguon
condensates,
\item[-]
the size of heavy diquark $r_{QQ'}\sim 1/(m_{Q,Q'} \cdot v)$, which is composed
of two nonrelativistic heavy quarks moving with a small relative velocity $v\ll
1$, for the interactions between the heavy quarks,
\item[-]
the scale of hard gluon corrections at the energies about the heavy quark
masses $m_Q$,
\end{description}
so that we assume that the leading approximation is valid if there is the
hierarchy of QCD interaction scales in $\Xi_{QQ'}$
\begin{equation}
\Lambda_{QCD}\ll m_Q \cdot v \ll m_Q.
\end{equation}
In this approach the doubly heavy diquark acts as a local heavy source of gluon
field for the light quark. This source is charged as the anti-triplet under the
color group of QCD, while the diquark itself is the system of two
nonrelativistic quarks in the low-frequency field of light quark. Thus, for the
motion of light quark and diquark we can use the effective theory of HQET, to
the moment for the motion of heavy quarks inside the diquark we should modify
NRQCD and pNRQCD in order to consider the nonrelativistic fields of heavy
quarks in the anti-triplet state, but the singlet one.

On the basis of quark-diquark representation for the interactions we consider
various physical aspects of baryons with two heavy quarks. In {\bf Chapter 1}
the mass spectrum of baryons $\Xi_{QQ'}$ is constructed in the potential
approach. We calculate the characteristics of groud state and its excitations
in the system of heavy diquark as well as in the system of light quark and
diquark. We show  that there is a family of $\Xi_{QQ'}$ levels with the
quasi-stable states for the heavy diquark composed of identical heavy quarks.
For these states the Pauli principle dictates quite definite values for the sum
of quark spins, so, the total spin equals unit for the P-even wave functions in
the configuration space, while it is equal to zero for the P-odd wave
functions, since the anti-triplet color state of diquark is anti-symmetric
under the permutation of color indices. Taking into account the small size of
diquark, the nonrelativistic motion of heavy quarks and a small ratio of
$\Lambda_{QCD}/m_Q$, the operators for the transitions of excited P-wave
diquark into the ground S-wave level with the emission of $\pi$ meson are
suppressed because in this transition both the spin state of diquark and
orbital state should change. We determine the region of consistency for the
quark-diquark approximation in the calculations of $\Xi_{QQ'}$ mass spectra in
the framework of potential approach.

In {\bf Chapter 2} the two-point sum rules of NRQCD are considered for the
baryonic currents with two heavy quarks. We discuss a criterium of stability
for the results of such the sum rules in estimates of masses and coupling
constants of $\Xi_{QQ'}$ baryons. We show that reliable results can be obtained
after account of the quark and gluon condensates as well as their product and
mixed consdensate, i.e. after the introduction of combined condensates of
higher dimensions. We calculate the masses and coupling constants for the
ground states of baryons $\Xi_{QQ'}$ and the doubly heavy baryons with
strangeness $\Omega_{QQ'}$, too. Reliable predictions for the mass splitting of
$M_{\Xi_{QQ'}}-M_{\Omega_{QQ'}}$ are presented. We get anomalous dimensions for
the baryonic currents in NQRCD up to two-loop approximation, that allows us to
estimate the coupling constants of baryons with the baryonic currents not only
in NRQCD but also in full QCD.

Mechanisms for the production of $\Xi_{QQ'}$ baryons are considered in {\bf
Chapter 3}. Following the factorization of interactions in the quark-diquark
approach, in $e^+e^-$ annihilation at high energies the inclusive production of
doubly heavy baryons can be described in the form of subsequent fragmentation
for both the heavy quark into the heavy diquark and the diquark into the
baryon. In this way the virtualities in the first process are determined by the
masses of heavy quarks, so that factorizing the soft movement of heavy quarks
inside the diquark we can use the perurbative QCD and get an analytic form of
fragmentation functions for the states with various spins and orbital quantum
numbers of diquark. Keeping in mind the difference between the color structures
of diquark and heavy quarkonium, these calculations repeat the consideration of
fragmentation into the doubly heavy mesons up to a color factor. Supposing the
representation of heavy diquark by the local field in its interactions with the
light quark and approximating the field components of baryon by its dominant
fast valence quarks, we develop a QCD-motivated perturbative model for the
fragmentation of diquark into the baryon and calculate an analytic form of
fragmentation function for the vector and scalar diquarks entering the baryon
$\Xi_{QQ'}$ with the spin $\frac{1}{2}$. The estimates for the pair production
of baryons in the $e^+e^-$ annihilation close to the threshold are presented,
too. The analysis turns to be more complicated in the consideration of
mechanism for the production of baryons $\Xi_{QQ'}$ in hadron collisions due to
subprocesses of quark-antiquark annihilation (low energies, fixed target
experiments) and gluon-gluon fusion (high energies, hadron colliders). This
complication is connected to a greater number of diagrams in the leading
approximation, i.e. in the fourth order of QCD coupling constant. Making use of
numerical method, we show that at high energies of partonic subprocess and at
transverse momenta much greater than the baryon mass the complete set of
diagrams in the given order of perturbation theory in QCD leads to the
factorization for the production of heavy quark and its fragmentation into the
doubly heavy diquark, expressed in the form of univerasl fragmetation function
analytically derived in the perturbative QCD. This fact implies the consistency
of approach used. The advantage of such considereation with the complete set of
diagrams dictated by the gauge invariance is a possibility to calculate not
only the leading term with respect to the transverse momentum $p_\perp$, that
gives the fragmentation dropping as $\sim \frac{1}{p^4_\perp}$, but also the
correction terms, the higher twists over the transverse momentum. In this way
we get a definite estimate for the transverse momentum determining the boundary
between the regions of fragmentation and recombination (the higher twists). We
point out that the statistics of events with the production of doubly heavy
baryons $\Xi_{QQ'}$ is dominantly integrated out at small transverse momenta
lying in the region of recombination. We represent estimates for the total and
differential cross sections of $\Xi_{QQ'}$ baryon production in hadronic
experiments at various energies and for the pair production of doubly heavy
baryons in the quark-antiquark annihilation.

The method of operator product expansion in the inverse powers of heavy quark
masses is exploited in {\bf Chapter 4} for the analysis of lifetimes and
inclusive width in decays of $\Xi_{QQ'}$ baryons. In this consideration of
mechanisms for the decays of heavy quarks entering the doubly heavy baryons,
the following three physical effects are essentially important:
\begin{description}
\item[-]
valuable corrections to the spectator width in decays of heavy quarks appear
because of movement of quarks in the heavy diquark, whose size is small and,
hence, the relative momenta of quarks $p\sim v\cdot m_Q$ are greater than the
momentum of heavy quark in the hadron with the single heavy quark, when $k\sim
\Lambda_{QCD}$, since we have $v\cdot m_Q \gg \Lambda_{QCD}$;
\item[-]
nonspectator contributions due to the Pauli interference between the products
of heavy quark decay and the valence quarks in the initial state can give a
fraction 30-50\% of the total width, so that the feature of anti-symmetric
color wave function for the baryons is a possibility of both a positive or
negative overall sign for the term with the interference\footnote{The overall
sign is determined by the anti-symmetric permutation of fermions multiplied by
the color factor.};
\item[-]
along with the Pauli interference the weak scattering of quarks in the initial
state appears in the OPE in the form of operator of higher dimension and it is
enhanced by a two-particle phase space in comparison with other operators with
the same dimension, since they have a three-particle phase space in the final
state, so that the three-particle phase space is suppressed in units of heavy
quark mass; the weak scattering gives about 30\% of total width for the baryons
with charmed quark.
\end{description}
In this way, we show how the arrangement of lifetimes is produced for the
baryons with two heavy quarks. Then we determine the parametric dependence of
estimates for the total and inclusive widths on the physical quantities of
hadron system. So, the small size of heavy diquark determines its wave function
yielding the factor in the evaluation of nonspectator decays. The masses of
heavy quarks in OPE are essentially correlated for the hadrons with different
quark contents, so that current experimental data on the semileptonic,
nonleptonic and total widths decrease uncertainties of estimates. Furthermore,
the experimental data are able to essentially improve the qualitative and
quantitative knowledges on the dynamics of heavy hadrons with the single or two
heavy quarks.

We analyze exclusive semileptonic decays and nonleptonic decays in the
assumption of factorization in the framework of three-point sum rules of NRQCD,
which allow us to derive relations for the formfactors of transitions given by
hadronic matrix elements from the spin symmetry of effective lagrangian. We
discuss uncertainties of calculations and compare the sum rule results with the
predictions of potential models for the exclusive decays.

In {\bf Conclusion} we summarize our results on the physics of baryons
containing two heavy quarks. The obtained predictions do not only point to a
way for a goal-recognized search of such baryons, but also make a basis for a
more accomplished and detailed theoretical analysis of physical effects in the
hadronic systems with two heavy quarks, which are, no doubts, of interest,
particularly, for the reliable predictions of total and exclusive widths.
Finally, we consider possibilities for an experimental observation of doubly
heavy baryons.

\setcounter{section}{0}
\makeatletter
\def\thesection {Chapter \arabic{section}.}
\def\thsection {\arabic{section}.}
\def\thesubsection {\thsection\arabic{subsection}.}
\def\thesubsubsection {\thesubsection\arabic{subsubsection}.}
 \@addtoreset{equation}{section}
  \@addtoreset{table}{section}
 \@addtoreset{figure}{section}
  \def\thetable{\thsection\arabic{table}}
  \def\thefigure{\thsection\arabic{figure}}

\def\theequation{\thsection\arabic{equation}}

\def\section{\@startsection {section}{1}{\z@}{-3.5ex plus -1ex minus
 -.2ex}{2.3ex plus .2ex}{\large\bf}}
\def\subsection{\@startsection{subsection}{2}{\z@}{-3.25ex plus -1ex minus
 -.2ex}{1.5ex plus .2ex}{\large\bf}}
\def\subsubsection{\@startsection{subsubsection}{3}{\z@}{-3.25ex plus
 -1ex minus -.2ex}{1.5ex plus .2ex}{\sl\bf}}
%
%
\def\l@section#1#2{\addpenalty{\@secpenalty} \addvspace{1.0em plus 1pt}
\@tempdima 
1in \begingroup
 \parindent \z@ \rightskip \@pnumwidth
 \parfillskip -\@pnumwidth
 \bf \leavevmode \advance\leftskip\@tempdima \hskip -\leftskip 
#1\nobreak\hfil
\nobreak\hbox to\@pnumwidth{\hss #2}\par
 \endgroup}

\makeatother

\newpage

\section{Spectroscopy of doubly heavy baryons:\\ the potential approach}

In this chapter we analyze basic spectroscopic characteristics for the families
of doubly heavy baryons $\Xi_{QQ'}=(QQ'q)$, where $q=u,\; d$ and
$\Omega_{QQ'}=(QQ's)$.

A general approach of potential models to calculate the masses of baryons
containing two heavy quarks was considered in refs.\cite{Wise}. The physical
motivation used the pair interactions between the quarks composing the baryon,
that was explored in the three-body problem. Clear implications for the mass
spectra of doubly heavy baryons were derived. So, for the given masses of heavy
charmed and beauty quarks, the approximation of factorization for the motion of
doubly heavy diquark and light quark is not accurate. It results in the ground
state mass and excitation levels, essentially deviating from the estimates in
the framework of appropriate three-body problem. For example, we can easily
find that in the oscillator potential of pair interactions an evident
introduction of Jacobi variables leads to the change of vibration energy
$\omega\to \sqrt{\frac{3}{2}} \omega$ in comparison with naive expectations
of diquark factorization.

\begin{figure}[th]
\hspace*{3.5cm}
\epsfxsize=10cm \epsfbox{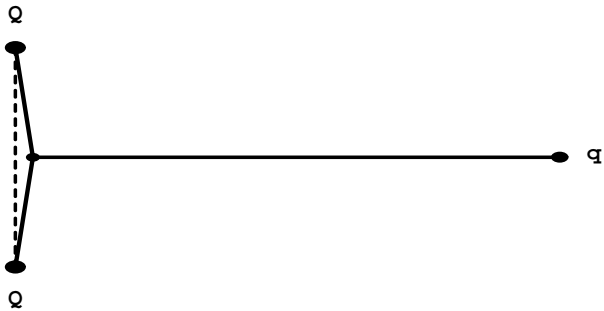}
\caption{The representation of doubly heavy baryon $QQq$ with the colored
fields forming the strings between the heavy and light quarks, that destroys
the picture of pair interactions and involves the additional `centre-of-mass'
point close to the centre of mass for the heavy-heavy system.}
\label{pic-str}
\end{figure}

There is another point of view to the problem of three-quark bound states in
QCD, i.e. the representation of quark-gluon string. In the string-like picture
of doubly heavy baryon shown in Fig.\ref{pic-str}, the above conclusions on the
structure of mass spectra for the doubly heavy baryons derived on the basis of
pair interactions should be essentially modified. Indeed, to the moment we have
to introduce the centre of string, which is very close to the centre of mass
for the doubly heavy diquark. Furthermore, the light quark interacts with the
doubly heavy diquark as a whole, i.e. with the string tension identical to that
in the heavy-light mesons $Q\bar q$. Therefore, two different assumptions on
the nature of interactions inside the doubly heavy baryons: pair interactions
or string-like picture, result in a distinct variation of predictions on the
mass spectra of these baryons for both the ground states and excitation levels.
The only criterion testing the assumptions is provided by an experimental
observation and measurements.

In this review we follow the approximation of doubly heavy diquark, which is
quite reasonable as we have clarified in the discussion given above. To enforce
this point we refer to the consideration of doubly heavy baryon masses in the
framework of QCD sum rules (see Chapter 2), which result in the masses of
ground states in a good agreement with the estimates obtained in the potential
approach with the factorization of doubly heavy diquark.

The qualitative picture for the forming of bound states in the system of
$(QQ'q)$ is determined by the presence of two scales of distances, which are
given by the size of $QQ'$-diquark subsystem, $r_{QQ'}$, in the anti-triplet
color state  as well as by the confinement scale, $\Lambda_{QCD}$, for the
light quark $q$, so that
$$
r_{QQ'}\cdot \Lambda_{QCD} \ll 1, \;\;\;\; \Lambda_{QCD}\ll m_Q.
$$
Under such conditions, the compact diquark $QQ'$ looks like a static source
approximated by the local colored QCD field interacting with the light quark.
Therefore, we can use a set of reliable results in models of mesons with a
single heavy quark, i.e. with a local static source belonging to the
anti-triplet representation of SU(3)$_c$ group. The successful approaches are
the potential models \cite{RQ} and the Heavy Quark Effective Theory (HQET)
\cite{HQET} in the framework of expansion in the inverse heavy quark mass. We
apply the nonrelativistic quark model with the potential by Buchm\"uller--Tye
\cite{BT}. Then {\it theoretically} we can talk on the rough approximation for
the light quark. Indeed, since $m_q^{QCD} \ll \Lambda_{QCD}$ the light quark is
relativistic. Nevertheless, we introduce the system with a finite number of
degrees of freedom and an instantaneous interaction $V({\bf r})$. This fact is
a disadvantage because the confinement supposes the following: a) the
generation of sea around the light quark, i.e. the presence of infinite number
of gluons and quark-antiquark pairs, and b) the nonperturbative effects with
the correlation time $\tau_{QCD}\sim 1/\Lambda_{QCD}$, that is beyond the
potential approach. However, {\it phenomenologically} the introduction of
constituent mass $m_q^{NP}\sim \Lambda_{QCD}$ as a basic parameter determining
the interaction with the QCD condensates, allows us to successfully adjust the
nonrelativistic potential model with a high accuracy ($\delta M\approx 30 - 40$
MeV) by fitting the existing experimental data, that makes the approach to be
quite a reliable tool for the prediction of masses for the hadrons, containing
the heavy and light quarks.

As for the diquark $QQ'$, it is completely analogous to the heavy quarkonium
$Q\bar Q'$ except the very essential peculiarities.
\begin{enumerate}
\item
$(QQ')_{\bar 3_c}$ is a system with the nonzero color charge.
\item
For the quarks of the same flavor $Q=Q'$ it is necessary to take into account
the Pauli principle for the identical fermions.
\end{enumerate}
The second item turns out to forbid the sum of quark spins S=0 for the
symmetric, spatial parity P-even wave functions of diquark, $\Psi_d({\bf
r})$ (the orbital momentum equals $L_d= 2n$, where $n=0,1,2\ldots$ ), as well
as S=1 is forbidden for the anti-symmetric, Р-odd functions $\Psi_d({\bf r})$
(i.e. $L_d= 2n+1$). The nonzero color charge leads to two problems.

First, we cannot generally apply the confinement hypothesis on the form of
potential (an infinite growth of energy with the increase of the system size)
for the object under consideration. However, it is unpossible to imagine a
situation, when a big colored object with a size $r > 1/\Lambda_{QCD}$ has a
finite energy of self-action, and, to the same moment, it is confined inside a
white hadron (the singlet over SU(3)$_c$) with $r \sim 1/\Lambda_{QCD}$ due to
the interaction with another colored source. In the framework of well-justified
picture of the hadronic string, the tension of such string in the diquark with
the external leg inside the baryons is only two times less than in the
quark-antiquark pair inside the meson $q\bar q'$, and, hence, the energy of
diquark linearly grows with the increase of its size. So, the effect analogous
to the confinement of quarks takes place in the similar way. In the potential
models we can suppose that the quark binding appears due to the effective
single exchange by a colored object in the adjoint representation of SU(3)$_c$
(the sum of scalar and vector exchanges is usually taken). Then, the potentials
in the singlet ($q\bar q'$) and anti-triplet ($qq'$) states differ by the
factor of 1/2, that means the confining potential with the linear term in the
QCD-motivated models for the heavy diquark $(QQ')_{\bar 3_c}$. In the present
chapter we use the nonrelativistic model with the Buchm\"uller--Tye potential
for the diquark, too.

Second, in the singlet color state $(Q\bar Q')$ there are separate
conservations of the summed spin S and the orbital momentum L, since the QCD
operators for the transitions between the levels determined by these quantum
numbers, are suppressed. Indeed, in the framework of multipole expansion in QCD
\cite{12}, the amplitudes of chromo-magnetic and chromo-electric dipole
transitions are suppressed by the inverse heavy quark mass, but in addition,
the major reason is provided by the following: a) the necessity to emit a white
object, i.e. at least two gluons, which results in the higher order in $1/m_Q$,
and b) the projection to a real phase space in a physical spectrum of massive
hadrons in contrast to the case of massless gluon. Furthermore, the probability
of a hybrid state, say, the octet sybsystem $(Q\bar Q')$ and the additional
gluon, i.e. the Fock state $|Q\bar Q^{\prime}_{8_c} g\rangle $, is suppressed
due to both the small size of system and the nonrelativistic motion of quarks
(for a more strict consideration see ref.\cite{NRQCD}). In the anti-triplet
color state, the emission of a soft nonperturbative gluon between the levels
determined by the spin $S_d$ and the orbital momentum $L_d$ of diquark, is not
forbidden, if there are no some other no-go rules or small order-parameters.
For the quarks of identical flavors inside the diquark, the Pauli principle
leads to that the transitions are possible only between the levels, which
either differ by the spin ($\Delta S_d= 1$) and the orbital momentum ($\Delta
L_d = 2n+1$), instantaniously, or belong to the same set of radial and orbital
excitations with $\Delta L_d = 2n$. Therefore, the transition amplitudes are
suppressed by a small recoil momentum of diquark in comparison with its mass.
The transition operator changing the diquark spin as well as its orbital
momentum, has the higher order of smallness because of either the additional
factor of $1/m_Q$ or the small size of diquark. These suppressions lead to the
existence of quasi-stable states with the quantum numbers of $S_d$ and $L_d$.
In the diquark composed by the quarks of different flavors, $bc$, the QCD
operators of dipole transitions with the single emission of soft gluon are not
forbidden, so that the lifetimes of levels can be about the times for
the forming of bound states or with the inverse distances between the levels
themselves. Then, we cannot insist on the appearance of excitation system for
such the diquark with definite quantum numbers of the spin and orbital
momentum\footnote{In other words, the presence of gluon field inside the baryon
$\Xi_{bc}$ leads to the transitions between the states with the different
excitations of diquark, like $|bc\rangle \to |bcg\rangle $ with $\Delta S_d =1$
or $\Delta L_d=1$, which are not suppressed.}.

Thus, in the present review we explore the presence of two physical scales in
the form of factorization for the wave functions of the heavy diquark and light
constituent quark. So, in the framework of nonrelativistic quark model the
problem on the calculation of mass spectrum and characteristics of bound states
in the system of doubly heavy baryon is reduced to two standard problems on the
study of stationary levels of energy in the system of two bodies. After that,
we take into account the relativistic corrections dependent of the quark spins
in two subsystems under consideration. The natural boundary for the region of
stable states in the doubly heavy system can be assigned to the threshold
energy for the decay into a heavy baryon and a heavy meson. As was shown in
\cite{13}, the appearance of such threshold in different systems can be
provided by the existence of an universal characteristics in QCD, a critical
distance between the quarks. At distances greater than the critical separation,
the quark-gluon fields become unstable, i.e. the generation of valence
quark-antiquark pairs from the sea takes place. In other words, the hadronic
string having a length greater than the critical one, decay into the strings of
smaller sizes with a high probability close to unit. In the framework of
potential approach this effect can be taken into account by that we will
restrict the consideration of excited diquark levels by the region, wherein the
size of diquark is less than the critical distance, $r_{QQ'}< r_c \approx 1.4 -
1.5$ fm. Furthermore, the model with the isolated structure of diquark looks to
be reliable, just if the size of diquark is less than the distance to the light
quark $r_{QQ'}<r_l$.

The peculiarity of quark-diquark picture for the doubly heavy baryon is the
possibility of mixing between the states of higher diquark excitations,
possessing the different quantum numbers, because of the interaction with the
light quark. Then it is difficult to assign some definite quantum numbers to
such excitations. We will discuss the mechanism of this effect.

In Section 1.1 we describe a general procedure for the calculation of masses
for the doubly heavy baryons in the framework of assumptions drawn above. We
take into account the spin-dependent corrections to the potential motivated in
QCD. The results of numerical estimates are presented in Section 1.2, and,
finally, our conclusions are discussed in the end of this chapter.

\subsection{Nonrelativistic potential model}

As we have mentioned in the Introduction, we solve the problem on the
calculation of mass spectra of baryons containing two heavy quarks,  in two
steps. First, we compute the energy levels of diquark. Second, we consider
the two-body problem for the light quark interacting with the point-like
diquark having the mass obtained in the first step. In accordance with the
effective expansion of QCD in the inverse heavy quark mass, we separate two
stages of such the calculations. So, the nonrelativistic Schr\" odinger
equation with the model potential motivated by QCD, is solved numerically.
After that, the spin-dependent corrections are introduced as perturbations
suppressed by the quark masses.

\subsubsection{Potential}

\begin{figure}[th]
\setlength{\unitlength}{1mm}
\begin{picture}(70,110)
\put(30,0){\epsfxsize=11cm \epsfbox{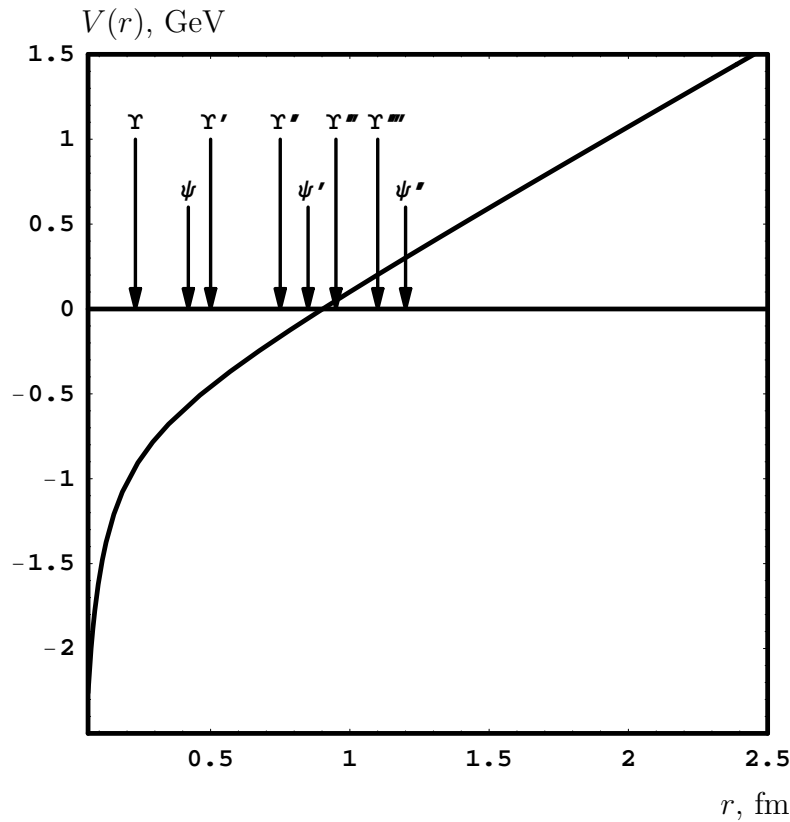}}
\put(130,3){$r$, fm}
\put(45,107){$V(r)$, GeV}
\end{picture}
\caption{The Corenell model of static potential and sizes of observed heavy
quarkonia with charmed quarks (the family $\psi$) and bottom quarks (the family 
$\Upsilon$).}
\label{gre}
\end{figure}

The potential of static heavy quarks illuminates the most important features of
QCD dynamics: the asymptotic freedom and confinement. In the leading order of
perturbative QCD at short distances and with a linear confining term in the
infrared region, the potential of static heavy quarks was considered in the
Cornell model \cite{Corn}, incorporating the simple superposition of both
asymptotic limits (the effective coulomb and string-like interactions). The
observed heavy quarkonia posed in the intermediate distances, where both terms
are important for the determination of mass spectra (see Fig. \ref{gre}). So,
the phenomenological approximations of potential (logarithmic one \cite{log}
and power law \cite{Mart}), taking into account the regularities of such the
spectra, were quite successful \cite{RQ}. 

The quantities more sensitive to the global properties of potential are the
wave functions at the origin as related to the leptonic constants and
production rates. So, the potentials consistent with the asymptotic freedom to
one and two loops as well as the linear confinement were proposed by Richardson
\cite{Richard}, Buchm\"uller and Tye \cite{BT}, respectively.

In QCD the static potential is defined in a manifestly gauge invariant way by
means of the vacuum expectation value of a Wilson loop \cite{Su},
\begin{eqnarray}
\label{def_WL}
V(r) &=& - \lim_{T\rightarrow\infty} 
\frac{1}{iT}\, \ln \langle{\cal W}_\Gamma\rangle \;, \nonumber\\
{\cal W}_\Gamma &=& \widetilde{\rm tr}\, 
{\cal P} \exp\left(ig \oint_\Gamma dx_\mu A^\mu\right) \;.
\end{eqnarray}
Here, $\Gamma$ is taken as a rectangular loop with time extension $T$ and
spatial extension $r$. The gauge fields $A_\mu$ are path-ordered along the
loop, while the color trace is normalized according to $\widetilde{\rm
tr}(..)={\rm tr}(..)/{\rm tr}1\!\!1\,$. This definition corresponds to the
calculation of effective action for the case of two external sources fixed at a
distance $r$ during an infinitely long time period $T$, so that the
time-ordering coincides with the path-ordering. Moreover, the contribution into
the effective action by the path parts, where the charges have been separated
to the finite distance during a finite time, can be neglected in comparison
with the infinitely growing term of $V(r)\cdot T$. Let us emphasize that the
defined static potential is, by construction, the renormalization invariant
quantity, since the action, by definition, doe not depend on the normalization
point.

Generally, one introduces the  V scheme of QCD coupling constant by the
definition of QCD potential of static quarks in momentum space as follows:
\begin{equation}
 V({\bf q}^2)  =  -C_F\frac{4\pi\alpha_{\Rsub V}({\bf q}^2)}{{\bf q}^2},
\end{equation}
so that for the such-way introduced value of $\alpha_{\Rsub V}$ one can derive
some results at large virtualities in the perturbative QCD as well as at low
transfer momenta in the approximation of linear term in the potential confining
the quarks.

In this section, first, we discuss two regimes for the QCD forces between the
static heavy quarks: the asymptotic freedom and confinement. Then we follow the
method by Buchm\"uller and Tye and formulate how these regimes can be combined
in a unified $\beta$ function for $\alpha_{\Rsub V}$ obeyed both limits of
small and large QCD couplings.

\subsubsection{Perturbative results at short distances}

Technically, using a given scheme of regularization, say, $\overline{\rm MS}$,
one has to calculate the perturbative expansion for the potential of static
quarks. This potential can be written down as the coulomb one with the running
coupling constant in the so-called  V scheme. Thus, the perturbative
calculations provide us with the matching of $\overline{\rm MS}$ scheme with
V-one. The calculations with the $n$ loop running of $\alpha_s^{\overline{\Rsub
MS}}$ requires the $n-1$ loop matching to $\alpha_{\Rsub V}$. Note, that
initial two coefficients of corresponding $\beta$ functions are scheme and
gauge independent, while others generally depend. The V scheme is defined for
the observed quantity, that implies its $\beta$ function to be gauge invariant.

In the perturbative QCD the quantity $\alpha_{\Rsub V}$ can be matched with
$\alpha_{\overline{\Rsub MS}}$
\begin{eqnarray}
\alpha_{\Rsub V}({\bf q}^2) & = & \alpha_{\overline{\Rsub MS}}(\mu^2)
\sum_{n=0}^\infty \tilde{a}_n(\mu^2/{\bf q^2})
\left(\frac{\alpha_{\overline{\Rsub MS}}(\mu^2)}{4\pi}\right)^n
\label{orig} = 
\alpha_{\overline{\Rsub MS}}({\bf q}^2)
\sum_{n=0}^\infty a_n\left(\frac{\alpha_{\overline{\Rsub MS}}({\bf q}^2)}
{4\pi}\right)^n. \label{vms}
\end{eqnarray}
Two loop results for the $\beta$ function and the one loop matching condition
for the potential were available to the moment of Buchm\"uller-Tye publication.
Recently, the progress in calculations has provided us with the two loop
matching of  V and $\overline{\rm MS}$ schemes \cite{Peter,Schroed}, that can
be combined with the three loop running of $\alpha_s^{\overline{\Rsub MS}}$. At
present, in expansion (\ref{vms}) the coefficients of tree approximation $a_0$,
the one loop contribution $a_1$ and new results for the two-loop term $a_2$
(see \cite{Peter,Schroed}) are known.

After the introduction of ${\EuFrak a} =\frac{\alpha}{4 \pi}$, the $\beta$
function is actually defined by 
\begin{equation}
\frac{d {\EuFrak a}(\mu^2)}{d\ln\mu^2} = \beta({\EuFrak a})
  = - \sum_{n=0}^\infty \beta_n \cdot {\EuFrak a}^{n+2}(\mu^2),
\end{equation}
so that $\beta_{0,1}^{\Rsub V}=\beta_{0,1}^{\overline{\Rsub MS}}$ and
$\beta_2^{\Rsub V} = \beta_2^{\overline{\Rsub MS}}-a_1\beta_1^{\overline{\Rsub
MS}} + (a_2-a_1^2)\beta_0^{\overline{\Rsub MS}}$. 

The Fourier transform results in the position-space potential \cite{Peter}
\begin{eqnarray}
 V(r) & = & -C_F\frac{\alpha_{\overline{\Rsub MS}}(\mu^2)}{r}\Bigg[
      1 + \frac{\alpha_{\overline{\Rsub MS}}(\mu^2)}{4\pi}\Big(
          2\beta_0\ln(\mu r^\prime)+a_1\Big) + \nonumber \\
   && + \Big(\frac{\alpha_{\overline{\Rsub MS}}(\mu^2)}{4\pi}\Big)^2 \Big(
      \beta_0^2(4\ln^2(\mu r^\prime)+\frac{\pi^2}{3}) \label{Vr}
      +2(\beta_1+2\beta_0a_1)\ln(\mu r^\prime)+a_2\Big) \Bigg],
\end{eqnarray}
with $r^\prime\equiv r\exp(\gamma_E)$. Defining the new running coupling
constant, depending on the distance,
\begin{equation}
  V(r) = -C_F\frac{\bar\alpha_{\Rsub V}(1/r^2)}{r}.
\end{equation}
we can calculate its $\beta$ function from (\ref{Vr}), so that \cite{Peter}
\begin{equation}
  \bar\beta_2^{\Rsub V} = \beta_2^{\Rsub V} + \frac{\pi^2}{3}\beta_0^3,
\end{equation}
and the minor coefficients $\bar\beta_{0,1}^{\Rsub V}$ are equal to the values
independent of the scheme. Note that the perturbative potential (\ref{Vr}), by
construction, is independent of normalization point, i.e. it is the
renormalization group invariant. However, in the problem under consideration
the truncation of perturbative expansion, wherein the coefficients do not
decrease\footnote{Moreover, according to the investigations of renormalon, the
coefficients in the series of perturbation theory for the potential increase in
the factorial power, so that the series has a meaning of asymptotic one.},
leads to a strong custodial dependence on the normalization point. So, putting
the normalization point $\mu$ in the region of charned quark mass, we find that
the two-loop potential with the three-loop running coupling constant
$\alpha_s^{\overline{\Rsub MS}}$ has an unremovable additive shift depending on
$\mu$. This shift has variation in wide limits. This fact illuminates the
presence of infrared singularity in the coupling constant of QCD, so that the
$\mu$-dependent shift in the potential energy has the form of a pole posed at
$\Lambda_{QCD}$ \cite{3loop}.

Thus, in order to avoid the ambiguity of static potential in QCD we have to
deal with infrared stable quantities. The motivation by Buchm\"uller and Tye
was to write down the $\beta$ function of $\alpha_{\Rsub V}$ consistent with
two known asymptotic regimes at short and long distances. They proposed the
function, which results in the effective charge determined by two parameters,
only: the perturbative parameter is the scale in the running of coupling
constant at large virtualities and the nonperturbative parameter is the string
tension. The necessary inputs are the coefficients of $\beta$ function. The
parameters of potential by Buchm\"uller and Tye were fixed by fitting the mass
spectra of charmonium and bottomonium \cite{PDG}. Particularly, in such the
phenomenological approach the scale $\Lambda_{\overline{\Rsub
MS}}^{n_f=4}\approx 510$ MeV was determined. It determines the asymptotic
behaviour of coupling constant at latge virtualities in QCD. This value is in a
deep contradiction with the current data on the QCD coupling constant
$\alpha_s^{\overline{\Rsub MS}}$ \cite{PDG}. In addition, one can easily find
that the three-loop coefficient $\beta_2^{\Rsub V}$ for the $\beta$ function
suggested by Buchm\"uller and Tye is not correct even by its sign and absolute
value in comparison with the exact coefficient recently calculated  in
\cite{Peter,Schroed}.

Thus, the modification of Buchm\"uller--Tye (BT) potential of static quarks as
dictated by the current status of perturbative calculations is of great
interest.

To normalize the couplings in deep perturbative region, we use (\ref{vms}) at
${\bf q}^2 = m_Z^2$.

\subsubsection{The quark confinement}

The nonperturbative behaviour of QCD forces between the static heavy quarks  at
long distances $r$ is usually represented by the linear potential (see
discussion in ref.\cite{simon})
\begin{equation}
V^{\rm conf}(r) = k\cdot r, \label{conf}
\end{equation}
which corresponds to the square-law limit for the Wilson loop. 

We can represent this potential in terms of constant chromo-electric field
between the sources posed in the fundamental representation of SU($N_c$). So,
in the Fock-Schwinger gauge of fixed point
$
x_\mu\cdot A^\mu(x) = 0,
$
we can represent the gluon field by means of strength tensor
$
A_\mu(x) \approx \frac{1}{2} x^\nu G_{\mu\nu}(0),
$
so that for the static quarks separated by the distance $\bf r$ we have
$
\bar Q_i(0)\; G^a_{m0}(0)\; Q_j(0) = \frac{{\bf r}_m}{r}\; E\; T^a_{ij},
$
where the heavy quark fields are normalized to unit. Then, the confining
potential is written down as
$$
V^{\rm conf}(r) = \frac{1}{2} g_s\, C_F\, E \cdot r.
$$
\begin{figure}[th]
\hspace*{2.5cm}
\epsfxsize=12cm \epsfbox{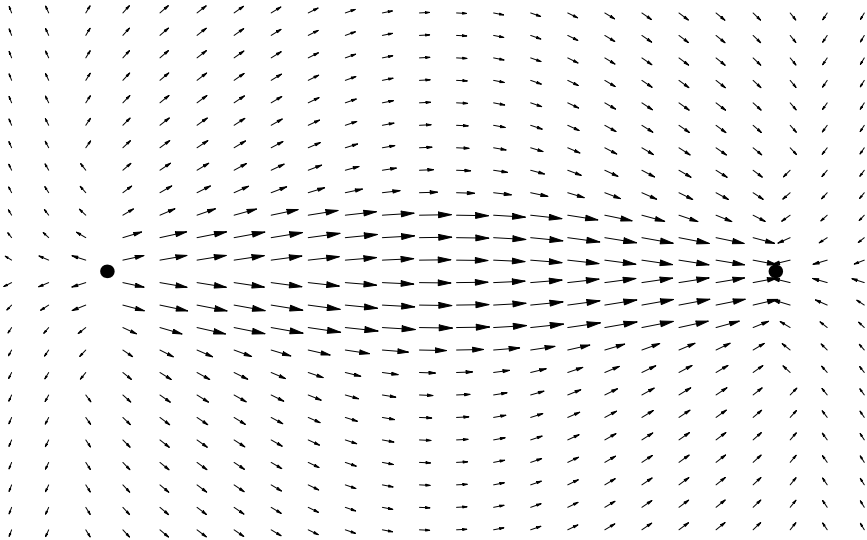}
\caption{Under the action of charged sources, the vacuum chromo-electric field
is aligned along the axis connecting the heavy quarks.}
\label{elect}
\end{figure}

\noindent
Supposing, that the same strength of the field is responsible for the forming 
of gluon condensate (see Fig. \ref{elect}), and introducing the colored sources
$n_i$, which have to be averaged in the vacuum, we can easily find
$$
\langle G^2_{\mu\nu}\rangle = 4\,C_F\, E^2 \langle \bar n n\rangle.
$$
For the linear term in the potential, the consideration in \cite{3loop} leads
to
\begin{equation}
k = \frac{\pi}{2 \sqrt{N_c}}\, C_F \sqrt{\langle \frac{\alpha_s}{\pi}
G^2_{\mu\nu}\rangle}.
\label{cond}
\end{equation}
The $k$ term is usually represented through a parameter $\alpha^\prime_P$ as
$$
k = \frac{1}{2 \pi \alpha^\prime_P}.
$$
Following Buchm\"uller and Tye, we put $\alpha^\prime_P= 1.04$ GeV$^{-2}$. This
value of tension, that is related with a slope of Regge trajectories, can be
compared with the estimate following from (\ref{cond}). At $\langle
\frac{\alpha_s}{\pi} G^2_{\mu\nu}\rangle = (1.6\pm 0.1) \cdot 10^{-2}$ GeV$^4$
\cite{SVZ} we have found
$$ 
\alpha^\prime_P = 1.04 \pm 0.03\;\; {\rm GeV}^{-2}, 
$$
which is in a good agreement with the well known value of Regge trajectory
slope.

The form of (\ref{conf}) corresponds to the limit, when at low virtualities
${\bf q}^2\to 0$ the coupling $\alpha_{\Rsub V}$ tends to
$$
\alpha_{\Rsub V}({\bf q^2}) \to \frac{K}{\bf q^2},
$$
so that
\begin{equation}
\frac{d \alpha_{\Rsub V}({\bf q^2})}{\ln \bf q^2} \to - \alpha_{\Rsub V}({\bf
q^2}),
\label{lim-c}
\end{equation}
which gives the confinement asymptotics for the $\beta_{\Rsub V}$ function.

\subsubsection{Unified $\beta$ function and potential}

Buchm\"uller and Tye proposed the procedure for the reconstruction of
$\beta$ function in the whole region of charge variation by the known limits of
asymptotic freedom to a given order in $\alpha_s$ and confinement regime.
Generalizing their method, the $\beta_{\rm PT}$ function found in the framework
of asymptotic perturbative theory (PT) to three loops, is transformed to the
$\beta$ function of effective charge as follows
\begin{eqnarray}
\displaystyle
\frac{1}{\beta_{\rm PT}({\EuFrak a})} &=& -\frac{1}{\beta_0 {\EuFrak a}^2} +
\frac{\beta_1+\left(\beta_2^{\Rsub V} - \frac{\beta_1^2}{\beta_0}\right)
{\EuFrak a}}{\beta_0^2 {\EuFrak a}} \Longrightarrow \nonumber \\
\frac{1}{\beta({\EuFrak a})} &=& -\frac{1}{\beta_0 {\EuFrak a}^2 \left(1-
\exp\left[-\frac{1}{\beta_0 {\EuFrak a}}\right]\right)}+
\frac{\beta_1+\left(\beta_2^{\Rsub V} - \frac{\beta_1^2}{\beta_0}\right)
{\EuFrak a}}{\beta_0^2 {\EuFrak a}}
\exp\left[-\frac{l^2 {\EuFrak a}^2}{2}\right],
\label{KKO}
\end{eqnarray}
where the exponential factor in the second term contributes to the
next-to-next-to-leading order at ${\EuFrak a}\to 0$. This function has the
essential peculiarity at ${\EuFrak a}\to 0$, so that the expansion is the
asymptotic series in ${\EuFrak a}$. At ${\EuFrak a}\to \infty$ the $\beta$
function tends to the confinement limit represented in (\ref{lim-c}). Remember
that the one and two-loop static potentials matched with the linear term of
confinement lead to the contradiction with the value of QCD coupling constant
extracted at the scale of $Z$ boson mass if we fit the mass spectra of heavy
quarkonia in such potentials. We will show that the static potential in the
three-loop approximation results in the consistent value of QCD coupling
constant at large virtualities.

In the perturbative limit the usual solution for the running coupling constant
\begin{eqnarray}
{\EuFrak a}(\mu^2) = \frac{1}{\beta_0 \ln
\frac{\mu^2}{\Lambda^2}}&&\left[1 - 
\frac{\beta_1}{\beta_0^2}\frac{1}{\ln
\frac{\mu^2}{\Lambda^2}} 
\ln \ln \frac{\mu^2}{\Lambda^2} + \right.
\nonumber \\
&&
\left.
\frac{\beta_1^2}{\beta_0^4}\frac{1}{\ln^2
\frac{\mu^2}{\Lambda^2}} 
\left( \ln^2 \ln \frac{\mu^2}{\Lambda^2} - \ln \ln \frac{\mu^2}{\Lambda^2} -1 +
\frac{\beta_2^{\Rsub V} \beta_0}{\beta_1^2}\right)\right],
\label{3pt}
\end{eqnarray}
is valid. Using the asymptotic limit of (\ref{3pt}), one can get the equation 
\begin{eqnarray}
\ln \frac{\mu^2}{\Lambda^2} & = & \frac{1}{\beta_0 {\EuFrak
a}(\mu^2)}+
\frac{\beta_1}{\beta_0^2} \ln \beta_0 {\EuFrak
a}(\mu^2)+\int_0^{{\EuFrak a}(\mu^2)} dx 
\left[\frac{1}{\beta_0 x^2}-\frac{\beta_1}{\beta_0^2
x}+\frac{1}{\beta(x)}\right], \label{ptlim} 
\end{eqnarray}
which can be easily integrated out, so that we get an implicit solution for the
charge depending on the scale. The implicit equation can be inverted by the
iteration procedure, so that well approximated solution has the form
\begin{equation}
{\EuFrak a}(\mu^2) = \frac{1}{\beta_0 \ln\left(1+\eta(\mu^2)
\frac{\mu^2}{\Lambda^2}\right)},
\label{eff}
\end{equation}
where $\eta(\mu^2)$ is expressed through the coefficients of perturbative
$\beta$ function and parameter $l$, which is related to the slope of Regge
trajectories and the integration constant, the scale $\Lambda$, by the relation
\begin{equation}
\ln 4 \pi^2 C_F \alpha^\prime_P \Lambda^2 = \ln \beta_0 +
\frac{\beta_1}{2
\beta_0^2} \left(\gamma_E + \frac{l^2}{2\beta_0^2}\right)-\frac{\beta_2^{\Rsub
V}\beta_0 -\beta_1^2}{\beta_0^3}\,
\frac{\sqrt{\frac{\pi}{2}}}{l},
\end{equation}
which completely fixes the parameters of $\beta$ function and the charge in
terms of scale $\Lambda$ and slope $\alpha^\prime_P$.

\subsubsection{Setting the scales}
As we have already mentioned the slope of Regge trajectories, determining the
linear part of potential, is fixed as
$ 
\alpha^\prime_P = 1.04\;{\rm GeV}^{-2}. 
$
We use also the measured value of QCD coupling constant \cite{PDG} and pose
$$
\alpha_s^{\overline{\Rsub MS}}(m_Z^2) = 0.123,
$$
as the basic input of the potential. Then, we evaluate
$
\alpha_{{\Rsub V}}(m_Z^2) \approx 0.1306,
$
and put it as the normalization point for ${\EuFrak a}(m_Z^2)=\alpha_{{\Rsub
V}}(m_Z^2)/(4\pi)$. Further, we find the values of $\Lambda$ for the
effective charge, depending on the number of active flavors \cite{3loop}. After
determining the momentum space dependence of the charge, we perform the Fourier
transform to get 
\begin{equation}
V(r) = k \cdot r - \frac{8 C_F}{r} u(r),
\label{r<}
\end{equation}
with the function
$$
u(r) = \int_0^\infty \frac{d q}{q}\, \left({\EuFrak
a}(q^2)-\frac{K}{q^2}\right) \sin
(q\cdot r),
$$
which is calculated numerically at $r>0.01$ fm, while at short distances the
behaviour of potential is purely perturbative, so that at $r<0.01$ fm we put 
\begin{equation}
V(r) = - C_F\, \frac{\bar \alpha_{{\Rsub V}}(1/r^2)}{r},
\label{r>}
\end{equation}
where the running $\bar\alpha_{{\Rsub V}}(1/r^2)$ is given by eq.(\ref{3pt})
with the appropriate value of $\bar \beta_2^{\Rsub V}$ at $n_f = 5$, and with
the matching with the potential (\ref{r<}) at $r_s=0.01$ fm, where we have
found
$
\bar\alpha_{\Rsub V}(1/r_s^2) = 0.22213,
$
which implies $\Lambda_{n_f=5}^{\overline{\Rsub V}}=617.42$ MeV.

\begin{figure}[th]
\setlength{\unitlength}{1mm}
\begin{center}
\begin{picture}(100,95)
\put(5,5){\epsfxsize=9cm \epsfbox{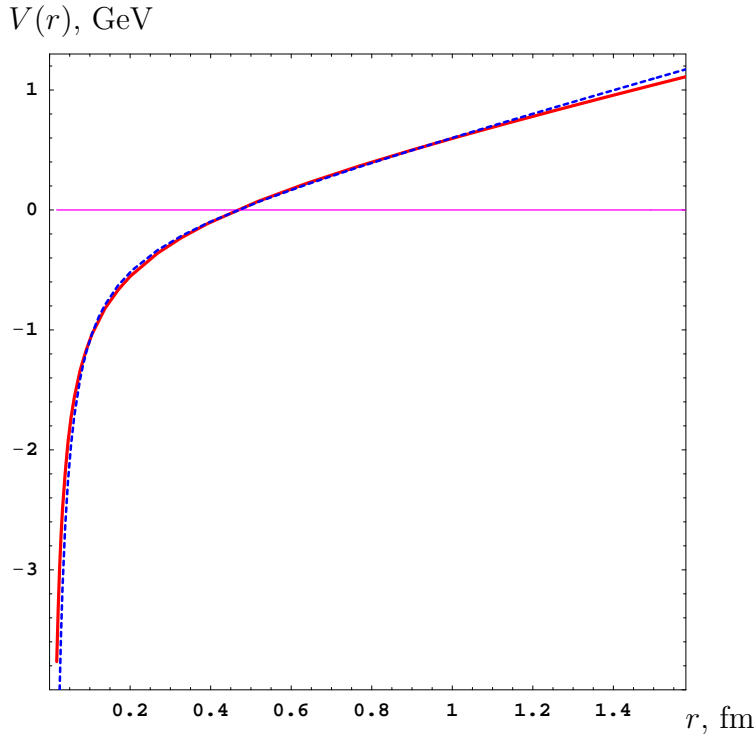}}
\put(95,5){$r$, fm}
\put(5,98){$V(r)$, GeV}
\end{picture}
\end{center}

\vspace*{-1cm}
\caption{The potential of static heavy quarks in QCD (solid line) in comparison
with the Cornell model (dashed line) (up to an additive shift of energy
scale).} 
\label{pot-fig}
\end{figure}

Thus, we have completely determined the potential of static heavy quarks in QCD
with the three-loop running of coupling constant. In Fig. \ref{pot-fig} we
present the potential versus the distance between the quarks. As we can see the
potential is very close to what was obtained in the Cornell model in the
phenomenological manner by fitting the mass spectra of heavy quarkonia. We can
draw the conclusion that accepting the normalization by the value of QCD
coupling constant at the virtuality $q^2 = m_Z^2$ and using the three-loop
evolution for the effective charge incorporating the confinement with the
linear term, we have got the static potential consistent with the
phenomenological models and, hence, with the calculations of mass spectra for
the heavy quarkonia in the nonrelativistic approximation.

Such the consistency of potential with the parameters of QCD has become
possible due to the fact that in the two-loop approximation for the coulomb
potential the calculations have led to the essential corrections to the
$\beta$ function of effective charge, so that $\Delta \beta/\beta \sim 10\%$.
This correction is important for the determination of critical values of
charge, i.e. the value in the intermediate region between the perturbative and
nonperturbative regimes. Moreover, the two-loop matching condition and the
three-loop running of coupling constant normalized by the data at the high
energy of $m_Z$ determine the region of energetic scale for changing the
regimes mentioned above. This scale strongly correlates with the data on the
mass spectra of heavy quarkonia. Indeed, it is connected with the splitting of
masses between the $1S$ and $2S$ levels. We stress that the two-loop
improvement gives the correct normalization of effective coulomb exchange at
the distances characteristic for the average separation between the heavy
quarks inside the heavy quarkonium and determines the evolution at short
distances $r < 0.08$ fm, that is important in the calculations of leptonic
constants related with the wave functions at the origin. 

The analysis of potential for the static quarks in the calculations of both the
mass spectra for the heavy quarkonia and the leptonic constants for the vector
states is presented in \cite{3loop}, where the heavy quark masses are
determined in the potential approach.

We emphasize that the potential by Buchm\"uller and Tye was obtained under
fitting the experimental mass spectra for the heavy quarkonia, and it is
numerically very close to the static potential under consideration consistent
with the normalization of QCD coupling constant at large virualities.
Therefore, the Buchm\"uller--Tye potential continues to keep its
phenomenological usage for the calculations of mass levels for the hadrons with
$c$ and $b$ quarks with the accuracy about 40-70 MeV, that is a systematic
uncertainty of potential approach.

\subsubsection{System of levels}

Following ref. \cite{PM1}, in the potential model we use the Buchm\"uller--Tye
ansatz, which takes into account the coulomb corrections at short distances
with the running coupling constant in two loops, while at large distances the
interaction energy linearly increases, that provides the confinement. In the
anti-triplet quark state we introduce the factor of 1/2 because of the color
structure of bound quark-quark state. For the interaction of diquark with the
light constituent quark, the corresponding factor is equal to unit.

As was shown in \cite{15}, the nonperturbative constituent term introduced into
the mass of nonrelativistic quark, exactly coincides with the additive
constant, subtracted from the coulomb potential. 

Thus, we extract the masses of heavy quarks by fitting the real spectra of
charmonium and bottomonium,
\begin{equation}
m_c = 1.486\; \mbox{GeV},\quad m_b = 4.88\; \mbox{GeV},
\end{equation}
so that the mass of the level in the heavy quarkonium has been calculated as,
say, $M(с\bar c)=2 m_c+E$, where $E$ is the energy of stationary Schr\" odinger
equation with the model potential $V$. Then, we have supposed that the mass of
meson with a single heavy quark is equal to $M(Q\bar q) = m_Q + m_q + E$, and
$E=\langle T\rangle + \langle V-\delta V\rangle$, whereas the additive term in
the potential is introduced because the constituent mass of light quark is
determined as a part of interaction energy $\delta V = m_q$. In accordance with
fitting the masses of heavy mesons, we get $m_q = 0.385$ GeV.

\begin{table}[ph]
\caption{The spectrum of $bb$-diquark levels without spin-dependent
splittings: masses and mean-squared radii.}
\label{bbt}
\begin{center}
\begin{tabular}{|c|c|c||c|c|c|}
\hline
diquark level  & М, GeV &  $\langle r^2\rangle ^{1/2}$, fm & 
diquark level  & М, GeV &  $\langle r^2\rangle ^{1/2}$, fm \\ 
\hline 
1S & 9.74 & 0.33 & 2P & 9.95 & 0.54\\
\hline 
2S & 10.02 & 0.69 & 3P & 10.15 & 0.86\\ 
\hline 
3S & 10.22 & 1.06 & 4P & 10.31 & 1.14\\
\hline 
4S & 10.37 & 1.26 & 5P & 10.45 & 1.39\\
\hline
5S & 10.50 & 1.50 & 6P & 10.58 & 1.61\\
\hline
3D & 10.08 & 0.72 & 4D & 10.25 & 1.01 \\
\hline 
5D & 10.39 & 1.28 & 6D & 10.53 & 1.51 \\
\hline 
4F & 10.19 & 0.87 & 5F & 10.34 & 1.15 \\
\hline 
6F & 10.47 & 1.40 & 5G & 10.28 & 1.01 \\
\hline 
6G & 10.42 & 1.28 & 6M & 10.37 & 1.15\\
\hline 

\end{tabular} 
\end{center} 
\end{table}
\begin{table}[ph]
\caption{The spectrum of $bc$-diquark levels without spin-dependent
splittings: masses and mean-squared radii.}
\label{bct}
\begin{center}
\begin{tabular}{|c|c|c||c|c|c|}
\hline
diquark level  & М, GeV &  $\langle r^2\rangle ^{1/2}$, fm & 
diquark level  & М, GeV &  $\langle r^2\rangle ^{1/2}$, fm \\ 
\hline 
1S & 6.48 & 0.48 & 3P & 6.93 & 1.16\\
\hline 
2S & 6.79 & 0.95 & 4P & 7.13 & 1.51\\ 
\hline 
3S & 7.01 & 1.33 & 3D & 6.85 & 0.96\\
\hline 
2P & 6.69 & 0.74 & 4D & 7.05 & 1.35\\
\hline 
4F & 6.97 & 1.16 & 5F & 7.16 & 1.52\\
\hline 
5G & 7.09 & 1.34 & 6H & 7.19 & 1.50\\
\hline 
\end{tabular} 
\end{center} 
\end{table}
\begin{table}[ph]
\caption{The spectrum of $cc$-diquark levels without spin-dependent
splittings: masses and mean-squared radii.}
\label{cct}
\begin{center}
\begin{tabular}{|c|c|c||c|c|c|}
\hline
diquark level  & М, GeV &  $\langle r^2\rangle ^{1/2}$, fm & 
diquark level  & М, GeV &  $\langle r^2\rangle ^{1/2}$, fm \\ 
\hline 
1S & 3.16 & 0.58 & 3P & 3.66 & 1.36\\
\hline 
2S & 3.50 & 1.12 & 4P & 3.90 & 1.86\\ 
\hline 
3S & 3.76 & 1.58 & 3D & 3.56 & 1.13\\
\hline 
2P & 3.39 & 0.88 & 4D & 3.80 & 1.59\\
\hline 
\end{tabular} 
\end{center} 
\end{table}

The results of calculations for the energy levels in the Schr\"odinger equation
with the Buch\-m\"ul\-ler--Tye potential for the various diquark systems are
presented in Tables \ref{bbt}--\ref{cct}, while the characteristics of
corresponding wave functions are shown in Tables \ref{fbbt}--\ref{fcct}. 

\begin{table}[ph]
\caption{The characteristics of radial wave function for the $bb$-diquark:
$R_{d(ns)}(0)$ (GeV$^{3/2}$), $R_{d(np)}^{'} (0)$ (GeV $^{5/2}$).}
\label{fbbt}
\begin{center}
\begin{tabular}{|c|c|c|c|}
\hline 
nL  & $R_{d(ns)}(0)$  & nL & 
$R_{d(np)}'(0)$  \\
\hline 
1S & 1.346 & 2P & 0.479  \\ 
\hline
2S & 1.027 & 3P & 0.539  \\
\hline 
3S & 0.782 & 4P & 0.585  \\
\hline 
4S & 0.681 & 5P & 0.343  \\
\hline 
\end{tabular}
\end{center}
\end{table} 

We have checked that with a good accuracy {\it the binding energy
and the wave function of light quark practically do not depend on the flavors
of heavy quarks}. Indeed, large values of diquark masses give small
contributions into the reduced masses. This fact leads to small corrections to
the wave functions in the Schr\"odinger equation.

\begin{table}[ph]
\caption{The characteristics of radial wave function for the $bc$-diquark:
$R_{d(ns)}(0)$ (GeV$^{3/2}$), $R_{d(np)}^{'} (0)$ (GeV $^{5/2}$).}
\label{fbct}
\begin{center}
\begin{tabular}{|c|c|c|c|}
\hline 
nL  & $R_{d(ns)}(0)$  & nL & 
$R_{d(np)}'(0)$  \\
\hline 
1S & 0.726 & 2P & 0.202  \\ 
\hline
2S & 0.601 & 3P & 0.240  \\
\hline 
\end{tabular}
\end{center}
\end{table} 

\begin{table}[ph]
\caption{The characteristics of radial wave function for the $cc$-diquark:
$R_{d(ns)}(0)$ (GeV$^{3/2}$), $R_{d(np)}^{'} (0)$ (GeV $^{5/2}$).}
\label{fcct}
\begin{center}
\begin{tabular}{|c|c|c|c|}
\hline 
nL  & $R_{d(ns)}(0)$  & nL & 
$R_{d(np)}'(0)$  \\
\hline 
1S & 0.530 & 2P & 0.128  \\ 
\hline
2S & 0.452 & 3P & 0.158  \\
\hline 
\end{tabular}
\end{center}
\end{table} 

So, for the states lying below the threshold of doubly heavy baryon decay into
the heavy baryon and heavy meson, the energies of levels of light constituent
quark are equal to
$$
E(1s) = 0.38\; {\rm GeV,}\;\;E(2s) = 1.09\; {\rm GeV,}\;\; E(2p) = 0.83\; {\rm
GeV,}
$$
where the energy has been defined as the sum of light quark constituent mass
and eigen-value of Schr\" odinger equation. In HQET the value of $\bar
\Lambda=E(1s)$ is generally introduced. Then we can draw a conclusion that our
estimate of $\bar \Lambda$ is in a good agreement with calculations in other
approaches. This fact confirms the reliability of such the
phenomenological predictions. For the light quark radial wave functions at the
origin we find
$$
R_{1S}(0) = 0.527\; {\rm GeV}^{3/2},\;\;
R_{2S}(0) = 0.278\; {\rm GeV}^{3/2},\;\;
R_{2P}'(0) = 0.127\; {\rm GeV}^{5/2}.\;\;
$$
The analogous characteristics of bound states of the $c$-quark
interacting with the $bb$-diquark, are equal to
$$
E(1s) = 1.42\; {\rm GeV,}\;\;E(2s) = 1.99\; {\rm GeV,}\;\; E(2p) = 1.84\; {\rm
GeV,}
$$
with the wave functions
$$
R_{1S}(0) = 1.41\; {\rm GeV}^{3/2},\;\;
R_{2S}(0) = 1.07\; {\rm GeV}^{3/2},\;\;
R_{2P}'(0) = 0.511\; {\rm GeV}^{5/2}.\;\;
$$

For the binding energy of strange constituent quark we add the current mass
$m_s\approx 100-150$ MeV.

\subsubsection{Spin-dependent corrections}

According to \cite{16}, we introduce the spin-dependent corrections causing
the splitting of $nL$-levels of diquark as well as in the system of light
constituent quark and diquark ($n=n_r+L+1$ is the principal number,
$n_r$ is the number of radial excitation, $L$ is the orbital momentum). For the
heavy diquark containing the identical quarks we have
\begin{eqnarray}
V_{SD}^{(d)}({\bf r}) &=& \frac{1}{2}\left(\frac{\bf L_d\cdot
S_d}{2m_Q^2}\right)
\left( -\frac{dV(r)}{rdr}+
\frac{8}{3}\alpha_s\frac{1}{r^3}\right)\nonumber \\
&& +\frac{2}{3}\alpha_s\frac{1}{m_Q^2}\frac{\bf L_d\cdot S_d}{r^3}+\frac{4}{3}
\alpha_s\frac{1}{3m_Q^2}{{\bf S}_{Q1}\cdot {\bf S}_{Q2}}[4\pi\delta({\bf r})]\\
&& -\frac{1}{3}\alpha_s\frac{1}{m_Q^2}\frac{1}{4{\bf L_d}^2 -3} [
6({\bf L_d\cdot S_d})^2+3({\bf L_d\cdot S_d})-2{\bf L_d}^2{\bf
S_d}^2]\frac{1}{r^3},
\nonumber     
\end{eqnarray}
where $\bf L_d,\; S_d$ are the orbital momentum in the diquark system and the
summed spin of quarks composing the diquark, respectively. Taking into
account the interaction with the light constituent quark gives
(${\bf S} = {\bf S_d}+{\bf S_l}$)
\begin{eqnarray}
V_{SD}^{(l)}({\bf r}) &=& \frac{1}{4}\left(\frac{\bf L\cdot S_d}{2m_Q^2}
+ \frac{2\bf L\cdot S_l}{2m_l^2}\right)
\left( -\frac{dV(r)}{rdr}+
\frac{8}{3}\alpha_s\frac{1}{r^3}\right)\nonumber \\
&& +\frac{1}{3}\alpha_s\frac{1}{m_Q m_l}\frac{(\bf L\cdot S_d + 
2L\cdot S_l)}{r^3}+ 
\frac{4}{3}\alpha_s\frac{1}{3m_Q m_l}{({\bf S_d}+{\bf L_d})\cdot {\bf S_l}}
[4\pi\delta({\bf r})]\\
&& -\frac{1}{3}\alpha_s\frac{1}{m_Q m_l}\frac{1}{4{\bf L}^2 -3} [
6({\bf L\cdot S})^2+3({\bf L\cdot S})-2{\bf L}^2{\bf S}^2\nonumber \\
&& -6({\bf L\cdot S_d})^2-3({\bf L\cdot S_d})+2{\bf L}^2{\bf S_d}^2]
\frac{1}{r^3}, \nonumber    
\end{eqnarray}
where the first term corresponds to the relativistic correction to the
effective scalar exchange, and other terms appear because of corrections to
the effective single-gluon exchange with the coupling constant $\alpha_s$. 

The value of effective parameter $\alpha_s$ can be determined in the following
way. The splitting in the $S$-wave heavy quarkonium $(Q_1\bar Q_2)$ is given by
the expression
\begin{equation}
\Delta M(ns) = \frac{8}{9}\alpha_s\frac{1}{m_1m_2}|R_{nS}(0)|^2,
\end{equation}
where $R_{nS}(r)$ is the radial wave function of quarkonium. From the
experimental data on the system of $c\bar c$ 
\begin{equation} 
\Delta M(1S,c\bar c) = 117\pm 2\; {\rm MeV,}
\label{delcc}
\end{equation} 
and $R_{1S}(0)$ calculated in the model, we can determine $\alpha_s(\Psi)$.

Let us take into account the dependence of this parameter on the reduced mass
of the system, $\mu $. In the framework of one-loop approximation for the
running coupling constant of QCD we have
\begin{equation} 
\alpha_s (p^2) = \frac{4\pi}{b\cdot\ln (p^2/\Lambda_{QCD}^2)}, 
\end{equation} 
whereas $b = 11 -2n_f/3$ and $n_f = 3$ at $p^2< m_c^2$. From the phenomenology
of potential models we know that the average kinetic energy of quarks
in the bound state practically does not depend on the flavors of quarks, and it
is given by the values
\begin{equation}
\langle T_{d}\rangle \approx 0.2\; {\rm GeV,}
\end{equation}
\begin{equation}
\langle T_{l}\rangle \approx 0.4\; {\rm GeV,}
\end{equation}
for the anti-triplet and singlet color states, correspondingly. Substituting
the definition of the nonrelativistic kinetic energy
\begin{equation}
\langle T\rangle  = \frac{\langle p^2\rangle }{2\mu },
\end{equation}
we get
\begin{equation}
\alpha_s(p^2) = \frac{4\pi}{b\cdot\ln (2\langle T\rangle \mu/\Lambda_{QCD}^2)},
\end{equation} 
whereas numerically $\Lambda_{QCD}\approx 113$ MeV.

For the identical quarks inside the diquark, the scheme of $LS$-coupling well
known for the corrections in the heavy quarkonium, is applicable. Otherwise,
for the interaction with the light quark we use the scheme of $jj$-coupling
(here, ${\bf LS_l}$ is diagonal at the given ${\bf J_l}$,  $({\bf J_l} =
{\bf L} + {\bf S_l}, {\bf J} = {\bf J_l} + {\bf \bar J})$, where $\bf J$
denotes the total spin of baryon, and $\bf\bar J$ is the total spin of
diquark, ${\bf\bar J}={\bf S_d}+{\bf L_d}$).

Then, to estimate various terms and mixings of states, we use the
transformations of bases (in what follows ${\bf S} = {\bf S_l} + {\bf\bar J}$)
\begin{equation}
|J;J_l\rangle  = \sum_{S} (-1)^{(\bar J+S_l+L+J)}\sqrt {(2S+1)(2J_l+1)}
\left\{\begin{array}{ccc} \bar J & S_l & S \\
                        L & J & J_l \end{array}\right\}|J;S\rangle 
\end{equation}
and 
\begin{equation}
|J;J_l\rangle  = \sum_{J_d} (-1)^{(\bar J+S_l+L+J)}\sqrt {(2J_d+1)(2J_l+1)}
\left\{\begin{array}{ccc} \bar J & L & J_d \\
                        S_l & J & J_l \end{array}\right\}|J;J_d\rangle .
\end{equation}
Thus, we have defined the procedure of calculations for the mass spectra of
doubly heavy baryons. This procedure leads to results presented in the next
section.

\subsection{Numerical results}
In this Section we present the results on the mass spectra with account for the
spin-dependent splitting of levels. As we have clarified in the Introduction,
the doubly heavy baryons with identical heavy quarks allow quite a reliable
interpretation in terms of diquark quantum numbers (the summed spin and the
orbital momentum). Dealing with the excitations of $bc$-diquark, we show the
results on the spin-dependent splitting of the ground 1S-state, since the
emission of soft gluon breaks the simple classification of levels for the
higher excitations of such diquark.

For the doubly heavy baryons, the quark-diquark model of bound states obviously
leads to the most reliable results for the system with the larger mass of heavy
quark, i.e. for $\Xi_{bb}$.

For the quantum numbers of levels, we use the notations $n_d L_d n_l l_l$, i.e.
we show the value of principal quantum number of diquark, its orbital momentum
by a capital letter and the principal quantum number for the excitations of
light quark and its orbital momentum by a lower-case letter. The splitting of
$\Xi_{bb}$ baryon levels was in detail considered in \cite{PM1}. The states
with the total spin $J = \frac{3}{2}$ (or $\frac{1}{2}$), can have different
values of $J_l$, and, hence, they have a nonzero mixing, when we perform the
calculations in the perturbation theory built over the states with the definite
total momentum $J_l$ of the light constituent quark. For $J =\frac{3}{2}$ the
mixing matrix can be approximated by the diaginal matrix with a high accuracy.
For $J=\frac{1}{2}$ the mixing of states with the different values of total
spin-orbital momentum of light quark is strong. The analysis for the $1S2p$ and
$2S2p$ levels in the system $\Xi_{bb}$ was done in ref. \cite{PM1}, 
where one can see that the difference between the wave functions because of the
slow change of diquark subsystem mass is not essential within the accuracy of
the method.

The splittings of $D$ and $G$ levels in the diquark are less than 11 MeV, so
that these corrections are small for the diquark excitations having the sizes
less than the distance to the light quark, i.e. for the states with low value
of principal quantum number, under the systematic uncertainty about $\delta
M\approx 30 - 40$ MeV.

For the hyper-fine spin-spin splitting in the system of quark-diquark, we have
\begin{equation}
\Delta_{h.f.}^{(l)} = \frac{2}{9}\bigg[J(J+1)-\bar J(\bar J +1 ) -
\frac{3}{4}\bigg]
\alpha_s(2\mu T)\frac{1}{m_bm_l} |R_l(0)|^2,
\end{equation}
where $R_l(0)$ is the radial wave function at the origin for the light
constituent quark, and for the analogous shift of diquark level, we find
\begin{equation}
\Delta_{h.f.}^{(d) }= \frac{1}{9}
\alpha_s(2\mu T)\frac{1}{m_b^2} |R_d(0)|^2.
\end{equation}

\begin{figure}[th]
\setlength{\unitlength}{1mm}
\begin{center}
\begin{picture}(150,90)
\put(0,0){\epsfxsize=14cm \epsfbox{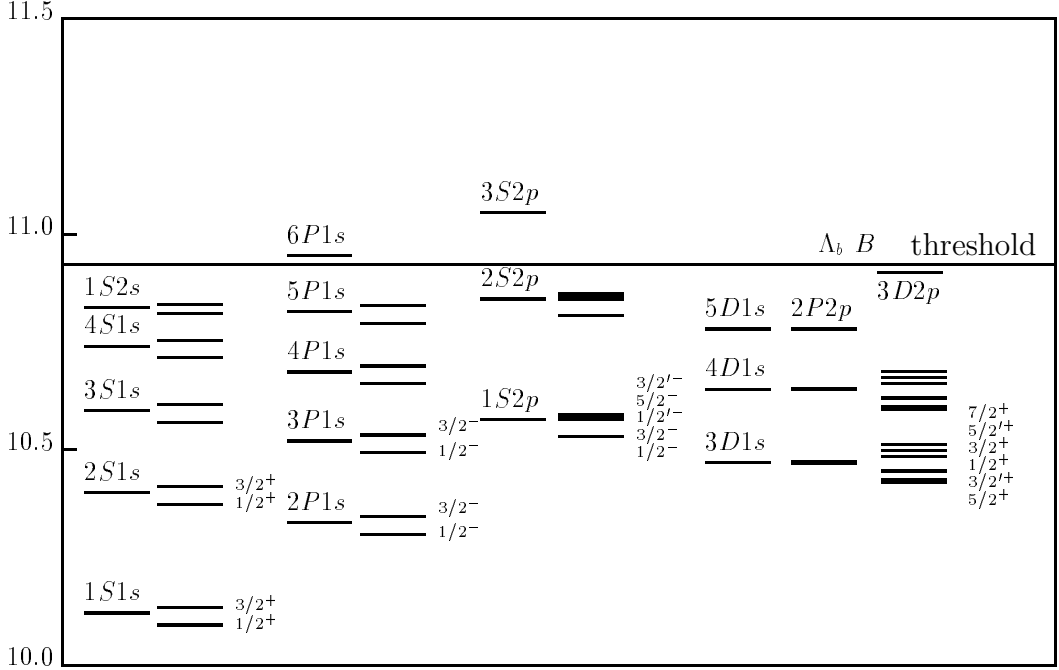}}
\put(120,55){threshold}
\end{picture}
\end{center}
\caption{The spectrum of baryons, containing two $b$-quarks: $\Xi_{bb}^{-}$ and
$\Xi_{bb}^{0}$, with account for the spin-dependent splittings of low-lying
excitations. The masses are given in GeV.}
\label{pic-bb}
\end{figure}

The mass spectrum of $\Xi_{bb}^{+}$ and $\Xi_{bb}^{0}$ baryons is shown in
Fig. \ref{pic-bb}, wherein we restrict ourselves by the presentation of S-,
P- and D-wave levels, while the table containing the numerical values of masses
for the $\Xi_{bb}$ baryons is presented in \cite{PM1}.

We can see in Fig. \ref{pic-bb} that the most reliable predictions are the
masses of baryons $1S1s\; (J^P=3/2^+,\; 1/2^+)$, $2P1s\; (J^P=3/2^-,\; 1/2^-)$
and $3D1s\; (J^P=7/2^+,\ldots 1/2^+)$. The $2P1s$-level is quasi-stable,
because the transition into the ground state requires the instantaneous change
of both the orbital momentum and the summed spin of quarks inside the diquark.
The analogous kind of transitions seems to be the transition between the states
of ortho- and para-hydrogen in the molecule of $H_2$. This transition take
place in a non-homogeneous external field due to the magnetic moments of other
molecules. For the transition of $2P1s\to 1S1s$, the role of such the external
field is played by the non-homogeneous chromo-magnetic field of the light
quark. The corresponding perturbation has the form
\begin{eqnarray}
\delta V & \sim & \frac{1}{m_Q} [{\bf S}_1\cdot {\bf H}_1 + {\bf S}_2\cdot {\bf
H}_2 - ({\bf S}_1+{\bf S}_2)\cdot \langle{\bf H}\rangle ] \nonumber\\
& = & \frac{1}{2m_Q}({\bf \nabla\cdot r_d})\; ({\bf S}_1-{\bf S}_2)\cdot {\bf
H} \sim \frac{1}{m_Q}\frac{{\bf r_l}\cdot {\bf r_d}}{m_q r_l^5}\; ({\bf
S}_1-{\bf S}_2)\cdot{\bf J_l}\; f(r_l),\nonumber
\end{eqnarray}
where $f(r_l)$ is a dimensionless nonperturbative function depending on the
distance between of the light quark and diquark. The $\delta V$
operator changes the orbital momentum of  light quark, too. It results in the
mixing between the states with the same values of $J^P$. If the splitting is
not small (for instance, $2P1s - 1S2p$, where $\Delta E\sim \Lambda_{QCD}$),
then the mixing is suppressed as $\delta V/\Delta E \sim \frac{1}{m_Q m_q}
\frac{r_d}{r_l^4} \frac{1}{\Delta E} \ll 1$. Since the admixture of $1S2p$ in
the $2P1s$-state is low, the  $2P1s$-levels are quasi-stable, i.e. their
hadronic transitions into the ground state with the emission of $\pi$-mesons 
are suppressed as we have derived, though an additional suppression is given by
a small value of phase space. Therefore, we have to expect the presence of
narrow resonances in the mass spectra of pairs $\Xi_{bb} \pi$, as they are
produced in the decays of quasi-stable states with $J^P=3/2^-,\; 1/2^-$. The
experimental observation of such levels could straightforwardly confirm the
existence of diquark excitations and provide the information on the character
of dependency in $f(r_l)$, i.e. on the non-homogeneous chromo-magnetic field in
the nonperturbative region.

Sure, the $3D1s\; J^P=7/2^+,\; 5/2^+$ states are also quasi-stable, since in
the framework of multipole expansion in QCD they transform into the ground
state due to the quadrupole emission of gluon (the E2-transition with the
hadronization $gq\to q' \pi$).

As for the higher excitations, the $3P1s$-states are close to the $1S2p$-levels
with $J^P=3/2^-,\; 1/2^-$, so that the operators changing both the
orbital momentum of diquark and its spin, can lead to the essential mixing with
an amplitude $\delta V_{nn'}/\Delta E_{nn'}\sim 1$, despite of suppression by
the inverse heavy quark mass and small size of diquark. We are sure that
the mixing slightly shifts the masses of states. The most important effect is
a large admixture of $1S2p$ in $3P1s$. It makes the state to be unstable
because of the transition into the ground $1S1s$-state with the emission of
gluon (the E1-transition). This transition leads to decays with the emission of
$\pi$-mesons\footnote{Remember, that the $\Xi_{QQ'}$-baryons are the
iso-dublets.}. 

The level $1S2p\; J^P=5/2^-$ has the definite quantum numbers of diquark and
light quark motion, because there are no levels with the same values of $J^P$
in its vicinity. However, its width of transition into the ground state and
$\pi$-meson is not suppressed and seems to be large, $\Gamma\sim 100$ MeV.

For the transitions we have
$$
{\frac{3}{2}}^- \to {\frac{3}{2}}^+ \pi \;\; {\rm in~ S-wave,}\;\;\;
{\frac{3}{2}}^- \to \frac{1^+}{2} \pi \;\; {\rm in~ D-wave,}
$$
$$
{\frac{1}{2}}^- \to {\frac{3}{2}}^+ \pi \;\; {\rm in~ D-wave,}\;\;\;
{\frac{1}{2}}^- \to {\frac{1}{2}}^+ \pi \;\; {\rm in~ S-wave.}
$$
The D-wave transitions are suppressed by the ratio of low recoil momentum to
the mass of baryon.

The width of state $J^P=3/2^+$ is completely determined by the
radiative electromagnetic M1-transition into the ground $J^P=1/2^+$ state.

The calculation procedure described above leads to the results for the
doubly charmed baryons as presented in Table \ref{ccqt}.
\begin{table}[th]
\caption{The mass spectrum of $\Xi_{cc}^{++}$ and $\Xi_{cc}^{+}$ baryons.}
\label{ccqt}
\begin{center}
\begin{tabular}{|p{40mm}|c||p{40mm}|c|}
\hline
$(n_d L_d n_l L_l)$, $J^{P}$ & mass, {\rm GeV}
 & $(n_d L_d n_l L_l)$, $J^{P}$ & mass, {\rm GeV}  \\
\hline
(1S 1s)$1/2^{+}$ & 3.478 & (3P 1s)$1/2^{-}$    & 3.972 \\
\hline
(1S 1s)$3/2^{+}$ & 3.61  & (3D 1s)$3/2^{\prime +}$ & 4.007 \\
\hline
(2P 1s)$1/2^{-}$ & 3.702  & (1S 2p)$3/2^{\prime -}$ & 4.034  \\
\hline
(3D 1s)$5/2^{+}$ & 3.781  & (1S 2p)$3/2^{-}$ & 4.039 \\
\hline
(2S 1s)$1/2^{+}$ & 3.812 & (1S 2p)$5/2^{-}$  & 4.047\\
\hline
(3D 1s)$3/2^{+}$ & 3.83 & (3D 1s)$5/2^{\prime +}$ & 4.05 \\
\hline
(2P 1s)$3/2^{-}$ & 3.834 & (1S 2p)$1/2^{\prime -}$ & 4.052 \\
\hline
(3D 1s)$1/2^{+}$ & 3.875 & (3S 1s)$1/2^{+}$   & 4.072\\
\hline
(1S 2p)$1/2^{-}$ & 3.927  & (3D 1s)$7/2^{+}$   & 4.089 \\
\hline
(2S 1s)$3/2^{+}$ & 3.944 & (3P 1s)$3/2^{-}$   & 4.104 \\
\hline
\end{tabular}
\end{center}
\end{table}

As we have already mentioned, the heavy diquark composed of
the quarks of different flavors, turns out to be unstable under the
emission of soft gluons. So, in the Fock state of doubly heavy baryon, there is
a sizable nonperturbative admixture of configurations
including the gluons and diquark with the various values of its spin $S_d$ and
orbital momentum $L_d$
$$
|B_{bcq}\rangle = O_B|bc_{\bar 3_c}^{S_d,L_d},q\rangle+
H_1|bc_{\bar 3_c}^{S_d\pm 1,L_d},g,q\rangle+
H_2|bc_{\bar 3_c}^{S_d,L_d\pm 1},g,q\rangle+\ldots ,
$$
whereas the amplitudes of $H_1$, $H_2$ are not suppressed with respect to
$O_B$. In the heavy quarkonium, the analogous operators for the octet-color
states are suppressed by the probability of emission by the nonrelativistic
quarks inside a small volume determined by the size of singlet-color system of
heavy quark and anti-quark. In the baryonic system under consideration, a soft
gluon is restricted only by the ordinary scale of confinement, and, hence,
there is no suppression.

We suppose that the calculations of masses for the excited $\Xi_{bc}$ baryons
are not so justified in the given scheme. Therefore, we present only the result
for the ground state with $J^P=1/2^+$
$$
M_{\Xi_{bc}^{\prime}} = 6.85\; {\rm GeV,}\;\;\;
M_{\Xi_{bc}} = 6.82\; {\rm GeV,}
$$
whereas for the vector diquark we have assumed that the spin-dependent
splitting due to the interaction with the light quark is determined by the
standard contact coupling of magnetic moments for the point-like systems. The
picture for the baryon levels with no account for the spin-dependent
perturbations suppressed by the heavy quark masses is shown in \cite{PM1}.

\subsubsection{The doubly heavy baryons with the strangeness $\Omega_{QQ'}$}
In the leading approximation, we suppose that the wave functions and the
excitation energies of strange quark in the field of doubly heavy diquark
repeat the characteristics for the analogous baryons containing the ordinary
quarks $u,\; d$. Therefore, the level system of baryons $\Omega_{QQ'}$
reproduces that of $\Xi_{QQ'}$ up to an additive shift of the masses by the
value of current mass of strange quark, $m_s \approx M(D_s)-M(D) \approx
M(B_s)-M(B) \approx 0.1$ GeV.

Further, we suppose that the spin-spin splitting of $2P1s$ and $3D1s$ levels of
$\Omega_{QQ'}$ is 20-30\% less than  in $\Xi_{QQ'}$ (the factor of
$m_{u,d}/m_s$). As for the $1S2p$-level, the procedure described above can be
applied. So, for $\Omega_{bb}$, the matrix of mixing for the states with the
different values of total momentum $J_l$ practically can be assigned to be
diagonal. This fact means that the following term of perturbation is dominant:
$$
\frac{1}{4}\left( \frac{2\bf L\cdot S_l}{2m_l^2}\right)
\left( -\frac{dV(r)}{rdr}+ \frac{8}{3}\alpha_s\frac{1}{r^3}\right).
$$
Therefore, we can think that the splitting of $1S2p$ is determined by the
factor of $m_{u,d}^2/m_s^2$ with respect to the splitting of corresponding
$\Xi_{bb}$, i.e. it is 40\% less than in $\Xi_{bb}$. Hence, the splitting is
very small.

For the baryon $\Omega_{сс}$, the factor of $m_s/m_c$ is not small. Hence,
for $1S2p$, the mixing matrix is not diagonal, so that the arrangement of
$1S2p$ states of $\Omega_{сс}$ can be slightly different from that of
$\Xi_{сс}$.

The following peculiarity of $\Omega_{QQ'}$ is of great interest: the low-lying
$S$- and $P$-excitations of diquark are stable. Indeed, even after taking into
account the mixing of levels, a gluon emission makes a hadronization into the
$K$-meson (the transitions of $\Omega_{QQ'}\to \Xi_{QQ'}+K$), while a single
emission of $\pi$-meson is forbidden because of the conservation of iso-spin
and strangeness. The hadronic transitions with kaons are forbidden because of
insufficient splitting between the masses of $\Omega_{QQ'}$ and $\Xi_{QQ'}$.
The decays with the emission of pion pairs belonging to the iso-singlet state,
are suppressed by a small phase space or even forbidden. Thus, the radiative
electromagnetic transitions into the ground state are the dominant modes of
decays for the low-lying excitations of $\Omega_{QQ'}$.

\subsubsection{$\Omega_{bbc}$ baryons}
In the framework of quark-diquark picture, we can build the model for the
baryons containing three heavy quarks, $bbc$. However, as we estimate, the size
of diquark turns out to be comparable with the average distance to the charmed
quark. So, the model assumption on the compact heavy diquark cannot be quite
accurate for the calculations of mass levels in this case. The spin-dependent
forces are negligibly small inside the diquark, as we have already pointed out
above. The spin-spin splitting of vector diquark interacting with the charmed
quark, is given by
$
\Delta (1s) =  33\; {\rm MeV,}\;\;
\Delta (2s) =  18\; {\rm MeV.}\;\;
$
For $1S2p$, the level shifts are small. So, for the state $J^P=1/2^-$ we have
to add the correction of $- 33$ MeV. For the  $3D1s$-state the splitting is
determined by the spin-spin interaction. The characteristics of excitations for
the charmed quark in the model with the potential by Buchm\"uller and Tye have
been presented above. Finally, we obtain the picture of $\Omega_{bbc}$ levels
presented in Table \ref{bbct}.

\begin{table}[th]
\caption{The mass spectrum of $\Omega_{bbc}^0$ baryons.}
\label{bbct}
\begin{center}
\begin{tabular}{|p{40mm}|c||p{40mm}|c|}
\hline
$(n_d L_d n_l L_l)$, $J^{P}$ & mass, {\rm GeV}
 & $(n_d L_d n_l L_l)$, $J^{P}$ & mass, {\rm GeV}  \\
\hline
(1S 1s)$1/2^{+}$ & 11.12 & (3D 1s)$3/2^{\prime +}$ & 11.52 \\
\hline
(1S 1s)$3/2^{+}$ & 11.18  & (3D 1s)$5/2^{\prime +}$ & 11.54 \\
\hline
(2P 1s)$1/2^{-}$ & 11.33  & (1S 2p)$1/2^{-}$ & 11.55 \\
\hline
(2P 1s)$3/2^{-}$ & 11.39  & (3D 1s)$7/2^{+}$   & 11.56 \\
\hline
(2S 1s)$1/2^{+}$ & 11.40 & (1S 2p)$3/2^{\prime -}$ & 11.58 \\
\hline
(3D 1s)$5/2^{+}$ & 11.42  & (1S 2p)$3/2^{-}$ & 11.58 \\
\hline
(3D 1s)$3/2^{+}$ & 11.44 & (1S 2p)$1/2^{\prime -}$ & 11.59 \\
\hline
(3D 1s)$1/2^{+}$ & 11.46 & (1S 2p)$5/2^{-}$  & 11.59\\
\hline
(2S 1s)$3/2^{+}$ & 11.46  & (3P 1s)$3/2^{-}$   & 11.59 \\
\hline
(3P 1s)$1/2^{-}$ & 11.52 & (3S 1s)$1/2^{+}$   & 11.62 \\
\hline
\end{tabular}
\end{center}
\end{table}

Further, the excitations of ground $\Omega_{bbc}^0$ state can strongly mix with
large amplitudes because of small splittings between the levels, but they have
small shifts of masses. This effect takes place for $3P1s$ -- $1S2p$ with $J^P=
1/2^-,\; 3/2^-$, and for $2S1s$ -- $3D1s$ with $J^P= 1/2^+,\; 3/2^+$. We
suppose the prediction to be quite reliable for the states of $1S1s$ with
$J^P=1/2^+,\; 3/2^+$, $1S2p$ with $J^P=5/2^-$ and $3D1s$ with $J^P=5/2^+,\;
7/2^+$.  For these excitations, we might definitely predict the widths of their
radiative electromagnetic transitions into the ground state in the framework of
multipole expansion in QCD. The widths for the transitions will be essentially
determined by the amplitudes of admixtures, which have a strong model
dependence. Therefore, the experimental study of electromagnetic transitions in
the family of $\Omega_{bbc}^0$ baryons could provide a significant information
on the mechanism of mixing between the different levels in the baryonic
systems. The electromagnetic transitions combined with the emission of pion
pairs, if not forbidden by the phase space, saturate the total widths of
excited $\Omega_{bbc}^0$ levels. The characteristic value of total width is
about $\Gamma \sim 10 - 100$ keV, in the order of magnitude.

Thus, the system of $\Omega_{bbc}^0$ can be characterized by a large number
of narrow quasi-stable states.

\subsection{Discussion}

In this paper we have calculated the spectroscopic characteristics of baryons
containing two heavy quarks, in the model with the quark-diquark factorization
of wave functions. We have explored the nonrelativistic model of constituent
quarks with the potential by Buchm\"uller and Tye. The region of applicability
of such the approximations has been pointed out.

We have taken into account the spin-dependent relativistic corrections to the
potential in the subsystems of diquark and light quark-diquark. Below the
threshold of decay into the heavy baryon and heavy meson, we have found the
system of excited bound states, which are quasi-stable under the hadronic
transitions into the ground state. We have considered the physical reasons for
the quasi-stability taking place for the baryons with two identical quarks. In
accordance with the Pauli principle, the operators responsible for the
hadronic decays and the mixing between the levels, are suppressed by the
inverse heavy quark mass and the small size of diquark. This suppression is
caused by the necessity of instantaneous change in both the spin and the
orbital momentum of compact diquark. In the baryonic systems with two heavy
quarks and the strange quark, the quasi-stability of diquark excitations is
provided by the absense of transitions with the emission of both a single kaon
and a single pion. These transitions are forbidden because of small splitting
between the levels and the conservation of iso-spin and strangeness.

The characteristics of wave functions can be used in calculations of
cross sections for the doubly heavy baryons in the framework of quark-diquark
approximation.

The quark-diquark factorization in calculating the masses of ground states for
the baryon systems with two heavy quarks was also considered in ref.
\cite{faust}, where the quasi-potential approach \cite{quasipot} was explored.
There is a numerical difference in the choice of heavy quark masses, that leads
to that in \cite{faust} the mass of doubly charmed diquark, say, about 100 MeV
greater that the mass used in the above calculations. This difference
determines the discrepacy of estimates for the masses of ground states
presented in this paper and in \cite{faust}. We believe that this deviation
between the quark masses is caused by the use of Cornell potential with the
constant value of effective coulomb exchange coupling in contrast to the above
consideration with the running coupling constant, that cancels the uncertainty
in the arbitrary additive shift of energy. Furthermore, in the potential
approach the masses of heavy quarks depend on the mentioned additive shift,
which adjusted in the phenomemological models by comparing, say, the leptonic
constants of heavy quarkonium calculated in the model with the values known
from experiments. In the QCD motivated potential such the ambiguity of
potential because of the additive shift is absent, so that the estimates of
heavy quark masses have less uncertainities. Let us stress that in the Cornell
model the leptonic constants were calculated by taking into account the
one-loop corrections caused by the hard gluons. This correction is quite
essential, in part, for the charmed quarks. The two-loop corrections are also
important for the consideration of leptonic constants in the potential approach
\cite{3loop}. Moreover, in \cite{faust} the constituent mass of light quark is
posed with no correlation with the normalization of potential, while we put the
constituent mass to be a part of nonperturbative energy in the potential. This
can lead to an additional deviation between the estimates of baryon masses
about 50 MeV. Taking into account the above notes on the systematic
differences, we can claim that the estimates of ground states masses for the
baryons with two heavy quarks in \cite{faust} agree with the values obtained in
the presented approach (see Table \ref{compar}).

\begin{table}[th]
\caption{The masses of ground states $M$ (in GeV) for the baryons with two
heavy quarks calculated in various approaches (* denotes the results of authors
in this review). The accuracy of predictions under the variation of model
parameters is about 30-50 MeV. The systematic unceratinties are discussed in
the text.}
\label{compar}
\begin{center}
\begin{tabular}{|c|c|c|c|c|c|c|c|}    \hline
baryon & * & \cite{guo} & \cite{faust} & \cite{Ron} & \cite{Korner} &
\cite{itoh} & \cite{kaur}\\
\hline
$\Xi_{cc}$ & 3.48 & 3.74 & 3.66 & 3.66 & 3.61 & 3.65 & 3.71
\\  \hline
$\Xi_{cc}^\ast$& 3.61 & 3.86 & 3.81 &3.74 & 3.68 & 3.73 & 3.79
\\  \hline
$\Omega_{cc}$ & 3.59 & 3.76 & 3.76 & 3.74 & 3.71 & 3.75 & 3.89
\\   \hline
$\Omega_{cc}^\ast$ & 3.69 & 3.90 & 3.89 & 3.82 & 3.76 & 3.83 & 3.91
\\   \hline
$\Xi_{bb}$ & 10.09 & 10.30 & 10.23 & 10.34 & - & - & 10.43
\\  \hline
$\Xi_{bb}^\ast$& 10.13 & 10.34 & 10.28 & 10.37 & - & - & 10.48
\\  \hline
$\Omega_{bb}$  & 10.18 & 10.34 & 10.32 & 10.37 & - & - & 10.59 
\\   \hline
$\Omega_{bb}^\ast$& 10.20  & 10.38 & 10.36 & 10.40 & -  & - & 10.62
\\   \hline
$\Xi_{cb}$     & 6.82 & 7.01 & 6.95 & 7.04 & - & - & 7.08 
\\  \hline
$\Xi'_{cb}$    & 6.85 & 7.07 & 7.00 & 6.99 & - & - & 7.10
\\  \hline
$\Xi_{cb}^\ast$& 6.90 & 7.10 & 7.02 & 7.06 & - & - & 7.13 
\\  \hline
$\Omega_{cb}$  & 6.91 & 7.05 & 7.05 & 7.09 & - & - & 7.23
\\  \hline
$\Omega'_{cb}$ & 6.93 & 7.11 & 7.09 & 7.06  & - & - & 7.24 
\\  \hline
$\Omega_{cb}^\ast$& 6.99 & 7.13 & 7.11 & 7.12 & -  & - & 7.27
\\  \hline
\end{tabular}
\end{center}
\end{table}

In ref. \cite{guo}, following \cite{faust} in the framework of quasi-potential
approach, the analysis of spin-dependent relativistic corrections was performed
so that the overestimated, to our opinion, value of heavy diquark from
\cite{faust} was used. Unfortunately, there is an evident mistake in the
description of calculations in \cite{guo}, because both the parameter giving
the relative contribution of scalar and vector part in the potential and the
anomalous chromo-magnetic moment of heavy quark are denoted by the same symbol,
that leads to numerical errors, since in \cite{faust} it was shown that these
quantities have different values. This mistake enlarges the unceratinty about
100 MeV into th estimates of \cite{guo}, so that we can consider that the
results of \cite{guo} do not contradict with the presented description (see
Table \ref{compar}).

The estimates based on the hypothesis of pair interactions were presented in
ref. \cite{Ron}, so that in the light of discussion given in the beggining of
this chapter the difference about 200-300 MeV, that follows from values in
Table \ref{compar}, is not amazing. This deviation is, in general, related with
the different character of interquark forces in the doubly heavy baryon, though
the uncertainty in the heavy quark masses is also important.

In ref. \cite{Korner} simple speculations based on the HQET with the heavy
diquark were explored, so that the estimates depend on the supposed mass of
diquark composed of two heavy quarks. In this way, if we neglected the binding
energy in the diquark, that is evidently related with the choice of heavy quark
masses, then we got the estimates of ground state masses shown in Table
\ref{compar}.

Finally, in \cite{kaur} the analysis given in \cite{khanna} was modified on the
basis of interpolation formulae for the mass of ground state with account for
the dependence of spin forces on both the wave functions and the effective
coupling constant, which were changed under the quark contents of hadrons. In
this way, the parameter of energy shift enters the fitting function, so that
this parameter essentially changes under the transition from the description of
mesons to baryons: $\delta_M\approx 80$ MeV $\longrightarrow$ $\delta_B\approx
210$ MeV. This shift of energy provides a good agreement of fitting with the
mass values for the mesons and baryons observed experimentally. However, if we
suggest that the doubly heavy baryon is similar with the meson containing the
local heavy source in the picture of strong interactions, then we should use
the energy shift prescribed to the heavy mesons but the heavy baryons, wherein
the presense of system with two light quarks leads to the essential difference
in the calculation of bound state masses, hence, to the energy shift different
from the mesonic one. Such the substitution of parameters would lead to a more
good agreement between the results of \cite{kaur} (see Table \ref{compar}) and
the values obtained in this review.

Summarizing, we can claim that, first of all, in the framework of potential
approach in the calculations of masses for the doubly heavy baryons the
dominant uncertainty is caused by the choice of heavy quark masses, so that due
to the adjustment on the systems with heavy quarks, the analysis presented in
the QCD-motivated model of potential with the running coupling constant at
short distances and the linear nonperturbative term confining quarks at large
distances, gives the most reliable predictions.

A new field of interest for the investigations is radiative, electromagnetic or
hadronic, transitions between the quasi-stable states in families of baryons
with two heavy quarks. A first step in the study of this problem was recently
done in \cite{guo2}, wherein some preliminary results were obtained on the
electromagnetic transitions between the levels of $\Xi_{bc}$.

\newpage
\section{Sum rules of nonrelativistic QCD:\\ two-point correlators}

In the framework of potential models in Chapter 1 we have described the
families of baryons with two heavy quarks, which form a set of narrow excited
states in addition to the ground states, so that the mass spectrum is very
similar with the system of levels in the heavy quarkonium. In the method of QCD
sum rules  \cite{SVZ} for the two-point correlators of baryonic currents, the
masses and coupling constants for the baryons with two heavy quarks were
calculated in \cite{QCDsr}. However, the analysis in \cite{QCDsr} has some
disadvantages connected to an unstable divergency of sum rules in the range of
parameters determining the baryonic currents. This leads to quite large
uncertainties in calculations.

In this chapter we investigate the sum rules of NRQCD for the two-point
correlators of currents corresponding to the baryons with two heavy quarks. The
main physical argument of such the consideration is the nonrelativistic motion
of heavy quarks inside the diquark of small size, that interacts with the light
quark. This fact leads to some definite expressions for the structure of
baryonic currents written down in terms of nonrelativistic heavy quark fields.
In the leading order of inverse heavy quark mass and relative velocity of heavy
quarks inside the diquark, we have to take into account the hard gluon
corrections in order to derive the relations between the correlators of
nonrelativistic quarks in NRQCD and the correlators in full QCD. The
corresponding anomalous dimensions for the baryonic currents were calculated up
to two loops in ref. \cite{anom}. The structure of currents in NRQCD
corresponds to some choice of parameters in expressions of full QCD. These
values of parameters are posed in the range of significant uncertainty observed
in the analysis performed in \cite{QCDsr}. We find a simple physical reason for
failure of stability in this case: the behaviour of quantities versus the
scheme parameters of sum rules (the Borel variable or the number of spectral
density moment) is determined by forming the doubly heavy diquark inside the
baryon and, hence, the difference between the masses of baryon and diquark.
This difference of masses determines the basic characteristics of correlators
if we do not take into account the corrections connected to the nonperturbative
interaction of doubly heavy diquark with the light quark in the baryon. In the
NRQCD sum rules, the introduction of such the interaction is related with the
nonperturbative condensates caused by operators of higher dimensions. We show
that the better stability and lower uncertainty of sum rules can be achieved by
taking into account  the product of quark and gluon condensates in addition to
quark, gluon and mixed condensates. Moreover, we accurately introduce the
coulomb $\alpha_s/v$-corrections inside the heavy diquark. These corrections
enforce the relative contribution of perturbative part of sum rules in
comparison with the contribution of condensates into the correlators under
consideration.

Further, we comparatively analize the sum rules for the doubly heavy baryons
with the both strange and light massless quarks.

In Section 2.1 we define the currents and represent the spectral densities in
the NRQCD sum rules for various operators included into the consideration.
Section 2.2 is devoted to numerical estimates. We calculate the masses of
ground states, which are in a good agreement with the values obtained in
potential models. Finally, we briefly summarize the results.

\newpage
\subsection{Sum rules for doubly heavy baryons}

\subsubsection{Baryonic currents}

The currents for the baryons with two heavy quarks $\Xi_{cc}^{\diamond}$,
$\Xi_{bb}^{\diamond}$ and $\Xi^{\prime \diamond}_{bc}$, where $\diamond$
denotes the electric charge of baryon depending on the flavor of light quark,
corespond to the quantum numbers of spin and parity $j^P_d=1^+$ and $j^P_d=0^+$
for the systems of heavy diquark with the symmetric and anti-symmteric
structure of flavor matrix, correspondingly (if the identical heavy
quarks form the diquark then the scalar ground state $j^P_d=0^+$ is forbidden).
Adding the light quark to the system of heavy quarks gives $j^P=\frac{1}{2}^+$
for the $\Xi^{\prime \diamond}_{bc}$ baryons and the couple of degenerate
states $j^P=\frac{1}{2}^+$ and $j^P=\frac{3}{2}^+$ for the baryons
$\Xi_{cc}^{\diamond}$, $\Xi_{bc}^{\diamond}$, $\Xi_{bb}^{\diamond}$ and
$\Xi_{cc}^{*\diamond}$, $\Xi_{bc}^{*\diamond}$, $\Xi_{bb}^{*\diamond}$.
Generally, the structure of baryonic currents with two heavy quarks is written
down in the form
\begin{equation}
J = [Q^{iT}C\Gamma\tau Q^j]\Gamma^{'}q^k\varepsilon_{ijk}.
\end{equation}
Here $T$ denotes the transposition, $C$ is the charge conjugation matrix with
the properties $C\gamma_{\mu}^TC^{-1} = -\gamma_{\mu}$ and $C\gamma_5^TC^{-1} =
\gamma_5$, $i,j,k$ are color indices and $\tau$ is a matrix in the flavor
space. The effective static field of the heavy quark is denoted by $Q$. To the
leading order over both the relative velocity of heavy quarks and their inverse
masses, this field contains the ``large'' component only in the hadron rest
frame.

Here, unlike the case of baryons with a single heavy quark \cite{yakov}, there
is the only independent current component $J$ for each of the ground state
baryon currents. They equal
\begin{eqnarray}
J_{\Xi^{\prime \diamond}_{QQ^{\prime}}} &=& [Q^{iT}C\tau\gamma_5
Q^{j\prime}]q^k\varepsilon_{ijk},\nonumber\\
J_{\Xi_{QQ}^{\diamond}} &=& [Q^{iT}C\tau\boldsymbol{\gamma}^m
Q^j]\cdot\boldsymbol{\gamma}_m\gamma_5
q^k\varepsilon_{ijk},
\label{def}\\
J_{\Xi_{QQ}^{*\diamond}}^n &=& [Q^{iT}C\tau\boldsymbol{\gamma}^n
Q^j]q^k\varepsilon_{ijk}+\frac{1}{3}\boldsymbol{\gamma}^n
[Q^{iT}C\boldsymbol{\gamma}^m
Q^j]\cdot\boldsymbol{\gamma}_m q^k\varepsilon_{ijk},\nonumber
\end{eqnarray}
where $J_{\Xi_{QQ}^{*\diamond}}^n$ satisfies the spin-3/2 condition
$\boldsymbol{\gamma}_n J_{\Xi_{QQ}^{*\diamond}}^n = 0$. The flavor matrix
$\tau$ is anti-symmetric for $\Xi^{\prime \diamond}_{bc}$ and symmetric for
$\Xi_{QQ}^{\diamond}$ and $\Xi_{QQ}^{*\diamond}$. The currents written down in
Eq. (\ref{def}) are taken in the rest frame of hadrons. The corresponding
expressions in a general frame moving with a velocity $v^{\mu}$ can
be obtained by the substitution of $\boldsymbol{\gamma}^m\to
\gamma_{\perp}^{\mu}=\gamma^{\mu}-\slashchar{v} v^{\mu}$.

Similar expressions can be written down for the doubly heavy baryons with
the strange quark.

To compare with the full QCD analysis we represent the expression for the
$J_{\Xi^{\prime \diamond}_{bc}}$ current given in \cite{QCDsr}
\begin{equation}
J_{\Xi^{\prime \diamond}_{bc}} = \{r_1 [u^{iT}C\gamma_5 
c^{j}]b^k + r_2 [u^{iT}C c^{j}]\gamma_5 b^k + r_3 [u^{iT}C\gamma_5
\gamma_mu c^{j}]\gamma^\mu b^k\}\varepsilon_{ijk},
\label{choice}
\end{equation}
so that the NRQCD structure can be obtained by the choice of $r_1=r_2=1$ and
$r_3=0$ and the anti-symmetric permutation of $c$ and $b$ flavors. This
connection can be achieved by the nonrelativistic limit of full QCD spinors
of heavy quarks, so that in the leading order of $1/m_Q$-expansion the
``large'' components of spinors contribute only. Therefore, for the ground
states of douby heavy baryons containing the heavy quarks with the identical
flavors, the leading approximation of NRQCD leads to the only structure of
baryonic current expressed in terms of nonrelativistic spinors of heavy quark,
since the total spin of heavy diquark is fixed by $S=1$ because of the Pauli
principle. The corrections of the first $1/m_Q$-order can contribute with the
other Lorentz structures, of course. However, we deal with the leading
approximation of NRQCD in the present paper. For the ground states of doubly
heavy baryons containing the heavy quarks with the different flavors, the
$1/2$-spin state of baryon can contain the mixture of diquark states with $S=1$
and $S=0$, as it does in full QCD. We perform the separate consideration of
these two currents in NRQCD, and such the approach can generally be not optimal
in full QCD currents. Nevertheless, as we show, in the leading order of NRQCD
there are relations between the masses and coupling constants of baryons
because of the spin symmetry, so that the NRQCD does not distinguish these spin
states untill the spin dependent $1/m_Q$-corrections are taken into account. In
addition, as we have already mentioned, the analysis in the sum rules of full
QCD was done with a large uncertainty beacuse of difference in the evaluations
of coupling constants from two correlation functions \cite{QCDsr}, and the
authors noted that this uncertainty became large in a region of parameters
$r_{1,2}$ defined above, so that this region of bad accuracy is placed in
the vicinity of point gining the NRQCD choice in (\ref{choice}). To get
rid this disadvantage and to find its reason, we analyze the NRQCD sum rules in
details.

\subsubsection{Description of the method}

In this section we describe steps required for the evaluation of two-point
correlation functions in the NRQCD approximation and the connection to physical
characteristics of doubly heavy baryons. We start from the correlator of two
baryonic currents with the half spin
\begin{equation}
\Pi(w)=i \int d^4 x e^{i p x} \langle 0|T{J(x), \bar J(0)}|0
\rangle=\slashchar v F_1(w)+F_2(w), 
\label{2pcor}
\end{equation}
where $w$ is defined by $p^2=({\cal{M}}+w)^2$, while
${\cal{M}}=m_Q+m_{Q'}+m_s$, $m_{Q,Q'}$ are the heavy quark masses, and $m_s$ is
the strange quark mass. Sure, the correlators for the baryonic currents with
the light quarks instead of the strange one can be obtained, if we put $m_s\to
m_{u,d}\approx 0$ in expressions below. The appropriate definitions of scalar
formfactors for the 3/2-spin baryon are given by the following:
\begin{equation}
\Pi_{\mu\nu}(w) = i\int d^4x e^{ipx}\langle 0|T\{J_\mu (x),\bar
J_\nu (0)\}|0\rangle = -g_{\mu\nu}[\slashchar{v} \tilde F_1(w) +
\tilde F_2(w)]+\ldots,\label{2vcor}
\end{equation}
where we do not concern for distinct Lorentz structures. The scalar correlators
$F$ can be evaluated in a deep euclidean region by employing the Operator
Product Expansion (OPE) for the chronological product of baryonic currents in
the framework of NRQCD, for instance, in Eqs. (\ref{2pcor}), (\ref{2vcor}),
\begin{equation}
F_{1,2}(w) = \sum_d C^{(1,2)}_d(w)O_d,
\end{equation}
where $O_d$ denotes the local operator with a given dimension $d$:
$O_0 = \hat 1$, $O_3 = \langle\bar qq\rangle$, $O_4 =
\langle\frac{\alpha_s}{\pi}G^2\rangle$, $O_{d>4}=\ldots$,
and the functions $C_d(w)$ are the corresponding Wilson coefficients of OPE.

In this review we take into accouont the nonperturbative terms connected to the
quark and gluon condensates, their product and mixed condensates. For the
contribution of quark condensate operator we explore the following OPE for the
correlator of two quark fields:
\cite{Smilga}:
\begin{eqnarray}
\langle 0|T\{q_i^a(x)\bar q_j^b(0)|0\rangle &=& -\frac{1}{12}
\delta^{ab}\delta_{ij}\langle\bar q  q \rangle\cdot [1 + \frac{m_0^2x^2}{16}
+ \frac{\pi^2x^4}{288}\langle\frac{\alpha_s}{\pi}G^2\rangle + ... ],
\label{qq}
\end{eqnarray}
where the value of mixed condensate is parametrized by introducing the
variable $m_0^2$, which is numerically determined as $m_0^2 \approx 0.8$
GeV$^2$. Taking into account the nonzero mass of strange quark in the framework
of OPE we get the following expression for the quark condensate up to terms of
fourth order in $x$ \cite{KK}:
\begin{eqnarray}
 \langle 0|T{s_i^a(x)\bar s_j^b(0)}|0
 \rangle = &-&\frac{1}{12}\delta^{ab}\delta_{ij}\langle \bar s s\rangle\cdot
 \left[1+\frac{x^2(m_0^2-2
 m_s^2)}{16}+\frac{x^4(\pi^2\langle
 \frac{a_s}{\pi}G^2\rangle-\frac{3}{2}m_s^2(m_0^2-m_s^2))}{288}\right]\nonumber
 \\
 &+& i m_s \delta^{ab}  x_{\mu}\,\gamma^{\mu}_{ij}\cdot\langle \bar s
 s\rangle\cdot\left[\frac{1}{48}+\frac{x^2}{24^2}\left(\frac{3
 m_0^2}{4}-m_s^2\right)\right] = \label{Exp} \\
 &&
 -\delta^{ab}\langle \bar s s \rangle \cdot 
 ({\cal P}_0 \delta_{ij}+{\cal P}_1 x_{\alpha}
 \gamma^{\alpha}_{ij}+{\cal P}_2 \delta_{ij}  x^2+{\cal P}_3
x_{\alpha}\gamma^{\alpha}_{ij} x^2+ {\cal P}_4
\delta_{ij} x^4). \nonumber
\end{eqnarray}
Note that at $m_s\neq0$ the expansion of quark condensate (\ref{Exp}) gives
contributions in both correlators in contrast with the sum rules for
$\Xi_{QQ^\prime}$ \cite{KO}, where putting $m_s = 0$ and neglecting the higher
condensates, the authors found the factorization of diquark correlator in $F_2$
and full baryonic correlator in $F_1$. This factorization led to the systematic
instability for the estimates in the sum rules.

We write down the Wilson coefficients in front of unity and quark-gluon
operators by making use of the dispersion relation over $w$,
\begin{equation}
C_d(w)=\frac{1}{\pi}\int_0^{\infty}\frac{\rho_d(\omega)
d\omega}{\omega-w},
\end{equation}
where $\rho$ denotes the imaginary part of corresponding Wilson coefficient in
the physical region of NRQCD. Thus, the calculation of Wilson coefficients for
the operators under consideration is reduced to the problem on the derivation
of corresponding spectral densities.

To relate the NRQCD correlators to the real hadrons, we use the dispersion
representation for the two-point function with the physical spectral density
given by the appropriate resonance and continuum part. The coupling constants
of baryons are defined by the following expressions:
\begin{eqnarray}
\langle 0| J(x)|{\Xi(\Omega)_{QQ}^{\diamond}}(p)\rangle & = & i 
Z_{\Xi(\Omega)_{QQ}^{\diamond}} u(v,M_{\Xi(\Omega)}) e^{ip x}, \nonumber
\\
\langle 0| J^m(x)|{\Xi(\Omega)_{QQ}^{*\diamond}}(p,\lambda)\rangle & = &
i Z_{\Xi(\Omega)_{QQ}^{*\diamond}} u^m(v,M_{\Xi(\Omega)}) e^{ip x}, \nonumber 
\end{eqnarray}
where the spinor field with the four-velocity $v$ and mass $M$ satisfies the
equation
$$
\slashchar{v} u(v,M) = u(v,M),
$$
and 
$u^m(v,M)$ denotes the transversal spinor, so that
$(\gamma^m-v^m\slashchar{v}) u^m(v,M) =0$.

We suppose that the continuum densities starting from the threshold
$\omega_{cont}$, is modelled by the NRQCD expressions. Then, in the sum rules
equalizing the correlators in NRQCD and those of given by the physical states,
we assume the model of continuum given by the calculated perturbative term.
This model cannot be exact because of binding effects as well as the truncation
of perturbative expansion in the given order of $\alpha_s$. Therefore, the
integration above $\omega_{cont}$ cannot be strictly cancelled, and the model
introduces the implicit dependence of masses and couplings on the choice of
value $\omega_{cont}$. This dependece causes an uncertainty, which is not
essential in comparison with uncertainties following from another methodics and
the variation of quark masses.

Then we use the nonrelativistic expressions for the physical spectral
functions 
\begin{equation}
\rho^{phys}_{1,2}(\omega)=\frac{M}{2{\cal{M}}} |Z|^2
\delta(\bar \Lambda-\omega),
\end{equation}
where we have performed the substitution $\delta(p^2-M^2)\to\frac{1}{2
{\cal{M}}} \delta(\bar \Lambda-w)$, here $\bar \Lambda$ is the binding energy
of baryon and $M={\cal{M}}+\bar \Lambda$. The nonrelativistic dispersion
relation for the hadronic part of sum rules has the form 
\begin{equation}
\int \frac{\rho^{phys}_{1,2} d\omega}{\omega-w}=\frac{1}{2{\cal{M}}}
\frac{|Z|^2
}{\bar \Lambda-w}.
\end{equation}
Further, we write down the correlators in the deep underthreshold point
of $w=-{\cal{M}}+t$ at $t\to0$, which corresponds to the limit of $p^2\to 0$.
The approximation of hadronic part by the only bound state leads to the
expression, which can be expanded in series of $t$. Thus, the sum rules result
in the equality of coefficients at the same powers of $t$
\begin{equation}
\frac{1}{\pi}\int_0^{\omega_{cont}} \frac{\rho_{1,2}\,
d\omega}{(\omega+{\cal{M}})^n}=\frac{M }{2{\cal{M}}}\frac{|Z|^2}{M^n},
\end{equation}
where $\rho_j$ contains the contributions given by various operators in OPE for
the corresponding scalar functions $F_j$. Introducing the following notation
for the $n$-th moment of two-point correlation function:
\begin{equation}
{\cal M}_n	=\frac{1}{\pi}\int_0^{\omega_{cont}}\frac{\rho (\omega
)\,d\omega}
{(\omega +{\cal{M}})^{n+1}},
\end{equation}
for the baryon mass $M_{\Xi}$ we have the following estimate:
\begin{equation}
M[n] = \frac{{\cal M}_n}{{\cal M}_{n + 1}},
\end{equation}
and the coupling constants are determined by the expression
\begin{equation}
|Z[n]|^2 =\frac{{2 \cal{M}}}{ M} {\cal M}_n M^{n + 1},
\end{equation}
where we see the dependence of sum rule results on the scheme parameter.
Therefore, we have to find the region of parameters, wherein, first, the
results are stable with respect to variation of $n$, and, second, the both
correlation functions $F_1$ and $F_2$ repeats the same values of physical
quantities: the masses and coupling constants. The problem of analysis given in
the full QCD was the existance of significant difference between the masses and
coupling constants calculated under different $F$.

\subsubsection{Calculating the spectral densities}
\noindent
In this subsection we present analytical expression for the perturbative
spectral functions in the NRQCD approximation. The evaluation of spectral
densities involves the standard use of Cutkosky rules \cite{Cutk} with
some modifications motivated by NRQCD. We use the rule under which the jump of
two-point function under study is calculated by the following substitutions for
the propogators of heavy and light quarks, respectively:
\begin{eqnarray}
 {\rm heavy\; quark:}\;\;&& \frac{1}{p_0-(m+\frac{{\bf p}^2}{2m})}\to 2\pi
 i\cdot\delta (p_0-(m+\frac{{\bf p}^2}{2m})), \nonumber\\
 {\rm light\; quark:}\;\;&& \frac{1}{p^2-m^2} \to 2\pi i\cdot\delta
 (p^2-m^2).\nonumber
\end{eqnarray}
We derive the spin symmetry relations for all the  spectral densities due to
the fact that in the leading order of the heavy quark effective theory the
spins of heavy quarks are decoupled, so 
\begin{equation}
\rho_{1,\Omega(\Xi)_{QQ}^{\diamond}}=3 \rho_{1,\Omega(\Xi)_{QQ'}^{'\diamond}}=3
\rho_{1,\Omega(\Xi)_{QQ}^{*\diamond}},
\end{equation}
\begin{equation}
\rho_{2,\Omega(\Xi)_{QQ}^{\diamond}}=3 \rho_{2,\Omega(\Xi)_{QQ'}^{'\diamond}}=3
\rho_{2,\Omega(\Xi)_{QQ}^{*\diamond}},
\end{equation}
and we have the following relation for the baryon couplings in NRQCD:
\begin{equation}
|Z_{\Omega(\Xi)}|^2=3|Z_{\Omega'(\Xi)}|^2=3|Z_{\Omega(\Xi)^*}|^2.
\end{equation}

For the perturbative spectral densities $\rho_{1,H} (\omega)$ and $\rho_{2,H}
(\omega)$ in front of unit operator for $F_1$ and $F_2$, respectively, we
explore the smallness of strange quark current mass with respect to the masses
of heavy quarks and make the expansion in series of $m_s$. So, we get the
following expressions:
\begin{equation}
\rho_{1,\Omega^{'\diamond}_{QQ'}}(\omega)=\frac{\sqrt{2} (m_{QQ'}
\omega)^{3/2} }{15015 \pi^3
({\cal {M}}_{diq}+\omega)^3}(\eta_{1,0}(\omega)+m_s
\eta_{1,1}(\omega)+m_s^2 \eta_{1,2}(\omega)),
\label{ro1}
\end{equation}
so that $m_{QQ'}=m_Q m_{Q'}/(m_Q+m_{Q'})$ is the reduced diquark mass,,
${\cal{M}}_{diq}=m_Q+m_{Q'}$, and the coefficients of spectral densities $\eta$
are presented in Appendix I\footnote{In what follows the coefficients of
spectral densities, which are not given explicitly, are also given in Appendix
I.}. The first term of this expansion reproduces the result obtained in
\cite{KO} for the zero mass of light quark. For the strange baryons the
perturbative density $\rho_{2,\Omega^{'\diamond}_{QQ'}}$ is prportional to
$m_s$, and it is not equal to zero
\begin{equation}
\rho_{2,\Omega^{'\diamond}_{QQ'}}(\omega)=\frac{2 \sqrt{2} \omega (m_{QQ'}
\omega)^{3/2} m_s}{105 \pi^3
({\cal {M}}_{diq}+\omega)^2} (\eta_{2,0}+m_s \eta_{2,1}+m_s^2 \eta_{2,2}).
\label{ro2}
\end{equation}
In the leading order of perturbative NRQCD the correlators $F_2$ are equal to
zero for the massless light quark. This fact is caused by the absense of
interaction between the light quark and the heavy diquark in this order, and,
therefore, there is no massive term in this correlator.

The coulomb interaction inside the diquark can be taken into account by the
introduction of Sommerfeld factor $\bf C$ for the spectral density of diquark
before the integration over the invariant mass of diquark in order to get the
baryonic spectral densities, so that
\begin{equation}
\rho_{diquark}^{\bf{C}}=\rho_{diquark}^{bare} \cdot {\bf{C}}
\end{equation}
whereas
\begin{equation}
{\bf C} = \frac{2\pi\alpha_s}{3v_{QQ'}}\left [ 1- \exp\left
(-\frac{2\pi\alpha_s}{3v_{QQ'}} \right)\right ]^{-1},
\end{equation}
where we have taken into account the anti-triplet color structure of diquark,
and $v_{QQ'}$ denotes the relative velocity of heavy quarks inside the
diquark:
\begin{equation}
v_{QQ'} = \sqrt{1-\frac{4m_Q m_{Q'}}{Q^2-(m_Q-m_{Q'})^2}},
\end{equation}
where $Q^2$ is the square of heavy diquark four-momentum. In NRQCD we take the
limit of low velocities, so that denoting the diquark invariant mass squared by
$Q^2=({\cal{M}}_{diq}+\epsilon)^2$, we find
$$
{\bf C} = \frac{2\pi\alpha_s}{3v_{QQ'}},\;\;\;
v_{QQ'}^2 = \frac{\epsilon}{2m_{QQ'}},
$$
at $\epsilon\ll m_{QQ'}$. The modified spectral densities are equal to
\begin{equation}
\rho_{1}^{{\bf C}}(\omega)=\frac{m_{QQ'}^2 \alpha_s \omega (2
{\cal M}_{diq}+\omega)}{6
\pi^2 ({\cal M}_{diq}+\omega)^3}(\eta_{1,0}^{{\bf C}}+m_s\eta_{1,1}^{{\bf
C}}+m_s^2\eta_{1,2}^{{\bf C}}).
\label{ro1c}
\end{equation}
The leading term gives the result for the zero mass of light quark \cite{KO}.
For $\rho_{2,\Omega^{'\diamond}_{QQ'}}^{\bf {C}}$ we have
\begin{equation}
\rho_{2}^{\bf {C}}=\frac{m_s m_{QQ'}^2 (2 {\cal M}_{diq}+\omega)\omega \alpha_s
}{2 \pi({\cal M}_{diq}+\omega)^2}(\eta_{2,0}^{{\bf C}}+m_s \eta_{2,1}^{{\bf
C}}+m_s^2 \eta_{2,2}^{{\bf C}}).
\label{ro2c}
\end{equation}
The truncated expansion in the mass of light quark leads to a small deviation
about 0.5\% from the exact integral representation, so that this approximation
is quite justified with the initial three terms in the explicit analytic form.

Further, the spectral functions, connected to the condensates of light quarks 
and gluons, can be derived in the analogous way. For the moments of
coefficients in front of quark condensate terms we have the following 
expressions:
\begin{eqnarray}
{\cal M}_{\bar q q}^{(1)}(n)&=&-\frac{(n+1)!}{n!} {\cal{P}}_1 {\cal
M}^{diq}(n+1)+\frac{(n+3)!}{n!} {\cal{P}}_3 {\cal M}^{diq}(n+3)\nonumber\\
{\cal {\cal M}}_{\bar q q}^{(2)}(n)&=&{\cal{P}}_0 {\cal
M}^{diq}(n)-\frac{(n+2)!}{n!} {\cal{P}}_2 {\cal M}^{diq}(n+2)+\frac{(n+4)!}{n!}
{\cal{P}}_4 {\cal M}^{diq}(n+4),
\end{eqnarray}
where we have used the coefficients of expansion in $x$, ${\cal{P}}_i$, in Eq.
(\ref{Exp}), while the n-th moment of two-point correlation function for the
diquark is denoted by ${\cal M}^{diq}(n)$, and it is derived under the
integration of spectral density
\begin{equation}
\rho_{diq}=\frac{12 \sqrt 2 m_{QQ'}^{3/2} \sqrt\omega}{\pi}, 
\end{equation}
which should be multiplied by the Sommerfeld factor $\bf C$, where the variable
$\epsilon$ is substituted by $\omega$, since in this case there is no
integration over the invariant mass of diquark. The modified density is equal
to
\begin{equation}
\rho_{diq}^{{\bf{C}}}=\frac{48 \pi \alpha_s m_{QQ'}^2}{3},
\end{equation}
and it does not depend on $\omega$.

It is interesting to stress that in NRQCD the light quark condensate
contributes to the $F_2$ correlators, only. This fact has a simple physical
explanation: to the leading order the light quark operator can be factorized in
the expression for the correlator of baryonic currents. Indeed, we can write
down for the condensate contribution
$$
\langle 0|T\{J(x),\bar J(0)\}|0\rangle \Rightarrow \langle 0|T\{q_i^a(x)\bar
q_i^a(0)|0\rangle\cdot \frac{\hat 1}{12}\cdot \langle 0|T\{J^j_d(x),\bar
J^j_d(0)\}|0\rangle +\ldots,
$$
where $J^j_d(x)$ denotes the appropriate diquark current with the color index
$j$, as it is defined by the baryon structure in Eqs. (\ref{def}). So, we see
that the restriction by the first term independent of $x$ in the expansion for
the quark correlator in (\ref{qq}) results in the independent contribution of
diquark correlator to the baryonic one. Then, since the diquark correlator is
isolated in $F_2$ from the baryonic formfactor $F_1$, the NRQCD sum rules lead
to the evaluation of diquark masses and couplings from $F_2$, and estimation of
baryon masses and couplings from $F_1$. These masses and couplings are
different. The positive point is the possibility to calculate the binding
energy for the doubly heavy baryons $\bar \Lambda = M_\Xi - {\cal M}_{diq}$.
The disadvantage is the instability of NRQCD sum rules at this stage, since the
various formfactors or correlators lead to the different results. In  sum rules
of full QCD various choices of parameters in the definitions of baryonic
currents result in an admixture of pure diquark correlator in various
formfactors, so that the estimations acquire huge uncertainties. Say, the
characteristic ambiguity in the evaluation of baryon mass in full QCD is about
300 MeV, i.e. the value close to the expected estimate of $\bar \Lambda$. The
analysis in the framework of NRQCD makes this result to be not unexpectable.
Moreover, it is quite evident that the introduction of interactions between the
light quark and the doubly heavy diquark destroys the factorization of diquark
correlator. Indeed, we see that due to the higher terms in expansion
(\ref{qq}), the diquark factorization is explicitly broken, which has to result
in the convergency of estimates obtained from $F_1$ and $F_2$. Below we show
numerically that this fact is valid. Technically, we point out that the
contribution to the moments of spectral density, determined by the light quark
condensate including the mixed condensate and the product of quark and gluon
condensates, can be calculated after the exploration of (\ref{qq}), so that
\begin{equation}
{\cal M}_n^{q\bar q} = {\cal M}_n^{\langle\bar q q\rangle}
-\frac{(n+2)!}{n!}\frac{m_0^2}{16} {\cal M}_{n+2}^{\langle\bar q q\rangle}+
\frac{(n+4)!}{n!} \frac{\pi^2}{288} \langle\frac{\alpha_s}{\pi}G^2\rangle
{\cal M}_{n+4}^{\langle\bar q q\rangle}.
\end{equation}
For the corrections determined by the gluon condensate and written down for the
operator $O_4 = \langle\frac{\alpha_s}{\pi}G^2\rangle$, we have
\begin{equation}
\rho_{1}^{G^2}(\omega)=\frac{(m_Q^2+m_{Q'}^2+11 m_Q m_{Q'}) m_{QQ'}^{5/2}\sqrt
\omega}{21\cdot2^{10} \sqrt 2 \pi 
m_Q^2 m_{Q'}^2 ({\cal{M}}_{diq}+\omega)^2} (\eta^{G^2}_{1,0}+m_s
\eta^{G^2}_{1,1}+m_s^2 \eta^{G^2}_{1,2}).
\label{ro1g}
\end{equation}
For the nonzero mass of light quark the density $\rho_{2}^{G^2}(\omega)$ is
proportional to $m_s$ and equal to
\begin{equation}
\rho_{2}^{G^2}(\omega)=\frac{m_s (m_Q^2+m_{Q'}^2+11 m_Q m_{Q'})
m_{QQ'}^{5/2}\sqrt
\omega}{3\cdot2^{9} \sqrt 2 \pi 
m_Q^2
m_{Q'}^2 ({\cal{M}}_{diq}+\omega)} (\eta^{G^2}_{2,0}+m_s
\eta^{G^2}_{2,1}),
\end{equation}
so that
\begin{eqnarray}
\eta^{G^2}_{2,0} &=& -(9 {\cal{M}}_{diq}+\omega),\;\;\;\;
\eta^{G^2}_{2,1} = \frac{9 {\cal{M}}_{diq}+\omega}{{\cal{M}}_{diq}+\omega}.
\end{eqnarray}

For the product of condensates $\langle\bar qq\rangle\langle
\frac{\alpha_s}{\pi}G^2\rangle$, wherein the gluon fields are connected to the
heavy quarks in contrast to the light quark, we have computed the contribution
to the two-point correlation function itself. It has the following form:
\begin{equation}
F^{\bar q q G^2}_2(\omega)=-\frac{m_{QQ'}^{5/2}(m_Q^2+m_{Q'}^2+11 m_Q
m_{Q'})}{ 2^{9} \sqrt 2 \pi  m_{Q}
m_{Q'}(-\omega)^{5/2}},
\end{equation}
and $F^{\bar q q G^2}_1(\omega)=0$, so that we restrict the consideration by
the operators with the dimension not greater than 7, while the nonzero
contribution to $F_1$ appears in the fifth order of expansion (\ref{Exp}). The
obtained result is presented in the form, which allows the analytic
continuation over $\omega=-{\cal{M}}+w$.

Thus, we provide the NRQCD sum rules, where we take into account the
perturbative terms and the vacuum expectations of quark-gluon operators up to
the contributions by the light quark condensate, gluon condensate, their
product and the mixed condensate. Note, that the product of condensates is
essential for the doubly heavy baryons, and we present the full NRQCD
expression for this term, including the interaction of nonperturbative gluons
with both the light and heavy quarks. The correct introduction of coulomb-like
interactions is done for the perturbative spectral densities of heavy diquark,
which is important for the nonrelativistic heavy quarks. Finally, we find the
spin-symmetry relation for the baryon couplings in NRQCD
$$
|Z_\Xi|^2 = 3 |Z_{\Xi^\prime}|^2 = 3 |Z_{\Xi^*}|^2.
$$

\subsubsection{Anomalous dimensions for the baryonic currents}

To connect the NRQCD sum rules to the quantities in full QCD we have to take
into account the anomalous dimensions of effective baryonic currents with the
nonrelativistic quarks. They determine the factors, which have to multiply the
NRQCD correlators to obtain the values in full QCD. Indeed, to the leading
order of NRQCD we have the relation
$$
J^{QCD} = {\cal K}_J(\alpha_s,\mu_{\rm soft},\mu_{\rm hard}) \cdot J^{NRQCD},
$$
where the coefficient ${\cal K}_J(\alpha_s,\mu_{\rm soft},\mu_{\rm hard})$
depends on the normalization scale $\mu_{\rm soft}$ and obeys the matching
condition at the starting point of $\mu_{\rm hard}={\cal{M}}_{diq}$. The
anomalous dimensions of NRQCD currents are independent of the diquark spin
structure in the leading order. They are equal to \cite{anom}
\begin{eqnarray}
\gamma &=& \frac{d\ln C_J(\alpha_s,\mu)}{d\ln (\mu)} =
\sum_{m=1}^\infty\left(\frac{\alpha_s}{4\pi}\right)^m\gamma^{(m)},
	   \nonumber\\
\gamma^{(1)} &=& \Big(-2C_B(3a-3)+3C_F(a-2)\Big), \label{gamma}\\
\gamma^{(2)} &=& \frac{1}{6}(-48(-2+6\zeta (2))C_B^2+C_A((104-240\zeta
(2))C_B-101C_F) \nonumber \\
&& -64C_Bn_fT_F+C_F(-9C_F+52n_fT_F)),\nonumber
\end{eqnarray}
where $C_F=(N_c^2-1)/2N_c$, $C_A=N_c$, $C_B=(N_c+1)/2N_c$, and $T_F=1/2$ for
$N_c=3$, $n_f$ being the number of light quarks. In Eq.(\ref{gamma}) we give
the one-loop result with the arbitrary gauge parameter $a$, and the two-loop
anomalous dimension is represented in the Feynman gauge $a=1$. 
So, numerically at $n_f=3$ and $a=1$ we find
\begin{equation}
\gamma^{(1)} = -4,\;\;\;
\gamma^{(2)} \approx -188.24.
\end{equation}
In the leading logarithmic approximation and to the one-loop accuracy, the
coefficient ${\cal K}_J$ is given by the expression
\begin{equation}
{\cal K}_J(\alpha_s,\mu_{\rm soft},\mu_{\rm hard}) = \left(
\frac{\alpha_s(\mu_{\rm hard})}{\alpha_s(\mu_{\rm
soft})}\right)^{\frac{\gamma^{(1)}}{2
\beta_0}},
\end{equation}
where $\beta_0 = 11N_c/3 - 2 n_f/3 =9$. To evaluate the two-loop expression for
$C_J$ we have to know subleading corrections in the first $\alpha_s$ order in
addition to the anomalous dimensions. These corrections are not available yet,
so we restrict ourselves by the one-loop accuracy.

Further, we have to determine the normalization point for the NRQCD estimates
$\mu_{\rm soft}$. We put it to the average momentum transfer inside the doubly
heavy diquark, so that $\mu_{\rm soft}^2=2 m_{QQ'}T_{diq}$, where $T_{diq}$
denotes the kinetic energy in the system of two heavy quarks, which is
phenomenologically independent of the quark flavors and approximately equal to
0.2 GeV. Then, the coefficients ${\cal K}_J$ are equal to 
\begin{equation}
{\cal K}_{\Omega(\Xi)_{cc}} \approx 1.95,\;\;\;
{\cal K}_{\Omega(\Xi)_{bc}} \approx 1.52,\;\;\;
{\cal K}_{\Omega(\Xi)_{bb}} \approx 1.30,
\end{equation}
with the characteristic uncertainty about 10\% because of the variation of
initial and final points $\mu_{\rm hard,soft}$.

Finally, we emphasize that the values of ${\cal K}_J$ do not change the
estimates of baryon masses calculated in the sum rules of NRQCD. However, they
are essential in the evaluation of baryon couplings, which acquire these
multiplicative factors.

\subsection{Numerical estimates}

Evaluating the two-point sum rules, we explore the scheme of moments. We point
out the well-known fact that an essential part of uncertainties is caused by
the variation of heavy quark masses. In the analysis we chose the following
region of mass values:
\begin{equation}
m_b = 4.6-4.7\; {\rm GeV,}\;\;\;\; 
m_c = 1.35-1.40\; {\rm GeV,}
\end{equation}
which is ordinary used in the sum rule estimates for the heavy quarkonia. Next
critical point is the value of QCD coupling constant determining the
coulomb-like interactions inside the doubly heavy diquark. Indeed, it stands
linearly in front of the perturbative functions of diquark contributions. Thus,
the introduction of $\alpha_s/v$-corrections is essential for both the baryon
couplings and the relative contributions of perturbative terms and condensates
to the baryon masses. To decrease the uncertainty we impose the same approach
to the heavy quarkonia, where it is well justified, and then, we extract the
characteristic values for the heavy-heavy systems from the comparison of
calculations with the current data on the leptonic constants of heavy
quarkonia, which are known experimentally for $c\bar c$ and $b\bar b$ or
evaluated in various approaches for $\bar b c$. So, our calculations give the
following couplings of coulomb interactions
\begin{equation}
\alpha_s(b\bar b) = 0.37,\;\;\;
\alpha_s(c\bar b) = 0.45,\;\;\;
\alpha_s(c\bar c) = 0.60.\;\;\;
\end{equation}
Since the squared size of diquark is two times larger than that of heavy
quarkonium composed of the same heavy quarks (see the dependence of average
square of relative momentum on the kinetic energy of heavy quarks), the
effective coulumb constants have to be rescaled according to the equation
of evolution in QCD. Since we use the one-loop approximation, we explore the
evolution equation
$$
\alpha_s(QQ')=\frac{\alpha_s(Q\bar Q')}{1-\frac{9 }{4 \pi}\alpha_s(Q\bar Q')
\ln 2}.
$$
So,
\begin{equation}
\alpha_s(b b) = 0.45,\;\;\;
\alpha_s(b c) = 0.58,\;\;\;
\alpha_s(c c) = 0.85.\;\;\;
\end{equation}
As for the dependence of results on inputs for the quark masses, we have to
remark that so called pole masses are not well defined due to infrared
problems, usually mentioned as the renormalon ambiguity \cite{Benekerev}. Thus,
it is important to fix the definition of mass \cite{Ben,Hoang,Mel}. 

To the given order in $\alpha_s$ for the NRQCD sum rules, we use the leading
quark loop approximation with account for the coulomb exchange between the
heavy quarks. At this stage the heavy quark masses and coulomb coupling
constants are strictly fixed by the data on the charmonium and bottomonium
leptonic constants and masses as described by the QCD sum rules to the same
accuracy. The stability or convergency of sum rule method applied to these
heavy quarkonia\footnote{We have required that the ratio of initial moments for
the spectral densities calculated over the data and in QCD sum rules was
stable.} results in the following masses of quarks:
$$
m_c = 1.40\pm 0.03\; {\rm GeV,}\;\;\; m_b = 4.60\pm 0.02\; {\rm GeV,}
$$
which well agree with the values of heavy quark masses defined as free of
infrared contributions: the potential subtracted mass $m_b^{\rm PS} = 4.60\pm
0.11$ GeV and the kinetic mass $m_b^{\rm kin} = 4.56\pm 0.06$ GeV both obtained
in the QCD sum rules for the bottomonium within the two-loop accuracy
\cite{Ben,Mel}. The corresponding 1S-mass defined in \cite{Hoang} has a
slightly larger value. We think that in the leading order over $\alpha_s$ the
PS and kinetic masses above determine the threshold of quark contribution and
can be taken as the appropriately defined heavy quark masses in the
calculations of characteristics for the doubly heavy baryons. The mass values
are dependent of normalization point, which was chosen in the range of $1-2$
GeV. Nevertheless, we slightly enlarge the region of mass variation. 

The QCD sum rules for the bottomonium and charmonium fix the values of coulomb
couplings, too, since the momentum stability yields the heavy quark masses,
while the leptonic constants linearly determine the corresponding values of
$\alpha_s$ shown above (see Fig. \ref{coulfig}). Note that the dependence of
coulomb coupling constant on the quark contents of quarkonia well agree with
the renormalization group evolution with the change of size for the system
composed of two heavy quarks. The uncertainty of further estimates on the
supposed values of coulomb couplings is about 5\%.

\setlength{\unitlength}{1mm}
\begin{figure}[th]
\begin{center}
\begin{picture}(100, 80)
\put(0,0){\epsfxsize=10 cm \epsfbox{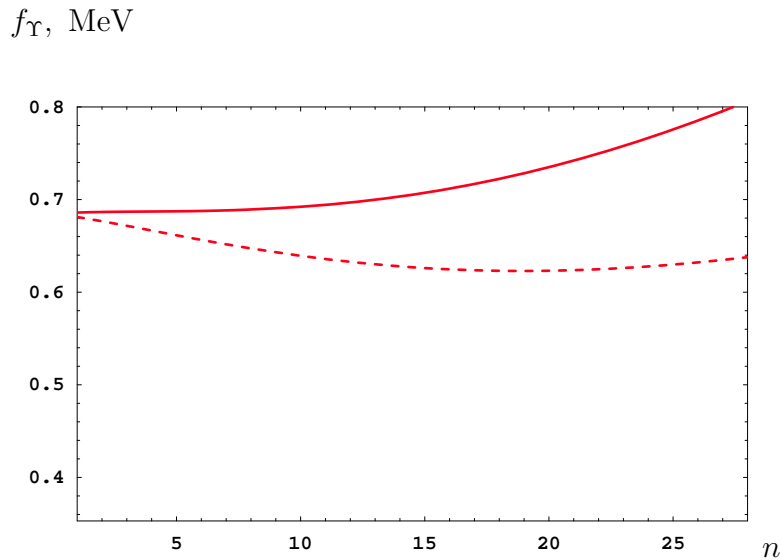}}
\put(0, 70){$f_{\Upsilon},~$MeV}
\put(100, 0){$n$}
\end{picture}
\end{center}
\caption{The leptonic constant of $\Upsilon$ in the two-point sum rules
formulated in the scheme of spectral density moments. The dashed line
represents the result at $m_b=4.63~$ GeV, the solid curve gives the estimate
at $m_b=4.59~$ GeV.}
\label{coulfig}
\end{figure}

As we have already mentioned, in the coupling constants of baryons the
uncertainty connected to the coefficients for the matching the NRQCD with full
QCD, is about 10\%.

The dependence of estimates on the value of thershold for the continuum
contribution is not so significant as on the quark masses. We fix the region of
$\omega_{cont}$ as
\begin{equation}
\omega_{cont} = 1.3-1.4\; {\rm GeV.}
\end{equation}
For the condensates of quarks and gluons the following regions are under
consideration:
\begin{equation}
\begin{array}{rcl}
\langle\bar q  q \rangle &=& -(250-270\;{\rm MeV})^3,\\
m_0^2 &=& 0.75- 0.85\;{\rm GeV}^2,\\
\langle\frac{\alpha_s}{\pi}G^2\rangle &=& (1.5-2)\cdot 10^{-2}\;{\rm GeV}^4.
\end{array}
\end{equation}
The main source of uncertainties in the ratios of the baryonic couplings is the
ratio of the condensates with the strange and light quarks. We use ${\langle
\bar s s \rangle}/{\langle \bar q q\rangle }=0.8\pm0.2$ that corresponds to the
variations of sum $(m_u+m_d)[1\, {\rm GeV}] = 12\div14~$ MeV \cite{Pivss}.

We suppose the strange quark mass equal to $m_s=150\pm 30$ MeV, that is the
wide-accepted estimate consistent with both the sum rules and the quark current
algebra, wherein the current mass of quark operates.

So, we have described the set of parameters entering the scheme of
calculations.

Fig. \ref{mbc} represents the calculated difference of masses
extracted from the $F_1$ and $F_2$ correlators\footnote{In these figures we
have fixed the value of gluon condensate $\langle\frac{\alpha_s}{\pi}G^2\rangle
= 1.7\cdot 10^{-2}\;{\rm GeV}^4$ and arranged $m_0^2$ in the above region to
reach zero differences between the masses, though the variation of parameters
leads to errors in the estimates quoted below.} for the baryon $\Xi_{bc}$ (we
have not shown the similar figures for the $\Xi_{cc}$ and $\Xi_{bb}$ baryons,
since they qualitatively and quantitatively repeat the picture clearly given by
Fig. \ref{mbc}). 

\setlength{\unitlength}{1mm}
\begin{figure}[th]
\begin{center}
\begin{picture}(100,80)
\put(0,0){\epsfxsize=10cm \epsfbox{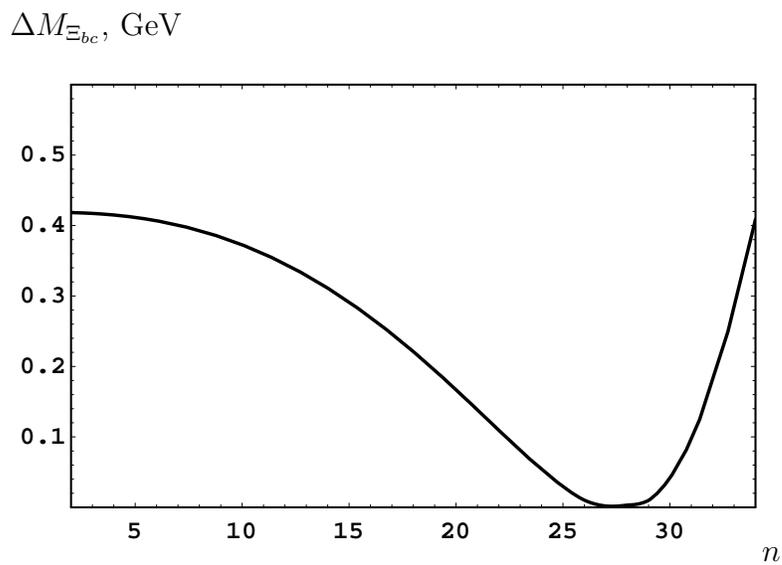}}
\put(100,0){$n$}
\put(0,70){$\Delta M_{\Xi_{bc}}$, GeV}
\end{picture}
\end{center}
\caption{The difference between the $\Xi_{bc}$-baryon masses calculated in the
NRQCD sum rules for the formfactors $F_1$ and $F_2$ in the scheme of moments
for the spectral densities.}
\label{mbc}
\end{figure}

We certainly see that at low numbers of moments for the spectral densities, the
baryon-diquark mass difference can be evaluated as
\begin{equation}
\bar \Lambda = 0.40\pm 0.03\; {\rm GeV,}
\end{equation}
which is quite a reasonable value, being in a good agreement with the estimates
in the heavy-light mesons. In the region of mass difference stability we can
fix the number of moment for the spectral density, say, $n=27\pm 1$ for
$\Xi_{bc}$, and calculate the corresponding masses of baryons, which are equal
to
\begin{equation}
M_{\Xi_{cc}} = 3.47\pm 0.05\; {\rm GeV},\;\;\;
M_{\Xi_{bc}} = 6.80\pm 0.05\; {\rm GeV},\;\;\;
M_{\Xi_{bb}} = 10.07\pm 0.09\; {\rm GeV},\;\;\;
 \label{met1}
\end{equation}
where we do not take into account the spin-dependent splitting caused by the
$\alpha_s$-corrections to the heavy-light interactions, which are not available
yet. The uncertainties in the mass values are basically given by the variation
of heavy quark masses. The convergency of NRQCD sum rules allows one to improve
the accuracy of estimates in comparison with the previous analysis in full QCD
\cite{QCDsr}. The obtained values are in agreement with the calculations in the
framework of nonrelativistic potential models (see Chapter 1).

\setlength{\unitlength}{1pt}
\begin{figure}[th]
\begin{center}
\begin{picture}(100,200)
\put(-100,0){\epsfxsize=10cm \epsfbox{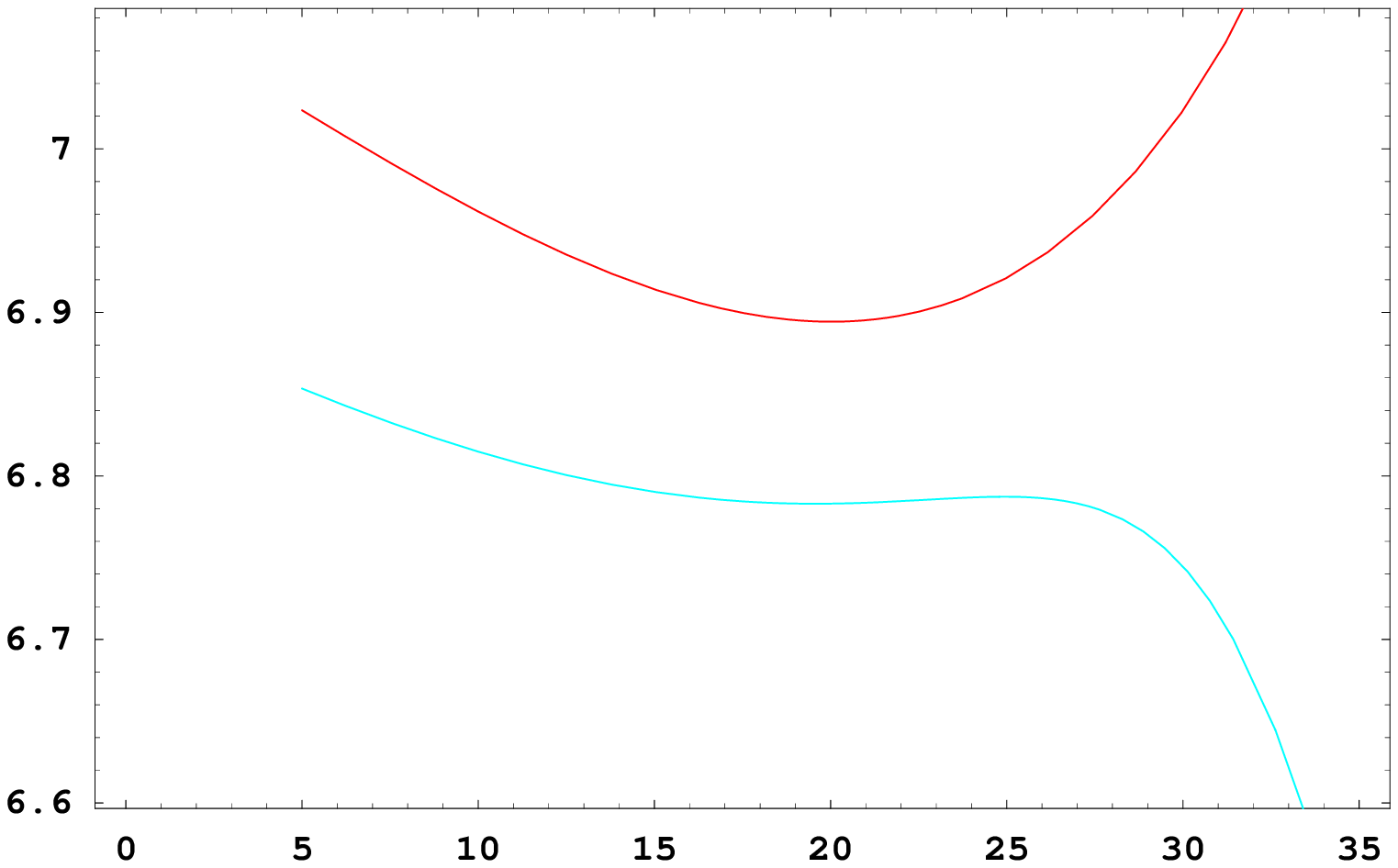}}
\put(200,5){$n$}
\put(-110,200){$M_{\Xi,\Omega_{bc}}$, GeV}
\end{picture}
\end{center}
\caption{The $\Xi_{bc}$ (lower curve) and $\Omega_{bc}$ (upper curve) masses
obtained in the NRQCD sum rules by averaging the results from two correlators
$F_{1,2}$.}
\label{M}
\end{figure}

In the two-point sum rules for the mass of $\Omega_{bc}$ (the conclusions for
other doubly heavy baryons are similar) we can observe the stability of
estimates with respect to changing the moment numbers in both correlators $F_1$
and $F_2$. We suppose that this fact is connected to the destroying of
mentioned factorization for the correlators of diquark and baryon in the
perturbative limit in contrast to the case of $\Xi$-baryons. The stability
regions for $F_1$ and $F_2$ do not coincide because the contributions of higher
dimension operators become valuable at the different numbers of moments.
However, the quantity 
$$
\frac{1}{2}\, (M_1[n]+M_2[n])
$$
has
the larger stability region, and we explore this fact to determine the $\Omega$
baryons masses as well as that of $\Xi$ baryons (see Fig. \ref{M}). Thus, in
the present review we consider two criteria for the stability of baryon mases:
the first is based on the study of mass difference obtained from two
correlators $F_{1,2}$, the second investigates the half sum of masses extracted
from two correlators. The second way is especially reliable for the doubly
heavy baryons with the strangeness, because the both correlators have the
stability regions at various numbers of moments. In this way, the difference
between the masses in two correlators at the stable points determines the
accuracy of estimates in the framework of NRQCD sum rules.

The second method results in the following values of baryon masses:
\begin{equation}
 \begin{array}{lcrrlcrr}
M_{\Omega_{cc}} &=& 3.65\pm0.05  &\mbox{GeV}, &  
   M_{\Xi_{cc}} &=& 3.55\pm0.06  &\mbox{GeV}, \\
M_{\Omega_{bc}} &=& 6.89\pm0.05 & \mbox{GeV}, &
   M_{\Xi_{bc}} &=& 6.79\pm0.06 & \mbox{GeV},\\
M_{\Omega_{bb}} &=& 10.09\pm0.05 &\mbox{GeV}, &
   M_{\Xi_{bb}} &=& 10.00\pm0.06 &\mbox{GeV},
 \end{array}
 \label{met2}
\end{equation}
We can see that the estimates of $\Xi_{QQ'}$ masses in both methods of
(\ref{met2}) and (\ref{met1}) well agree with each other.

\begin{figure}[t]
\begin{center}
\begin{picture}(100,200)
\put(-100,0){\epsfxsize=10cm \epsfbox{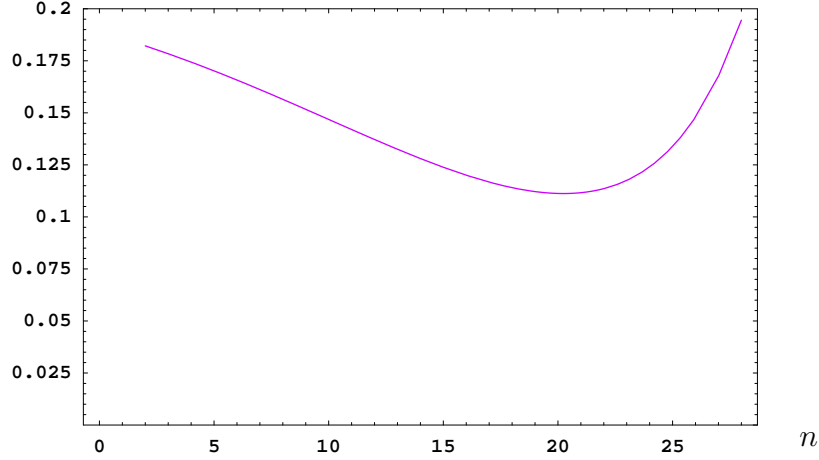}}
\put(200,5){$n$}
\put(-110,200){$\Delta M$, GeV}
\end{picture}
\end{center}
\caption{The mass difference $\Delta M=M_{\Omega_{bc}}-M_{\Xi_{bc}}$ obtained
from the results shown in Fig. \ref{M}.}
\label{Mres}
\end{figure}

Then, we investigate the difference between the masses of doubly heayy baryons
with strangeness and wothout it: $1/2((M_{1,\Omega} + M_{2,\Omega}) -
(M_{1,\Xi}+M_{2,\Xi})) $, shown in Fig. \ref{Mres}. In our scheme of baryon
masses determination  this quantity has the meaning of average difference
between the masses, for which we observe a wide interval of stability
indicating a good systematic accuracy of estimate. We have obtained
$$
\Delta M=M_{\Omega_{bb}}-M_{\Xi_{bb}} = M_{\Omega_{cc}}-M_{\Xi_{cc}} =
M_{\Omega_{bc}}-M_{\Xi_{bc}}=100\pm 30\; {\rm MeV.}
$$

\begin{figure}[th]
\begin{center}
\begin{picture}(100,200)
\put(-100,0){\epsfxsize=10cm \epsfbox{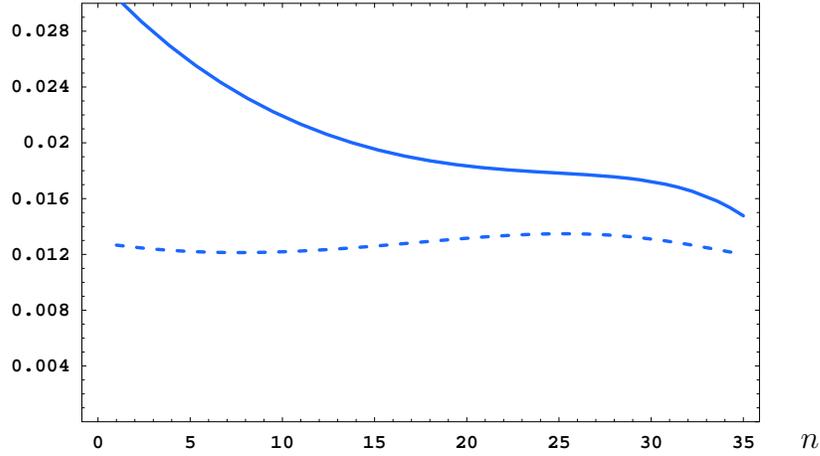}}
\put(200,5){$n$}
\put(-110,200){$|Z_{\Omega_{bc}}|^2$, $\mbox {GeV}^6$}
\end{picture}
\end{center}
\caption{The couplings $|Z_{\Omega_{bc}}^{(1,2)}|^2$ calculated in the NRQCD
sum rules for the formfactors $F_1$ and $F_2$ (solid and dashed lines,
correspondingly) in the scheme of moments for the spectral densities.}
\label{Zbcs}
\end{figure}

\begin{figure}[ph]
\begin{center}
\begin{picture}(100,200)
\put(-100,0){\epsfxsize=10cm \epsfbox{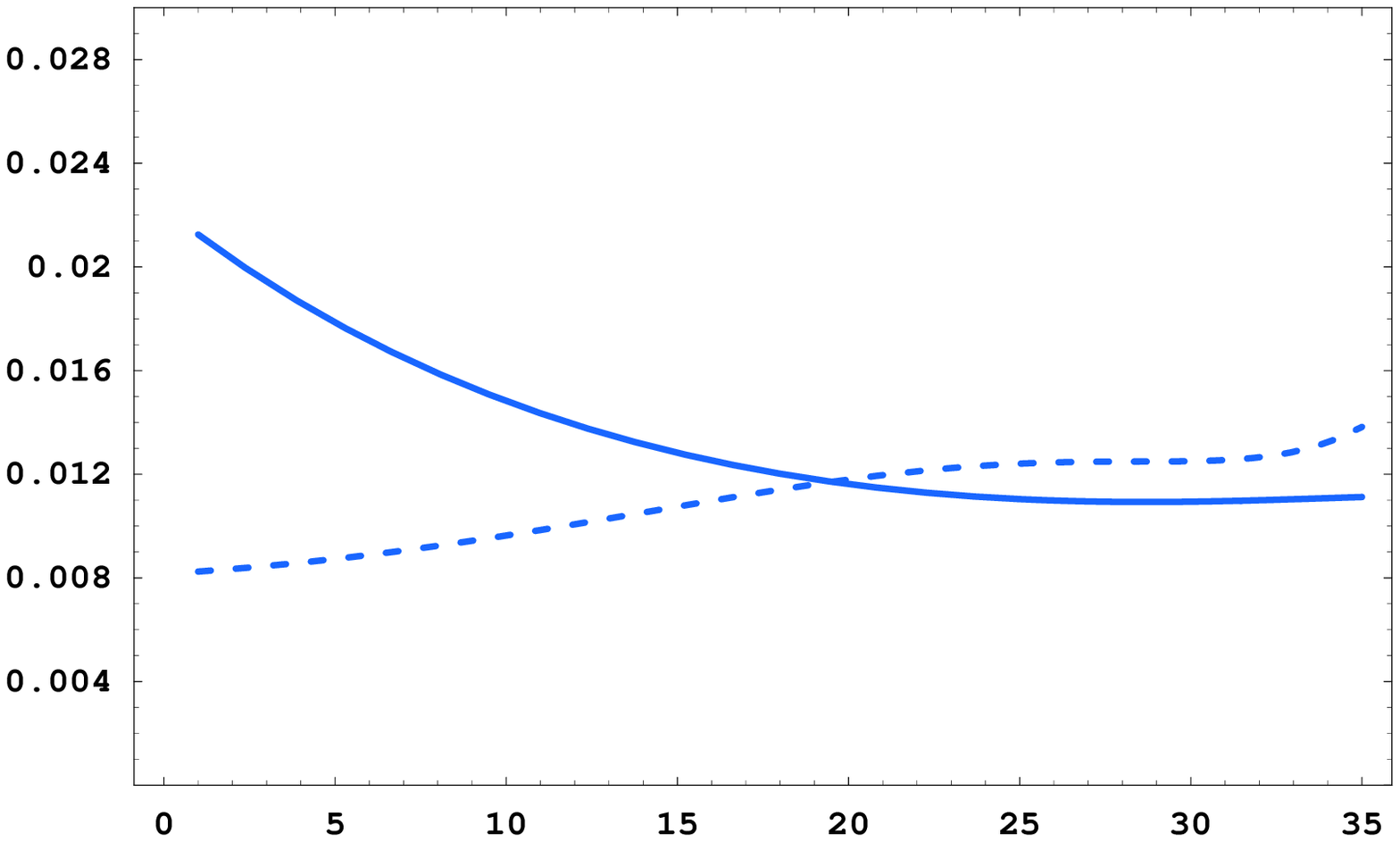}}
\put(200,5){$n$}
\put(-110,200){$|Z^{NR}_{\Xi_{bc}}|^2$, $\mbox {GeV}^6$}
\end{picture}
\end{center}
\caption{The couplings $|Z_{\Xi_{bc}}^{(1,2)}|^2$ calculated in the NRQCD sum
rules for the formfactors $F_1$ and $F_2$ (solid and dashed lines,
correspondingly) in the scheme of moments for the spectral densities.}
\label{Zbc}
\end{figure}

Figs. \ref{Zbcs}, \ref{Zbc} show the dependence of baryon couplings calculated
in the moment scheme of NRQCD sum rules for the doubly heavy baryons with the
strangeness and without it, respectively. Numerically, we find
\begin{equation}
 \begin{array}{lcrrlcrr}
|Z_{\Omega_{cc}}|^2 &=& (10.0\pm1.2)\cdot10^{-3} &\mbox{GeV}^6, &
   |Z_{\Xi_{cc}}|^2 &=& (7.2\pm0.8)\cdot10^{-3}  &\mbox{GeV}^6,\\[1mm]
|Z_{\Omega_{bc}}|^2 &=& (15.6\pm1.6)\cdot10^{-3} &\mbox{GeV}^6, &
   |Z_{\Xi_{bc}}|^2 &=& (11.6\pm1.0)\cdot10^{-3} &\mbox{GeV}^6,\\[1mm]
|Z_{\Omega_{bb}}|^2 &=& (6.0\pm0.8)\cdot10^{-2}  &\mbox{GeV}^6, &
   |Z_{\Xi_{bb}}|^2 &=& (4.2\pm0.6)\cdot10^{-2}  &\mbox{GeV}^6.
\end{array}
\label{nrqcd}
\end{equation}

\begin{figure}[ph]
\begin{center}
\begin{picture}(100,200)
\put(-100,0){\epsfxsize=10cm \epsfbox{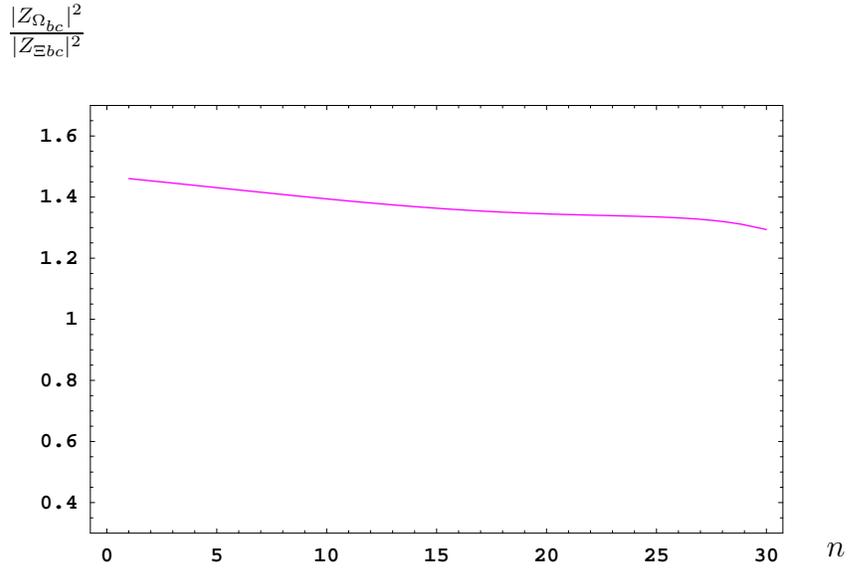}}
\put(200,5){$n$}
\put(-110,200){$\frac{|Z_{\Omega_{bc}}|^2}{|Z_{\Xi{bc}}|^2}$}
\end{picture}
\end{center}
\caption{The ratio $\frac{|Z_{\Omega_{bc}}|^2}{|Z_{\Xi{bc}}|^2}$ calculated in
the NRQCD sum rules in the scheme of moments for the spectral densities at
${\langle \bar s s \rangle}/{\langle \bar q q\rangle }=0.8$.}
\label{Ratio}
\end{figure}

In Fig. \ref{Ratio} we present the sum rules results for the ratio of baryonic
constants $|Z_{\Omega_{bc}}|^2/|Z_{\Xi_{bc}}|^2$. We have also found
$$
|Z_{\Omega_{bc}}|^2/|Z_{\Xi_{bc}}|^2=|Z_{\Omega_{cc}}|^2/|Z_{\Xi_{cc}}|^2=
|Z_{\Omega_{bb}}|^2/|Z_{\Xi_{bb}}|^2=1.3\pm0.2.
$$
The uncertainty of this result as was mentioned above is mainly connected with
the pourly known ratio of $\langle \bar s s \rangle /\langle \bar q q
\rangle=0.8\pm0.2$.

We can see in figures that the region of stability for the baryonic constants
coincides with the region of stability for the average mass described above.

For the sake of comparison, we derive the relation between the baryon coupling
and the wave function of doubly heavy baryon evaluated in the framework of
potential model, where we have used the approximation of quark-diquark
factorization. So, we find
\begin{equation}
|Z^{\rm PM}| = 2 \sqrt{3} |\Psi_d(0)\cdot \Psi_{l,\,s}(0)|,
\end{equation}
where $\Psi_d(0)$ and $\Psi_{l,\,s}(0)$ denote the wave functions at the origin
for the doubly heavy diquark and light (strange) quark-diquark systems,
respectively. In the approximation used, the values of $\Psi(0)$ were
calculated in the potential by Buchm\"uller and Tye \cite{BT}, so that
\begin{eqnarray}
\sqrt{4\pi}\;|\Psi_l(0)|    &=& 0.53\; {\rm GeV}^{3/2},\;\;\;
\sqrt{4\pi}\;|\Psi_s(0)|    = 0.64\; {\rm GeV}^{3/2},\nonumber\\
\sqrt{4\pi}|\Psi_{cc}(0)| &=& 0.53\; {\rm GeV}^{3/2},\;\;\;
\sqrt{4\pi}|\Psi_{bc}(0)| = 0.73\; {\rm GeV}^{3/2},\nonumber\\
\sqrt{4\pi}|\Psi_{bb}(0)| &=& 1.35\; {\rm GeV}^{3/2}.\nonumber
\end{eqnarray}
In the static limit of potential models, these parameters result in the
estimates
\begin{eqnarray}
|Z^{\rm PM}_{\Omega_{cc}}|^2~ =~ 8.8\cdot 10^{-3}\; {\rm GeV}^6,\;
|Z^{\rm PM}_{\Xi_{cc}}|^2 &=& 6.0\cdot 10^{-3}\; {\rm GeV}^6,~\nonumber\\
|Z^{\rm PM}_{\Omega_{bc}}|^2 ~=~ 1.6\cdot 10^{-2}\; {\rm GeV}^6,\;
|Z^{\rm PM}_{\Xi_{bc}}|^2 &=& 1.1\cdot 10^{-2}\; {\rm GeV}^6,~\label{pot}\\
|Z^{\rm PM}_{\Omega_{bb}}|^2~ =~ 5.6\cdot 10^{-2}\; {\rm GeV}^6,\;
|Z^{\rm PM}_{\Xi_{bb}}|^2 &=& 3.9\cdot 10^{-2}\; {\rm GeV}^6.~\nonumber
\end{eqnarray}
The estimates in the potential model (\ref{pot}) are close to the values
obtained in the sum rules of NRQCD (\ref{nrqcd}). We also see that the
SU(3)-flavor splitting for the baryonic constants $|Z_{\Omega}|^2/|Z_{\Xi}|^2$
is determined by the ratio $|\Psi_s(0)|^2/|\Psi_l(0)|^2=1.45$ which is in
agreement with the sum rules result.

The values obtained in the NRQCD sum rules
have to be multiplied by the Wilson coefficients coming from the expansion of
full QCD operators in terms of NRQCD fields, as they have been estimated by use
of corresponding anomalous dimensions. This procedure results in the estimates
\begin{eqnarray}
|Z_{\Omega_{cc}}|^2&=&(38\pm5)\cdot10^{-3}~\mbox{GeV}^6,\;\;
|Z_{\Xi_{cc}}|^2=(27\pm3)\cdot10^{-3}~\mbox{GeV}^6, \nonumber\\
|Z_{\Omega_{bc}}|^2&=&(36\pm4)\cdot10^{-3}~\mbox{GeV}^6,\;\;
|Z_{\Xi_{bc}}|^2=(27\pm3)\cdot10^{-3}~\mbox{GeV}^6, \\
|Z_{\Omega_{bb}}|^2&=&(10\pm1)\cdot10^{-2}~\mbox{GeV}^6,\;\;
|Z_{\Xi_{bb}}|^2=(70\pm 8)\cdot10^{-3}~\mbox{GeV}^6. \nonumber
\end{eqnarray}

Thus, we have got the reliable estimates for the masses and coupling constants
of baryons containing two heavy quarks in the framework of NRQCD sum rules.

\subsection{Discussion}

We have considered the NRQCD sum rules for the two-point correlators of
baryonic currents with two heavy quarks. The nonrelativistic approximation for
the heavy quark fields allows us to fix the structure of baryonic currents and
to take into account the coulomb-like interactions inside the doubly heavy
diquark. Moreover, we have introduced into the consideration the operators of
higher dimensions, which are responsible for the quark-gluon condensates in
order to reach the convergency of the sum rule method for two scalar
correlation functions. To the leading approximation, including the
perturbative term and the contributions of quark and gluon condensates, the
correlators of three-quarks state and the doubly heavy diquark are factorized
in separate functions, so that the sum rules result in the different values of
masses and couplings. This fact indicates the divergency of approach unless the
product of quark and gluon condensates and the mixed condensate are taken into
account. Then, the interaction of two heavy quarks and light quark destroy the
factorization, which allows one to get meaningful estimates of masses and
couplings. Moreover, we have also calculated the binding energy of doubly heavy
diquark, which is in a good agreement with the estimates in the framework of
potential models. In the doubly heavy baryons with the strangeness the
factorization of diquark correlator is already destroyed in the approximation
of quark loop in the perturbative QCD, so that the better convergency of sum
rule approach takes place. So, the both correlators have the intervals of
stability under the variation of number for the moment of spectral density. We
have calculated the splitting of masses between the strange baryons
$\Omega_{QQ'}$ and baryons with the massless light quark $\Xi_{QQ'}$. We have
also estimates the ratio of baryonic coupling constants
$|Z_{\Omega_{QQ'}}|^2/|Z_{\Xi_{QQ'}}|^2$ for the baryons under consideration.

Thus, the NRQCD sum rules allow us to improve the analysis of masses and
couplings for the doubly heavy baryons and to obtain reliable results.

\newpage

\section{Production processes}

For the production of baryons with two heavy quarks $\Xi_{QQ'}$ a small ratio
of $\Lambda/m_Q$ and, hence, a small value of quark-gluon coupling constant in
QCD $\alpha_s \sim 1/ln(m_Q/\Lambda) \ll 1$ allow us not only to consider the
production of two pairs of heavy quarks $Q\bar Q$ and $Q'\bar Q'$ in the
framework of perturbative QCD, but also to factorize the contributions caused
by the perturbative production of heavy quarks and the nonperturbative forming
of heavy diquark by the quarks enetering the $\Xi_{QQ'}$-baryons. So, in order
to calculate the cross sections for the production of S-wave $\Xi_{bc}$ states
at the $Z$-boson peak we have to evaluate the matrix elements for the
associated production of $b\bar b$ and $c\bar c$ pairs with the anti-triplet
color state of $b c$ pair having got a definite sum of quark spins ($S=0,1$),
while the quarks are assigned to the same velocity equal to the velocity of
diquark composed by these quarks. Then we have to multiply these matrix
elements by a nonperturbative factor determined by some spectroscopic
characteristics of bound state, i.e. by both the wave function of diquark, that
gives the probability to find the quarks at short distances between them inside
the bound state, and the quark masses. This approach is justified due to the
following: the characteristic virtualities of heavy quarks inside the heavy
diquark are much less than their masses, since the heavy quarks
nonrelativistically move in the bound state, while the quark virtualities are
about their masses in the hard production. Therefore, considering the
production of $\Xi_{bc}$, we can suppose that the quarks $b$ and $c$ are on the
mass shell in the diquark, and they are at rest with respect to each other.
Thus, after the isolation of nonperturbative factor the analysis of
$\Xi_{bc}$-baryon production is reduced to the consideration of matrix elements
calculated in the perturbative QCD, if we suppose that the total and
differential cross sections of baryon repeat the corresponding quantities for
the doubly heavy diquark.

Note, first of all, that in the electromagnetic and strong interactions of
colliding particles the associated production of two heavy quark pairs
necessary for the hadronization into $\Xi_{QQ'}$, can be done provided the
leading order of perturbative QCD for such processes has the additional factor
of $\alpha_s^2$ in comparison with the leading order for the production of
single pair of heavy quarks $Q\bar Q$, so that $\sigma(\Xi_{QQ'})/\sigma(Q\bar
Q)\sim \alpha_s^2|\Psi(0)|^2/m_{Q'}^3$. This suppression causes a small
relative yield of $\Xi_{QQ'}$ in comparison with the inclusive production of
heavy mesons $M_Q$. 

So, we have to analyze the leading approximation of perturbative QCD for the
production of $\Xi_{QQ'}$, that allows us to get some analytic expressions for
the cross sections of $\Xi_{QQ'}$, particularly, for the fragmentation
functions of both the heavy quark into the heavy diquark and the diquark into
the baryon in the scaling limit of $M^2/s \to 0$. Thus, the production of
$\Xi_{QQ'}$ in the regime of fragmentation can be reliably described in terms
of analytic expressions, that opens new possibilities in the study of QCD
dynamics, which is significant in the complete picture of heavy quark physics.

The mechanism for the production of baryons with two heavy quarks in the
hadronic collisions involves the consideration of complete set of diagrams in
the fourth order of perturbative QCD because the fragmentation regime does not
dominate in the total cross section determined by other nonfragmentational
contributions rapidly decreasing at large transverse momenta. We investigate a
role of these higher twists over the transverse momentum of doubly heavy baryon
in the associated hadronic production. We numerically determine the limits for
the consistent use of factorization regime in the hard production of heavy
quarks with the consequent fragmentation.

\subsection{Production of doubly heavy baryons in $e^+e^-$-annihilation}

The analysis of mechanisms for the production of hadrons with two heavy quarks
shows that the expected production yield of such hadrons with respect to the
number of hadrons with a single heavy quark is of the order of $10^{-(3-4)}$.
For example, at the $Z^0$-boson pole the number of events with heavy quarks is
about $10^6$, consequently, the number of hadrons with two heavy quarks is
expected to be $\sim 100-1000$. Taking into account specific decay modes of
hadrons with two heavy quarks one could expect the detection of several events
with such hadrons, which makes their observation problematic at LEP. 

In this section we consider the doubly charmed baryon production
($\Sigma^{(*)}_{cc}$) under the conditions of a $B$-factory with high
luminosity $L=10^{34}$~cm${}^{-2}$s${}^{-1}$, where the number of
$\Sigma^{(*)}_{cc}$ is two orders of magnitude greater than the yield at the
$Z^0$-boson pole.

\subsubsection{Fragmentation mechanism}
In \cite{falk} the production cross sections for $\Xi^{(*)}_{cc}$,
$\Xi^{(*)}_{bc}$ and $\Xi^{(*)}_{bb}$ were evaluated in the region of heavy
quark fragmentation at high energies. These estimations are based on the exact
analytical calculations for the heavy quarkonium production in the QCD
perturbation theory in the limit of small $M^2/s$ ratio and nonrelativistic
potential model. In \cite{falk} the $cc$-diquark momentum spectrum was
considered to be equal to that of heavy vector quarkonium $(\bar c
c)$\footnote{There is a wrong additional factor of 2 in \cite{falk}.}
\begin{equation}
D_{c\to cc}(z) = \frac{2}{9\pi}\; \frac{|R_{cc}(0)|^2}{m_c^3}\;
\alpha_s^2(4m_c^2)F(z),
\label{g3b1}
\end{equation}
where
$$
F(z) = \frac{z(1-z)^2}{(2-z)^6}\; (16-32z+72z^2-32z^3+5z^4),
$$
and $R_{cc}(0)$ is a radial wave function of the bound diquark at the origin.

Let us note that identical quarks $cc$ in the color anti-triplet state can have
the symmetrical spin wave function in the $S$-wave, i.e. they must be in the
total spin $S=1$ state. The normalization of the fragmentation function
$D_{c\to cc}(z)$ is determined by the model dependent value of $R_{cc}(0)$. In
\cite{falk} a rather rough approximation with the pure coulomb potential in the
system of heavy quarks was used. This factor gives a huge
uncertainty\footnote{The calculation of diquark wave function in the model with
the Martin potential accounting for the factor of $1/2$ due to the anti-triplet
color state of quarks enforces the corresponding factor by an order of
magnitude.} in the estimation of $\Xi^{(*)}_{cc}$. Moreover, expression
(\ref{g3b1}) obtained in the scaling limit $M^2/s\to 0$, is not justified for
estimating the $\Xi^{(*)}_{cc}$ production at the $B$-factory, where the
$M^2/s$ ratio is not small.

We propose another method to estimate the production of 
hadrons containing two heavy quarks, on the basis of 
quark-hadron duality.

\subsubsection{Calculations under the quark-hadron duality}
A production cross section of the $B_c$-meson $S$-wave states at the
$Z^0$-boson pole calculated in the fragmentation model is in a good agreement
with the cross section estimations for the production of quark pair $({\bar
b}c)$ in the color singlet state with small invariant masses
\begin{equation}
m_b+m_c< M(\bar b c)<M_{\rm th}=M_B+M_D+\Delta M,
\label{g3b3}
\end{equation}
where $\Delta M=0.5-1$ GeV.

In the same range of duality (\ref{g3b3}) the $bc$-diquark production
cross section is approximately equal to that of ${\bar b}c$-pair. Selecting the
color anti-triplet state $({ b}c)$ and multiplying by the factor of 2/3, we
obtain the estimate for the the $\Xi^{(*)}_{bc}$-baryon production
cross section $\sigma(\Xi^{(*)}_{bc})/\sigma(b\bar b)\simeq 6\cdot 10^{-4}$,
i.e. 6 times greater than the estimate made in \cite{falk} for the $1S$-state
production. This difference is caused, first, by taking into account the
contribution by higher excitations of diquark in the framework of quark-hadron
duality and, second, by the strong suppression due to supposed low value of
$R_{bc}(0)$, evidently underestimated in the pure coulomb approximation.

Let us consider the $\Xi^{(*)}_{cc}$-baryon production at the energy of
$B$-factory $(\sqrt{s} = 10.58$ GeV). Remember that expression (\ref{g3b1}) may
not be used at the given energy because the power corrections over $M^2/s$ are
substantial. The method of calculations in the leading order of QCD
perturbation theory was described in \cite{d9,d12,d30,d31}.

In the method of quark-hadron duality the cross section for the associated
production of quarkonium bound states can be estimated by using the formula
\begin{equation}
\sum_{nL,J}\sigma(e^+e^-\to (nL(c\bar c)_J)c\bar c) =
\int_{M_i}^{M_{\rm th}}dM_{c\bar c}\;\frac{d\sigma(e^+e^-\to
(\bar c c)_{singlet}c\bar c)}{dM_{c\bar c}},
\label{g3b4}
\end{equation}
where $M_i=2m_c$ is the kinematical threshold of $c\bar c$-pair production,
$M_{\rm th}=2M_D+\Delta M$, $\Delta M=0.5-1$ GeV.
In Table \ref{g3bt1} the results for the numerical calculations of the QCD
perturbation theory diagrams are presented for the production of the bound
$1S$- and $2S$-levels of charmonium at the energy $\sqrt{s}=10.58$ GeV and
$\alpha_s=0.2$. The values of the radial wave functions at the origin
$R_{nS}(0)$ have been determined from the experimental data on the lepton decay
widths of charmonia $\psi (nS)$ \cite{PDG}. As we can conclude from Table
\ref{g3bt1}, below the threshold for the decay into the pair of $D\bar D$
mesons the sum over the $S$-wave states of charmonia is equal to
\begin{equation}
\sigma(\sum \eta_c,\psi) = 0.093\;\; {\rm pb.}
\label{g3b5}
\end{equation}
Note that the ratio of the vector and pseudoscalar state yields at the energy
under consideration is equal to $\omega_V/\omega_P \simeq 2.2$ in contrast to
the value $\omega_V/\omega_P \simeq 1$ obtained in the fragmentation mechanism
\cite{falk}.

\begin{table}[t]
\caption{The production cross sections of $\eta_c$ and $\psi$ mesons 
in $e^+e^-$ annihilation at the B-factory.}
\label{g3bt1}
\begin{center}
\begin{tabular}{||c|c||}
\hline
meson & $\sigma$, pb \\
\hline
$\eta_c(1S)$ & 0.025 \\
$\eta_c(2S)$ & 0.003 \\
$\psi(1S)$   & 0.055 \\
$\psi(2S)$   & 0.010 \\
\hline
\end{tabular}
\end{center}
\end{table}
Our estimations of the integral in the r.h.s. of expression (\ref{g3b4}) give
\begin{eqnarray}
\sigma_{c\bar c}(\Delta M=0.5\;{\rm GeV}) & = & 0.093\;\;{\rm pb,}
\label{g3b6}\\
\sigma_{c\bar c}(\Delta M=1\;{\rm GeV}) & = & 0.110\;\;{\rm pb,}
\label{g3b7}
\end{eqnarray}
where we set $m_c=1.4$ GeV.

From (\ref{g3b5}) and (\ref{g3b6}), (\ref{g3b7}) we see that the relation of
quark-hadron duality (\ref{g3b4}) is well satisfied for the bound states of
$(\bar c c)$.

Further, the calculations show that the invariant mass distributions for the
$c\bar c$ and $cc$ pairs coincide with each other in the region of small
invariant masses. Hence, in the same duality region, the estimates for the
production cross sections of $c\bar c$ pair and $cc$-diquark are approximately 
equal to each other, so that comparing with (\ref{g3b6}) and (\ref{g3b7}) we
have
\begin{eqnarray}
\sigma_{cc}(\Delta M=0.5\;{\rm GeV}) & = & 0.086\;\;{\rm pb,}
\label{g3b8}\\
\sigma_{cc}(\Delta M=1\;{\rm GeV}) & = & 0.115\;\;{\rm pb.}
\label{g3b9}
\end{eqnarray}
Selecting the anti-triplet color state, we can obtain the summed total cross
section for the production of $\Xi_{cc}^{(*)}$ baryons
\begin{equation}
\sigma(\Xi_{cc}^{(*)})=(70\pm 10)\cdot 10^{-3}\;\; {\rm pb,}
\end{equation}
so that the yield fraction of doubly charmed baryons approximately equals 
\begin{equation}
\sigma(\Xi_{cc}^{(*)})/\sigma(c\bar c) = 7\cdot 10^{-5}.
\end{equation}
For the luminosity equal to $L=10^{34}\mbox{cm}^{-2}\mbox{s}^{-1}$ the number 
of events with the production of $\Xi_{cc}^{(*)}$ is equal to
$N(\Xi_{cc}^{(*)}) = 7\cdot 10^3$ per year, so it is by two orders 
of magnitude greater than the yield of $\Xi_{cc}^{(*)}$ at LEP.

In ref. \cite{d18} the distribution over the $cc$-diquark momentum is presented
for the conditions of anti-symmetric collider KEK.

Thus, in this section we have presented the calculations of the doubly charmed
$\Xi_{cc}^{(*)}$ baryon production on the basis of the quark-hadron duality in
the leading order of QCD perturbation theory. We have evaluated the
$\Xi_{cc}^{(*)}$ production cross section at the energy of $B$ factory, where
the fragmentation model \cite{falk} does not work.

The main theoretical uncertainty in the estimations for the production
cross section of the double charmed baryons is related with the description of
the process for the heavy $cc$-diquark hadronization. First of all, a
considerable fraction of the diquarks (1/3) is produced in the color sixtet
state and can transmit into both the exotic four quark states ($cc{\bar q}{\bar
q}$) and the  $DD$-meson pair. As in ref. \cite{falk}, we assume that color 
anti-triplet state hadronizes into the $\Xi_{cc}^{(*)}$ baryon with  100\% 
probability. Thus, at the $B$-factory one could expect $10^4$ events per year
with the production of $\Xi_{cc}^{(*)}$ at the luminosity $L=10^{34}$ см$^{-2}$
с$^{-1}$.

\subsubsection{Exclusive production of diquark pairs}
Close to the threshold for the production of doubly heavy baryons in $e^+e^-$
annihilation a significant contribution could be given by the pair production.
In order to estimate a yield of such kind events the calculation of cross
sections was done in \cite{brag} for the exclusive production of doubly heavy
diquark pairs. Supposing the 100\% probability for the diquark fragmentation
into the baryons we can put the pair yield equal to the yield of baryons.
Authors of \cite{brag} considered both the axial-vector states and the scalar
ones for the $S$-wave diquarks, so that the amplitudes, differential and total
cross sections were presented for the scalar-scalar, scalar-vector and
vector-vecotr pairs. The details are given in the original ref. \cite{brag}.
For the representativity we show the expression for the total cross section of
scalar pairs versus the square of total energy $s$
\begin{equation}
\sigma_{ss}=256 \pi^3 \frac {f_{ss}^2} {9 s^3}~|\Psi_s(0)|^4	
\left(1-\frac {4M^2} {s}\right)^{3/2},
\end{equation}
where $\Psi_s(0)$ is the wave function of diquark at the origin, and the
formfactor has the form
\begin{eqnarray}
f_{ss} &=& \alpha_s \alpha_{em} M \left[\left( \frac{q_2 }{m_1^2} + \frac{q_1
}{m_2^2}\right) - \frac {2M^2}{s} \left(\frac{q_2 m_2}{m_1^3} +
\frac{q_1m_1}{m_2^3}\right) \right],
\end{eqnarray}
so that $q_{1,2}$ are the heavy quark charges, $m_{1,2}$ are their masses, and 
$M=m_1+m_2$. 

Numerical estimates show that the production of axial vector pairs dominates,
so that in comparison with the yield of heavy quark pairs the fraction of
diquark pairs is about $(2-6)\cdot 10^{-6}$. This fact implies that, say, at
the $B$ factory among the charm events one should try to observe the events
with the production of doubly charmed baryon pairs yielding a 10\% fraction of
their inclusive production.

\subsection{Perturbative fragmentation of diquark}
In this section we investigate the production of baryons in the fragmentation
of vector and scalar particles interacting with quarks. From the QCD point of
view the doubly heavy diquark of small size is a local color-triplet field,
therefore, the results of this section can be used for the calculation of
fragmentation into the baryons with doubly heavy vector and scalar diquarks. In
this approach we explore the perturbative QCD for the calculation of hard
amplitude in the fragmentation, which is factorized in front of soft amplitude
for the forming of bound state. Sure, this method is quite accurate, if the
hardness is provided by a large value of mass for the quark, which composes the
hadronic state of baryon by coupling together with the diquark, for example, in
the fragmentation of $bb$ into $bbc$. However, the obtained expressions could
be used as the QCD-motivated parametrizations for the processes with the light
quarks, too.

The fragmentation of scalar color-triplet local field was considered in
\cite{ScLep}. A new problem arising in the case of vector particle is a choice
of the lagrangian for the vector diquark interaction with gluons. Indeed, to
the lagrangian of a free vector field $-{\rm tr}\, [H_{\mu \nu} \bar{H}^{\mu
\nu}]$, where $H_{\mu \nu}= \partial_{\mu} U_{\nu}-\partial_{\nu} U_{\mu}$,
$U_{\mu}$ is the vector complex field with derivatives substituted by covariant
ones, we can add the gauge invariant term proportional to $S^{\alpha
\beta}_{\mu \nu} {\rm tr}\, [G^{\mu \nu} U_{\beta} \bar{U}_{\alpha}]$, where
$S^{\alpha \beta}_{\mu \nu}=1/2(\delta^{\alpha}_{\mu}
\delta^{\beta}_{\nu}-\delta^{\alpha}_{\nu} \delta^{\beta}_{\mu})$ is the tensor
of spin, $G^{\mu \nu}$ is the gluon field strength tensor. It leads to the
appearance of a parameter in the gluon--diquark vertex (the so-called anomalous
chromo-magnetic moment). In this section we discuss the production of a
$1/2$-spin bound state containing the heavy vector particle at various values
of this parameter.

At high transverse momenta, the dominant production mechanism for the heavy
baryon bound states is the diquark fragmentation, which can be calculated in
perturbative QCD \cite{Bra-rev} after the isolation of soft binding factor
extracted from the nonrelativistic potential models \cite{RQ}. The
corresponding fragmentation function is universal for any high energy process
for the direct production of baryon.

In the leading $\alpha_s$-order, the fragmentation function has a scaling form,
which is the initial condition for the perturbative QCD evolution caused by the
emission of hard gluons by the diquark before the hadronization. The
corresponding splitting function differs from that for the heavy quark because
of the spin structure of gluon coupling to the diquark, which is the vector or
scalar color-triplet particle.

In this work the scaling fragmentation function is calculated in the leading
order of perturbation theory.  The limit of infinitely heavy diquark,
$m_{diq}\to \infty$, is obtained from the full QCD consideration for the
fragmentation. The distribution of bound state over the transverse momentum
with respect to the axis of fragmentation is calculated in the scaling limit of
perturbative QCD. The splitting kernel of the
Dokshitzer--Gribov--Lipatov--Altarelli--Parisi-evolution (DGALAP) is derived,
while the one-loop equations of renormalization group for the moments of
fragmentation function are obtained and solved. These equations are universal,
since they do not depend on whether the diquark will bound or free at low
virtualities, where the perturbative evolution stops. The integrated
probabilities of diquark fragmentation into the doubly heavy baryons are
evaluated.

\subsubsection{Fragmentation function in the leading order}

The contribution of fragmentation into direct production of heavy
baryon has the form
$$
d\sigma[\Xi_H(p)] = \int_0^1 dz\; d\hat \sigma[diq(p/z), \mu]\; 
D_{diq\to \Xi_H}(z, \mu), 
$$
where $d\sigma$ is the differential cross section for the production of
baryon with the 4-momentum $p$, $d\hat \sigma$ is that of the hard
production of diquark with the scaled momentum $p/z$, and $D(z)$ is interpreted
as the fragmentation function depending on the fraction of momentum carried out
by the bound state $z$. The value of $\mu$ determines the factorization scale.
In accordance with the general DGLAP-evolution, the $\mu$-dependent
fragmentation function satisfies the equation
\begin{equation}
\frac{\partial D_{diq\to \Xi_H}(z, \mu)}{\partial \ln \mu} =
\int_z^1 \frac{dy}{y} \; P_{diq\to diq}(z/y, \mu)\; D_{diq\to \Xi_H}(y, \mu), 
\label{DGLAP}
\end{equation}
where $P$ is the kernel caused by the emission of hard gluons by the diquark
before the production of heavy quark pair. Therefore, the initial form of
fragmentation function is determined by the diagram shown in Fig.\ref{diag},
and, hence, the corresponding initial factorization scale is equal to $\mu=
2m_Q$. Furthermore, this function can be calculated as an expansion in
$\alpha_s(2m_Q)$. The leading order contribution is evaluated in this section.

\setlength{\unitlength}{1mm}
\begin{figure}[th]
\begin{center}
\begin{picture}(100,50)
\put(0,0){\epsfxsize=9cm \epsfbox{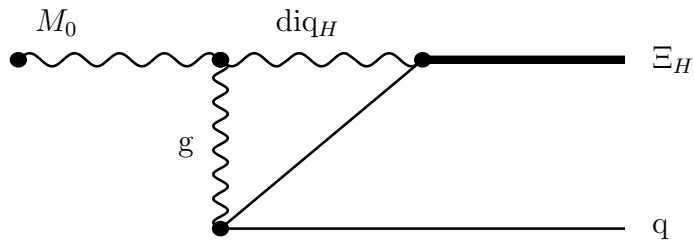}}
\put(90,8){q}
\put(90,30){$\Xi_H$}
\put(40,35){diq$_H$}
\put(27,19){g}
\put(8,35){$M_0$}
\end{picture}
\end{center}
\caption{The diagram for the fragmentation of diquark diq$_H$ into the heavy
baryon $\Xi_H$.}
\label{diag}
\end{figure}

Consider the fragmentation diagram in the system, where the momentum of initial
diquark has the form $q=(q_0, 0, 0, q_3)$ and the baryon momentum is denoted by
$p$, so that 
$$
q^2=s,  \;\; p^2=M^2.\;\; 
$$
In the static approximation for the bound state of diquark and heavy quark,
the quark mass is expressed as $m_Q= r M$, and the diquark mass equals
$m=(1-r)M=\bar r M$. The gluon--vector diquark vertex has the form
\begin{equation}
T_{\alpha\mu\nu}^{VVg}=-i g_{s} t^{a} [g_{\mu\nu} (q+\bar r
p)_{\alpha}-g_{\mu\alpha} ((1+\ae )\bar r p-\ae  q)_{\nu}-g_{\nu\alpha}
((1+\ae )q -\ae  \bar r p)_{\mu}],
\label{VgV}
\end{equation}
where $\ae $ is the anamalous chromo-magnetic moment, $t^a$ is the QCD group
generator in the fundamental representation. The sum over the vector diquark
polarizations with the momentum $q\; (q^2=s)$ depends on the choice of the
gauge of free field lagrangian (for example, the Stueckelberg gauge), but the
fragmentation function is a physical quantity and has not to depend on the
gauge parameter changing the contribution of longitudinal components of the
vector field. So, in a general case, the sum over polarization can be taken in
the form
$$
P(q)_{\mu\nu}= -g_{\mu\nu}+\frac{q_\mu q_\nu}{s}.
$$ 
The matrix element of the fragmentation into the baryon with the spin of $1/2$
has the form 
\begin{equation}
{\cal M} = -\frac{2\sqrt{2\pi}\alpha_s}{9\sqrt{M^3}}
\frac{R(0)}{r\bar r (s-m^2)^2}
P(q)_{\nu\delta}P(\bar r p)_{\mu\eta}T_{\alpha\mu\nu}^{VVg}
\rho_{\alpha\beta}\;
\bar q \gamma^\beta(\hat p-M) \gamma^\eta \gamma^5 \xi_H \; {\cal M}_0^\delta, 
\label{Matr}
\end{equation}
where the sum over the gluon polarization is written down in the axial gauge
with $n=(1, 0, 0, -1)$
$$
\rho_{\mu\nu}(k) = -g_{\mu\nu}+\frac{k_\mu n_\nu+k_\nu n_\mu}{k\cdot n}, 
$$
and $k=q-(1-r)p$. The spinors $\xi_H$ and $\bar q$ correspond to the baryon and
heavy quark associated to the fragmentation. ${\cal M}_0$ denotes the matrix
element for the hard production of diquark at high energy, $R(0)$ is the radial
wave function at the origin. The matrix element squared and summed over the
helicities of particles in the final state has the following structure:
$$
|\overline{\cal M}|^2=W_{\mu\nu} M_0^\mu M_0^\nu .
$$
In the limit of high energies $q\cdot n \to \infty$ $ W_{\mu\nu}$ behaves like
\begin{equation}
W_{\mu\nu}=-g_{\mu\nu}W+R_{\mu\nu}, 
\label{W}
\end{equation}
where $R_{\mu\nu}$ can depend on the gauge parameters. After the separation of
Lorentz structures it leads to scalar formfactor terms, which are small in
comparison with $W$ in the limit of $q\cdot n \to \infty$. Define 
$$
z=\frac{p\cdot n}{q\cdot n}. 
$$
The fragmentation function is determined by the expression \cite{Bra-frag}
$$ 
D(z) = \frac{1}{16\pi^2}\int ds\, \theta\biggl(s-\frac{M^2}{z}-\frac{m_Q^2}
{1-z}\biggr)\;W , 
$$ 
where $W$ is defined in (\ref{W}). The integral in the expression for the
fragmentation function diverges logarithmically at a constant value of
anamalous chromo-magnetic moment if $\ae$ does not equal $-1$. We consider two
sets for the behaviour of anamalous chromo-magnetic moment. The first is $\ae
=-1$. Here we observe that the obtained fragmentation function coincides with
that for the scalar diquark up to the factor of $1/3$
\begin{eqnarray}
D(z) = \frac{8\alpha_s^2}{243\pi}\;
\frac{|R(0)|^2}{M^3 r^2\bar r^2}\;
\frac{z^2(1-z)^2}{[1-\bar r z]^6} &\cdot&
\bigg\{3+3r^2-(6-10r+2r^2+2r^3)z+\nonumber \\
&&+(3-10r+14r^2-10r^3+3r^4)z^2)\bigg\}, 
\label{frscal}
\end{eqnarray}
which tends to
\begin{equation}
\tilde D(y)= \frac{8\alpha_s^2}{243\pi\ y^6}\;
\frac{|R(0)|^2}{m_Q^3}\; \frac{(y-1)^2}{r}\bigg\{8+4y+3y^2\bigg\}, 
\label{tilde}
\end{equation}
at $r\to 0$ and $y=(1-(1-r)z)/(rz)$.  
\setlength{\unitlength}{1mm}
\begin{figure}[th]
\begin{center}
\begin{picture}(100,80)
\put(0, 0){\epsfxsize=11cm \epsfbox{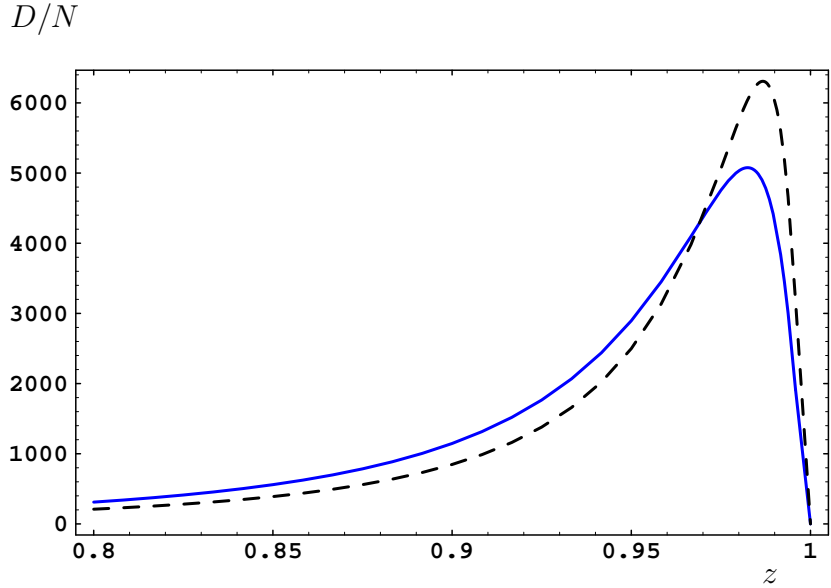}}
\put(100, 6){$z$}
\put(0, 80){$D/N$}
\end{picture}
\end{center}
\caption{The fragmentation function of diquark into the heavy
baryon, the $N$-factor is determined by $N=\frac{8\alpha_s^2}{243\pi}\;
\frac{|R(0)|^2}{M^3 r^2(1-r)^2}$, the fragmentation function at $\ae=-1$ is
shown by the dashed line, the fragmentation function at
$1+\ae=3M^2/(s-m_{diq}^2)$ is given by the solid line ($r=0.02$).} 
\label{d-fig}
\end{figure}

The limit of $\tilde D(y)$ is in agreement with the general consideration of
$1/m$-expansion for the fragmentation function \cite{JR}, where 
$$
\tilde D(y) = \frac{1}{r}a(y) +b(y). 
$$
The dependence of $a(y)$ on $y$ has the same form as for the fragmentation of
heavy quark into the quarkonium \cite{Bra-frag}.

The consideration of fragmentation at $\ae = -1+A M^2 /(s-m_{diq}^2)$ was done
in \cite{kova}.

The perturbative functions in the leading $\alpha_s$-order are shown in
Fig.\ref{d-fig} at $r=0.02$. We see hard distributions, which become softer
with the evolution (see ref.\cite{ScLep}).

\subsubsection{Transverse momentum of baryon}

In the system with an infinite momentum of the fragmentating diquark its
invariant mass is expressed by the fraction of longitudinal diquark
momentum $z$ and transverse momentum with respect to the fragmentation axis
$p_T$ (see Fig. \ref{diag}) as 
$$
s = m^2 + \frac{M^2}{z(1-z)}[(1-(1-r)z)^2+t^2], 
$$
where $t=p_T/M$. The calculation of diagram in Fig. \ref{diag} gives the double
distribution for the fragmentation probability
$$
\frac{d^2 P}{ds\; dz} = {\cal D}(z, s), 
$$
where at $(\ae =-1)$ the function ${\cal D}$ has the form 
\begin{eqnarray}
{\cal D}(z, s) &=& \frac{256\alpha_s^2}{81 \pi} \;
\frac{|R(0)|^2}{r^2\bar r^2}\;
\frac{M^3}{[1-\bar r z]^2 (s-m^2)^4} \nonumber\\ &&
\biggl\{r\bar r^2+\bar r (1+r-z(1+4r-r^2))\frac{s-m^2}{M^2}-
 z(1-z) \biggl(\frac{s-m^2}{M^2}\biggr)^2\biggr\}. 
\label{dtdz}
\end{eqnarray}
\begin{figure}[th]
\setlength{\unitlength}{0.8mm}
\begin{center}
\begin{picture}(110,80) 
\put(0, 0){\epsfxsize=11cm \epsfbox{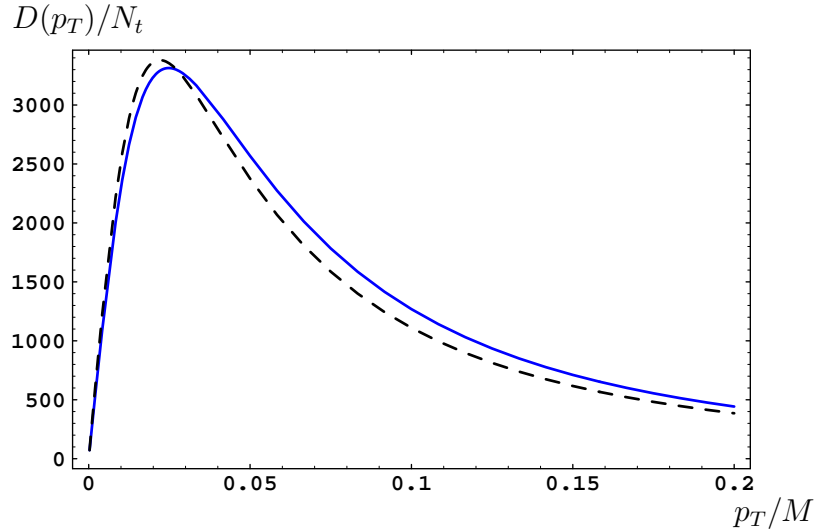}}
\put (120, 0){$p_T/M$}
\put (0,82){$D(p_T)/N_t$}
\end{picture}
\end{center}
\caption{The distributions over the transverse momentum with respect to the
axis of diquark fragmentation into the baryon, $N_t$-factor is determined by
$N_t=\frac{8\alpha_s^2}{81\pi}\; \frac{|R(0)|^2}{M^4 r^2(1-r)^7}$ at $r=0.02$.
The dashed line represents the result at $\ae=-1$, the solid line does it at
$\ae=-1+3M^2/(s-m_{diq})$.}
\label{dt-fig}
\end{figure}
The distribution of baryon over the transverse momentum can be
obtained by the integration of (\ref{dtdz}) over $z$
$$
D(t) = \int_0^1 dz\; {\cal D}(z, s)\; \frac{2M^2 t}{z(1-z)}. 
$$
Finally, we get quite a cumbersome expression presented in Appendix II. The
characteristic form of distribution over the transverse momentum of baryon with
respect to the axis of diquark fragmentation is shown in Fig. \ref{dt-fig}.

\subsubsection{Hard gluon emission}

The one-loop contribution can be calculated in the way described in the
previous sections. This term does not depend on the part of diquark-gluon
vertex with the anomalous chromo-magnetic moment, therefore the splitting
kernel coincides with that for the scalar diquark. It equals
\begin{equation}
P_{diq\to diq}(x, \mu) = \frac{4\alpha_s(\mu)}{3\pi}\;
\bigg[\frac{2x}{1-x}\biggr]_+, 
\label{P}
\end{equation}
where the "plus" denotes the ordinary action: $\int_0^1 dx f_+(x)\cdot g(x)=
\int_0^1 dx f(x)\cdot [g(x)-g(1)]$. The diquark splitting function can be
compared with that of the heavy quark 
$$
P_{Q\to Q}(x, \mu) = \frac{4\alpha_s(\mu)}{3\pi}\;
\bigg[\frac{1+x^2}{1-x}\biggr]_+, 
$$
which has the same normalization factor at $x\to 1$.

Furthermore, multiplying the evolution equation by $z^n$ and integrating over
$z$, one can get from (\ref{DGLAP}) the $\mu$-dependence of moments $a_{(n)}$
up to the one-loop accuracy of renormalization group
\begin{equation}
\frac{\partial a_{(n)}}{\partial \ln \mu} = - \frac{8\alpha_s(\mu)}{3\pi}\;
\bigg[\frac{1}{2}+\ldots +\frac{1}{n+1}\biggr]\; a_{(n)},  \;\;\; n\ge 1.
\label{da}
\end{equation}
At $n=0$ the right hand side of (\ref{da}) equals zero, which means that the
integrated probability of diquark fragmentation into the heavy baryon does not
change during the evolution, and it is determined by the initial fragmentation
function calculated perturbatively \cite{ScLep}. 

The solution of equation (\ref{da}) has the form 
\begin{equation}
a_{(n)}(\mu) = a_{(n)}(\mu_0)\; \biggl[\frac{\alpha_s(\mu)}
{\alpha_s(\mu_0)}\biggr]^{\frac{16}{3\beta_0}
\bigg[\frac{1}{2}+\ldots +\frac{1}{n+1}\biggr]}, 
\label{a}
\end{equation}
where one has used the one-loop expression for the QCD coupling constant
$$
\alpha_s(\mu) = \frac{2\pi}{\beta_0\ln(\mu/\Lambda_{QCD})}, 
$$
where $\beta_0= 11-2 n_f/3$, with $n_f$ being the number of quark flavors with
$m_q<\mu<m_{diq}$. 

Relation (\ref{a}) is universal one, since it is independent of whether the
diquark is free or bound at the virtualities less than $\mu_0$. In this paper
we take into account the evolution for the fragmentation into the heavy baryon.
The diquark can lose about 20\% of its momentum before the
hadronization \cite{ScLep}.

\subsubsection{Integrated probabilities of fragmentation}

\begin{figure}[th]
\setlength{\unitlength}{0.8mm}
\begin{center}
\begin{picture}(110,90)
\put(0, 0){\epsfxsize=10cm \epsfbox{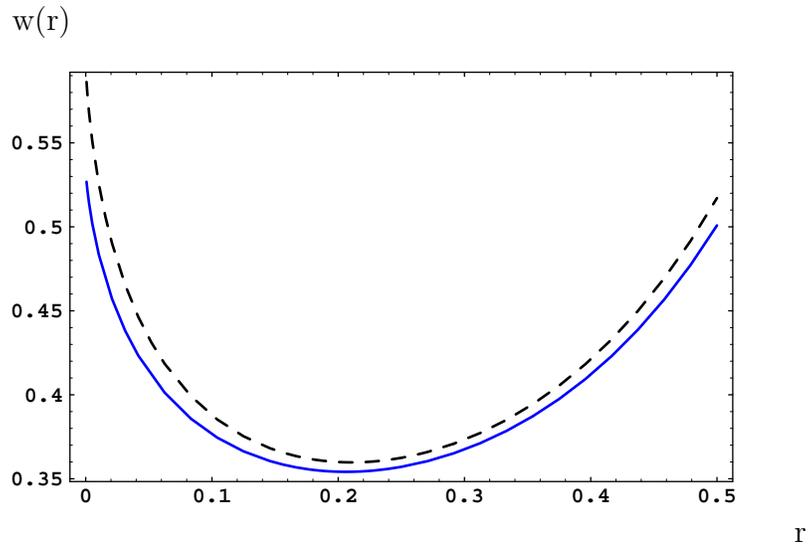}}
\put (130, 5){r}
\put (0, 90){$\rm{w(r})$}
\end{picture}
\end{center}
\caption{The w functions for the diquark fragmentation into the heavy
baryon versus the fraction $r=m_Q/M$. The dashed curve corresponds to the case
of $\ae=-1$, while the solid line does to $\ae=-1+3M^2/(s-m_{diq}^2)$.}
\label{w-fig}
\end{figure}

As has been mentioned above, the evolution conserves the integrated probability
of fragmentation which can be calculated explicitly
\begin{eqnarray}
\int dz\; D(z) &=& \frac{8\alpha_s^2}{81\pi}\; \frac{|R(0)|^2}{16 m_Q^3}\; 
{\rm w}(r),
\end{eqnarray}
so that
\begin{equation}
{\rm w}(r)=  \frac{16[(8+15r-60r^2+100r^3-60r^4-3r^5)+
30r(1-r+r^2+r^3)\ln r]}{15(1-r)^7}. 
\end{equation}
The ${\rm w}(r)$ functions are shown in Fig. \ref{w-fig} at low $r$. 

Thus, we have considered the dominant mechanism for the production of bound
spin $1/2$ states composed by a local color-triplet vector field with a heavy
anti-quark for high energy processes at large transverse momenta, where the
fragmentation contributes as the leading term. We have investigated two cases
for the behaviour of anamalous chromo-magnetic moment. At $\ae =-1$ we
observe\footnote{The expression for the fragmentation function diverges
logarithmically at a constant value of anamalous chromo-magnetic moment if
$\ae$ is not equal to $-1$.} that the obtained fragmentation function coincides
with that for the scalar diquark up to a factor. In the infinitely heavy
diquark limit, $D(z)$ has the form, which agrees with what expected from
general consideration of $1/m$-expansion for the fragmentation function. The
distribution of bound state over the transverse momentum with respect to the
axis of diquark fragmentation is calculated in the scaling limit of
perturbative QCD. The hard gluon corrections caused by the splitting of vector
diquark has been taken into account in the perturbation theory, that has led to
the corresponding one-loop equations of renormalization group for the moments
of fragmentation function (see  (\ref{da}),  (\ref{a})). 

The numerical estimates show that the probabilities of fragmentation into the
bound states containing the heavy vector diquark with the mass from 3 to 10 GeV
depend on the effective mass of quark entering the baryon together with the
diquark. In this way the ratio of yields for the baryons with the strangeness
and without it, is approximately equal to $\sigma(\Omega_{QQ'}) /
\sigma(\Xi_{QQ'}) \approx 0.2$. Certainly, the introduction of light and
strange quarks into the consideration makes the obtained results to be not
strictly justified, since the calculations suggest that the constituent masses
lead to the effective and correct description of dominant contributions caused
by the infrared dynamics. However, we can use the perturbative expressions as
models for the fragmentation into the hadrons containing light and strange
quarks because in such processes we describe the "fast" valence degrees of
freedom in the baryon. In the limit of small dispersion these degrees of
freedom can be approximated by the introduction of ratio for the assigned
fractions of longitudinal moment for the parton inside the hadron, while we do
not provide the consideration of contribution by a soft sea of light quarks and
gluons.

Another approach to the fragmentation mechanism for the production of doubly
heavy baryons was considered in ref. \cite{saleev}, wherein the perturbative
formfactors of doubly heavy and heavy-light diquarks were calculated and the
heavy quark fragmentation function into the baryons was derived in the process
with the production of vector diquark pairs: $Q\to H_{Q(Q'q)_{diq}}+ (\bar
Q'\bar q)_{diq}$. In this way the estimates strictly mean the minimal
expectations, since they are based on the elastic formfactor of diquark. We
also believe that the hierarchy of scales $m_Q\gg m_Q\cdot v\gg \Lambda_{QCD}$
in the relevant strong interactions provides that after the hard production of
heavy quark (the virtualities about $m_Q$) quite a fast forming of the doubly
heavy diquark occures (the virtualities of the order of $m_Q\cdot v$), while a
more slow hadronization of diquark into the baryons (the virtualities sloe to
$\Lambda_{QCD}$) takes place in the final stage. V.Saleev analyzed the
fragmentation into the thriply charmed baryon $\Omega_{ccc}$ \cite{ccc} due to
the cascade process of quark into the diquark and diquark into the baryon in
comparison with the direct fragmentation of quark into the baryon due to the
elastic production of vector diquark. Unfortumately, the general conclusion of
\cite{ccc} on the significant dominance of direct fragmentation is not correct,
because, first, the use of elastic formfactor for the vector diquark in the
cascade process leads to the double counting of suppressing factor due to the
charmed quark fusion into the diquark in the fragmetation of quark into the
diquark as well as in the elastic formfactor, where one again took the
projection of initial quark state on the bound diquark. This procedure gives
the overcoming factor of suppression $\alpha_s^2 |\Psi_{cc}(0)|^2/m_c^3 \sim
10^{-3}$. Second, the idea on the forming the $cc$-diquark with the consequent
hard production of charmed quark off diquark is not correct because the time
period (the diquark size) for forming the diquark is much greater than the time
interval (the compton length) for the production of charmed quark.

\subsection{Hadronic production}
Recent years are marked by a rapid increase of charmed particles observed in
modern experiments. So, the study of about $10^6$ charmed particles is expected
at fixed target FNAL facilities of E831 and E781. An increase of this value by
two orders of magnitude is proposed in experiments of next generation. Along
with standard problems of CP-violation in the charmed quark sector and a
measuring of rare decays etc., an investigation of processes with more than one
$c\bar c$-pair production becomes actual. The production of  additional $c\bar
c$-pair strongly decreases a value of  cross section for  such processes. This
fact must be especially taken into account in fixed target experiments, where
the quark-partonic luminosities are strongly suppressed in the region of  heavy
mass production.

An interesting process of the mentioned kind is the doubly charmed baryon
production. The doubly charmed $\Xi_{cc}^{(*)}$-baryon represents an absolutely
new type of objects in comparison with the ordinary baryons containing light
quarks only. The ground state of such baryon is analogous to a $\bar Q
q$-meson,
which contains a single heavy anti-quark $\bar Q$ and a light quark $q$. In the
doubly heavy baryon the role of heavy anti-quark is played by the $cc$-diquark,
which is in anti-triplet color state \cite{1c}. It has a small size in
comparison with the scale of the light quark confinement. The study of
mechanism for the production of these states are of interest. The $ccq$-baryon
production was discussed in \cite{falk,d31,d18,6c,7c}. The main problem of
calculations is reduced to an evaluation of the production cross section for
the diquark in the $\bar 3$-triplet color state. One assumes further that the
$cc$-diquark nonperturbatively transforms into the $ccq$-baryon with a
probability close to unit.

\begin{figure}[th]
\hspace*{3.5cm}
\epsfxsize=10cm \epsfbox{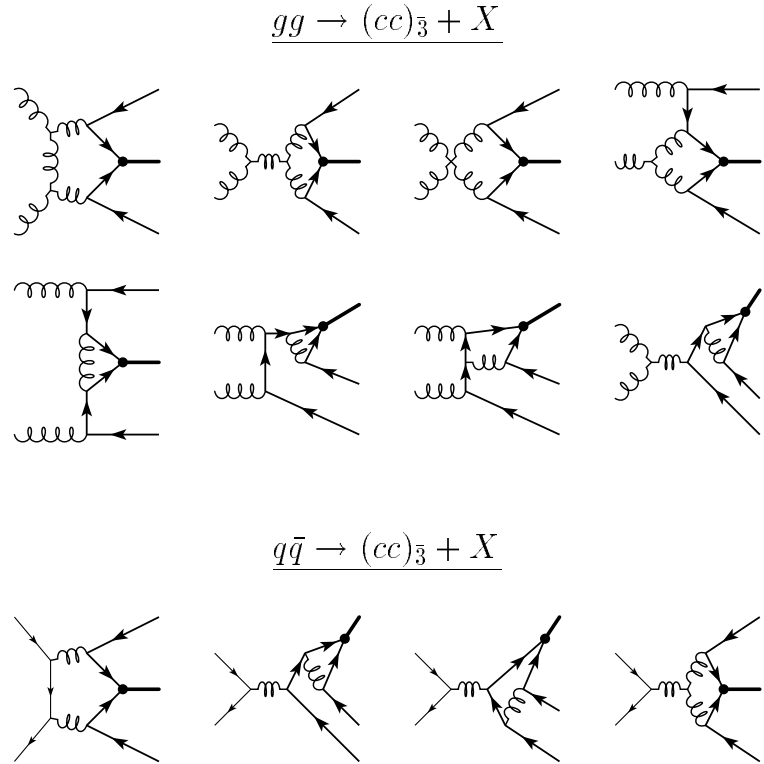}

\caption{The examples of diagrams for the gluon-gluon and quark-antiquark
production of $cc$-diquark. The initial quarks are denoted by the thin fermion
lines, the final quarks are denoted by the bold fermion lines and the gluons
are denoted by the helical lines.}
\label{ccf1}
\end{figure}

The hadronic production of diquark is subdivided into two parts. The first
stage is the hard production of two $c\bar c$-pairs  in the processes of $gg
\to c \bar c c \bar c$ and $q \bar q \to c \bar c c \bar c$ described by the
Feynman diagrams in the fourth order of $\alpha_s$ coupling constant in the
perurbative QCD (see Fig. \ref{ccf1}).

The second step is the nonperturbative fusion of two $c$-quarks with a small
relative momentum into the $cc$-diquark. For the $S$-wave states this process
is characterized by  the radial wave function at the origin, $R(0)$.

The main difference between the existing evaluations of the doubly charmed
baryon cross section consists in the methods used for the hard subprocess
calculation. In \cite{8c} a part of diagrams connected with the
$c$-fragmentation  into the $(c c)$-diquark is only used instead of the
complete set of diagrams. As was shown in \cite{6c} this estimation is not
absolutely correct, because it becomes true only at large transverse momenta
greater than $p_T^{min} >35$ GeV, where the fragmentation mechanism is
dominant. In other kinematical regions the application of fragmentational
approximation is not justified and it leads to wrong results, especially at
$\sqrt{\hat s}$ being not much greater than $p_T^{min}$. 
     
However, even after taking into account the complete set of diagrams considered
in \cite{6c,7c}, essential uncertainties in the estimations of the
$ccq$-baryon production remain. The basic parameters determining these
uncertainties are the values of $\alpha_s$, $m_c$ and $R_{cc}(0)$. In addition,
it is not clear, to what extent the hypothesis on the hadronization of
$cc$-diquark into the $ccq$-baryon with the unit probability is correct or not.
The matter is that the interaction between the diquark and gluons is not
suppressed in contrast to the $c \bar c$-pair production in the color singlet
state, when the quarkonium dissociation supposes exchange with the
quark-gluonic sea by two hard gluons with virtualities, which are greater than
the inverse size of quarkonium.

A decrease of the uncertainty in the $ccq$-baryon cross section would be
possible by means of comparing the process of baryon production with the
analogous process of $J/\Psi + D \bar D$ production\footnote{We calculate the
$J/\psi+c\bar c$ production and assume that the $c\bar c$-pair transforms to
$D\bar D+$ some light hadrons with the probability very close to unit. So, we
neglect the production of charmed baryons as well as bound states of
charmonium in the hadronization of associated $c\bar c$ pair.}. The associated
production of $J/\Psi$ is described by practically
the same diagrams of fourth order with the well-known wave function of $J/\Psi$
at the origin\footnote{The value of $|R_{\Psi}(0)|$ is determined by the width
of leptonic decay, $J/\Psi\to l^+l^-$ with taking into account the hard gluonic
correction, so, numerically, $|R_{\psi}(0)|=\sqrt{\pi M/3}\tilde f_{\psi}$,
where $\tilde f_{\psi}=540$ MeV.}. In this way of connection to the
$J/\Psi+D\bar D$ process we could remove the part of uncertainties following
from $\alpha_s$ and $m_c$ in the $cc$-diquark production process.

In the following sections of the paper the joint cross section calculations of
these processes in $\pi^-p$ and $pp$ interactions are performed, the
description of production models for the $ccq$-baryons and $J/\Psi +D \bar D$
is given. We presents the calculation results for the production cross section
of $ccq$-baryons and $J/\Psi +D \bar D$ in the fixed target experiments E781
and HERA-B as well as at the collider energies of Tevatron and LHC. We
discuss possibilities of searching for the baryons $\Xi_{cc}^{(*)}$. 

\subsubsection{Production mechanism}
As was mentioned, we suppose that the diquark production can be subdivided by
two stages. On the first stage the production amplitude of four free quarks is
calculated for the following processes
\begin{eqnarray}
&& gg \to cc \bar c \bar c, \nonumber \\
&& q\bar q \to cc \bar c \bar c. \nonumber
\end{eqnarray}

The calculation technique applied in this work is analogous to that for the
hadronic production of $B_c$ \cite{d30,9c}, but in this case the bound state is
composed by two quarks ($Q_1$ and $Q_2$) instead of the quark and anti-quark
\cite{d31,d18,6c}.

One assumes that the binding energy in the diquark is much less than the masses
of constituent quarks and, therefore, these quarks are on the mass shells. So,
the quark four-momenta are related to the $(Q_1 Q_2)$ diquark momentum $P_{(Q_1
Q_2)}$ in the following way:
\begin{equation}
p_{Q_1}=\frac{m_{Q_1}}{M_{(Q_1 Q_2)}}P_{(Q_1 Q_2)}
\;,\;\; \quad p_{Q_2}=\frac{m_{Q_2}}{M_{(Q_1 Q_2)}}P_{(Q_1 Q_2)},
\label{kin}
\end{equation}
where $M_{(Q_1 Q_2)}=m_{Q_1}+m_{Q_2}$ is the diquark mass,
$m_{Q_1}, m_{Q_2}$ are the quark masses.

In the given approach the diquark production is described by 36 Feynman
diagrams of the leading order, corresponding to the production of four free
quarks with the combining of two quarks into the color anti-triplet diquark
with the given quantum numbers over the Lorentz group. The latter procedure is
performed by means of the projection operators
\begin{equation}
\displaystyle
{\cal N}(0,0)=
\sqrt{\frac{2M_{(Q_1 Q_2)}}{2m_{Q_1}2m_{Q_2}}}
\frac{1}{\sqrt{2}}\{ \bar u_1(p_{Q_1},+)\bar u_2(p_{Q_2},-)
- \bar u_1(p_{Q_1},-)\bar u_2(p_{Q_2},+) \}
\end{equation}
for the scalar state of diquark (the corresponding baryon is denoted by
$\Xi_{Q_1 Q_2}'(J =1/2)$);
\begin{eqnarray}
\displaystyle
{\cal N}(1,-1) &=&
\sqrt{\frac{2M_{(Q_1 Q_2)}}{2m_{Q_1}2m_{Q_2}}}
\bar u_1(p_{Q_1},-)\bar u_2(p_{Q_2},-), \nonumber\\
{\cal N}(1,0) &=&
\sqrt{\frac{2M_{(Q_1 Q_2)}}{2m_{Q_1}2m_{Q_2}}}
\frac{1}{\sqrt{2}}\{ \bar u_1(p_{Q_1},+)\bar u_2(p_{Q_2},-) +
\bar u_1(p_{Q_1},-)\bar u_2(p_{Q_2},+) \}, \nonumber\\
{\cal N}(1,+1) &=&
\sqrt{\frac{2M_{(Q_1 Q_2)}}{2m_{Q_1}2m_{Q_2}}}
\bar u_1(p_{Q_1},+)\bar u_2(p_{Q_2},+) 
\end{eqnarray}
for the vector state of diquark (the baryons are denoted as  $\Xi_{Q_1 Q_2}(J
=1/2)$ and $\Xi_{Q_1 Q_2}^*(J =3/2)$).

To produce the quarks, composing the diquark in the $\bar 3_c$ state, one has
to introduce the color wave function as $\varepsilon_{ijk}/\sqrt{2}$, into the
diquark production vertex, so that $i=1,2,3$ is the color index of the first
quark, $j$ is that of the second one, and $k$ is the color index of diquark.
For the identical quarks $Q_1=Q_2$ possessing the momenta equal to each other,
the anti-triplet color state can have the summed spin $S=1$, only.

The diquark production amplitude $A^{Ss_z}_k$ is expressed through the
amplitude $T^{Ss_z}_k(p_i)$ for the free quark production in kinematics
(\ref{kin}) under the substitution for the product $\bar u_1 \bar u_2$ by the
projection operators as well as under the condition that two heavy quarks in
the $\bar 3$-color state, so that
\begin{equation}
A^{Ss_z}_k=\frac{R_{Q_1 Q_2}(0)}{\sqrt{4\pi}}T^{Ss_z}_k(p_i),
\end{equation}
where $R_{Q_1 Q_2}(0)$ is the diquark radial wave function at the origin, $k$
is the color state of diquark, $S$ and $s_z$ are the diquark spin and its
projection on the $z$-axis, correspondingly.

In numerical calculations under discussion we suppose the following values of
parameters:
\begin{eqnarray}
&&\alpha_s=0.2,\;\;\;\;
m_c=1.7\; {\rm GeV},\;\;\;m_b=4.9\; {\rm GeV}, \nonumber \\
&&R_{cc(1S)}(0)=0.601\; {\rm GeV}^{3/2}, \;\;\;
R_{bc(1S)}(0)=0.714\; {\rm GeV}^{3/2}, 
\end{eqnarray}
where the value of $R_{cc}(0)$ has been calculated by means of numerical
solution of the Schr\"odinger equation with the Martin potential  multiplied by
the 1/2 factor caused by the color anti-triplet state of quarks instead of the
singlet one.

To calculate the production cross section of diquarks composed of two
$c$-quarks, one has to account for their identity. One can easily find, that
the anti-symmetrization  over the identical fermions leads to the scalar
diquark amplitude equal to zero, and it results in the amplitude of the vector
$cc$-diquark production being obtained by the substituting of equal masses in
the production amplitude of vector diquark composed of two quarks with the
different flavors, and taking into account the 1/2 factor for the identical
quarks and anti-quarks.

In this approach we suppose that the produced diquark forms the baryon with the
unit probability by catching up the light quark from the quark-antiquark sea at
small $p_T$ or having the fragmentation into the baryon at large $p_T$.

The typical diagrams of fourth order describing the parton processes 
are shown in Fig. \ref{ccf1}. One can subdivide them into two groups. The first
group contains the diagrams of fragmentation type, wherein the $c\bar c$-pair
emits another one. The second group corresponds to the independent dissociation
of gluons into the $c\bar c$-pairs with the consequent fusion into the diquark.
The diagrams of second group belong to the recombination type.  

As was mentioned above, the authors of some papers \cite{8c} restricted
themselves by the consideration of fragmentation diagrams, only. In this way
they reduced the cross-section formulae to the $c\bar c$-pair production
cross section multiplied  by the fragmentation function of $c$-quark into the
$cc$-diquark.

\begin{figure}[th]
\hbox to 1.5cm {\hfil\mbox{$N\, d\hat \sigma_{gg}^{cc}/dp_T$, pb/GeV}}
\hspace*{3cm} {\epsfxsize=10cm \epsfbox{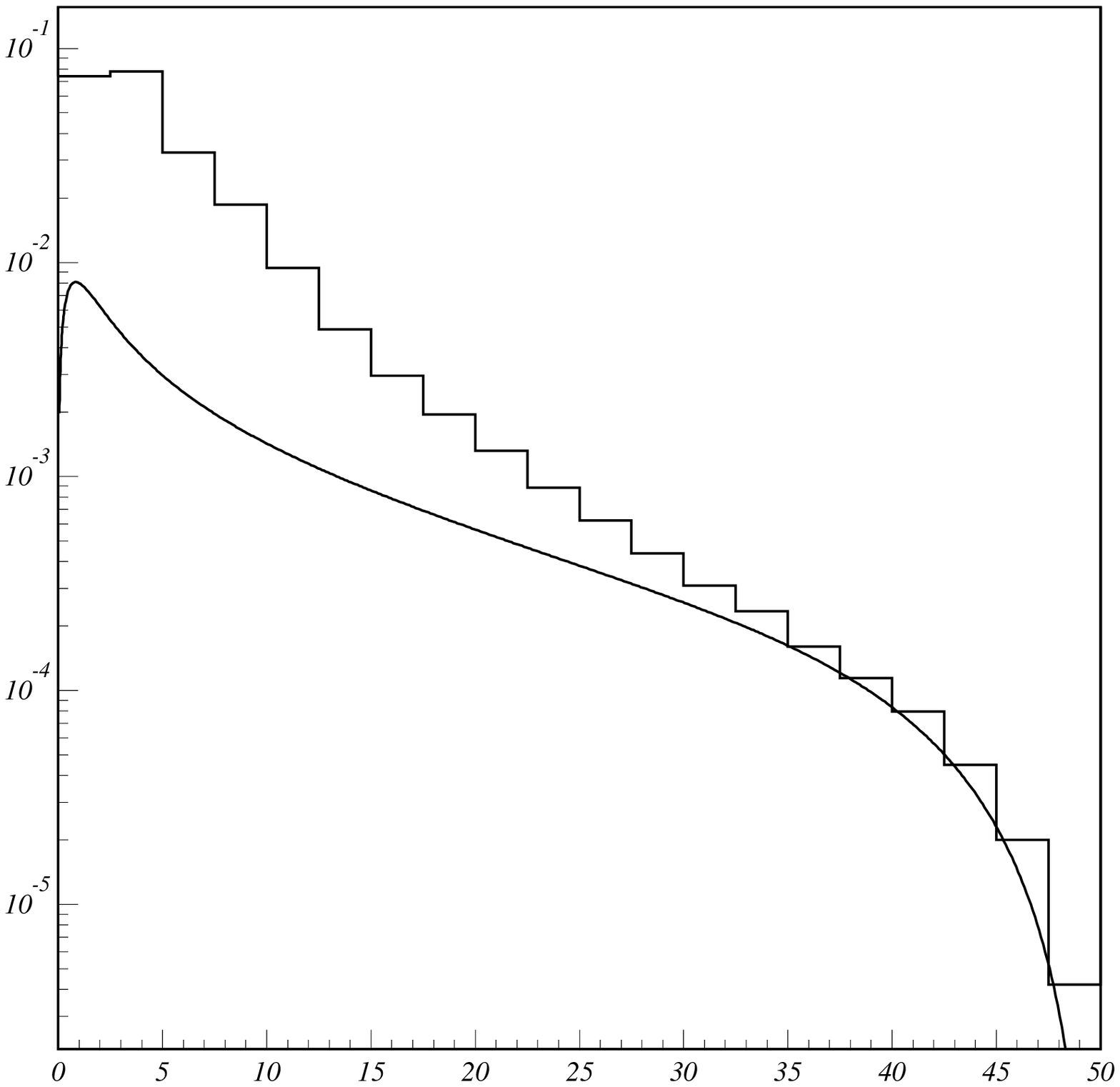}}

\vspace*{-0.5cm}
\hbox to 14.cm {\hfil \mbox{$p_T$, GeV}}
\caption{The differential cross section for the associated production of $cc$
diquarkin the gluon-gluon subprocess at 100 GeV (histogram) in comparison with
the prediction of fragmentation model (solid curve), correspondingly.}
\label{ccfr}
\end{figure}

As we shown in \cite{6c}, this approximation is correct only under the two
following conditions: $M^2_{QQ}\ll \hat s$ and $p_T  \gg M_{QQ}$. In other
kinematical regions, the contribution of recombination diagrams dominates.
The typical value of $p_T$, wherefrom the fragmentation begins to dominate in
the production of $сс$-diquark, is $p_T > 35$ GeV (see Fig. \ref{ccfr}). It is
clear, that at realistic $p_T$ one has to take into account all contributions
including the recombination. For the first time, the complete set of diagrams
was taken into account in \cite{6c} and confirmed in \cite{7c}. In the both
papers the calculations are performed only for the gluon-gluon production,
which is a rather good approximation at collider energies. For the fixed target
experiments the value of total energy strongly decreases, and, hence, the
values of energy in the parton subprocesses decrease too. The contribution of
quark-antiquark annihilation becomes essential at fixed target energies,
especially for the processes with initial valence anti-quarks. In the following
consideration we take into account the quark-antiquark annihilation into four
free charmed quarks in the estimation of yield for the doubly charmed baryon. 

\subsubsection{Doubly charmed baryon production in fixed target experiments}
\begin{figure}[th]
\hbox to 1.5cm {\hfil\mbox{$\hat \sigma_{gg}^{cc}$,
 $\hat \sigma_{gg}^{J/\Psi+D\bar D}$, pb}}
\hspace*{2.5cm} {\epsfxsize=10cm \epsfbox{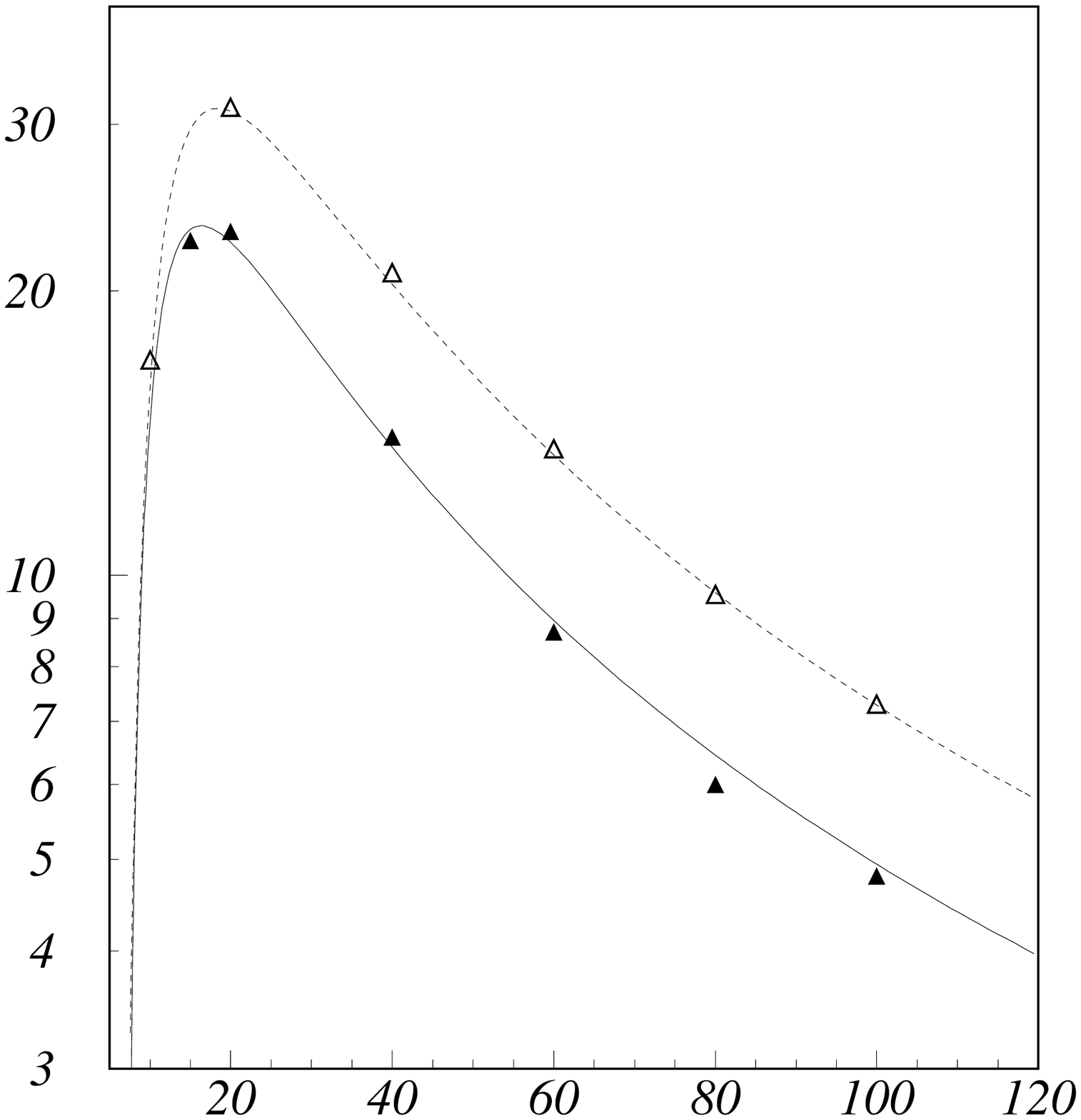}}

\vspace*{-1.cm}
\hbox to 14.cm {\hfil \mbox{$\sqrt{s_{gg}}$, GeV}}
\caption{The total cross sections for the gluon-gluon production of $cc$
diquark ({\scriptsize$\blacktriangle$}) and $J/\Psi+D\bar D$
({\tiny$\tiny\triangle$}) in comparison with the approximations of (\ref{ccgg})
and (\ref{cacgg}) (solid and dashed lines, correspondingly).}
\label{fig2}
\end{figure}

\begin{figure}[th]
\hbox to 1.5cm {\hfil\mbox{$\hat \sigma_{q \bar q}^{cc}$,
 $\hat \sigma_{q \bar q}^{J/\Psi+D\bar D}$, pb}}
\hspace*{2.5cm} {\epsfxsize=10cm \epsfbox{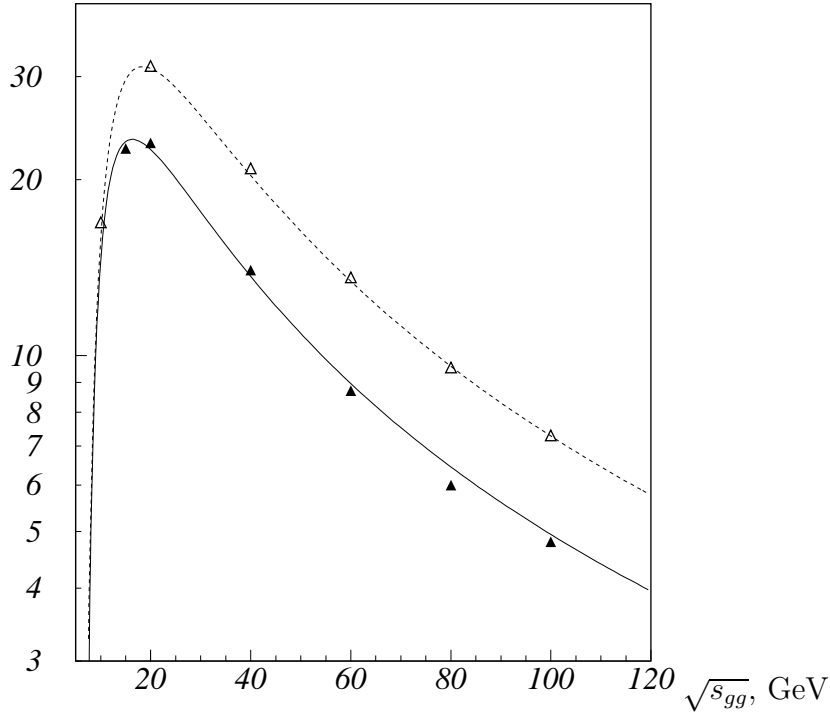}}

\vspace*{-1.cm}
\hbox to 14.cm {\hfil \mbox{$\sqrt{s_{q \bar q}}$, GeV}}
\caption{The total cross sections for the gluon-gluon production of $cc$
diquark ({\scriptsize$\blacktriangle$}) and $J/\Psi+D\bar D$
({\tiny$\tiny\triangle$}) in comparison with the approximations of
(\ref{ccqq}) and (\ref{cacqq}) (solid and dashed lines, correspondingly).}
\label{fig3}
\end{figure}

In Figs \ref{fig2} and \ref{fig3} the calculation results for the total
cross section of the diquark-production subprocesses versus the total energy
$\sqrt{\hat s}$ for the given values of $\alpha_s$, $m_c$ and $R_{cc}(0)$. We
have also shown  the parametrizations for the dependencies of total cross
sections versus the process energy for the $cc$-diquark as given by 
\begin{eqnarray}
&& \hat\sigma_{gg}^{(cc)}=213.\left (1-\frac{4m_c}{\sqrt{\hat s}}
\right )^{1.9} \left (\frac{4m_c}
{\sqrt{\hat s}}\right )^{1.35}\quad {\rm pb}, \label{ccgg}\\
&& \hat\sigma_{q\bar q}^{(cc)}=
206.\left (1-\frac{4m_c}{\sqrt{\hat s}}\right )^{1.8}
\left (\frac{4m_c} {\sqrt{\hat s}} \right )^{2.9}\quad {\rm pb}.
\label{ccqq}
\end{eqnarray}
We have to stress that the numerical coefficients  depend on the
model parameters, so that $\hat \sigma\sim \alpha_s^4|R(0)|^2/m_c^5$.

\begin{figure}[th]
\hbox to 1.5cm {\hfil\mbox{$d\hat \sigma_{gg}^{J/\Psi+c\bar
c}/dp_T$, pb/GeV}}
\hspace*{3cm} {\epsfxsize=10cm \epsfbox{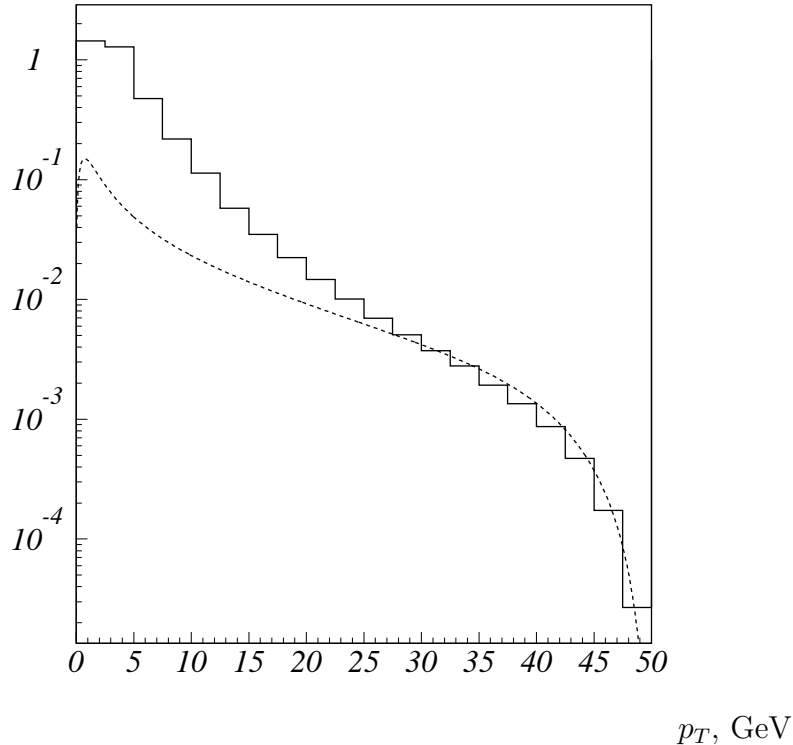}}

\vspace*{-0.5cm}
\hbox to 14.cm {\hfil \mbox{$p_T$, GeV}}
\caption{The differential cross section for the associated production of
$J/\Psi+c\bar c$  in the gluon-gluon subprocess at 100 GeV (solid
histogram) in comparison with the prediction of fragmentation model
(dashed curve), correspondingly.}
\label{add}
\end{figure}

As was mentioned, the production of $J/\Psi$ in the subprocesses of $gg \to
J/\Psi + c\bar c$ and $q\bar q \to J/\Psi +c\bar c$ is also calculated in this
work. The numerical results of such consideration are shown  in Figs.
\ref{fig2} and \ref{fig3}. The parameterization of these results versus the
energy $\sqrt{\hat s }$ are presented below
\begin{eqnarray}
&& \hat\sigma_{gg}^{J/\Psi}=518.\left (1-
\frac{4m_c}{\sqrt{\hat s}}\right )^{3.0} \left (\frac{4m_c}
{\sqrt{\hat s}} \right )^{1.45}\quad {\rm pb}, \label{cacgg}\\
&& \hat\sigma_{q\bar q}^{J/\Psi}=699.\left( 1- 
\frac{4m_c}{\sqrt{\hat s}}\right )^{1.9}
\left ( \frac{4m_c}{\sqrt{\hat s}} \right )^{2.97}\quad {\rm pb}.
\label{cacqq}
\end{eqnarray}
As well as for the associated $B_c+b\bar c$ and $\Xi_{cc}+\bar c\bar c$, we see
the following regularity for the partonic production of $J/\psi+c\bar c$: the
fragmentation regime occures at $p_T>25-30$ GeV. This fact can be certainly
observed in Fig. \ref{add} for the differential cross section of $gg\to
J/\psi+c\bar c$ at $\sqrt{\hat s}=100$ GeV. Thus,  for the associated
production of $J/\psi+c\bar c$ and $\Xi_{cc}+\bar c \bar c$ the fragmentation
works at $p_T\gg m_c$.

\begin{figure}[th]
\hbox to 1.5cm {\mbox{$ \sigma_{\pi^- p}^{cc}$,
$\hat \sigma_{\pi^- p}^{J/\Psi+D\bar D}$, nb}
\hspace*{4cm} \mbox{$ \sigma_{p p}^{cc}$,
 $\hat \sigma_{p p}^{J/\Psi+D\bar D}$, nb}}
\hbox to 0cm {\epsfxsize=7cm \epsfbox{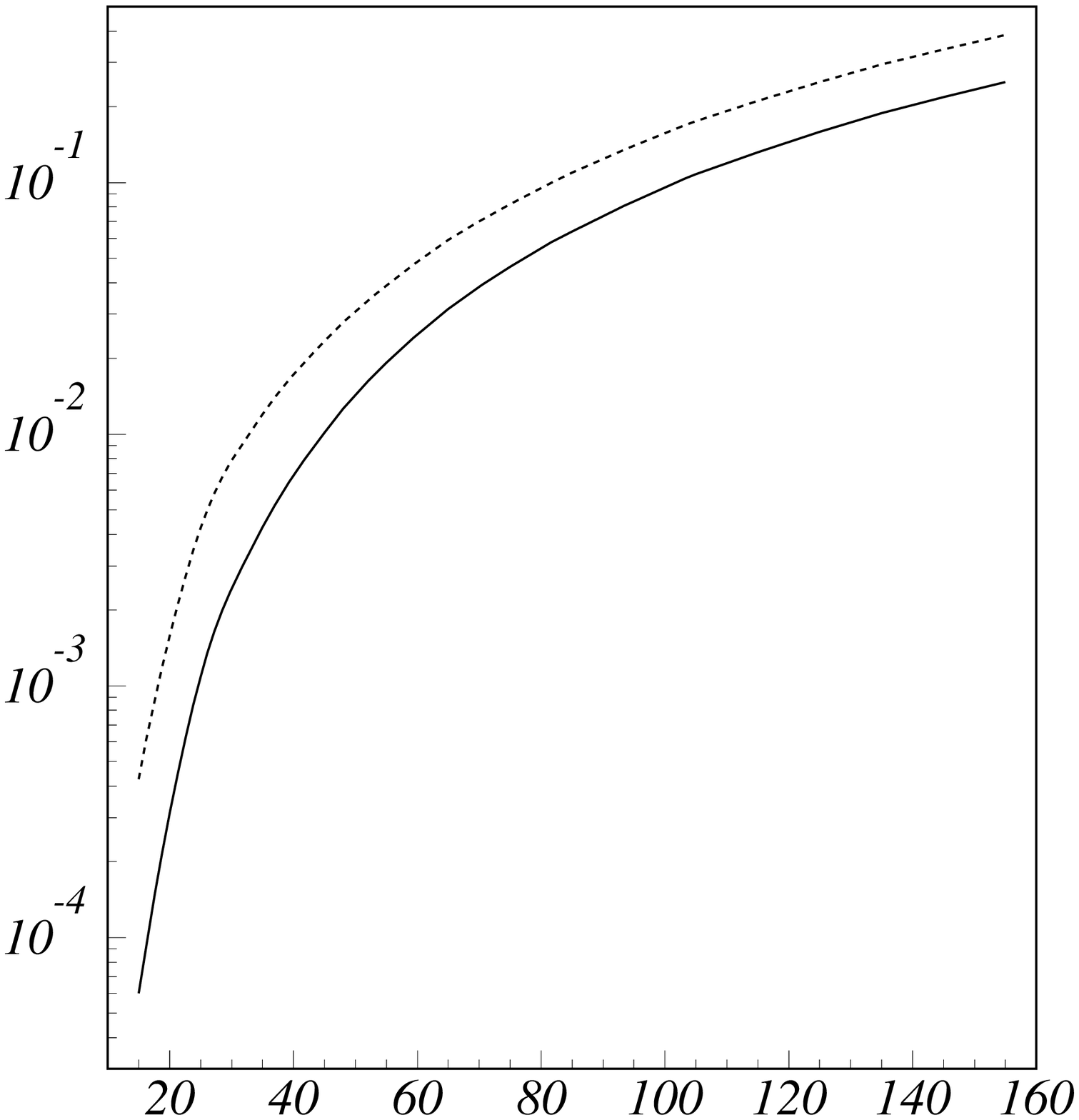}
\hspace*{1cm}
\epsfxsize=7cm \epsfbox{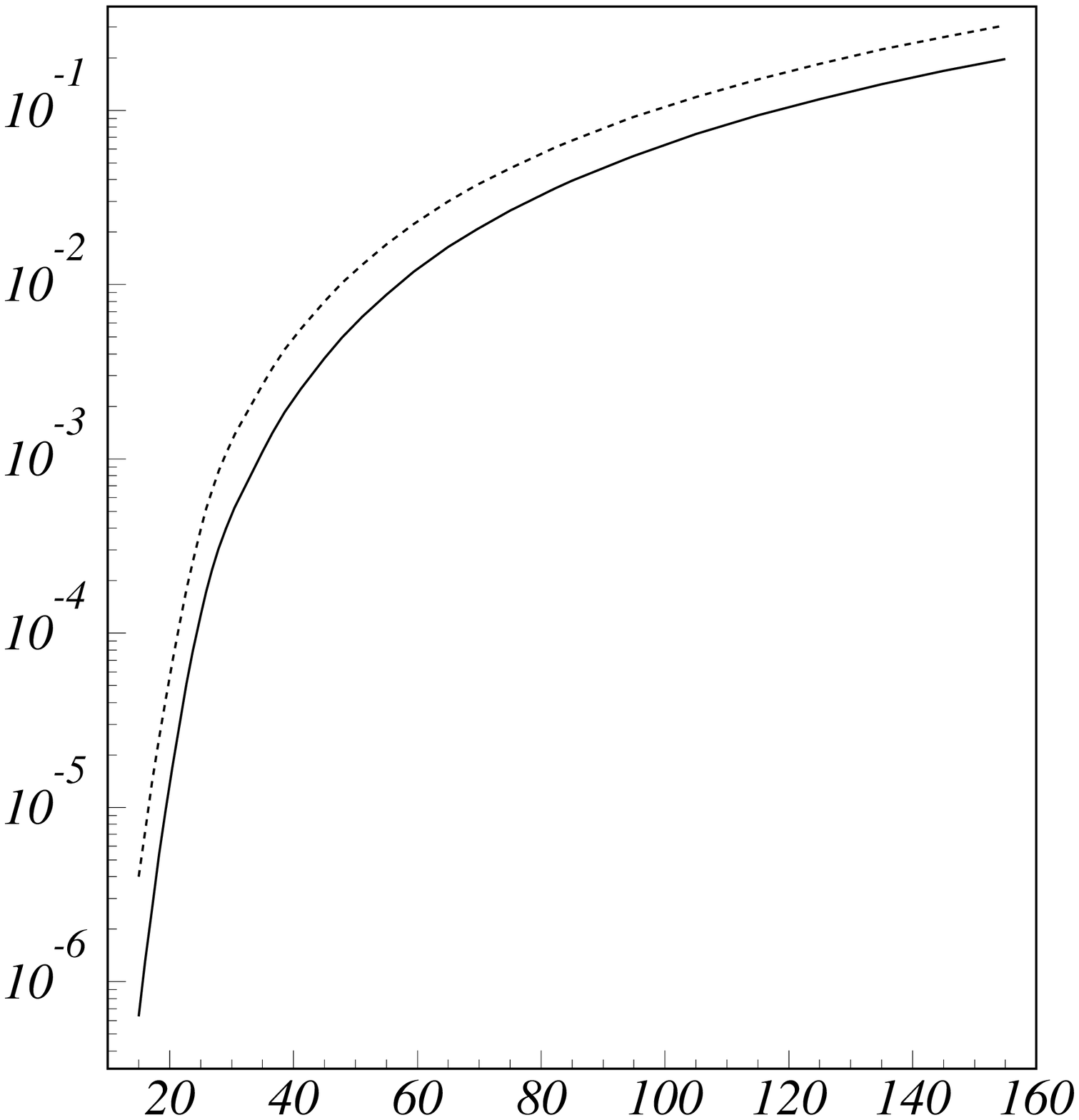}}

\hspace*{5cm} {\mbox{$\sqrt{s_{\pi^- p}}$, GeV}\hspace*{6cm}
\mbox{$\sqrt{s_{pp}}$, GeV}}
\caption{The total cross sections for the production of $cc$-diquark and
$J/\Psi+D\bar D$ (solid and dashed curves, correspondingly) in $\pi^- p$ and
$pp$ collisions.}
\label{pip}
\end{figure}

The above parametrizations quite accurately reconstruct the results of precise
calculations at $\sqrt{\hat s}<150$ GeV, and that is why they can be used for
the approximate estimation of total hadronic production cross section for the
$cc$-diquark and $J/\Psi$ by means of their convolution with the partonic
distributions
\begin{equation}
\sigma=\sum_{i,j}\int dx_1 dx_2 f_{i/A}(x_1,\mu) f_{j/B}(x_2,\mu)
\hat\sigma ,
\end{equation}
where $f_{i/A}(x,\mu)$ is the distribution of $i$-kind parton in the
$A$-hadron. The parton distributions  used for the proton are the CTEQ4 
parameterizations \cite{10c}, and those of used for the $\pi^-$-meson are the
Hpdf ones \cite{11c}. In both cases the virtuality scale is fixed at 10 GeV. As
for the choice of fixed scale in the structure functions, this is caused by the
fact that the cross section of subprocesses is integrated in the region of low
$\hat s$ close to the fixed scale, so that the account of "running" scale
weakly changes the estimate of $\Xi_{cc}$-baryon yield in comparison with the
mentioned uncertainty of diquark model\footnote{We have found the
scale-dependent variation to be at the level of $\delta \sigma/\sigma \sim $
10\%.}. The cross sections convoluted with the gluon and quark luminosities are
presented in Fig. \ref{pip} for both the $cc$-diquark and $J/\Psi$ production
in $\pi^-p$ and $pp$ collisions. 

As we can see in these figures, the cross section of $cc$-diquark as well as
the cross section of $J/\Psi+c\bar c$ are strongly suppressed at low energies
in comparison with the values at the collider energies. The ratio for the
$cc$-diquark production and total charm production is
$\sigma_{cc}/\sigma_{charm} \sim 10^{-4} -10^{-3}$  in  the collider
experiments and $\sim 10^{-6} -10^{-5}$ in the fixed target experiments. The
same situation is observed for the hadronic $J/\Psi+D \bar D$ production.

The distributions for the $ccq$-baryon and $J/\Psi+D \bar D$
production are shown in \cite{hprod} for the $\pi^-p$-interaction at 35 GeV and
for the $pp$-interaction at 40 GeV, correspondingly. The rapidity distributions
point to the central character of $ccq$-baryon production as well as $J/\Psi
+D\bar D$. The $p_T$-distributions  of  these processes are alike to each other
also\footnote{We assume that at the given energies the $cc$-diquark has no
fragmentational transition into the  baryon, but it catches up the light quark
from the quark-antiquark pair sea.}. The form of differential distributions
leads to the conclusion that the process of $J/\Psi +D\bar D$ production can be
used to normalize the estimate of $ccq$-baryon yield, wherein the following
additional uncertainties appear: 
\begin{enumerate}
\item
the unknown value of $|R_{(cc)}(0)|^2$,
\item
uncertainties related with the hadronization of $cc$-diquark.
\end{enumerate}

We see from the given estimates that in the experiments  with the number of
charmed events at the level about $10^6$ (for example, in the E781 experiment,
where $\sqrt{s}=35$ GeV), we have to expect about one event with the doubly
charmed baryon.  The situation is more promising and pleasant for the
$pp$-interaction at 800 GeV (HERA-B). The considered processes yield about
$10^5\; \Xi_{cc}^{(*)}$-baryons and a close number  of $J/\Psi +D\bar D$ in the
experiment specialized for the detection of about $10^8$ events with the
$b$-quarks.

\subsubsection{Production of $ccq$-baryon at colliders}

As one can see in the previous Section, the observation of
$\Xi_{cc}^{(*)}$-baryons  presents a rather difficult problem in the
experiments specialized for the study of charmed particles.  As a rule, such
experiments are carried out at fixed targets, so that the effective value of
subprocess energy is strongly decreased. So, the relative contribution of
doubly charmed baryons into the total charm yield is of the order of
$10^{-6}-10^{-5}$. The production of $ccq$-baryons at colliders with large
$p_T$ is more effective. In this case the cross section is determined by the
region of quark-antiquark and gluon-gluonic energy, where the threshold effect
becomes negligible  and the partonic luminosities are quite large at $x\sim
M/\sqrt{s}$. So, the suppression factor with respect to the single production
of $c\bar c$-pairs is much less and it is in the range of $10^{-4}-10^{-3}$.

In \cite{hprod} the $p_T$-distributions for $\Xi_{cc}^{(*)}$ and $J/\Psi$
associated with $D$ and $\bar D$ are shown at the energies of Tevatron and LHC
with the rapidity cut $|y|<1$.

One can easily understand that the presented $\Xi_{cc}^{(*)}$ cross sections
are the upper estimates for the real cross sections because of the possible
dissociation of  heavy diquark into the $D D$-pair.

Further, even if the $cc$-diquark being the color object, transforms into the
baryon with the unit probability, one has to introduce the fragmentation
function describing  the hadronization of diquark into the baryon at quite
large $p_T$ values. The simplest form of this function can be chosen by the
analogy with that for the heavy quark
\begin{equation}
D(z)\sim \frac{1}{z}\frac{1}{(M^2-\frac{m_{cc}^2}{z}-\frac{m_q^2}{1-z})^2},
\label{d}
\end{equation}   
where $M$ is the mass of  baryon $\Xi_{cc}^{(*)}$, $m_{cc}$ is the mass of
diquark, $m_q$ is the mass of light quark (we suppose it to be equal to 300
MeV). This function practically repeats the form of fragmentation function if
diquark into the baryon, that has been derived above in this chapter in the
framework of perturbative QCD. In \cite{hprod} the $p_T$-distributions of
doubly charmed baryon production are calculated with the use of (\ref{d}). One
has to mention, that in the leading order over the inverse heavy quark mass,
the relative yield of $\Xi_{cc}$ and $\Xi_{cc}^*$ is determined by the simple
counting rule for the spin states, and it equals $\sigma (\Xi_{cc}):\sigma
(\Xi_{cc}^*)=1:2$. In this approach one does not take into account a possible
difference between the fragmentation functions for the baryons with the
different spins. The corresponding difference is observed in the perturbative
fragmentation functions for the heavy mesons and quarkonia \cite{d9,Bra-frag}.

\subsubsection{Hadronic production of $\Xi_{bc}$}

\begin{figure}[th]
\hbox to 2.5cm {\hfil\mbox{$\hat \sigma_{gg}^{\Xi_{bc}}$, pb}}
\hspace*{2.5cm} {\epsfxsize=10cm \epsfbox{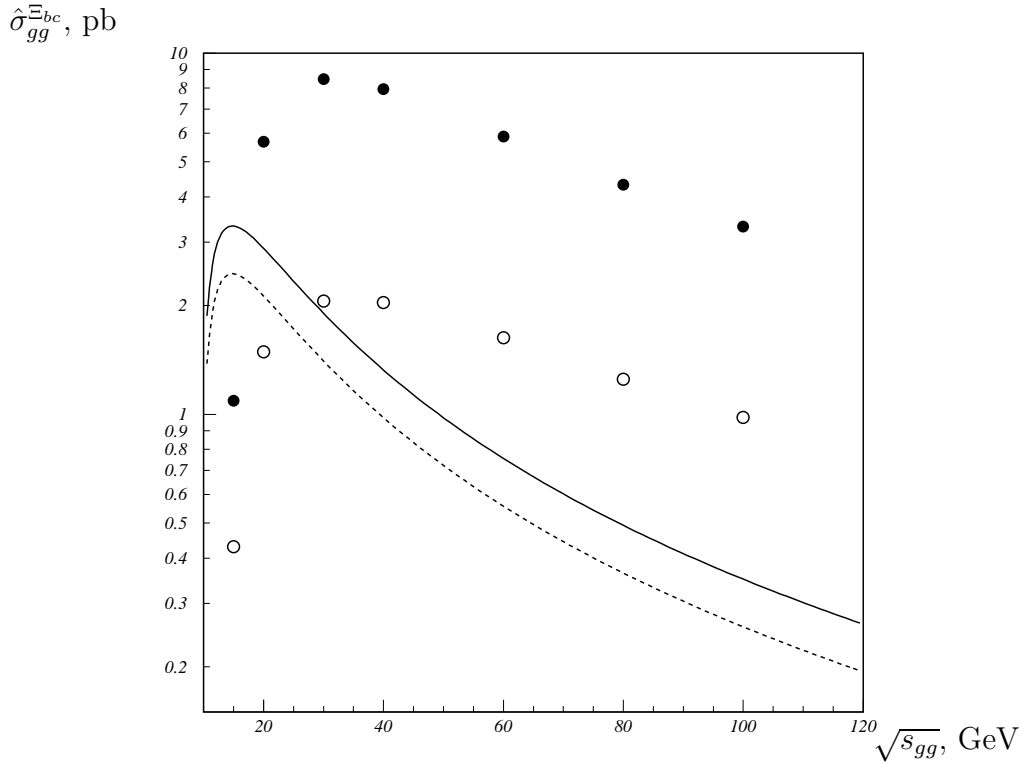}}

\vspace*{-1.cm}
\hbox to 14.5cm {\hfil \mbox{$\sqrt{s_{gg}}$, GeV}}
\caption{The gluonic production cross sections of $\Xi_{bc}'$ ($\circ$) and
$\Xi_{bc}^{(*)}$ ($\bullet$) in comparison with the predictions of the
fragmentational mechanism for $\Xi_{bc}'$ (dashed line) and $\Xi_{bc}^{(*)}$
(solid line).}
\label{bcgg}
\end{figure}

The total energy dependence of the gluonic production cross sections of
$\Xi_{bc}'$ ($\circ$) and $\Xi_{bc}^{(*)}$ ($\bullet$) baryons is shown in Fig.
\ref{bcgg}. For the sake of comparison, the predictions of the fragmentational
mechanism for  $\Xi_{bc}^{(*)}$ (solid line) and $\Xi_{bc}'$ (dashed line) are
also presented. One can see from the figure, that the fragmentational
production mechanism assuming the validity of factorization in the cross
section at $ M^2/s\ll 1$ and at large transverse momenta in accordance with the
formula
\begin{equation}
\frac{d\sigma_{gg\rightarrow\Xi_{bc}'(\Xi_{bc}^{(*)}) \bar b \bar c}}
{dz}=\sigma_{gg \rightarrow b \bar b } \cdot 
D_{b \rightarrow \Xi_{bc}'(\Xi_{bc}^{(*)})}(z),  
\end{equation}
where $z=2|\vec P|/\sqrt{s}$, does not work at low gluon energies, where it 
overestimates the cross section, because of the incorrect evaluation of the
phase space, and it is also not valid at large energies, where the predictions
of the fragmentational mechanism are essentially less that the exact results.

So, the fragmentational values underestimate the $\Xi_{bc}^{(*)}$ and
$\Xi_{bc}'$ cross sections by 10 and 3 times, respectively, at $\sqrt{\hat
s}=100$ GeV. When the fragmentational predictions give the ratio
$\sigma_{\Xi_{bc}^{(*)}}/ \sigma_{\Xi_{bc}'}\simeq 1.4$, the exact perturbative
calculations result in $\sigma_{\Xi_{bc}^{(*)}}/ \sigma_{\Xi_{bc}'}\simeq 3.5$
even at $\sqrt{\hat s}=100$ GeV.

\begin{figure}[th]
\hbox to 2.5cm {\hfil\mbox{$\frac{d\hat \sigma_{gg}^{\Xi_{bc}}}{d\, p_T}$,
pb/GeV}}
\hspace*{2.5cm} {\epsfxsize=10cm \epsfbox{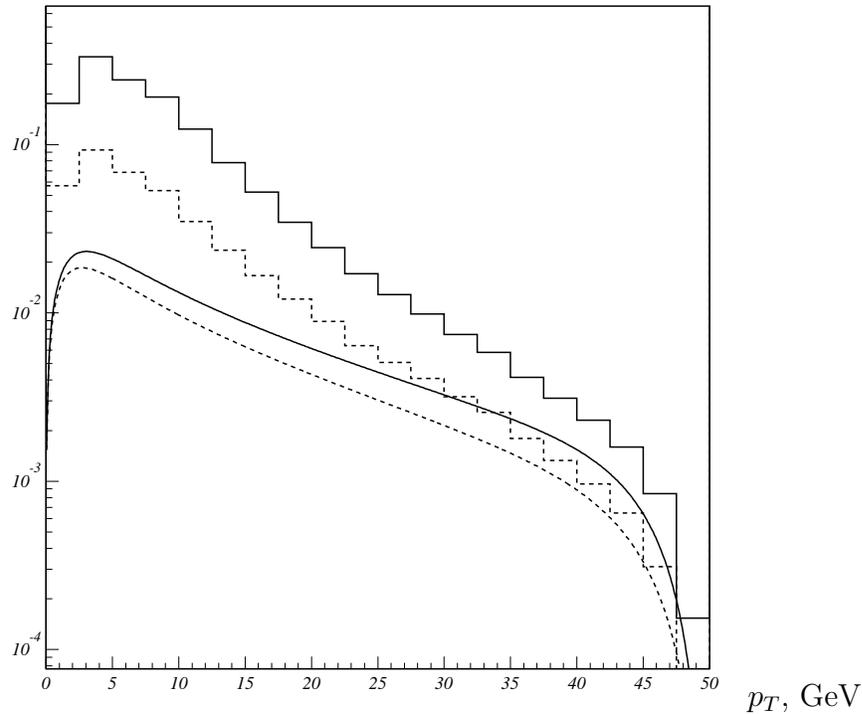}}

\vspace*{-1.cm}
\hbox to 14.5cm {\hfil \mbox{$p_T$, GeV}}
\caption{The distributions over the transverse momentum in the gluonic 
production of $\Xi_{bc}'(\Xi_{bc}^{(*)})$ in comparison with the fragmentation 
result at the interaction energy of 100 GeV. The solid line corresponds to the
production of $\Xi_{bc}^{(*)}$, the dashed curve represents the yield of
$\Xi_{bc}'$, while the complete results and fragmentation estimates are given
by the histogram and smooth curves, correspondingly.}
\label{bcpt}
\end{figure}

The agreement with the fragmentational production at $\sqrt{\hat s}=100$ GeV
takes place at large transverse momenta of the baryon, as one can see from the
distributions over $p_T$ for the $\Xi_{bc}^{(*)}$ and $\Xi_{bc}'$ production,
shown in Fig. \ref{bcpt}. Note, that in contrast to the doubly heavy baryon
production, the exact perturbative calculations of the gluonic production of
$B_c(B_c^*)$ mesons with $p_T > 35$ GeV at $\sqrt{\hat s}=100$ GeV agree with
the fragmentational predictions. For the baryon production, a visible deviation
is observed up to the largest values of $p_T$, though it is evident that at
larger energies of gluon-gluon subprocess the region of justified use of
fragmentation regime will become wider in the direction of large transverse
momenta.

The differential cross section $d\sigma /d p_T$ of the 
$\Xi_{bc}'$ and $\Xi_{bc}^{(*)}$ production in $p \bar p$ interactions
at $\sqrt{s}=1.8$ TeV is presented in \cite{hprod} in comparison with the 
fragmentational predictions. So, we can also conclude that the fragmentation
approach gives rather rough estimate for the yield and momentum distribution of
$\Xi_{bc}$ baryons.

At the chosen values of parameters and with the account for the cuts over the
transverse momentum and rapidity of the baryons ($p_T>5$ GeV and $|y|<1$), the
production cross section of the $1S$-wave $bcq$-baryons and its anti-particles
is evaluated as $\sigma_{bcq} \simeq 1 $ nb (without cuts the value of
$\sigma_{bcq}$ is about two times greater). After the expected end of Run Ib at
Tevatron with the integral luminosity $100\div 150\; \mbox{pb}^{-1}$, one has
the yields of $1.0\div 1.5 \cdot 10^5$ of $bcq$-baryons.

\subsubsection{Pair production}
At the energies of fixed target experiments the luminosity of parton
subprocesses with the valence quarks is not suppressed in comparison with the
luminosity of gluon-gluon collisions in the region of large invariant masses.
In this way a significant fraction of total cross section for the production of
baryons with two heavy quarks is given by the pair production of such baryons.
The total and differential cross sections for the pair production were
considered in \cite{brag}, wherein the contributions of scalar and axial-vector
diquarks were taken into account. So, the expression for the total cross
section of scalar pairs versus the square of total energy $s$ has the form 
\begin{equation}
\sigma_{ss}=\frac{ 8 \pi^3}{81s^3}~|\Psi_s(0)|^4
\left(1-\frac{4M^2}{s}\right)^{3/2}\left(\frac{16}{3}(\tilde
f^{[1]}_{ss})^2+\frac{11} {20} \left(1-\frac{4M^2}{s}\right)(\tilde
f^{[2]}_{ss})^2\right),
\end{equation}
where $\Psi_s(0)$ is the diquark wave function at the origin, while the
formfactors are determined by the functions
\begin{eqnarray}
\tilde f^{[1]}_{ss} &=& M \alpha_s^2 \left[\left(\frac{1}{m_1^2}
+ \frac{1}{m_2^2}\right) - \frac {2M^2}{s} \left(\frac{m_2}{m_1^3} + 
\frac{m_1}{m_2^3}\right) \right],
\\
\tilde f^{[2]}_{ss} &=& \frac{M^5}{m_1^3m_2^3}
\alpha_s^2,
\end{eqnarray}
so that $m_{1,2}$ are the heavy quark masses, and $M=m_1+m_2$. 

Numerical estimates show that the pair production of vector diquarks dominates,
so that it gives about 10\% of single doubly heavy baryon production in the
parton subprocess.

\subsubsection{Discussion}
On the basic of perturbative calculations for the hard production of doubly
charmed diquark fragmentating in the baryon we have shown that the
observation of doubly charmed baryons is a difficult problem, because the ratio
of $\sigma ( \Xi_{cc}^{(*)}) /\sigma ( charm)$ for these baryons and charmed
particles yields the value of $10^{-6}-10^{-3}$ depending on the process
energy. The suppression of doubly charmed baryon yield at low energies of fixed
target experiments is explained by the threshold effect.
\begin{table}[th]
\caption{The total cross section of doubly charmed baryon production 
at different facilities.}
\begin{center}
\begin{tabular}{|p{5cm}|c|c|c|c|}
\hline
facility & HERA-B & E781 & Tevatron & LHC\\
\hline
\parbox{3.5cm}{ \vspace*{1mm} total cross section, nb/nucleon}
 & $2\cdot 10^{-3}$ & $4.6\cdot 10^{-3}$
 & $12$ & $122$\\
\hline
\end{tabular}
\end{center}
\label{totcc}
\end{table}
\noindent
As one can see in Table \ref{totcc}, the low value of cross section for the
production of doubly charmed baryons in the fixed target experiments allows us
to expect about $10^5$ events with the production of these baryons at HERA-B.
Practically the same number of events at $p_T>5$ GeV and  $|y|<1$ is expected
at Tevatron with the integrated luminosity of $100\ {\rm pb}^{-1}$. The large
luminosity and large interaction energy allow one to increase the yield of the
doubly charmed baryon by $10^4$ times at LHC.

Taking into account the increase of luminosity of collider at FNAL, we can
believe that the experimental search for the events with the baryons $\Xi_{bc}$
and $\Xi_{cc}$ is the actual task.

Under conditions of large yield of the doubly charmed baryons, the problem of
their registration is under challenge.

First of all it is interesting to estimate the lifetimes of the lightest states
of $\Xi_{cc}^{++}$ and $\Xi_{cc}^{+}$. The simple study of quark diagrams shows
that in the decay of $\Xi_{cc}^{++}$-baryon the Pauli interference for the
decay products of charmed quark and valence quark in the initial state takes
place as well as in the case of $D^+$-meson decay. In the decay of $\Xi_{cc}^+$
the exchange by the $W$-boson  between the valence quarks plays an important
role as well as in the annihilation of $D^0$. Therefore we suppose that the
mentioned mechanisms\footnote{We present quite naive estimates for the
lifetimes, while a strict consideration of operator product expansion for the
inclusive decays of doubly heavy baryons is addressed in Chapter 4.} give the
same ratio for the both baryon and $D$-meson lifetimes
$$
\tau(\Xi_{cc}^+)\approx 0.4\cdot \tau(\Xi_{cc}^{++}).
$$
Further, the presence of two charmed quark in the initial state results in
following expressions:
\begin{eqnarray}
\tau(\Xi_{cc}^{++})&\approx & \frac{1}{2} \tau(D^+)\simeq 0.53\; {\rm ps,}
\nonumber \\
\tau(\Xi_{cc}^{+})&\approx & \frac{1}{2} \tau(D^0)\simeq 0.21\;{\rm ps.}
\nonumber
\end{eqnarray}
Following the analogy with the decays of baryon containing a single heavy
quark, we can expect the decay modes
\begin{eqnarray}
&&{\rm BR}(\Xi_{cc}^{++}\to K^{0(*)}\Sigma_c^{++(*)})\approx
{\rm BR}(\Xi_{cc}^{+}\to K^{0(*)}(\Sigma_c^{+(*)}+\Lambda_c^+))\approx
\nonumber \\
&& \quad \approx {\rm BR}(\Lambda_c\to K^{0(*)}p)\simeq 4\cdot 10^{-2}.
\nonumber
\end{eqnarray}

One can observe $4\cdot 10^{3}$ events in these decay modes at HERA-B and
Tevatron without taking into account a detection efficiency. One has to expect
the yield of $4\cdot 10^7$ such decays at LHC. Among other decay modes, the
$\Xi_{cc}^{++}\to \pi^+\Xi_c^+$ and $\Xi_{cc}^{+}\to \pi^+\Xi_c^0$ processes
taking place with branching ratios of about 1\%, can be essential.

The excited $\Xi_{cc}^*$ states always decay into $\Xi_{cc}$ by the emission of
$\gamma$-quantum, so that the branching fraction of transition is equal to
100\%, since the emission of $\pi$-meson is impossible in the $\Xi_{cc}^*$
decay  because of the small value of splitting between the ground state and the
excited one, in contrast to the charmed  meson decay.

In conclusion we mention another possibility to increase the yield of
doubly charmed baryons in fixed  target experiments. In the model of intrinsic
charm \cite{13c} one assumes, that the nonperturbative admixture of exotic
hybrid state $|c\bar c uud \rangle$ presents in the proton along with the
ordinary state $|uud\rangle$ including three light valence quarks.The
probability $P_{ic}$ of $|c\bar c uud\rangle$-state is suppressed at the level
of 1\%. The valence charmed quark from that state can recombine with the
charmed quark produced in the hard partonic process with the $c\bar c$-pair
production. The energy dependence for such doubly charmed baryon production
repeats one for the single charmed quark production in the framework of pQCD up
to the factor of exotic state suppression and the factor of fusion of two
charmed quarks into the diquark, $K\sim 0.1$. This mechanism has no threshold
of four quark state production in contrast to the discussed perturbative one.
Therefore at low energies of fixed target experiments, where the threshold
suppression of perturbative mechanism is strong, the model of intrinsic charm
would yield the dominant contribution into the $\Xi_{cc}^{(*)}$ production. So,
the number of events in this model would be increased by three orders of
magnitude, and ratio of $\Xi_{cc}^{(*)}$ and charmed particle yields would
equal $\sigma(\Xi_{cc}^{(*)})/\sigma(\rm{charm})\sim 10^{-3}$. At high energies
the perturbative production is comparable with the intrinsic charm
contribution. We have to note that the $|c\bar c c\bar c uud\rangle$-state
suppressed at the level of $3\cdot 10^{-4}$, also could increase the doubly
charmed baryon production at low energies of hadron-hadron collisions.

Thus, the observation of $\Xi_{cc}^*$-baryons in hadronic interactions is a
quite realistic problem, whose solution opens new possibilities to research the
heavy quark interactions. The observation of $\Xi_{cc}^{(*)}$-baryons at fixed
target experiments \cite{14c} would allow one to investigate the contributions
of different mechanisms in the doubly charmed baryon production, as the
contribution of the perturbative mechanism and that of the intrinsic charm,
which strongly increases the yield of these baryons.

\newpage
\section{Lifetimes and decays of $\Xi_{QQ'}$ baryons}

In the framework of Operator Product Expansion (OPE) over the inverse powers of
heavy quark mass there is a definite scheme for consistent calculations of QCD
effects, that was developed for the evaluation of various characteristics in
decays of hadrons containing the heavy quarks \cite{HQET,NRQCD,BB}. The
consideration based on the predictions of this approach allows one to extract
the parameters assigned to the electroweak interactions of heavy quarks in the
background of dynamics with strong interactions of quarks and gluons, which
form the hadrons observed in the experimental measurements. The accuracy of QCD
description in the sector of heavy quarks has an important significance in the
search for subtle effects such as the violation of {\sf CP}-invariance,
deviations from the predictions of Standard Model as well as in clarifying the
mechanisms for contributions of virtual corrections caused by a ``new'' physics
operating on characteristic energy scales of TeV range. Therefore, the study of
OPE over the inverse powers of heavy quark mass is quite an informative problem
of interest serving a manyfold attention. An importamt challenge is a complex
investigation of system containing the heavy quarks with the analysis and
comparison of various characteristics such as {\sc i)} the convergency of
expansion in both the inverse mass and the QCD coupling constant, {\sc ii)}
relative and absolute values of various contributions and their dependence on
the quark contents of systems, {\sc iii)} qualitative conclusions on the role
of some mechanisms \cite{Okun}, and {\sc iv)} uncertainties of numerical
estimates.  

The efficiency of approach under discussion was convincingly shown in the
description of weak decays for the hadrons having a single heavy quark, carried
out in the framework of the heavy quark effective theory (HQET) \cite{HQET}, in
the annihilation and radiative decays of heavy quarkonia $Q\bar Q$, using the
framework of nonrelativistic QCD (NRQCD) \cite{NRQCD}, and in the weak decays
of long-lived heavy quarkonium with mixed heavy flavours\footnote{The first
experimental observation of $B_c$-meson has been reported by the CDF
Collaboration \cite{cdf}, while a review on the theoretical status of $B_c$ is
given in \cite{bc-rev}.} $B_c^+$ \cite{BB}. We emphasize that the experimental
data on the weak decays of doubly heavy hadrons could be able to bring a
significant improvement of numerical accuracy in the parameters entering the
description of systems with the heavy quarks. The advantage is caused by the
presence of an additional small parameter in the NRQCD expansion in contrast to
HQET. This parameter is a relative velocity of heavy quark motion $v$.
Moreover, there is an essential variability of characteristics assigned to the
bound states with the heavy quarks, so that these changes of
properties allow us to investigate the dependence of OPE on the nonperturbative
parameters, which can be evaluated, for example, in the potential approach.

The baryons with two heavy quarks, $QQ^\prime q$, provide a new insight in the
description of systems containing the heavy quarks. For these baryons we can
apply a method based on the combined HQET-NRQCD techniques
\cite{HQET,NRQCD,BB}, if we use the quark-diquark factorization for the bound
states. The expansion in the inverse powers of heavy quark mass for the heavy
diquark $QQ^\prime$ is a straightforward generalization of these techniques in
the mesonic decays \cite{NRQCD,BB} with the difference that we deal with the
color anti-triplet system of heavy quarks with the appropriate account of
interaction with the light quark instead of the color singlet systems. The HQET
methods have to reliably work for describing the interaction of local diquark
with the light quark.

In this chapter, we present the consistent calculation of lifetimes for the
doubly heavy baryons $\Xi_{bc}^{+}$, $\Xi_{bc}^{0}$, $\Xi_{cc}^{+}$ and
$\Xi_{cc}^{++}$. Taking into account necessary generalizations to the case of
hadrons with two heavy quarks and other corrections in the description of
inclusive decays for the baryons, we follow the general consideration of heavy
hadron decays in \cite{vs,BB}, where the decays of the hadrons with a single
heavy quark and the doubly heavy $B_c$ meson were discussed. The justified
basis for such calculations is the optical theorem for the inclusive decay
width combined with the Operator Product Expansion (OPE) for the transition
currents in accordance with the consequent nonrelativistic expansion of
hadronic matrix elements derived in OPE. Using OPE at the first step, we
exploit the fact, that, due to the presence of heavy quarks in the initial
state, the energy release in the decay of both quarks is large enough in
comparison with the binding energy in the state. Thus, we can use the expansion
over the ratio of these scales. Technically, this step repeats an analogous
procedure for the inclusive decays of heavy-light mesons as it was reviewed in
\cite{bigi}. Exploring the nonrelativistic expansion of hadronic matrix
elements at the second step, we use the approximation of nonrelativistic QCD
\cite{NRQCD}, which allows one to reduce the evaluation of matrix elements for
the full QCD operators, corresponding to the interaction of heavy quarks inside
the diquark, to the expansion in powers of $\frac{p_c}{m_c}$, where $p_c
=m_cv_c\sim 1~GeV$ is a typical momentum of the heavy quark inside the baryon.
The same procedure for the matrix elements, determined by the strong
interaction of heavy quarks with the light quark, leads to the expansion in
powers of $\frac{\Lambda_{QCD}}{m_c}$.

We note, that keeping just the leading term in the OPE, the inclusive widths
are  determined by the mechanism of spectator decays involving free quarks,
wherein the corrections in the perturbative QCD are taken into account. The
introduction of subleading terms in the expansion over the inverse heavy quarks
mass\footnote{It was shown in \cite{bigi} that the leading order correction in
$1/m_Q$ is absent and the corrections begin with $1/m_Q^2$.} allows one to
take into account the corrections due to the quark confinement inside the
hadron. In this way, an essential role is played by the following
nonperturbative characteristics: the motion of heavy quark inside the hadron
and the corresponding time dilation in the hadron rest frame with respect to
the quark rest frame, and the influence of the chromo-magnetic interaction of
the quarks. The important ingredient of such corrections in the baryons with
two heavy quarks is the presence of a compact heavy diquark. The next
peculiarity of baryons with two heavy quarks is the significant numerical
influence on the lifetimes by the quark contents of hadrons, since in the third
order of inverse heavy quark mass, $1/m_Q^3$, the four-quark correlations
in the total width are enforced in the effective lagrangian due to the
two-particle phase space in the intermediate state (see discussion in
\cite{vs}). In this situation, we have to add the effects of the Pauli
interference between the products of heavy quark decays and the quarks in the
initial state as well as the weak scattering involving the quarks composing the
hadron. Through such terms we introduce the corrections involving the masses of
light and strange quarks in the framework of nonrelativistic models with the
constituent quarks. We include the corrections to the effective weak lagrangian
due to the evolution of Wilson coefficients from the scale of the order of
heavy quark mass to the energy, characterizing the binding of quarks inside the
hadron.

Following the picture given above, we describe the general scheme for the
construction of OPE for the total widths of baryons with two heavy quarks with
account of the corrections to the spectator widths in Section 4.1. In Section
4.2, the procedure for the estimation of nonperturbative matrix elements over
the states of doubly heavy baryons is considered for the operators of
nonrelativistic heavy quarks. Section 4.3 is devoted to the numerical
evaluation of lifetimes for the baryons with two heavy quarks and their partial
decay rates, as well as to the discussion of underlying uncertainties. We
conclude the chapter by summarizing our results.

\subsection{Operator Product Expansion for the heavy baryons}

In this section we describe the general scheme for the construction of OPE for
the total widths of baryons containing two heavy quarks.

Now let us start the description of our approach for the calculation of 
lifetimes for the doubly charmed baryons. The optical theorem, taking into
account the integral quark-hadron duality, allows us to relate the total decay
width of the heavy hadron $\Gamma$ with the imaginary part of its forward
scattering amplitude. This relationship, applied to the
$\Xi_{cc}^{\diamond}$-baryon total decay width $\Gamma_{\Xi_{cc}^{\diamond}}$,
can be written down as
\begin{equation}
\Gamma_{\Xi_{cc}^{\diamond}} =
\frac{1}{2M_{\Xi_{cc}^{\diamond}}}\langle \Xi_{cc}^{\diamond}|{\cal
T}|\Xi_{cc}^{\diamond}\rangle ,
\label{optic}
\end{equation}
where the state $\Xi_{cc}^{\diamond}$ in (\ref{optic}) has the ordinary
relativistic normalization
$$\langle \Xi_{cc}^{\diamond}|\Xi_{cc}^{\diamond}\rangle  = 2EV,$$
and the transition operator ${\cal T}$ is determined by the expression
\begin{equation}
{\cal T} = \Im m\int d^4x~\{{\rm T}H_{eff}(x)H_{eff}(0)\},
\end{equation}
where $H_{eff}$ is the standard effective hamiltonian describing the low
energy weak interactions of initial quarks with the decay products. For the
transition of $c$-quark into the $u$-quark and the quarks $q_{1,2}$ with the
charge $-1/3$, the lagrangian has the form
\begin{equation}
H_{eff} = \frac{G_F}{2\sqrt 2}V_{uq_1}V_{cq_1}^{*}[C_{+}(\mu)O_{+} +
C_{-}(\mu)O_{-}] + {\rm h.c.}
\label{heff}
\end{equation}
where $V$ is the matrix of mixing between the charged currents, and
$$
O_{\pm} = [\bar q_{1\alpha}\gamma_{\nu}(1-\gamma_5)c_{\beta}][\bar
u_{\gamma}\gamma^{\nu}(1-\gamma_5)q_{2\delta}](\delta_{\alpha\beta}\delta_{
\gamma\delta}\pm\delta_{\alpha\delta}\delta_{\gamma\beta}),
$$
so tat the indices $\alpha,\beta$ denote the color states of quarks, and
$$
C_+ = \left [\frac{\alpha_s(M_W)}{\alpha_s(\mu)}\right ]^{\frac{6}{33-2n_f}},
\quad
C_- = \left [\frac{\alpha_s(M_W)}{\alpha_s(\mu)}
\right]^{\frac{-12}{33-2n_f}},\\
$$
$n_f$ is the number of flavors.      

Assuming that the energy release in the heavy quark decay is large, we can
perform the operator product expansion for the transition operator ${\cal T}$
in (\ref{heff}). In this way we find a series of local operators with
increasing dimensions over the energy scale, wherein the contributions to
$\Gamma$ are suppressed by the increasing inverse powers of the heavy quark
masses. This formalism was already applied to calculate the total decay
rates for the hadrons, containing a single heavy quark (see refs. \cite{bigi}
and \cite{vs,7d}). Here we would like to stress that the expansion, applied in
this paper, is simultaneously in the powers of the inverse heavy quark mass and
the relative velocity of heavy quarks inside the hadron. Thus, this fact
points to the difference from the description of both the heavy-light mesons
(the expansion in powers of $\frac{\Lambda_{QCD}}{m_c}$) and the heavy-heavy
mesons \cite{BB} (the expansion in powers of relative velocity of heavy quarks
inside the hadron, where one can apply the scaling rules of nonrelativistic QCD
\cite{NRQCD}).

In this work we will extend this approach to the treatment of baryons 
containing two heavy quarks. The operator product expansion applied has the
following form:
\begin{equation}
{\cal T} = C_1(\mu)\bar cc + \frac{1}{m_c^2}C_2(\mu)\bar
cg\sigma_{\mu\nu}G^{\mu\nu}c
+ \frac{1}{m_c^3}O(1).
\end{equation}
The leading contribution in OPE is determined by the operator $\bar cc$,
corresponding to the spectator decays of $c$-quarks. The use of
the equation of motion for the heavy quark fields allows one to eliminate some
redundant operators, so that no operators of dimension four contribute. There
is a single operator of dimension five, $Q_{GQ} = \bar Q g \sigma_{\mu\nu}
G^{\mu\nu} Q$. As we will show below, significant contributions come from the
operators of dimension six $Q_{2Q2q} = \bar Q\Gamma q\bar q\Gamma^{'}Q$,
representing the effects of Pauli interference and weak scattering for
$\Xi_{cc}^{++}$ and $\Xi_{cc}^{+}$, correspondingly, which are enforced by the
two-particle phase space. Furthermore, there are also other operators of
dimension six $Q_{61Q} = \bar Q \sigma_{\mu\nu} \gamma_{l} D^{\mu}G^{\nu l}Q$
and $Q_{62Q} = \bar Q D_{\mu} G^{\mu\nu} \Gamma_{\nu}Q$. In what follows, we
neglect the corresponding contributions for the latter two operators, since
they are suppressed by the mentioned smallness of three-particle phase space,
so that the expansion is certainly complete up to the second order of $\frac
{1}{m}$, only. The reason for the restriction of dimension-six operators is
twofold. First, the operators $Q_{61Q}$ and $Q_{62Q}$ do not contribute to the
lifetime difference between the doubly charmed baryons under consideration,
since they are independent of the quark contents of hadrons. Second, the
four-quark operators are enhanced in comparison with the quark-gluon terms with
the same dimension because the two-particle phase space integrated in the
calculation of coefficients in front of Pauli interference and weak scattering
has the additional factor of $16 \pi^2$ in contrast to the the three-particle
phase space expressed in the units of heavy quark mass as it occurs for the
coefficients of operators $Q_{61Q}$ and $Q_{62Q}$, so that they are suppressed.

Further, the different contributions to OPE are given by the following:
$$
{\cal T}_{\Xi_{cc}^{++}} = {\cal T}_{35c} + {\cal T}_{6,PI},
$$
$$
{\cal T}_{\Xi_{cc}^{+}} = {\cal T}_{35c} + {\cal T}_{6,WS},
$$
where the first terms account for the operators of dimension three $O_{3Q}$ and
five $O_{GQ}$, the second terms correspond to the effects of Pauli interference
and weak scattering. The explicit formulae for these contributions have the
following form:
\begin{equation}
{\cal T}_{35c} = 2 (\Gamma_{c,spec}\bar cc - \frac{\Gamma_{0c}}{m_c^2}[(2 +
K_{0c})P_{1}
+
K_{2c}P_{2}]O_{Gc}), 
\label{cc35}
\end{equation}
where $\Gamma_{0c} = \frac{G_F^2m_c^2}{192{\pi}^3}$
and $K_{0c} = C_{-}^2 + 2C_{+}^2$,  $K_{2c} = 2(C_{+}^2 - C_{-}^2)$.
This expression has been derived in \cite{9d} (see also \cite{10d}), and it is
also discussed in \cite{bigi}. The phase space factors $P_i$ \cite{bigi,11d}
have the form
$$
P_1 = (1-y)^4,\quad P_2 = (1-y)^3
$$
where $y = \frac{m_s^2}{m_c^2}$. $\Gamma_{c,spec}$ denotes the contribution to
the total decay width
of the free decay for one of the two $c$-quarks, which is explicitly 
expressed in \cite{ltcc}. 

Quite cumbersome expressions for the contributions of Pauli interference and
weak scattering terms in the inclusive widths are given in Appendix III.

In the numerical estimates for the evolution of coefficients $C_{+}$ and
$C_{-}$, we have taken into account the threshold effects, connected to the
$b$-quark, as well as the threshold effects, related to the $c$-quark mass in
the Pauli interference and weak scattering.

In the expressions for $C_{\pm}$ the scale $\mu$ is approximately equal to
$m_c$. For the Pauli interference and weak scattering, this scale in the factor
$k$ is chosen in the way to obtain an agreement between the experimental
differences of lifetimes for the $\Lambda_c$, $\Xi_c^{+}$ and
$\Xi_c^{0}$-baryons and the theoretical predictions based on the effects
mentioned above. This problem is discussed below. Anyway, the choice of these
scales allows some variations, and a complete answer to this question requires
calculations in the next order of perturbative theory.

\setlength{\unitlength}{1mm}
\begin{figure}[th]
\begin{center}
\begin{picture}(90.,35.)
\put(0,0){\epsfxsize=9cm \epsfbox{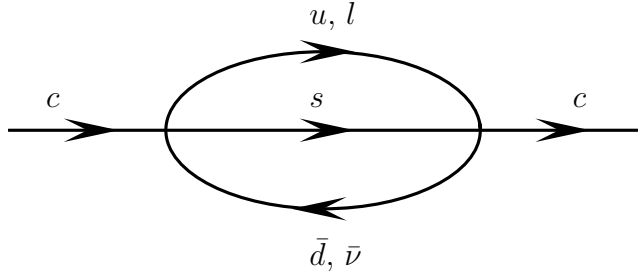}}
\put(5,24){$c$}
\put(75,24){$c$}
\put(40,24){$s$}
\put(40,35){$u$, $l$}
\put(40,3){$\bar d$, $\bar \nu$}

\end{picture}
\end{center}
\vspace*{-8mm}
\caption{The spectator contribution into the total widths of doubly charmed
baryons.}
\label{f35}
\end{figure}

The contribution of the leading operator $\bar cc$ corresponds to the imaginary
part of the diagram in Fig. \ref{f35}, as it stands in expression (\ref{cc35}).
The coefficient in front of $\bar cc$ can be obtained in the usual way by
projection of contribution due to the diagram in Fig. \ref{f35} to the operator
$\bar cc$. This coefficient is equivalent to the free quark decay rate, and it
is known in the next-to-leading logarithmic approximation of QCD 
\cite{12d,13d,14d,15d,16d}, including the strange quark mass effects in the
final state \cite{16d}. To calculate the next-to-leading logarithmic effects,
the Wilson coefficients in the effective weak lagrangian are required at the
next-to-leading accuracy, and the single gluon exchange corrections to
the diagram in Fig. \ref{f35} must be considered. In our numerical estimates we
use the expression for $\Gamma_{spec}$, including the next-to-leading order
corrections, $s$-quark mass effects in the final state, but we neglect the
Cabibbo-suppressed decay channels for the $c$-quark. The bulky explicit
expression for the spectator $c$-quark decay is placed in the Appendix of 
\cite{ltcc}.

\begin{figure}[th]
\begin{center}
\begin{picture}(90.,35.)
\put(0,0){\epsfxsize=9cm \epsfbox{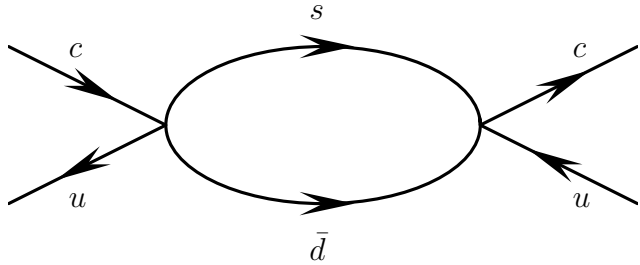}}
\put(8,30){$c$}
\put(8,10){$u$}
\put(75,30){$c$}
\put(75,10){$u$}
\put(40,35){$s$}
\put(40,3){$\bar d$}

\end{picture}
\end{center}
\vspace*{-8mm}
\caption{The Pauli interference of $c$-quark decay products with the valence
quark in the initial state for the $\Xi_{cc}^{++}$-baryon.}
\label{fpi}
\end{figure}

\begin{figure}[th]
\begin{center}
\begin{picture}(90.,35.)
\put(0,0){\epsfxsize=9cm \epsfbox{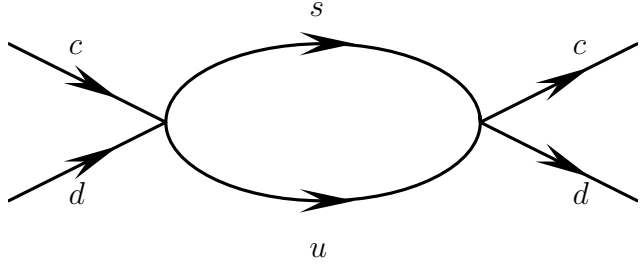}}
\put(8,30){$c$}
\put(8,10){$d$}
\put(75,30){$c$}
\put(75,10){$d$}
\put(40,35){$s$}
\put(40,3){$u$}

\end{picture}
\end{center}
\vspace*{-8mm}
\caption{The weak scattering of the valence quarks in the initial
state for the $\Xi_{cc}^{+}$-baryon.}
\label{fws}
\end{figure}

Similarly, the contribution by $O_{GQ}$ is obtained, when an external gluon
line is attached to the inner quark lines in Fig. \ref{f35} in all possible
ways. The corresponding coefficients are known in the leading logarithmic
approximation. Finally, the dimension six operators and their coefficients
arise due to those contributions, wherein one of the internal $u$ or $\bar d$
quark line is "cut". The resulting graphs are depicted in Figs. \ref{fpi} and
\ref{fws}. These contributions correspond to the effects of Pauli interference
and weak scattering. We have calculated the expressions for these effects with
account for both the $s$-quark mass in the final state and the logarithmic
renormalization of effective electroweak lagrangian at low energies. This
effective lagrangian is shown in Appendix III.

To calculate the contribution of semileptonic modes to the total decay width of
$\Xi_{cc}^{\diamond}$-baryons (we have taken into account the electron and muon
decay modes only) we use the following expressions \cite{10d} (see also
\cite{16d}):
\begin{eqnarray}
\Gamma_{sl} &=& 4\Gamma_c(\{1-8\rho+8\rho^3-\rho^4-12\rho^2\ln\rho\}
+\nonumber\\
&&
E_c\{5-24\rho+24\rho^2-8\rho^3+3\rho^4-12\rho^2\ln\rho\}+\\
&& K_c\{-6+32\rho-24\rho
^2-2\rho^4+24\rho^2\ln\rho\}+\nonumber\\
&& G_c\{-2+16\rho-16\rho^3+2\rho^4+24\rho^2\ln\rho\}), \nonumber 
\end{eqnarray}
where $\Gamma_c = |V_{cs}|^2G_F^2\frac{m_c^5}{192\pi^3}$, $\rho
=\frac{m_s^2}{m_c^2}$, $E_c = K_c + G_c$, while $K_c$ and $G_c$ are given by
the
expressions
\begin{eqnarray}
K_c &=& -\langle \Xi_{cc}^{\diamond}(v)|\bar
c_v\frac{(i{\bf D})^2}{2m_c^2}c_v|\Xi_{cc}^{\diamond}(v)\rangle ,\nonumber\\
G_c &=& \langle \Xi_{cc}^{\diamond}(v)|\bar
c_v\frac{gG_{\alpha\beta}\sigma^{\alpha\beta}}{4m_c^2}c_v|\Xi_{cc}^{\diamond}(v
)
\rangle ,
\end{eqnarray}
where the spinor field $c_v$ in the effective heavy quark theory is defined by
the form
\begin{equation}
c(x) = e^{-im_cv\cdot x}\Bigl[1+\frac{i D_\mu \gamma^\mu}{2m_c}\Bigr]c_v(x),
\end{equation}
where $D_\mu$ denotes the partial derivative over coordinates.

The analogous scheme for the calculation of inclusive widths
$\Gamma_{\Xi_{bc}^{\diamond}}$ for the baryon $\Xi_{bc}^{\diamond}$, where
$\diamond$ denotes the electric charge of the system, was developed in
\cite{ltbc}. Then, for the total widths we have the expressions
\begin{eqnarray}
{\cal T}_{\Xi_{bc}^{+}} &=& {\cal T}_{35b} + {\cal T}_{35c} + 
{\cal T}_{6,PI}^{(1)} + {\cal T}_{6,WS}^{(1)},\nonumber\\
{\cal T}_{\Xi_{bc}^{0}} &=& {\cal T}_{35b} + {\cal T}_{35c} + 
{\cal T}_{6,PI}^{(2)} + {\cal T}_{6,WS}^{(2)},
\nonumber
\end{eqnarray}
where two initial terms denote the contributions into the decays of quark $Q$
by the operators with the dimensions 3 and 5, and the forthcoming terms are the
interference and scattering of constituents. Various contributions were
explicitly presented in \cite{ltbc}. 

The calculation of both the Pauli interference effect for the products of heavy
quark decays with the quarks in the initial state and the weak scattering of
quarks composing the hadron, results in summming up the various channels of
decays, so that 
\begin{eqnarray}
{\cal T}_{6,PI}^{(1)} &=& {\cal T}_{PI,u\bar d}^c + {\cal T}_{PI,s\bar c}^b +
{\cal T}_{PI,d\bar u}^b + \sum_l{\cal T}_{PI,l\bar\nu_l}^b, \nonumber\\
{\cal T}_{6,PI}^{(2)} &=& {\cal T}_{PI,s\bar c}^b + {\cal T}_{PI,d\bar u}^b +
{\cal T}_{PI,d\bar u}^{'b} + \sum_l{\cal T}_{PI,l\bar\nu_l}^b,\\
{\cal T}_{6,WS}^{(1)} &=& {\cal T}_{WS,bu} + {\cal T}_{WS,bc},\nonumber\\
{\cal T}_{6,WS}^{(2)} &=& {\cal T}_{WS,cd} + {\cal T}_{WS,bc}. \nonumber
\end{eqnarray}
\noindent
and 
\begin{eqnarray}
{\cal T}_{PI,d\bar u}^b &=& {\cal T}_{PI,s\bar c}^b~(z_{-}\to 0),\nonumber\\
{\cal T}_{PI,e\bar\nu_e}^b &=& {\cal T}_{PI,\mu\bar\nu_{\mu}}^b = 
{\cal T}_{PI,\tau\bar\nu_{\tau}}^b~(z_{\tau}\to 0).
\end{eqnarray}
Further, we have taken into account the known $\alpha_s$-corrections to the
semileptonic widths of quarks. 

Thus, the calculation of lifetimes for the baryons with two heavy quarks is
reduced to the problem of evaluating the matrix elements of operators, that is
the subject of next section.

\subsection{Hadronic matrix elements}
According to the equation of motion for the heavy quarks, the matrix element of
local operator $\bar Q^jQ^j$ can be expanded in the following series over the
powers of ${1}/{m_{Q^j}}$, so that
\begin{eqnarray}
\frac{\langle \Xi_{QQ'}^{\diamond}|\bar Q^jQ^j|\Xi_{QQ'}^{\diamond}\rangle}{2M}
= 1 - \frac{\langle \Xi_{QQ'}^{\diamond}|\bar
Q^j[(i\boldsymbol{D})^2-(\frac{i}{2}\sigma
G)]Q^j|\Xi_{QQ'}^{\diamond}\rangle}{4 M m_{Q^j}^2} +
O(\frac{1}{M m_{Q^j}^3}).
\end{eqnarray}
Thus, we have to make numerical estimates for the following list of operators:  
\begin{eqnarray}
&& \bar Q^j(i\boldsymbol{ D})^2Q^j,\quad (\frac{i}{2})\bar Q^j\sigma GQ^j,\quad
\bar
Q^j\gamma_{\alpha}(1-\gamma_5)Q^j\bar q\gamma^{\alpha}(1-\gamma_5)q,\nonumber\\
&& \bar Q^j\gamma_{\alpha}\gamma_5Q^j\bar q\gamma^{\alpha}(1-\gamma_5)q,\quad 
\bar Q^j\gamma_{\alpha}\gamma_5Q^j\bar
Q^k\gamma^{\alpha}(1-\gamma_5)Q^k,\\ 
&& \bar Q^j\gamma_{\alpha}(1-\gamma_5)Q^j\bar
Q^k\gamma^{\alpha}(1-\gamma_5)Q^k.\nonumber
\end{eqnarray}
The first operator corresponds to the time dilation connected to the motion of
heavy quarks inside the hadron. The second is related to the spin interaction
of heavy quarks with the chromo-magnetic field of light quark and the 
other heavy quark. The further operators are the four-quark operators
representing the effects of Pauli interference and weak scattering.

Next, following the general methods of effective theories, we introduce the
effective filed $\Psi_Q$, which represents the nonrelativistic spinor of heavy
quark, so that we integrate out the contributions with the virtualities $\mu$
in the range $m_{Q} >  \mu >  m_{Q}v_{Q}$ in the framework of perturbative QCD,
while the nonperturbative effects in the matrix elements could be expressed in
terms of effective nonrelativistic fields. So, we have 
\begin{eqnarray}
\bar QQ &=& \Psi_Q^{\dagger}\Psi_Q -
\frac{1}{2m_Q^2}\Psi_Q^{\dagger}(i\boldsymbol{
D})^2\Psi_Q +
\frac{3}{8m_Q^4}\Psi_Q^{\dagger}(i\boldsymbol{ D})^4\Psi_Q -\nonumber\\
&& \frac{1}{2m_Q^2}\Psi_Q^{\dagger}g\boldsymbol{\sigma}\boldsymbol{ B}\Psi_Q -
\frac{1}{4m_Q^3}\Psi_Q^{\dagger}(\boldsymbol{ D}g\boldsymbol{ E})\Psi_Q + ...
\label{32}\\
\bar Qg\sigma_{\mu\nu}G^{\mu\nu}Q &=&
-2\Psi_Q^{\dagger}g\boldsymbol{\sigma}\boldsymbol{ B}\Psi_Q -
\frac{1}{m_Q}\Psi_Q^{\dagger}(\boldsymbol{ D}g\boldsymbol{ E})\Psi_Q + ...
\label{33}
\end{eqnarray}
In these expressions we have omitted the term
$\Psi_Q^{\dagger}\boldsymbol{\sigma} (g\boldsymbol{E} \times 
\boldsymbol{ D})\Psi_Q$, corresponding to the spin-orbital interaction, because
it vanishes in the ground states of baryons under consideration. For the
normalization we take
\begin{equation}
\int d^3x\Psi_Q^{\dagger}\Psi_Q = \int d^3x Q^{\dagger}Q.
\end{equation} 
Then for $Q$ determined by
\begin{equation}
Q\equiv e^{-imt}\left(\phi\atop \chi\right)
\end{equation}
we have
\begin{equation}
\Psi_c = \left( 1 + \frac{(i\boldsymbol{ D})^2}{8m_c^2}\right)\phi.
\end{equation}

Let us emphasize the difference in the descriptions of interactions between the
heavy quarks and light quark and interactions between two heavy quarks. As we
have mentioned, there is an additional small parameter in the heavy subsystem.
It is the relative velocity of quark motion, that introduces the energetic
scale $m_Q v$. Therefore, for example, the Darvin term ($\boldsymbol{
D}\boldsymbol{ E}$) in the heavy subsystem turns out to be the same order of
magnitude in comparison with the chromo-magnetic term
($\boldsymbol{\sigma}\boldsymbol{ B}$) in the expansion over the inverse power
of heavy quark mass, since it has the same degree of smallness in $v$. This
fact becomes clear if we explore the scaling relations derived in the
nonrelativistic QCD \cite{NRQCD}
$$
\Psi_Q\sim (m_Qv_Q)^{\frac{3}{2}},\quad|\boldsymbol{ D}|\sim m_Qv_Q,\quad
g|\boldsymbol{E}|\sim m_Q^2v_Q^3,\quad g|\boldsymbol{B}|\sim m_Q^2v_Q^4,\quad
g\sim v_Q^{\frac{1}{2}}.
$$
For the interaction of heavy quark with the light one there is no small
parameter defined by the relative velocity, so that the Darvin term is
suppressed by the factor of $k/m_Q\sim \Lambda_{QCD}/m_Q$.

\subsubsection{Baryons $\Xi_{cc}^{+}$ and $\Xi_{cc}^{++}$}
Let us now start the calculation of matrix elements with the use of potential
models for the bound states of hadrons. While estimating the matrix element
value of the kinetic energy, we note, that the heavy quark kinetic energy
consists of two parts: the kinetic energy of the heavy quark motion inside the
diquark and the kinetic energy related to the diquark motion inside the
hadron. According to the phenomenology of meson potential models, in the range
of average distances between the quarks: $0.1\; {\rm fm} < r < 1\; {\rm fm}$,
the average kinetic energy of quarks is constant and independent of both the
quark flavors, composing the meson, and the quantum numbers, describing the
excitations of the ground state. Therefore, we determine $T = m_dv_d^2/2 +
m_lv_l^2/2$ as the average kinetic energy of diquark and light quark with
$m_l=m_q$, and $T/2 = m_{c1}v_{c1}^2/2 + m_{c2}v_{c2}^2/2$ as the average
kinetic energy of heavy quarks inside the diquark (the coefficient 1/2 takes
into account the anti-symmetry of color wave function for the diquark).
Finally, we have the following expression for the matrix element of the heavy
quark kinetic energy:
\begin{equation}
\frac{\langle \Xi_{cc}^{\diamond}|\Psi_c^{\dagger}(i{\bf 
D})^2\Psi_c|\Xi_{cc}^{\diamond}\rangle }{2M_{\Xi_{cc}^{\diamond}}m_c^2}\simeq
v_c^2\simeq
\frac{m_lT}{2m_c^2+m_cm_l}+\frac{T}{2}.
\end{equation}
We use the value $T\simeq 0.4$ GeV, which results in $v_c^2 = 0.146$, where the
dominant contribution comes from the motion of heavy quarks inside the diquark.

Now we would like to estimate the matrix element of chromo-magnetic operator,
corresponding to the interaction of heavy quarks with the chromo-magnetic field
of the light quark. For this purpose, we will use the following definitions:
$O_{mag} = \sum_{i=1}^2 \frac{g_s}{4m_c}\bar c^i\sigma_{\mu\nu}G^{\mu\nu}c^i$
and $O_{mag}\sim \lambda (j(j+1) - s_d(s_d+1) - s_l(s_l+1))$, where $s_d$ is
the diquark spin (as was noticed by the authors earlier, there is only the
vector state of the $cc$-diquark in the ground state of such baryons), $s_l$ is
the light quark spin and $j$ is the total spin of the baryon. Since both
$c$-quarks additively contribute to the total decay width of baryons, we can
use the diquark picture and substitute for the sum of $c$-quark spins the
diquark spin. This leads to the parametrization for $O_{mag}$, as it is given
above, and, moreover, it allows us to relate the value of the matrix element
for this operator to the mass difference between the excited and ground state
of baryons
\begin{equation}
O_{mag} = -\frac{2}{3}(M_{\Xi_{cc}^{*\diamond}}-M_{\Xi_{cc}^{\diamond}}).
\end{equation}
The account for the interaction of heavy quarks inside the diquark leads to the
following expressions for the chromo-magnetic and Darwin terms:
\begin{equation}
\frac{\langle \Xi_{cc}^{\diamond}|\Psi_c^{\dagger}g{\boldsymbol{\sigma}}\cdot
{\bf 
B}\Psi_c|\Xi_{cc}^{\diamond}\rangle }{2M_{\Xi_{cc}^{\diamond}}} =
\frac{2}{9}g^2\frac{|\Psi^d(0)|^2}{m_c},
\end{equation}
\begin{equation}
\frac{\langle \Xi_{cc}^{\diamond}|\Psi_c^{\dagger}({\bf  D}\cdot g{\bf 
E})\Psi_c|\Xi_{cc}^{\diamond}\rangle }{2M_{\Xi_{cc}^{\diamond}}} =
\frac{2}{3}g^2|\Psi^d(0)|^2,
\end{equation}
where $\Psi^d(0)$ is the diquark wave function at the origin. Collecting the
results given above, we find the matrix elements of dominant operators
determining the spectator decay
\begin{eqnarray}
\frac{\langle \Xi_{cc}^{\diamond}|\bar cc|\Xi_{cc}^{\diamond}\rangle
}{2M_{\Xi_{cc}^{\diamond}}} &=& 1 -
\frac{1}{2}v_c^2 -
\frac{1}{3}\frac{M_{\Xi_{cc}^{*\diamond}}-M_{\Xi_{cc}^{\diamond}}}{m_c}
- \frac{g^2}{9m_c^3}|\Psi^d(0)|^2 - 
\frac{1}{6m_c^3}g^2|\Psi^d(0)|^2 + ... \nonumber\\
&\approx & 1 - 0.074 -0.004 - 0.003 -
0.005
\end{eqnarray}
We can see that the largest contribution to the decrease of the
decay width comes from the time dilation, connected to the motion of heavy
quarks inside the baryon. For the matrix element of the operator $\bar
cg\sigma_{\mu\nu}G^{\mu\nu}c$, we get
\begin{eqnarray}
\frac{\langle \Xi_{cc}^{\diamond}|\bar
cg\sigma_{\mu\nu}G^{\mu\nu}c|\Xi_{cc}^{\diamond}\rangle
}{2M_{\Xi_{cc}^{\diamond}}m_c^2}
&=&
-\frac{4}{3}\frac{(M_{\Xi_{cc}^{*\diamond}} - 
M_{\Xi_{cc}^{\diamond}})}{m_c} - 
\frac{4g^2}{9m_c^3}|\Psi^d(0)|^2 - \frac{g^2}{3m_c^3}|\Psi^d(0)|^2.
\end{eqnarray}
Now let us continue with the calculation of the matrix elements for the
four-quark operators, corresponding to the effects of Pauli interference and
weak scattering. The straightforward calculation in the framework of
nonrelativistic QCD gives the following expressions for the operators $\langle
\bar c\gamma_{\mu}(1-\gamma_5)c\cdot \bar q\gamma^{\mu}(1-\gamma_5)q\rangle$
and $\langle\bar c\gamma_{\mu}\gamma_5c\cdot\bar q\gamma^{\mu} (1-\gamma_5)
q\rangle$:
\begin{eqnarray}
\langle\bar c\gamma_{\mu}(1-\gamma_5)c\cdot\bar
q\gamma^{\mu}(1-\gamma_5)q\rangle &=&
2m_cV^{-1}(1-4S_cS_q), \\
\langle\bar c\gamma_{\mu}\gamma_5c\cdot\bar q\gamma^{\mu}(1-\gamma_5)q\rangle
&=& -4S_cS_q\cdot 2m_cV^{-1},
\end{eqnarray}
where $V^{-1} = |\Psi_l(0)|^2$, $\Psi_l(0)$ is the light quark wave function at
the origin of two $c$-quarks. We suppose that, for estimates, $|\Psi_l(0)|$ has
the same value as that in the $D$-meson. So, we find
\begin{equation}
|\Psi_l(0)|^2\approx\frac{f_D^2m_D^2}{12m_c}.
\end{equation}
We would like to note that in the numerical estimate the supposed value of
light quark wave function obtained from the leptonic constant of charmed meson
$f_D\approx 200$ MeV, is about two times less than the wave function calculated
in Chapter 1 in the approximation of quark-diquark factorization. This is
related with the fact that the leptonic constants of charmed hadrons have large
corrections of both logarithmic form and powers of inverse charmed quark mass.
So, the leptonic constant of $D$ meson calculated in the potential models is
two times greater than the value obtained in the QCD sum rules with account for
the mentioned corrections. Therefore, we think that the supposed approximation
for the wave function of light quark in the baryon is justified enough.

Then, again remembering that both $c$-quarks additively contribute to
the total decay width and using the diquark picture, we can substitute
for $S_{c_1} + S_{c_2}$ by $S_d$, where $S_d$ is the diquark spin. Thus, for
the matrix elements of operators under study we have the following expressions:
\begin{equation}
\langle \Xi_{cc}^{\diamond}|(\bar c\gamma_{\mu}(1-\gamma_5)c)(\bar
q\gamma^{\mu}(1-\gamma_5)q)|\Xi_{cc}^{\diamond}\rangle  = 12m_c\cdot
|\Psi_l(0)|^2,
\end{equation}
\begin{equation}
\langle \Xi_{cc}^{\diamond}|(\bar c\gamma_{\mu}\gamma_5c)(\bar
q\gamma^{\mu}(1-\gamma_5)q)|\Xi_{cc}^{\diamond}\rangle  = 8m_c\cdot
|\Psi_l(0)|^2.
\end{equation}
The color anti-symmetry of the baryon wave function results in relations
between the matrix elements of operators with the different sums over the color
indices: $\langle \Xi_{cc}^{\diamond}|(\bar c_iT_{\mu}c_k)(\bar
q_k\gamma^{\mu}(1-\gamma_5)q_i|\Xi_{cc}^{\diamond}\rangle = -\langle
\Xi_{cc}^{\diamond}|(\bar cT_{\mu}c)(\bar q\gamma^{\mu}(1-\gamma_5)q
|\Xi_{cc}^{\diamond}\rangle $, where $T_{\mu}$ is any spinor structure.

Thus, we have formally constructed the procedure for the evaluation of matrix
elements obtained in the OPE of ${\cal T}$ for the baryon with two identical
heavy quarks.

\subsubsection{Baryons $\Xi_{bc}^{+}$ and $\Xi_{bc}^{0}$}
Considering the baryons with two heavy quarks of different flavors, we point
out the modifications, which have to be introduced in the estimation of
hadronic matrix elements for the quark operators determining the inclusive
decay widths.

First of all, according to the quar-diquark factorization we get the
following expressions for the kinetic terms:
\begin{equation}
\frac{\langle
\Xi_{bc}^{\diamond}|\Psi_c^{\dagger}(i\boldsymbol{D})^2\Psi_c|\Xi_{bc}^{
\diamond}
\rangle }{2M_{\Xi_{bc}^{\diamond}}m_c^2}\simeq
v_c^2\simeq
\frac{2m_lT}{(m_l+m_b+m_c)(m_b+m_c)}+\frac{m_bT}{m_c(m_c+m_b)}.
\end{equation}
\begin{equation}
\frac{\langle
\Xi_{bc}^{\diamond}|\Psi_b^{\dagger}(i\boldsymbol{D})^2\Psi_b|\Xi_{bc}^{
\diamond}
\rangle }{2M_{\Xi_{bc}^{\diamond}}m_b^2}\simeq
v_b^2\simeq
\frac{2m_lT}{(m_l+m_b+m_c)(m_b+m_c)}+\frac{m_cT}{m_b(m_c+m_b)}.
\end{equation}
Numerically, we assign $T\simeq 0.4$ GeV, which results in $v_c^2 =0.195$ and
$v_b^2 =0.024$, where the dominant contribution is provided by the motion
inside the diquark.

Let us define
\begin{eqnarray}
O_{mag} &=& \frac{g_s}{4m_c}\bar c\sigma_{\mu\nu}G^{\mu\nu}c +
\frac{g_s}{4m_b}\bar b\sigma_{\mu\nu}G^{\mu\nu}b,\\
\langle O_{mag} \rangle &=& \frac{\lambda}{m_c}(S_{cl}(S_{cl}+1) - S_c(S_c+1) -
S_l(S_l+1)) +\\
&& \frac{\lambda}{m_b}(S_{bl}(S_{bl}+1) - S_b(S_b+1) - S_l(S_l+1)),\nonumber
\end{eqnarray}
where $S_{bl} = S_b + S_l$, $S_{cl} = S_c + S_l$ , $S_b$ is the $b$-quark spin,
$S_c$ is that of $c$-quark, and $S_l$ is the light quark spin. The operator
under study is related to the hyperfine splitting in the baryon system
\begin{eqnarray}
&& \langle S_{bc} = 1, S = \frac{3}{2}|O_{mag}|S_{bc} = 1, S =
\frac{3}{2}\rangle
- \langle S_{bc} = 1, S = \frac{1}{2}|O_{mag}|S_{bc} = 1, S =
\frac{1}{2}\rangle~~~\nonumber\\
&& = \langle S_{bc} = 1, S = \frac{3}{2}|V_{hf}|S_{bc} = 1, S =
\frac{3}{2}\rangle
- \langle S_{bc} = 1, S = \frac{1}{2}|V_{hf}|S_{bc} = 1, S =
\frac{1}{2}\rangle,~~~
\end{eqnarray}
so that $S$ denotes the total spin of the system, and $S_{bc}$ is the diquark
spin. Further, the perturbative term, depending on the spins, is equal to
\begin{equation}
V_{hf} = \frac{8}{9}\alpha_s\frac{1}{m_lm_c}{\bf S}_l{\bf S}_c|R^{dl}(0)|^2 + 
\frac{8}{9}\alpha_s\frac{1}{m_lm_b}{\bf S}_l{\bf S}_b|R^{dl}(0)|^2,
\end{equation}
where $R^{dl}(0)$ is the radial wave function at the origin of quark-diquark
system. In contrast to the diquark system with the identical quarks, this
operator is not diagonal in the basis of $S$ and $S_{bc}$. To proceed, we use
the change of basises
\begin{equation}
|S;S_{bc}\rangle  = \sum_{S_{bl}} (-1)^{(S+S_l+S_c+S_b)}\sqrt
{(2S_{bl}+1)(2S_{bc}+1)}
\left\{\begin{array}{ccc} \bar S_l & S_b & S_{bl}\label{44} \\
                        S_c & S & S_{bc} \end{array}\right\}|S;S_{bl}\rangle 
\end{equation}
and 
\begin{equation}
|S;S_{bc}\rangle  = \sum_{S_{cl}} (-1)^{(S+S_l+S_c+S_b)}\sqrt
{(2S_{cl}+1)(2S_{bc}+1)}
\left\{\begin{array}{ccc} \bar S_l & S_c & S_{cl}\label{45} \\
                        S_b & S & S_{bc} \end{array}\right\}|S;S_{cl}\rangle 
\end{equation}
The result of substitutions gives
\begin{equation}
\lambda = \frac{4|R^{dl}(0)|^2\alpha_s}{9m_l},
\end{equation}
however, for the state with the zero spin of heavy diquark, which is considered
in what follows, we have
\begin{equation}
\frac{\langle\Xi_{bc}^{\diamond}|O_{mag}|\Xi_{bc}^{\diamond}\rangle}{2M_{\Xi_{b
c}^{\diamond}}}
= 0
\end{equation}
The account of Darwin and chromo-magnetic terms results in
\begin{eqnarray}
\frac{\langle \Xi_{bc}^{\diamond}|\Psi_c^{\dagger}g\boldsymbol{\sigma}\cdot
\boldsymbol{B}\Psi_c|\Xi_{bc}^{\diamond}\rangle }{2M_{\Xi_{bc}^{\diamond}}} &=&
-\frac{2}{3}g^2\frac{|\Psi^d(0)|^2}{m_b}, \\
\frac{\langle \Xi_{bc}^{\diamond}|\Psi_c^{\dagger}(\boldsymbol{ D}\cdot
g\boldsymbol{E})\Psi_c|\Xi_{bc}^{\diamond}\rangle }{2M_{\Xi_{bc}^{\diamond}}}
&=&
\frac{2}{3}g^2|\Psi^d (0)|^2.
\end{eqnarray}
where $\Psi^d (0)$ is the wave function at the origin of diquark. The analogous
matrix elements\footnote{The obtained expressions differ from those for the
$B_c$ meson \cite{BB} because of the color structure of the state, providing
the factor of $\frac{1}{2}$.} for the operators of beauty quarks can be written
down by the permutation of heavy quark masses.

Combining the results above, we find
\begin{eqnarray}
\frac{\langle \Xi_{bc}^{\diamond}|\bar cc|\Xi_{bc}^{\diamond}\rangle
}{2M_{\Xi_{bc}^{\diamond}}} &=& 1 -
\frac{1}{2}v_c^2 + \frac{g^2}{3m_bm_c^2}|\Psi^d (0)|^2 - 
\frac{1}{6m_c^3}g^2|\Psi^d (0)|^2 + \ldots \nonumber\\
&\approx& 1 - 0.097 +0.004 - 0.007\ldots
\end{eqnarray}
The dominant role in the corrections is played by the term, connected to the
time dilation because of the quark motion inside the diquark. Next, for the
operator $cg\sigma_{\mu\nu}G^{\mu\nu}c$ we have
\begin{eqnarray}
\frac{\langle \Xi_{bc}^{\diamond}|\bar
cg\sigma_{\mu\nu}G^{\mu\nu}c|\Xi_{bc}^{\diamond}\rangle
}{2M_{\Xi_{bc}^{\diamond}}m_c^2} &=&
\frac{4g^2}{3m_bm_c^2}|\Psi^d (0)|^2 - \frac{g^2}{3m_c^3}|\Psi^d (0)|^2 \approx
0.002.
\end{eqnarray}
The permutation of quark masses lead to the required expressions for the
operators of $\bar bb$ and $\bar bg\sigma_{\mu\nu}G^{\mu\nu}b$.

Making use of (\ref{44}) and (\ref{45}) for the ground states of baryons
results in 
\begin{eqnarray}
\langle \Xi_{bc}^{\diamond}|(\bar b\gamma_{\mu}(1-\gamma_5)b)(\bar
c\gamma^{\mu}(1-\gamma_5)c)|\Xi_{bc}^{\diamond}\rangle  &=& 8(m_b+m_c)\cdot
|\Psi^{d}(0)|^2,\\
\langle \Xi_{bc}^{\diamond}|(\bar b\gamma_{\mu}\gamma_5b)(\bar
c\gamma^{\mu}(1-\gamma_5)c)|\Xi_{bc}^{\diamond}\rangle  &=& 6(m_b+m_c)\cdot
|\Psi^{d}(0)|^2,\\
\langle \Xi_{bc}^{\diamond}|(\bar b\gamma_{\mu}(1-\gamma_5)b)(\bar
q\gamma^{\mu}(1-\gamma_5)q)|\Xi_{bc}^{\diamond}\rangle  &=& 2(m_b+m_l)\cdot
|\Psi^{dl}(0)|^2,\\
\langle \Xi_{bc}^{\diamond}|(\bar b\gamma_{\mu}\gamma_5b)(\bar
q\gamma^{\mu}(1-\gamma_5)q)|\Xi_{bc}^{\diamond}\rangle  &=& 0,\\
\langle \Xi_{bc}^{\diamond}|(\bar c\gamma_{\mu}(1-\gamma_5)c)(\bar
q\gamma^{\mu}(1-\gamma_5)q)|\Xi_{bc}^{\diamond}\rangle  &=& 2(m_c+m_l)\cdot
|\Psi^{dl}(0)|^2,\\
\langle \Xi_{bc}^{\diamond}|(\bar c\gamma_{\mu}\gamma_5c)(\bar
q\gamma^{\mu}(1-\gamma_5)q)|\Xi_{bc}^{\diamond}\rangle  &=& 0.
\end{eqnarray}
Thus, we can make numerucal estimates of inclusive decay widths.

\subsection{Numerical estimates}

Now we are ready to collect the contributions, described above, and to estimate
the lifetimes of baryons $\Xi_{cc}^{++}$ and $\Xi_{cc}^{+}$. For the
beginning, we list the values of parameters, which we have used in our
calculations, and give some comments on their choice.
\begin{equation}
\begin{array}{rclrclrcl}
m_c &=& 1.6~~{\rm {\rm GeV}}, & m_s &=& 0.45~~{\rm {\rm GeV}},& |V_{cs}|
&=& 0.9745,\\
M_{\Xi_{cc}^{++}} &=& 3.56~~{\rm {\rm GeV}},& M_{\Xi_{cc}^{+}} &=&
3.56~~{\rm {\rm GeV}},&
M_{\Xi_{cc}^{*\diamond}} - M_{\Xi_{cc}^{\diamond}} &=& 0.1~~{\rm {\rm
GeV}},\\ 
T &=& 0.4~~{\rm {\rm GeV}},& |\Psi (0)| &=& 0.17~~{\rm {\rm
GeV}}^{\frac{3}{2}},& m_l &=&
0.30~~{\rm {\rm GeV}}.
\end{array}
\end{equation}
For the parameters $M_{\Xi_{cc}^{++}}$, $M_{\Xi_{cc}^{+}}$ and
$M_{\Xi_{cc}^{*\diamond}} - M_{\Xi_{cc}^{\diamond}}$ we use the mean values,
given in the literature. Their evaluation has been also performed by the
authors in the potential model for the doubly charmed baryons with the
Buchm\"uller--Tye potential \cite{PM1}, and also in refs.
\cite{faust,Korner,Ron,QCDsr}. For $f_D$ we use the value, given in
\cite{vs,bigi}, and for $T$ we take it from \cite{25}.  The mass $m_c$
corresponds to the pole mass of the $c$-quark. For its determination we have
used a fit of theoretical predictions for the lifetimes and semileptonic width
of the $D^0$-meson from the experimental data. This choice of $c$-quark mass
seems to effectively include unknown contributions of higher orders in
perturbative QCD to the total decay width of baryons under consideration.  

The renormalization scale $\mu$ is chosen in the following way: $\mu_1 = m_c$
in the estimate of Wilson coefficients $C$ for the effective four-fermion weak
lagrangian with the $c$-quarks at low energies and $\mu_2 = 1.2~~GeV$ for the
Pauli interference and weak scattering. The latter value of renormalization
scale has been obtained from the fit of theoretical predictions for the
lifetimes differences of baryons $\Lambda_c$, $\Xi_c^{+}$, $\Xi_c^{0}$ over the
experimental data. Here we would like to note, that the theoretical
approximations used in \cite{5d} include the effect of logarithmic
renormalization and do not take into account the mass effects, related to the
$s$-quark in the final state. For the corresponding contributions to the decay
widths of baryons with the different quark contents we have
\begin{eqnarray}
\triangle\Gamma_{nl}(\Lambda_c) &=& c_d\langle O_d\rangle _{\Lambda_c} +
c_u\langle O_u\rangle _{\Lambda_c},\nonumber\\
\triangle\Gamma_{nl}(\Xi_c^{+}) &=& c_s\langle O_s\rangle _{\Xi_c^{+}} +
c_u\langle O_u\rangle _{\Xi_c^{+}},\\
\triangle\Gamma_{nl}(\Xi_c^{0}) &=& c_d\langle O_d\rangle _{\Xi_c^{0}} +
c_s\langle O_s\rangle _{\Xi_c^{0}},\nonumber
\end{eqnarray}
where $\langle O_q\rangle _{X_c} = \langle X_c|O_q|X_c\rangle $, $O_q = (\bar
c\gamma_{\mu}c)(\bar q\gamma^{\mu}q)$ with $q = u, d$ or $s$. The coefficients 
$c_q(\mu)$ are equal to
\begin{eqnarray}
c_d &=& \frac{G_F^2m_c^2}{4\pi}[C_{+}^2 + C_{-}^2 +
\frac{1}{3}(4k^{\frac{1}{2}} - 1)(C_{-}^2 - C_{+}^2)],\nonumber\\
c_u &=& -\frac{G_F^2m_c^2}{16\pi}[(C_{+} + C_{-})^2 + \frac{1}{3}(1 -
4k^{\frac{1}{2}})(5C_{+}^2 + C_{-}^2 - 6C_{+}C_{-})],\\
c_s &=& -\frac{G_F^2m_c^2}{16\pi}[(C_{+} - C_{-})^2 + \frac{1}{3}(1 -
4k^{\frac{1}{2}})(5C_{+}^2 + C_{-}^2 + 6C_{+}C_{-})].\nonumber
\end{eqnarray}
We use the spin averaged value of the $D$-meson mass for the estimation of 
the effective light quark mass $m_l = \bar\Lambda$ as it stands below
\begin{equation}
m_D = m_c + \bar\Lambda + \frac{\mu_{\pi}^2}{2m_c} = m_c + m_l + \frac{T\cdot
m_l}{m_c+m_l} = 1.98~~{\rm {\rm GeV}}.
\end{equation}
The $s$-quark mass can be written down in terms of $m_l$ introduced above, so
that
\begin{equation}
m_s = m_l + 0.15~~{\rm {\rm GeV}}.
\end{equation}
As we have already mentioned, the spectator decay width of $c$-quark
$\Gamma_{c,spec}$ is known in the next-to-leading order of perturbative QCD
\cite{12d,13d,14d,15d,16d}. The most complete calculation, including the mass
effects, connected to the $s$-quark in the final state, is given in \cite{16d}.
In the present work we have used these results for the calculation of the
spectator contribution to the total decay width of doubly charmed baryons. In
the calculation of the semileptonic decay width, we neglect the electron and
muon masses in the final state. Moreover, we neglect the $\tau$-lepton mode.

Now, let us proceed with the numerical analysis of contributions by the
different decay modes into the total decay width. In Table \ref{cont} we have
listed the results for the fixed values of parameters, described above. From
this table one can see the significance of effects caused by both the Pauli
interference and the weak scattering in the decays of doubly charmed baryons.
The Pauli interference gives the negative correction about $63\%$ for the
$\Xi_{cc}^{++}$-baryons, and the weak scattering increases the total width by
$61\%$ for $\Xi_{cc}^{+}$. As we have already noted above, these effects take
place separately in these baryons, and, thus, they enhance the difference of
lifetimes. 

\begin{table}[th]
\caption{The contributions of different modes to the total decay width
of doubly charmed baryons.}
\label{cont}
\begin{center}
\begin{tabular}{|p{37mm}|r|p{33mm}|p{33mm}|}
\hline
Mode or decay mechanism & Width, $ps^{-1}$ & Contribution,$\%$
($\Xi_{cc}^{++}$) & Contribution,$\%$ ($\Xi_{cc}^{+}$) \\
\hline
$c\to s\bar du$ & $2.648$~~ & ~~~127 & ~~~31\\
\hline
$c\to se^{+}\nu$ & $0.380$~~ & ~~~~18 & ~~~~4.2\\
\hline
PI & $-1.317$~~ & ~~~-63 & ~~~~--\\ 
\hline
WS & $5.254$~~ & ~~~~-- & ~~~60.6\\
\hline
 $\Gamma_{\Xi_{cc}^{++}}$ & $2.089$~~ & ~~~100 &~~~--\\
\hline
$\Gamma_{\Xi_{cc}^{+}}$ & $8.660$~~ & ~~~--  &~~100\\
\hline
\end{tabular}
\end{center}
\end{table}

It is worth here to recall that the lifetime difference of $D^{+}$ and
$D^{0}$-mesons is generally explained by the Pauli interference of $c$-quark
decay products with the anti-quark in the initial state, while in the current
consideration, we see the dominant contribution of weak scattering. This could
not be surprising, because under a more detailed consideration we will find, 
that the formula for the Pauli interference operator  for the $D$-meson 
coincides with that for the weak scattering in the case of baryons, 
containing, at least, a single $c$-quark. 

Finally, collecting the different contributions for the total lifetimes of
doubly charmed baryons, we obtain the following values:
$$\tau_{\Xi_{cc}^{++}} = 0.48~~{\rm ps},\;\;\;\;
\tau_{\Xi_{cc}^{+}} = 0.12~~{\rm ps}.$$
Note that the supposed exploration of fitting the data on the semileptonic
decay width of $D$ mesons, the difference between the widths of baryons
containing the charmed quark as well the spectroscopic characteristics allow us
to significantly decrease the variations of model parameters, i.e. the quark
masses, the scale for the normalization of Wilson coefficients, and the wave
function of light quark in the nonrelativistic model. In this way we decrease
the uncertainties of theoretical predictions. The variation of charmed quark
mass in the limits of $1.6 - 1.65~~{\rm GeV}$ and difference between the
masses of strange quark and light quark in the range $0.15 - 0.2~~{\rm GeV}$
results in the following uncertainties for the lifetimes of baryons under
consideration: $\delta\tau_{\Xi_{cc}^{++}} = 0.1~~{\rm ps}$,
$\delta\tau_{\Xi_{cc}^{+}} = 0.01~~{\rm ps}$, whereas we see that the abolute
values of uncertainties in the widths are equal to
$\delta\Gamma_{\Xi_{cc}^{++}} =  0.4~~{\rm ps}^{-1}$,
$\delta\Gamma_{\Xi_{cc}^{+}} = 0.9~~{\rm ps}^{-1}$. However, since the total
width of ${\Xi_{cc}^{+}}$ is significantly enhanced by the contribution of weak
scattering for the constituent quarks, the relative uncertainty in the estimate
of lifetimes for this baryon is much less, i.e. it is 10\% in comparison with
20\% for ${\Xi_{cc}^{++}}$). 

Calculating the inclusive widths of decays for the $\Xi_{bc}^{+}$ and 
$\Xi_{bc}^{0}$ we have used
\begin{equation}
m_b  =  m_c+3.5\; {\rm GeV,}
\label{mbmc}
\end{equation}
in addition to the parameters given above.
The baryon mass has been put to 7 GeV. For the wave function in the subsystem
of diquark we have used the results of calculations in the nonrelativistic
model with the potential by Buchm\"uller and Tye \cite{BT} corrected for the
color structure of diquark, so that 
$$
\Psi^d(0) = 0.193\;\; {\rm GeV}^{3/2}.
$$
Further, it is quite evident that the estimates of spectator widths of free
heavy quarks do not depend on the system, wherein they are bound, so that we
can exploit the results of the calculations performed earlier. We have chosen
the quark masses to be the same as in \cite{BB}, and we have put the
corresponding values, presented in Table \ref{spectator}.
\begin{table}[th]
\caption{The widths of inclusive spectator decays for the $b$ and $c$ quarks,
in ps$^{-1}$.}
\begin{center}
\begin{tabular}{|c|c|c|c|}
\hline
mode    & $b\to c \bar u d$ & $b\to c \bar c s$ & $b\to c e^+ \nu$ \\
\hline
$\Gamma$ & 0.310 & 0.137 & 0.075 \\
\hline
mode    & $b\to c \tau^+ \nu$ & $c\to s \bar d u$ & $c\to s e^- \bar \nu$ \\
\hline
$\Gamma$ & 0.018 & 0.905 & 0.162 \\
\hline
\end{tabular}
\end{center}
\label{spectator}
\end{table}

Then the procedure, described above with the shown parameters, leads to the
lifetimes of the $\Xi_{bc}^{+}$ and $\Xi_{bc}^{0}$ baryons equal to
\begin{equation}
\tau_{\Xi_{bc}^{+}}  =  0.33\;\; {\rm ps},\;\;\;\;
\tau_{\Xi_{bc}^{0}}  =  0.28\;\; {\rm ps}.
\end{equation}
We can clearly see that the difference in the lifetimes caused by the decay
processes with the Pauli interference and weak scattering is about
20 \%. The relative contributions by various terms in the total
widths of the baryons under consideration are presented in Table \ref{br}.
\begin{table}[th]
\caption{The branching fractions (in \%) of various modes in the inclusive
decays of $\Xi_{bc}^{+}$ and $\Xi_{bc}^{0}$ baryons.}
\begin{center}
\begin{tabular}{|c|c|c|c|c|}
\hline
mode    & $\Gamma_b$ & $\Gamma_c$ & $\Gamma_{PI}$ & $\Gamma_{WS}$ \\
\hline
$\Xi_{bc}^+$ & 20 & 37 & 23 & 20\\
\hline
$\Xi_{bc}^0$ & 17 & 31 & 21 & 31\\
\hline
\end{tabular}
\end{center}
\label{br}
\end{table}

Note that the contributions by the Pauli interference and weak scattering
depending on the baryon contents, can be quite significant, up to $40-50\%$.
The corrections due to the quark-gluon operators of dimension 5 are numerically
very small. The most important corrections are those due to the operator of
dimension 3, where the role of time dilation is essential for the heavy quarks
in the hadron rest frame.

For the semileptonic decays, whose relative fractions are presented in Table
\ref{br-semi}, the largest corrections appear in the decays of $b$-quark
because of the Pauli interference, so that the corresponding widths practically
increase twice. This leads to the result that the semileptonic widths of $b$
and $c$ quarks in the electron mode are equal to each other, whereas for the
spectator decays, the width of the charmed quark is twice that of $b$.
\begin{table}[th]
\caption{The branching ratios for the inclusive semileptonic widths of
$\Xi_{bc}^{+}$ and $\Xi_{bc}^{0}$, in \%.}
\begin{center}
\begin{tabular}{|c|c|c|c|}
\hline
mode     & $\Gamma_с^{e\nu}$ & $\Gamma_b^{e\nu}$ & $\Gamma_b^{\tau\nu}$  \\
\hline
$\Xi_{bc}^+$ & 5.0 & 4.9 & 2.3\\
\hline
$\Xi_{bc}^0$ & 4.2 & 4.1 & 1.9\\
\hline
\end{tabular}
\end{center}
\label{br-semi}
\end{table}

As for the sign of terms, caused by the Pauli interference, it is basically
determined by the leading contribution, coming from the interference for the
charmed quark of the initial state with the charmed quark from 
the $b$-quark decay. In this way, the anti-symmetric color structure 
of baryon wave function leads to the positive sign for the Pauli interference.

Finally, concerning the uncertainties of the estimates presented, 
we note that they are mainly related to the following:

1) the spectator width of charmed quark, where the error can reach 
50 \%, reflecting the agreement of theoretical evaluation with the lifetimes
of charmed hadrons, so that for the baryons under
consideration this term produces an uncertainty of $\delta \Gamma/\Gamma
\approx 10 \%$,

2) the effects of Pauli interference in the decays of beauty quark and in its
weak scattering off the charmed quark from the initial state, wherein we use
the nonrelativistic wave function, which, being model-dependent, can lead to
an error estimated close to 30\%, producing an uncertainty of $\delta
\Gamma/\Gamma \approx 15\%$ in the total widths.

Thus, we estimate, that the uncertainty in the predictions of total widths for
the $\Xi_{bc}^{+}$ and $\Xi_{bc}^{0}$ baryons is about 20\%.

\subsubsection{Parametric dependence of results}

Despite of that we have already pointed out what is the expected accuracy of
predictions on the inclusive widths and lifetimes of baryons with two heavy
quarks, let us consider this important problem in a more detail way.

First of all, we have to investigate the dependence of widths on the masses of
heavy quarks composing the diquarks in the baryons. Indeed, the spectator decay
widths of heavy quarks are determined by the fifth power of masses, while the
dominant corrections due to the Pauli interference and weak scattering of
constituents depend as the third degree of the heavy quark masses. In this way,
a natural challenge is the justification of quark-hadron duality with the OPE
of quark operators, that leads to zero term of correction in the first inverse
power of heavy quark mass, $1/m_Q$. Such the nullification follows from the
theorem by Adomolo--Gato claiming that the introduction of term, which is
proportional to $\lambda$ and breaking a symmetry of lagrangian, leads to
corrections of second order, i.e. $\lambda^2$, in observable quantities, so
that under the introduction of $1/m_Q$ in the lagrangian of heavy quarks, the
decay widths of heavy hadrons do not contain contributions linear in $1/m_Q$,
if we suppose the quark-hadron duality. In the connection with the problem on
the lifetime of $\Lambda_b$ baryon, whose total width measured experimentally
is about 20\% greater than the widths of $B$ mesons, that is in a contradiction
with the predictions of heavy quark theory \cite{volosh}, authors of
\cite{altarelli} suggested the hypothesis on a strong violation of quark-hadron
duality, i.e. on the possibility of significant contribution by terms linear in
$1/m_Q$ into the inclusive decay widths of heavy hadrons. This assumption, in
fact, implies that the heavy quark effective mass determining the contribution
of leading term, can vary with the mass and contents of hadron. So, considering
a more large system of $\Lambda_b$ baryon with the light diquark, wherein the
string tension is twice less than the tension in the meson, we have to
introduce a larger effective mass of heavy quark, since it is determined at a
smaller energetic scale\footnote{The cloud of virtual gluons and quarks has a
larger size in the heavy baryon than in the heavy-light meson.}. Then, the
total width of $\Lambda_b$ increases. Such the approach is not acceptable in
the OPE with the quark-hadron duality, which operates with the heavy quark mass
being the same for all kinds of hadrons, otherwise the corrections linear in
$1/m_Q$ could appear in the widths. However, the hypothesis of the strong
violation of quark-hadron duality is practically removed by the experimental
measurement of $B_c$ meson lifetime. Indeed, the author of \cite{quigg}
following the ideology of \cite{altarelli} predicted the value of lifetime
$\tau[B_c]\approx 1.3-1.5$ ps, since the quarks inside the heavy quarkonium are
strongly bound, and their effective masses as well as the admissible phase
space of the final state in decays decrease. This fact leads to a significant
suppression of decay widths for both $\bar b$ and $c$ quarks in the $B_c$
meson. The experimental measurement yields $\tau[B_c]=0.48\pm 0.19$ ps, which
is in a good agreement with the estimates in the framework of OPE \cite{BB,bi},
QCD sum rules \cite{KKLO} and potential models \cite{bc-rev}. Thus, at present
we can claim that the OPE with the quark-hadron duality is a correct tool in
the calculations of inclusive decay widths and lifetimes for the hadrons with
heavy quarks. 

As we have mentioned, the estimates presented in the above sections have been
done under the assumption that the chosen value of charmed quark mass results
in quite the precise agreement of theoretical prediction on the semileptonic
widths of $D$ mesons with the experimental values. Behind our paper \cite{ltcc}
the analogous calculations were performed in ref. \cite{guber}, whose authors
supposed essentially lower value for the charmed quark mass $m_c=1.35$ GeV,
which certainly resulted in that the semileptonic widths of charmed mesons
could not be satisfactorily described in the same approach. Such the preference
was probably caused by the following: first, a low value of current mass of $c$
quark is usually obtained in the QCD sum rules for the charmonium, and, second,
the description of inclusive decay widths of $D$ mesons is generally considered
to be quite qualitative, but quantitative, because the charmed quark mass is
not large enough, therefore, the convergency of series in $1/m_Q$ could be
slow\footnote{Again, one does not take into account the data on $B_c$.}. Such
the suggestions do lead to just a qualitative predictions on the lifetimes of
baryons with two heavy quarks in \cite{guber}, so, their estimates are twice or
thrice different from the values obtained in previous sections. Indeed, the
significant decrease of leading term because of the variation of charmed quark
mass leads to that the negative contribution of Pauli interference strongly
diminishes the total widths of $\Xi_{cc}^{++}$, so that the lifetime is
essentially overestimated.

Another important source of uncertainty in the estimates of widths for the
decays of charmed quark is the mass of strange quark. Indeed, since the charmed
quark has the mass about $1.5$ GeV, the phase space of its decay strongly
depends on the chosen mass of strange quark: the current mass about $150-200$
MeV or the constituent mass close to the mass of $K$ meson. In previous
estimates we have supposed that the suppression of phase space is determined by
the constituent mass. The problem on the dependence of inclusive decay widths
on the masses of heavy and strange quarks was studied in detail in ref.
\cite{lion} under the consideration of $B_c$ meson lifetime, where the
uncertainty caused by the modelling the wave function is low\footnote{The heavy
quarkonium is quite precisely described due to lots of data on the charmonium
and bottomonium.}, and, moreover, the corresponding contribution of weak
annihilation (corrections in the second order of $1/m_Q$) is small (about
10\%).

Accepting (\ref{mbmc}), that follows from the analysis of data on the decays of 
$B$ mesons, authors of \cite{lion} get the estimates presented in Table
\ref{ltbc}. In this table we see that, first, the low value of charmed quark
mass supposed in \cite{guber} gives quite an overestimating value for the
lifetime of $B_c$. Second, the choice of current mass for the strange quark is
slightly preferable, since it leads to the value of $B_c$ lifetime, which well
agrees with the central value of experimental inteval, though the uncertainty
of data allows the description with the constituent mass of strange quark, too.
It is worth to note that this analysis yields the quark mass values consistent
with the choice of charmed quark mass under the semileptonic decay widths of
$D$ mesons.

\begin{table}[th]
\caption{The lifetime of $B_c$ meson and contributions of spectator widths as
well as of corrections caused by the Pauli interference (PI) and weak
annihilation (WA) at different  values of quark masses.}
\begin{center}
\begin{tabular}{|l|c|c|c|c|c|}
\hline
Parameters, GeV & $\bar b\to\bar c$, ps$^{-1}$ & $c\to s$, ps$^{-1}$&
PI, ps$^{-1}$  & WA, ps$^{-1}$ & $\tau_{B_c}$, ps \\
\hline
$m_b = 5.0, m_c = 1.5, m_s =0.20$ & 0.694 & 1.148 & -0.115 & 0.193 & 0.54 \\
\hline
$m_b = 4.8, m_c = 1.35, m_s =0.15$ & 0.576 & 0.725 & -0.132 & 0.168 & 0.75 \\
\hline
$m_b = 5.1, m_c = 1.6, m_s =0.45$ & 0.635 & 1.033 & -0.101 & 0.210 & 0.55 \\
\hline
$m_b = 5.1, m_c = 1.6, m_s =0.20$ & 0.626 & 1.605 & -0.101 & 0.210 & 0.43 \\
\hline
$m_b = 5.05, m_c = 1.55, m_s =0.20$ & 0.623 & 1.323 & -0.107 & 0.201 & 0.48 \\
\hline
$m_b = 5.0, m_c = 1.5, m_s =0.15$ & 0.620 & 1.204 & -0.114 & 0.193 & 0.53 \\
\hline
\end{tabular}
\end{center}
\label{ltbc}
\end{table}

Along with the quark masses, which, to the moment, could be generally
considered as quantities with quite low uncertainties, the variation of light
quark wave function plays a significant role in the calculations of inclusive
decay widths for the baryons with two heavy quarks. As we have explained above,
this quantity was estimated under the assumption that it is analogous to the
wave functions of $D$ mesons, i.e. we suggest that the corrections to the value
obtained in the framework of potential models are similar in the mesons and
baryons. In the analysis of \cite{guber,lion} the following relation for the
wave function of light quark in the doubly heavy baryon was supposed:
\begin{equation}
|\Psi_{l} (0) |^2 = \frac{2}{3}\frac{f_D^2 M_D k^{-\frac{4}{9}}}{12},
\end{equation}
where $f_D = 170$ MeV, and the factor of $k^{-\frac{4}{9}}$ is caused by the
so-called ``hybrid'' logarithms for the nonrelativistic heavy quarks. Such the
expression is derived if we suppose the scaling of hyperfine spin-spin
splitting in the charmed mesons and baryons and take into account corresponding
spin factors as well as the double mass of diquark composed by two heavy
quarks. The assumption on the independence of splitting on the mesonic or
baryonic state looks quite speculative, while, nevertheless, if we leave
physical motivations, the numerical effect is the reduction of wave function of
light quark by two or three times. In contrast, the calculations in the
framework of potential models result in the twice enhancement of wave function
factor. Thus, the numerical value accepted in previous sections gives the
central value for the widths under the variation of wave function of light
quark.

\begin{table}[th]
\caption{The lifetime and inclusive widths of $\Xi_{cc}^{++}$ baryon.}
\begin{center}
\begin{tabular}{|c|c|c|c|}
\hline
Parameters, GeV & $\sum c\to s$, ps$^{-1}$&
PI, ps$^{-1}$  & $\tau_{\Xi_{cc}^{++}}$, ps \\
\hline
$m_c = 1.35, m_s =0.15$ & 1.638 & -0.616  & 0.99 \\
\hline
$m_c = 1.6, m_s =0.45$ & 2.397 & -0.560  & 0.56 \\
\hline
$m_c = 1.55, m_s =0.2$ & 3.104 & -0.874  & 0.45 \\
\hline
\end{tabular}
\end{center}
\label{cc1}
\end{table}
\begin{table}[th]
\caption{The lifetime and inclusive widths of $\Xi_{cc}^{+}$ baryon.}
\begin{center}
\begin{tabular}{|c|c|c|c|}
\hline
Parameters, GeV & $\sum c\to s$, ps$^{-1}$&
WS, ps$^{-1}$  & $\tau_{\Xi_{cc}^{+}}$, ps \\
\hline
$m_c = 1.35, m_s =0.15$ & 1.638 & 1.297  & 0.34 \\
\hline
$m_c = 1.6, m_s =0.45$ & 2.397 & 2.563  & 0.20 \\
\hline
$m_c = 1.55, m_s =0.2$ & 3.104 & 1.776  & 0.20 \\
\hline
\end{tabular}
\end{center}
\label{cc2}
\end{table}
\begin{table}[th]
\caption{The lifetime and inclusive widths of $\Omega_{cc}^{+}$ baryon.}
\begin{center}
\begin{tabular}{|c|c|c|c|}
\hline
Parameters, GeV & $\sum c\to s$, ps$^{-1}$&
PI, ps$^{-1}$  & $\tau_{\Omega_{cc}}$, ps \\
\hline
$m_c = 1.35, m_s =0.15$ & 1.638 & 1.780  & 0.30 \\
\hline
$m_c = 1.6, m_s =0.45$ & 2.397 & 0.506  & 0.34 \\
\hline
$m_c = 1.55, m_s =0.2$ & 3.104 & 1.077  & 0.24 \\
\hline
\end{tabular}
\end{center}
\label{cc3}
\end{table}
For the sake of presentation on the degree of variations in the theoretical
predictions for the inclusive widths of doubly heavy baryons we show the
estimates from \cite{lion} exploring the underestimated value of wave function
of light quark, in Tables  \ref{cc1}-\ref{cc3}, which should be compared with
the results in Table \ref{cont}.

Remember that the estimates with the low value of charmed quark mass give the
illustrations, only, and they cannot be correct because of the contradiction
with the data on the lifetime of $B_c$.

Summimg up the analyzed uncertainties caused by the masses and wave function of
light quark in the baryon, we present our final estimates in Table \ref{total}.

\begin{table}[th]
\caption{The lifetimes of doubly heavy baryons.}
\begin{center}
\begin{tabular}{|c|c||c|c||c|c|}
\hline 
baryon & $\tau$, ps & baryon & $\tau$, ps & baryon & $\tau$, ps \\
\hline
$\Xi_{cc}^{++}$ & $0.46\pm 0.05$ & $\Xi_{bc}^{+}$ & $0.30\pm 0.04$  &
$\Xi_{bb}^{0}$ & $0.79\pm 0.02$ \\
\hline
$\Xi_{cc}^{+}$  & $0.16\pm 0.05$ & $\Xi_{bc}^{0}$ & $0.27\pm 0.03$ &
$\Xi_{bb}^{-}$ & $0.80\pm 0.02$ \\
\hline
$\Omega_{cc}^{+}$ &$0.27\pm 0.06$& $\Omega_{bc}^{0}$&$0.22\pm 0.04$ & 
$\Omega_{bb}^{-}$ & $0.80\pm 0.02$ \\
\hline 
\end{tabular}
\end{center}
\label{total}
\end{table}

In ref. \cite{ltscale} the analysis comparing the structures of OPE for the
heavy hadrons was done on the basis of symmetry in the hadronic matrix elements
determining the contributions by the Pauli interference and weak scattering of
constituents\footnote{In this way, we deal with the wave function of light
quark independent on the flavor of infinitely heavy quark being the source of
gluon field determining the motion of light quark.}, so that up to both
corrections in the inverse powers of heavy quark mass and logarithmic terms,
which are given by anomalous dimensions of corresponding operators, the
following scaling relations were derived:
\begin{equation}
\frac{\Gamma[B^-]-\Gamma[B^0]}{\Gamma[D^+]-\Gamma[D^0]} = 
\frac{\Gamma[\Xi_b^-]-\Gamma[\Xi_b^0]}{\Gamma[\Xi_c^+]-\Gamma[\Xi_c^0]} = 
\frac{\Gamma[\Xi_{bb}^-]-\Gamma[\Xi_{bb}^0]}{\Gamma[\Xi_{cc}^{++}]-\Gamma[\Xi_{
cc }^+]} = \frac{m_b^2}{m_c^2}\, \frac{|V_{cb}|^2}{|V_{cs}|^2}.
\label{guba}
\end{equation}
The accuracy of such relations should be of the order of 50\%, since, for
instance, according to the consideration of leptonic constants for the heavy
mesons, the hadronic matrix elements of quark currents with the charmed and
light quarks acquire large corrections about 50-90\% certainly due to both the
$1/m_Q$ terms and the logarithmic renormalization. Since the $cc$-diquark is
twice heavier than the charmed quark, in the leading order we can put that the
mentioned corrections could be twice smaller for the doubly heavy baryons. We
have explored the data of Table \ref{total} in order to test the last equality
in (\ref{guba}). First of all, as we see, the accuracy of theoretical
predictions does not allow us to make some convincing quantitative conclusions
on the difference of lifetimes for the baryons with two $b$ quarks. If we
restrict ourselves by the difference of central values, then the studied part
of relation (\ref{guba}) is really satisfied with the accuarcy of 50\%, which
points to a qualitative consistency of relation, whose quantitative accuracy is
sadly low.

\subsection{Exclusive decays in sum rules of NRQCD}
In this section we describe the calculations of exclusive semileptonic cascade
decays of doubly heavy baryons as well as two-paricle hadronic decays in the
approximation of factorization for the weak trasition current of quarks
\cite{exclus}.

In the framework of NRQCD sum rules, the following form of baryonic current was
chosen in \cite{exclus}:
\begin{eqnarray}
J_{\Xi_{QQ}^{\diamond}} = \varepsilon^{\alpha\beta\gamma }
:(Q^{T}_{\alpha}C\gamma_5 q_{\beta })Q_{\gamma }:,
\end{eqnarray}
which leads to the necessary anti-symmetrization in corresponding diagrams,
since two identical heavy quarks can enter the baryon. In this way, while
taking the matrix element of chosen current over the physical baryonic states
and the vacuum there is a component giving zero unphysical contribution, which
is not essential. In this approach, the baryonic coupling constants are
generally different from the values calculated above. Therefore, we have to
analyze the two-point correlation functions with the new choice of current
structure, that was done in \cite{exclus}. However, the new choice has a formal
advantage in the consideration of three-point correlators determining, for
instance, the formfactors of semileptonic decays (see Fig. \ref{diagram}).

\setlength{\unitlength}{1mm}
\begin{figure}[th]
\begin{center}
\begin{picture}(80,55)
\put(5,-5){\epsfxsize=8cm \epsfbox{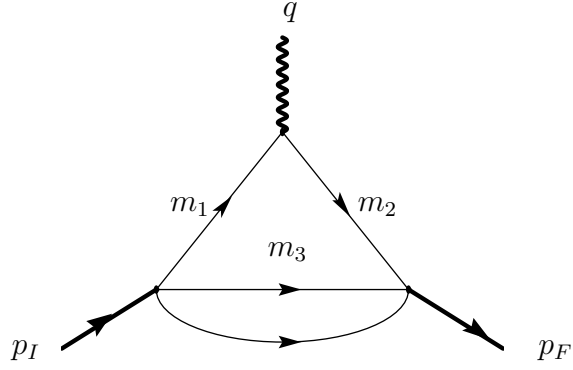}}
\put(28,31){$m_1$} \put(53,31){$m_2$} \put(41,25){$m_3$}
\put(7,12){$p_I$} \put(77,12){$p_F$} \put(43,57){$q$}
\end{picture}
\end{center}
\caption{The quark loop for the thre-point correlator in the decay of baryon.}
\label{diagram}
\end{figure}
\normalsize

Consider the correlator
\begin{eqnarray}
\Pi_{\mu } &=& i^2\int d^4xd^4y \langle
0|T \{J_{H_I}(y) J_{\mu}(0)\bar J_{H_F}(x) \}|0\rangle\; 
e^{i p_F\cdot x}e^{-i p_I\cdot y},
\end{eqnarray}
where $J_{\mu}$ is the weak current of quark decay. The theoretical part of sum
rules can be expressed in the form of dispersion relation
\begin{eqnarray}
&& \Pi_{\mu }^{(theor)}(s_1,s_2,q^2) = 
\frac{1}{(2\pi )^2}\int_{m_I^2}^{\infty}ds_1 \int_{m_F^2}^{\infty}ds_2
\frac{\rho_{\mu} (s_1,s_2,q^2)}{(s_1-s_1^0)(s_2-s_2^0)} + \ldots 
\end{eqnarray}
where dots denote possible subtractions providing the convergency of integrals,
and the spectral densities were calculated in \cite{exclus} in the limit of
spin symmetry in the effective lagrangian of heavy nonrelativistic quarks under
account of quark loop and corrections due to the condensate of light quarks. In
the approximation of symmetry the only scalar correlator should be calculated.
Indeed, the hadronic part od sum rules has the form
\begin{eqnarray}
\Pi_{\mu }^{(phen)}(s_1,s_2,q^2) &=& \sum_{spins}\frac{\langle
0|J_{H_F}|H_F(p_F)\rangle}{s_2^0-M_{H_F}^2}\; 
\langle H_F(p_F)|J_{\mu}|H_I(p_I)\rangle\; 
\frac{\langle H_I (p_I)|\bar J_{H_I}|0\rangle }{s_1^0-M_{H_I}^2},
\end{eqnarray}
where the formfactor of decay for the baryon with the spin $\frac{1}{2}$ into
the baryon with the spin $\frac{1}{2}$ is expressed in the general form
\begin{eqnarray}
\langle H_F(p_F)|J_{\mu}|H_I(p_I)\rangle = \bar u(p_F) &&\{
 \gamma_{\mu} G^V_1 + v_{\mu }^I G^V_2 + 
v_{\mu }^F G^V_3 + \nonumber \\ && \gamma_5 (\gamma_{\mu} G^A_1 +  
v_{\mu }^I G^A_2 + v_{\mu }^F G^A_3 )\}\;\, u(p_I).
\end{eqnarray}
In general, all of six formfactors are independent. However, in the leading
order the lagrangian of NQRCD possesses the symmetry, so that at small recoil
momenta restricting the virtualities of gluonic exchanges in the hadronic state
we can derive relations connecting the formfactors with each other, if they
give nonzero contributions. So, in this limit the 4-velocities of baryons in
the initial and final states slightly deviate from each other $v_I\neq v_F$,
while their scalar product is close to unit $w= (v_I\cdot v_F) \to 1$. Then,
the correlation function for the decay of heavy quark into the heavy quark has
the form 
\begin{equation}
\Pi_{\mu}^{(theor)} \sim
\xi^{IW}(w)(1+\slashchar{\tilde{v}}_F)\gamma_{\mu}(1-\gamma_5)
(1+\slashchar{\tilde{v}}_I),
\end{equation}
where
\begin{eqnarray}
\tilde{v}_I &=& v_I + \frac{m_3}{2m_1}(v_I-v_F),
\\ \tilde{v}_F &=& v_F + \frac{m_3}{2m_2}(v_F-v_I).
\end{eqnarray}
We see that the correlation function is determined by the only formfactor
$\xi^{IW}$ at the minimal recoil momentum. However, this formfactor is not
universal, i.e. it depends on the quark contents of baryons in the process of
decay.

\setlength{\unitlength}{1mm}
\begin{figure}[th]
\begin{center}
\begin{picture}(90,60)
\put(0,0){\epsfxsize=8cm \epsfbox{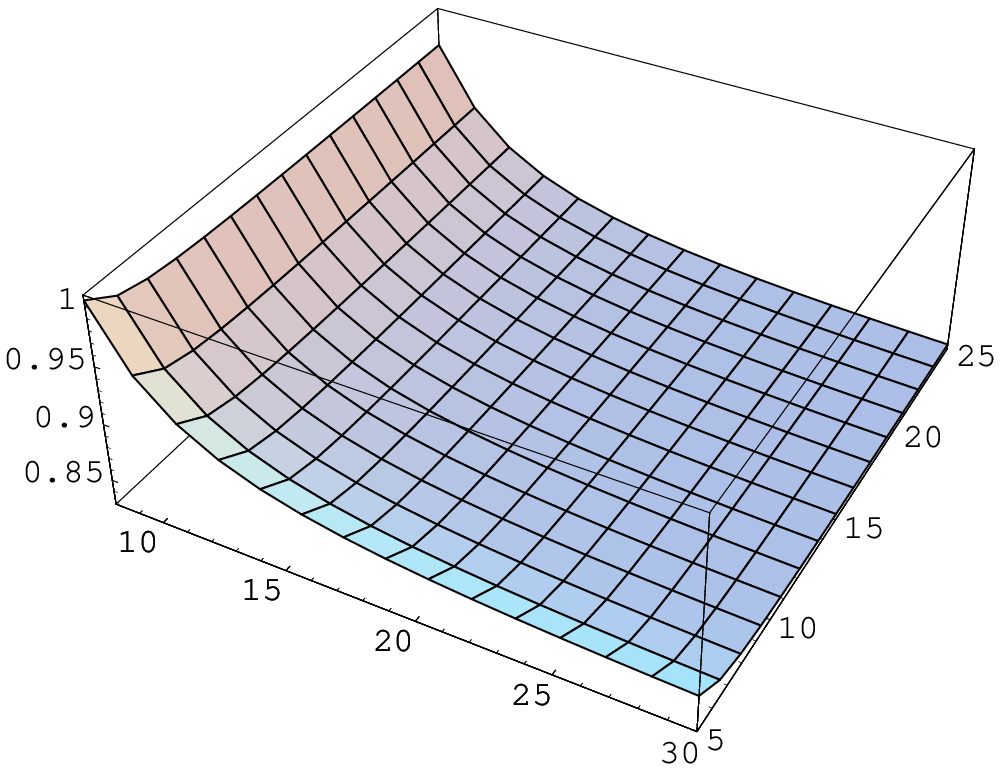}} 
\put(0,48){$\xi (1)$} 
\put(25,2){$B_I$, GeV$^2$} 
\put(73,20){$B_F$, GeV$^2$}
\end{picture}
\end{center}
\caption{The forfactor $\xi (1)$ for the transition of $\Xi_{bb}^{\diamond}\to
\Xi_{bc}^{\diamond}$ in the Borel scheme of sum rules.}
\label{fIW}
\end{figure}

For the decay of heavy quark into the light one we find
\begin{equation}
\Pi_{\mu}^{(theor)} \sim \{\xi_1(w)\slashchar{v}_I +
\xi_2(w)\slashchar{v}_F + \xi_3(w)\}\gamma_{\mu}(1-\gamma_5)
(1+\slashchar{\tilde{v}}_I),
\end{equation}
wherefrom we can get the spin-symmtry relation
\begin{eqnarray}
(G_1^V+G_2^V+G_3^V) &=& \xi^{IW}(w), \\ G_1^A &=& \xi^{IW}(w)
\end{eqnarray}
and the following connection between the functions $\xi_i(w)$ takes place:
$$
\xi_1(w) +
\xi_2(w) = \xi_3(w) = \xi^{IW}(w). 
$$ 
Exploring the presence of heavy hadrons in both the initial and final states,
we draw the conclusion that two form factors $G_1^V = G_1^A = \xi (w)$ are not
suppressed by the heavy quark mass, only.

The formfactor $\xi$ at nozero recoils is determined in accordance with 
\begin{eqnarray}
\xi^{IW}(w) &=& \frac{1}{(2\pi )^2}\frac{1}{8 M_I M_F Z_I Z_F }\,
\int_{(m_1+m_3)^2}^{s_I^{th}}
\int_{(m_1+m_2)^2}^{s_F^{th}} \rho(s_I,s_F,q^2)ds_Ids_F 
\nonumber \\ && 
\times\exp (-\frac{s_I-M_I^2}{B_I^2})\exp
(-\frac{s_F-M_F^2}{B_F^2}),
\end{eqnarray}
where $B_I$ and $B_F$ are the parameters of Borel transform over the invariant
squares of masses in the initial and final states of decay, while $Z_I$ and
$Z_F$ are the coupling constants of baryons with the appropriate currents of
quarks.

\subsubsection{Numerical estimates}
The estimates obtained in the sum rules for the function analogous to the
Isgur--Wise function are presented in Table \ref{tIW} for the formfactors of
semileptonic decays of doubly heavy baryons into the baryonic states with the
spin $\frac{1}{2}$, in comparison with the values calculated in the potential
model. The deviation between the values of $\xi(1)$ in these two approaches
does not lead to any discrepancy, since the corresponding systematic
uncertainty is about 10\%.

\begin{table}[th]
\caption{The normalization of Isgur--Wise formfactor at zero recoil in the sum
rules and potential model.}
\begin{center}
\begin{tabular}{|c|c|c|}
\hline Mode & $\xi (1)$, sum rules &  $\xi (1)$, pot.model \\ \hline
$\Xi_{bb}\to \Xi_{bc}$ & 0.85 & 0.91 \\\hline $\Xi_{bc}\to
\Xi_{cc}$ & 0.91 & 0.99 \\\hline $\Xi_{bc}\to \Xi_{bs}$ & 0.9 &
0.99 \\
\hline $\Xi_{cc}\to \Xi_{cs}$ & 0.99 & 1. \\
\hline
\end{tabular}
\end{center}
\label{tIW}
\end{table}

The result on the normaliztion of Isgur--Wise function in the transition of
$\Xi_{bb}\to \Xi_{bc}$ is presented in Fig. \ref{fIW} obtained in the Borel
scheme of NRQCD sum rules. 

In order to calculate the exclusive widths we suggest that the dependence of
formfactors on the transfer momentum has the form of pole
\begin{equation}
\xi^{IW} (w) = \xi_0\frac{1}{1-\frac{q^2}{m_{pole}^2}},
\end{equation}
with
\begin{eqnarray}
m_{pole} &=& 6.3 \mbox{~~~GeV for~~}  b\to c
\mbox{~~transitions,}\nonumber \\ m_{pole} &=& 1.85 \mbox{~~GeV for~~} c\to s
\mbox{~~transitions}.\nonumber
\end{eqnarray}
The results of calculation for the exclusive decay widths of doubly heavy
baryons in the framework of NRQCD sum rules are given in Table \ref{exwid},
where the total widths have been supposed equal to the mean values presented in
the previous section.
\begin{table}[th]
\caption{The branching ratios (Br) for the exclusive decays of baryons with two
heavy quarks.}
\begin{center}
\begin{tabular}{|l|r||l|r|}
\hline Mode & Br (\%) &  Mode & Br (\%) \\ \hline
$\Xi_{bb}^{\diamond}\to \Xi_{bc}^{\diamond}l\bar\nu_l$ & 14.9 &
$\Xi_{bc}^{+}\to \Xi_{cc}^{++}l\bar\nu_l$ & 4.9 \\\hline
$\Xi_{bc}^{0}\to \Xi_{cc}^{+}l\bar\nu_l$ & 4.6 & $\Xi_{bc}^{+}\to
\Xi_{b}^{0}\bar l\nu_l$ & 4.4 \\\hline $\Xi_{bc}^{0}\to
\Xi_{b}^{-}\bar l\nu_l$ & 4.1 & $\Xi_{cc}^{++}\to
\Xi_{c}^{+}\bar l\nu_l$ & 16.8 \\\hline $\Xi_{cc}^{+}\to
\Xi_{c}^{0}\bar l\nu_l$ & 7.5 & $\Xi_{bb}^{\diamond}\to
\Xi_{bc}^{\diamond}\pi^{-}$ & 2.2 \\
\hline 
$\Xi_{bb}^{\diamond}\to \Xi_{bc}^{\diamond}\rho^{-}$ & 5.7 & 
$\Xi_{bc}^{+}\to \Xi_{cc}^{++}\pi^{-}$ & 0.7 \\\hline
$\Xi_{bc}^{0}\to \Xi_{cc}^{+}\pi^{-}$ & 0.7 & $\Xi_{bc}^{+}\to
\Xi_{cc}^{++}\rho^{-}$ & 1.9 \\\hline $\Xi_{bc}^{0}\to
\Xi_{cc}^{+}\rho^{-}$ & 1.7 & $\Xi_{bc}^{+}\to
\Xi_{b}^{0}\pi^{+}$ & 7.7 \\\hline $\Xi_{bc}^{0}\to
\Xi_{b}^{-}\pi^{+}$ & 7.1 & $\Xi_{bc}^{+}\to
\Xi_{b}^{0}\rho^{+}$ & 21.7 \\\hline $\Xi_{bc}^{0}\to
\Xi_{b}^{-}\rho^{+}$ & 20.1 & $\Xi_{cc}^{++}\to
\Xi_{c}^{+}\pi^{+}$ & 15.7 \\\hline $\Xi_{cc}^{+}\to
\Xi_{c}^{0}\pi^{+}$ & 11.2 & $\Xi_{cc}^{++}\to
\Xi_{c}^{+}\rho^{+}$ & 46.8 \\\hline $\Xi_{cc}^{+}\to
\Xi_{c}^{0}\rho^{+}$ & 33.6 &  & \\\hline
\end{tabular}
\end{center}
\label{exwid}
\end{table}
The contribution of cascade decays into the baryons with the spin $\frac{3}{2}$
is also taken into account in the table. In this estimate the results of
\cite{Lozano} on decays $\Xi_{bc}\to \Xi_{cc}+l\bar\nu$ were used. In the
decays with $\Xi_{bb}^{\diamond }$ and $\Xi_{cc}^{\diamond }$ the factor caused
by the anti-symmetrization of identical quarks was taken into account. For the
decays of $\Xi_{cc}^{++}\to \Xi_{c}^{+} X$ the correction factor of $0.62$ was
introduced because of negative interference of Pauli. The author of
\cite{exclus} claims that the values obtained in these sum rules agree with the
results in the potential approach of \cite{Lozano} as well as with the above
estimates of inclusive decay widths if we sum up the corresponding exclusive
widths calculated in the sum rules.

\subsection{Discussion}

In this chapter we have calculated the lifetimes of baryons with two heavy
quarks in the framework of consistent consideration of OPE in the inverse
powers of heavy quark masses. In this expansion the leading constribution is
determined by the spectator widths of heavy quarks in the inclusive decays,
while the significant corrections appear under the taking into account the
effects of Pauli interference and weak scattering, which contribute about
20-30\% for the baryons $\Xi_{cc}$ and $\Xi_{bc}$. The measurement of lifetimes
for the doubly heavy baryons allows us to make the comparative analysis of
deacy mechanisms for the hadrons with heavy quarks, that is especially actual
in the light of searching for the fine effects with the violation of combined
СР-parity in the sector of heavy quarks, because the characteristics of quark
inteactions enter the measured quantities with the factors composed by the
hadronic matrix elements of quark currents. A reliable knowledge of properties
for such matrix elements could be essentially enriched by the study of decays
and lifetimes of baryons with two heavy quarks. Such the investigations allow
us to numerically analyze possible effects with the violation of quark-hadron
duality (these effects should be small, as we have stressed). An actual
challenge is studying the dependence of heavy quark decay widths on the
contents of hadron, which could be essential in the clarification of reasons
causing the large deviation from unit for the ration of $\Lambda_b$ and $B$
lifetimes. The measurement of doubly heavy baryon lifetimes allows us also to
investigate the characteristics of confinement for the heavy quarks in various
systems.

\begin{figure}[th]
\setlength{\unitlength}{1mm}
\begin{center}
\begin{picture}(120,57)
\put(0,0){\epsfxsize=11cm \epsfbox{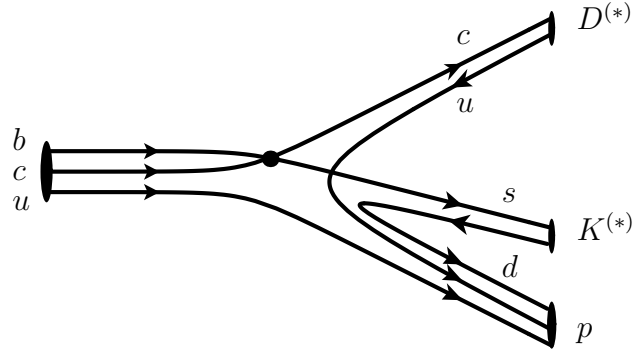}}
\put(17,29){$u$}
\put(17,33){$c$}
\put(17,37){$b$}
\put(76,51){$c$}
\put(76,42){$u$}
\put(82,30){$s$}
\put(82,20){$d$}
\put(92,25){$K^{(*)}$}
\put(92,53){$D^{(*)}$}
\put(92,12){$p$}
\end{picture}
\end{center}
\caption{The decays of $\Xi_{bc}\to D^{(*)}K^{(*)} p$ in the process with the
weak scattering of constituent $b$ and $c$ quarks.}
\label{exbc}
\end{figure}
An open field of activity is a calculation of exclusive decay widths for the
baryons with two heavy quarks. The presented estimates in the framework of
NQRCD sum rules and potential models are quite preliminary, of course, since
the challenge here is a role of corrections in the inverse powers of heavy
quark mass, which could be very significant for the hadrons containing the
charmed quark.

Moreover, the studies of exclusive hadronic decays for the doubly heavy baryons
are of interst for the experimental practice. Some channels have to be
distiguished among the hadronic decays. In contrast to the cascade decays of
two heavy quarks, for example, $\Xi_{bc}\to \Xi_{cc}+X\to \Xi_c+X\to \Xi+X$,
which require a reconstruction of three ``secondary'' vertices, there are the
processes with the weak scattering of constituents, in part, the heavy quarks,
that will lead to the only heavy quark in the final state as it happens in Fig.
\ref{exbc} for $\Xi_{bc}$, so that the only heavy hadron should be detected.
The contribution of weak scattering into the total width is large enough (about
20\%). Simple estimates of suppression factors show that the branching fraction
of such decay is about several tenth per cent. 

We believe that the experimental investigations of doubly heavy baryons are
quite actual problem, first of all, at the hadron colliders, and the
measurements of their decay characteristics could significantly enrich the
knowledges on the mechanisms of heavy quark decays.

\newpage
\makeatletter
\def\thesection {Conclusion}
\makeatother

\section{}

In this review we have presented the basic physical characteristics of baryons
containing two heavy quarks. The description of such hadrons is based on the
hierarchy of scales in the interactions determining the time intervals and
distances ascribed for the strong interactions in the subsystems composing
these baryons. So, because of the nonrelativistic motion of heavy quarks with
respect to each other the time interval for forming the system of two heavy
quarks is greater than the time period for the hard production of heavy quarks
or ``dressing'' the quarks by hard gluons. On the other hand, this time for the
forming the heavy diquark is less than the characteristic time interval for the
interactions providing the confinement of light quarks involving the
low-frequency strong-interacting fields. Due to the mentioned hierarchy of
strong interactions in the quark systems under consideration we have developed
the methods of heavy quark effective theory for the baryons with two heavy
quarks. In the effective theory we have isolated the leading approximation and
constructed the systematic way for the calculation of corrections to it.

Having started from the general approaches in the description of hadronic
systems with the heavy quarks, we have developed the following particular
methods for the consideration of baryons with two heavy quarks:
\begin{description}
\item[-]
the potential of static quarks and the potential model for the doubly heavy
baryons, the separation between the heavy quark motion inside the diquark and
the light quark motion in the field of diquark;
\item[-]
the formulation of two-point sum rules in NRQCD for the quark currents
corresponding to the baryons with two heavy quarks, and the calculations of
both ground state masses for such the baryons and their coupling constants with
the currents, taking into account the corrections to the local condensates of
light and strange quarks;
\item[-]
the calculation of anomalous dimensions for the baryonic currents with two
heavy quarks in NRQCD;
\item[-]
the scaling functions of diquark fragmentation;
\item[-]
the numerical calculations for the complete set of diagrams in the fourth order
of QCD and the analysis of higher twists over the transverse momentum;
\item[-]
the generalization of OPE for the inclusive widths of baryons with two heavy
quarks and a single light quark;
\item[-]
the formulation of three-point sum rules in NQRCD for the exclusive
semileptonic decays and hadronic decays in the approximation of factorization
for the transition current.
\end{description}

The most bright physical effects in the baryons containing two heavy quarks are
the following:
\begin{description}
\item[-]
there are the families of doubly heavy baryons, i.e. the systems of
quasi-stable
excited levels for the baryons with identical heavy quarks due to a suppression
of operators determining the transitions into the low-lying and ground states,
since the quantum numbers of diquark have to change in such transitions, so
that for some states the operators, which are not suppressed by the heavy quark
mass or small size of diquark, have to be equal to zero because of the Pauli
principle;
\item[-]
the cascade processes of fragmentation at high transverse momenta, wherein we
can get analytical results for the universal fragmentation functions in the
perturbative QCD for the heavy quark into the heavy diquark and for the heavy
diquark into the doubly heavy baryon, while the evolution of fragmentation
function because of emission of hard gluons can be described in the framework
of renornalization group in QCD;
\item[-]
the separation of regimes for the fragmentation and recombination in hadronic
processes by taking into account the higher twists over the transverse
momentum, which can be described in the framework of peturbative QCD by
calculating the complete set of diagrams in the given order of coupling
constant;
\item[-]
large, about 20-50\%, contributions of nonspectator terms into the lifetimes of
doubly heavy baryons depend on the valence quark contents of baryons and are
given by the effects of Pauli interference and weak scattering, especially in
the presense of charmed quark, that leads to a strong splitting of lifetimes:
$$
\begin{array}{lclcl}
\tau[\Xi_{cc}^{++}] &>& \tau[\Omega_{cc}^{+}] &>& \tau[\Xi_{cc}^{+}],\\
\tau[\Xi_{bc}^{+}] &>& \tau[\Xi_{bc}^{0}] &>& \tau[\Omega_{bc}^{0}],\\
\tau[\Xi_{bb}^{-}] &\approx& \tau[\Omega_{bb}^{-}] &>& \tau[\Xi_{bb}^{0}],
\end{array}
$$
\item[-]
cascade mechanisms of decays for the baryons with two heavy quarks, while there
are peculiar modes due to the weak scattering, which have sizable branching
ratios.
\end{description}

Sure, direct measuring the masses of ground states and excited levels
allows us to essentially move forward more deep understanding the dynamics of
forming the bound states with the heavy quarks. The theory for the radiative
transitions between the quasi-stable levels of doubly heavy baryons due to
either the electromagnetic or strong forces should be developed, wherein the
method of chiral lagrangian should be generalized for the soft emission of
Goldstone's mesons, particularly, pions.

A quite complete picture of production mechanisms has been constructed for the
doubly heavy baryons, so that the search for these baryons at hadron colliders
with high luminosities could be prospectively proposed. The distributions over
the transverse momentum of baryons could give the most important information on
the regimes of production and probably make a contribution into the
understanding of reasons resposible for the discrepance between the theory and
the experimental data on the yield of $b$-hadrons in hadron collisions.

The experimental data on the lifetimes of doubly heavy baryons could represent
the most interesting information, since they are strictly connected to the
whole system for describing the inclusive decays of heavy hadrons. The
branching ratios of semileptonic decays could be important for determining the
role of gluon corrections to the nonleptonic lagrangian for the weak charged
currents of quarks. The lifetimes would allow us to enrich the knowledge on the
masses of heavy quarks and the relative contributions of various decay
mechanisms, since the total widths are extremely sensitive to the above
physical characteristics.

We could suppose that the description of exclusive hadronic decays would
require extensive theoretical efforts for the baryons with two heavy quarks.

Thus, the physics of baryons containing two heavy quarks is sufficiently rich
and informative, so that it will justly occupy a honour place in the
experimental investigations, in the light of optimistic occurancies and
tendancies depicted by the modern theory, which deals with this field entering
the time of mature progress.

\noindent
\underline{~~~~~~~~~~~~~~~~~~~~~~~~~~}

The authors express their gratitude to the colleagues S.S.Gershtein,
A.V.Berezhnoy, A.E.Ko\-val\-sky and M.V.Shevlyagin, who took a part in getting
some results included in this review. We especially emphasize a large
contribution to the represented studies by A.I.Onishchenko. We thank
V.A.Rubakov, A.L.Kataev, A.Ali, W.Buchm\"uller, K.A.Ter-Martirosian and
V.A.Saleev for fruitfull discussions on the results of investigations, which
made the basis of review matter.

This work is in part supported by the Russian Foundation for Basic Research,
grants 99-02-16558 and 00-15-96645.

\newpage

\setcounter{section}{0}
\makeatletter
\def\thesection {Appendix \Roman{section}.}
\def\thsection {A.\Roman{section}.}
\def\thesubsection {\thsection\arabic{subsection}.}
\def\thesubsubsection {\thesubsection\arabic{subsubsection}.}
 \@addtoreset{equation}{section}
  \@addtoreset{table}{section}
 \@addtoreset{figure}{section}
  \def\thetable{\thsection\arabic{table}}
  \def\thefigure{\thsection\arabic{figure}}

\def\theequation{\thsection\arabic{equation}}

\def\section{\@startsection {section}{1}{\z@}{-3.5ex plus -1ex minus
 -.2ex}{2.3ex plus .2ex}{\large\bf}}
\def\subsection{\@startsection{subsection}{2}{\z@}{-3.25ex plus -1ex minus
 -.2ex}{1.5ex plus .2ex}{\large\bf}}
\def\subsubsection{\@startsection{subsubsection}{3}{\z@}{-3.25ex plus
 -1ex minus -.2ex}{1.5ex plus .2ex}{\sl\bf}}
%
%
\def\l@section#1#2{\addpenalty{\@secpenalty} \addvspace{1.0em plus 1pt}
\@tempdima 
1in \begingroup
 \parindent \z@ \rightskip \@pnumwidth
 \parfillskip -\@pnumwidth
 \bf \leavevmode \advance\leftskip\@tempdima \hskip -\leftskip 
#1\nobreak\hfil
\nobreak\hbox to\@pnumwidth{\hss #2}\par
 \endgroup}

\makeatother

\section{~~The coefficients of spectral densities}

The spectral densities in (\ref{ro1}) have the coefficients
\begin{eqnarray}
\eta_{1,0}(\omega) &=&16\omega^2(429 {\cal {M}}_{diq}^3+715 {\cal {M}}_{diq}^2
\omega+403{\cal {M}}_{diq} \omega^2+77\omega^3),\nonumber\\
\eta_{1,1}(\omega)&=& 104  \omega (231 {\cal {M}}_{diq}^3+297 {\cal
{M}}_{diq}^2
\omega+121 {\cal {M}}_{diq}
\omega^2+15
\omega^3),\nonumber\\
\eta_{1,2}(\omega)&=&\frac{10 }{({\cal {M}}_{diq}+\omega)^2}(3003
{\cal {M}}_{diq}^5+9009 {\cal {M}}_{diq}^4
\omega+9438
{\cal {M}}_{diq}^3 \omega^2\nonumber\\
&&+4290
{\cal {M}}_{diq}^2 \omega^3+871 {\cal {M}}_{diq} \omega^4 +77 \omega^5).
\end{eqnarray}
The coefficients of spectral densities in (\ref{ro2}) have the form
\begin{eqnarray}
\eta_{2,0} &=& 42 \omega ({\cal {M}}_{diq}^2+48 {\cal {M}}_{diq} \omega +14
\omega^2),\nonumber\\
\eta_{2,1} &=& 3(35 {\cal {M}}_{diq}^2+28 {\cal {M}}_{diq}
\omega+5\omega^2),\nonumber\\
\eta_{2,2} &=& \frac{1}{({\cal {M}}_{diq}+\omega)^2} (105
{\cal{M}}_{diq}^3+315{\cal {M}}_{diq}^2
\omega+279 {\cal {M}}_{diq} \omega^2+77 \omega^3).
\end{eqnarray}
The spectral densities taking into account the coulomb corrections in
(\ref{ro1c}) have the coefficients
\begin{eqnarray}
\eta_{1,0}^{{\bf C}}&=&(2 {\cal M}_{diq}+\omega)^2\omega^2,\nonumber\\
\eta_{1,1}^{{\bf C}}&=&\frac{3(2 {\cal M}_{diq}+\omega)\omega}{({\cal
M}_{diq}+\omega)}(4 {\cal M}_{diq}^3+6{\cal M}_{diq}^2 \omega+4 {\cal M}_{diq}
\omega^2+\omega^3),\nonumber\\
\eta_{1,2}^{{\bf C}}&=&\frac{1}{({\cal M}_{diq}+\omega)^2}(12 {\cal
M}_{diq}^4+24 {\cal M}_{diq}^3
\omega+32 {\cal M}_{diq}^2 \omega^2+20 {\cal M}_{diq} \omega^3+5 \omega^4).
\end{eqnarray}
The coefficients of spectral densities taking into account the coulomb
corrections in (\ref{ro2c}) have the form
\begin{eqnarray}
\eta_{2,0}^{{\bf C}}&=&(2 {\cal M}_{diq}+\omega)\omega ,\nonumber\\
\eta_{2,1}^{{\bf C}}&=& \frac{2}{{\cal M}_{diq}+\omega}(2 {\cal M}_{diq}^2+2
{\cal M}_{diq}
\omega+\omega^2), \nonumber\\
\eta_{2,0}^{{\bf C}}&=&\frac{2}{({\cal M}_{diq}+\omega)^2}(2 {\cal M}_{diq}^2+2
{\cal M}_{diq} \omega+\omega^2).
\end{eqnarray}
For the spectral densities with the gluon condensate in (\ref{ro1g}) we find
\begin{eqnarray}
\eta^{G^2}_{1,0}&=& 84 {\cal{M}}_{diq}^3+196 {\cal{M}}_{diq}^2
\omega+133{\cal{M}}_{diq} \omega^2+11 \omega^3,\nonumber\\
\eta^{G^2}_{1,1}&=& -\frac{2(210{\cal{M}}_{diq}^3+70 {\cal{M}}_{diq}^2
\omega+21{\cal{M}}_{diq}
\omega^2+3 \omega^3)}{{\cal{M}}_{diq}+\omega},\nonumber\\
\eta^{G^2}_{1,2}&=& \frac{2(210{\cal{M}}_{diq}^3+70 {\cal{M}}_{diq}^2
\omega+21{\cal{M}}_{diq}
\omega^2+3 \omega^3)}{({\cal{M}}_{diq}+\omega)^2}.
\end{eqnarray}

\section{~~The distribution over the transverse momentum}
For the fragmentation of vector diquark into the baryon state in the scaling
limit we get the following distribution over $t=p_T/M$ with respect to the
fragmentation axis:
\begin{eqnarray}
D(t) &=& \frac{64\alpha_s^2}{81 \pi} \; \frac{|R(0)|^2}{3(1-r)^5 M^3}\;
\frac{1}{t^6}\;
\nonumber \\ &&
\bigg\{t (30r^3-30r^4-61t^2r+45r^2t^2+33r^3t^2-
\nonumber \\ &&
-17r^4t^2+3t^4-9rt^4+15 r^2t^4-9r^3t^4)-
\nonumber\\ &&
(30r^4-99r^2t^2-54r^3t^2+27r^4t^2+9t^4+18rt^4-6r^2t^4+
\nonumber \\ &&
+18r^3t^4+3r^4t^4+3t^6-6rt^6+9r^2t^6){\rm arctg}
\biggl(\frac{(1-r)t}{r+t^2}\biggr)+
\nonumber \\ &&
24(2r^3t+rt^3+r^2t^3)\;\ln\biggl(\frac{r^2(1+t^2)}{r^2+t^2}\biggr)\bigg\}. 
\end{eqnarray}

\section{~~The spectator effects in the baryon decays}
For the effects of Pauli interference and electroweak scattering in the baryons
$\Xi_{cc}$ we have the following expresions:
\begin{eqnarray}
{\cal T}_{PI} &=& -\frac{2G_F^2}{4\pi}m_c^2(1-\frac{m_c}{m_u})^2\nonumber\\
&& ([(\frac{(1-z_{-})^2}{2}- \frac{(1-z_{-})^3}{4}) 
(\bar c_i\gamma_{\alpha}(1-\gamma_5)c_i)(\bar
q_j\gamma^{\alpha}(1-\gamma_5)q_j) + \nonumber\\ 
&& (\frac{(1-z_{-})^2}{2} -
\frac{(1-z_{-})^3}{3})(\bar c_i\gamma_{\alpha}\gamma_5 c_i)(\bar
q_j\gamma^{\alpha}(1-\gamma_5)q_j)][(C_{+} + C_{-})^2 + \\
&& \frac{1}{3}(1-k^{\frac{1}{2}})(5C_{+}^2+C_{-}^2-6C_{-}C_{+})]+ \nonumber\\
&& [(\frac{(1-z_{-})^2}{2} - \frac{(1-z_{-})^3}{4})(\bar
c_i\gamma_{\alpha}(1-\gamma_5)c_j)(\bar q_j\gamma^{\alpha}(1-\gamma_5)q_i) +
\nonumber\\
&&  (\frac{(1-z_{-})^2}{2} - \frac{(1-z_{-})^3}{3})(\bar
c_i\gamma_{\alpha}\gamma_5c_j)(\bar
q_j\gamma^{\alpha}(1-\gamma_5)q_i)]k^{\frac{1}{2}}(5C_{+}^2+C_{-}^2-6C_{-}C_{+}
)), \nonumber \\
{\cal T}_{WS} &=& \frac{2G_F^2}{4\pi}p_{+}^2(1-z_{+})^2[(C_{+}^2 + C_{-}^2 +
\nonumber\\
&& \frac{1}{3}(1 - k^{\frac{1}{2}})(C_{+}^2 - C_{-}^2))(\bar
c_i\gamma_{\alpha}(1
- \gamma_5)c_i)(\bar q_j\gamma^{\alpha}(1 - \gamma_5)q_j) + \\
&& k^{\frac{1}{2}}(C_{+}^2 - C_{-}^2)(\bar c_i\gamma_{\alpha}(1 - \gamma_5)c_j)
(\bar q_j\gamma^{\alpha}(1 - \gamma_5)q_i)],\nonumber 
\end{eqnarray}
where $p_{+} = p_c + p_q$, $p_{-} = p_c - p_q$ и $z_{\pm} =
\frac{m_c^2}{p_{\pm}^2}$. Further, for $p_{+}$ and $p_{-}$ we use their
threshold values
$$
p_{+}=p_c(1+\frac{m_q}{m_c}),\quad p_{-}=p_c(1-\frac{m_q}{m_c}).
$$
These expressions were derived with account of low-energy renormalization of
nonleptonic lagrangian for the weak interaction of nonrealtivistic heavy
quarks. This lagrangian \cite{vs,5d} has the form
\begin{eqnarray}
L_{eff,\, log} &=&
\frac{G_F^2m_c^2}{2\pi}\{\frac{1}{2}[C_{+}^2+C_{-}^2+\frac{1}{3}(1-k^{\frac{1}{
2}})(C_{+}^2-C_{-}^2)](\bar c\Gamma_{\mu}c)(\bar d\Gamma^{\mu}d) +\nonumber\\
&& \frac{1}{2}(C_{+}^2-C_{-}^2)k^{\frac{1}{2}}(\bar c\Gamma_{\mu}d)(\bar
d\Gamma^{\mu} c) +
\frac{1}{3}(C_{+}^2-C_{-}^2)k^{\frac{1}{2}}(k^{-\frac{2}{9}}-1)(\bar
c\Gamma_{\mu}t^ac)j_{\mu}^a - \\
&& \frac{1}{8}[(C_{+}+C_{-})^2+\frac{1}{3}(1-k^{\frac{1}{2}})(5C_{+}^2+C_{-}^2-
6C_{+}C_{-})](\bar c\Gamma_{\mu}c+\frac{2}{3}\bar c\gamma_{\mu}\gamma_5c)(\bar
u\Gamma^{\mu}u)-\nonumber\\
&& \frac{1}{8}k^{\frac{1}{2}}(5C_{+}^2+C_{-}^2-6C_{+}C_{-})(\bar
c_i\Gamma_{\mu}c_k+\frac{2}{3}\bar c_i\gamma_{\mu}\gamma_5c_k)(\bar
u_k\Gamma^{\mu}u_i)-\nonumber\\
&& \frac{1}{8}[(C_{+}-C_{-})^2+\frac{1}{3}(1-k^{\frac{1}{2}})(
5C_{+}^2+C_{-}^2+6C_{+}C_{-})](\bar c\Gamma_{\mu}c+\frac{2}{3}\bar
c\gamma_{\mu}\gamma_5c)(\bar
s\Gamma^{\mu}s)-\nonumber\\
&& \frac{1}{8}k^{\frac{1}{2}}(5C_{+}^2+C_{-}^2+6C_{+}C_{-})(\bar
c_i\Gamma_{\mu}c_k+\frac{2}{3}\bar c_i\gamma_{\mu}\gamma_5c_k)(\bar
s_k\Gamma^{\mu}s_i)-\nonumber\\
&& \frac{1}{6}k^{\frac{1}{2}}(k^{-\frac{2}{9}}-1)(5C_{+}^2+
C_{-}^2)(\bar c\Gamma_{\mu}t^ac+\frac{2}{3}\bar
c\gamma_{\mu}\gamma_5t^ac)j^{a\mu}\},
\nonumber
\end{eqnarray}
where $\Gamma_{\mu} = \gamma_{\mu}(1-\gamma_5)$,
$k=(\alpha_s(\mu)/\alpha_s(m_c))$ and $j_{\mu}^a = \bar u\gamma_{\mu}t^au +
\bar d\gamma_{\mu}t^ad + \bar s\gamma_{\mu}t^as$ is the color current of light
quarks ($t^a = \lambda^a/2$ are the color generators). 

Here we would like to make a note on the effective lagrangian terms containing
the colored current of light quarks. we have neglected those terms, since they
enter into the lagrangian with the factor of $k^{-\frac{2}{9}}-1$ numerically
about $0.054$.

\newpage

\makeatletter
\def\thesection {References}
\makeatother

\section{}

\end{document}